\documentclass[12pt,a4paper,nofootinbib,titlepage,final,twoside]{report}
\usepackage[Sonny]{fncychap}
\usepackage{graphicx}
\usepackage{amsmath}
\usepackage{amssymb}
\usepackage{bbm}
\usepackage{fancyheadings}
\usepackage[T1]{fontenc}
\usepackage[math]{iwona}
\usepackage{dsfont}
\usepackage{color}
\usepackage{slashed}
\usepackage{anysize}
\usepackage{float}
\usepackage{wrapfig}
\usepackage{mathabx}

\usepackage{geometry}
\usepackage{amsfonts}

\newcommand{\bg}{\bar{g}}
\newcommand{\Eqref}[1]{eq.~\eqref{#1}}

\newcommand{\tr}{{\rm tr}}
\newcommand{\Tr}{{\rm Tr}}
\newcommand{\p}{\phantom{x}}

\newcommand{\const}{{\rm const}}
\newcommand{\bare}{\bar{e}}
\begin{document}

\lhead[\fancyplain{}{\thepage}]{\fancyplain{}{\it\small\rightmark}}
\rhead[\fancyplain{}{\it\small\leftmark}]{\fancyplain{}{\thepage}}
\cfoot{\fancyplain{\bfseries\thepage}{}}
\cfoot{}

\makeatletter
\def\ps@headings{\let\@mkboth\markboth
\def\@oddfoot{} \def\@evenfoot{}
\def\@evenhead{\protect\underline{
\hbox to\hsize{ \thepage \it \hfil \leftmark}}}
\def\@oddhead{\protect\underline{
\hbox to\hsize{{\it \rightmark\hfil}\thepage}}}
\def\chaptermark##1{\markboth {\ifnum \c@secnumdepth>\m@ne
\thechapter. \ \fi ##1}{}}
\def\sectionmark##1{\markright{\ifnum \c@secnumdepth >\z@
\thesection \ \fi ##1}}}
\makeatother     

\makeatletter
\renewcommand*{\@makechapterhead}[1]{%
  \vspace*{1\p@}%
  {\parindent \z@ \raggedright \normalfont
    \ifnum \c@secnumdepth >\m@ne
      \if@mainmatter
        \DOCH
      \fi
    \fi
    \interlinepenalty\@M
    \if@mainmatter
      \DOTI{#1}%
    \else%
      \DOTIS{#1}%
    \fi
  }}
\renewcommand*{\@makeschapterhead}[1]{%
  \vspace*{1\p@}%
  {\parindent \z@ \raggedright
    \normalfont
    \interlinepenalty\@M
    \DOTIS{#1}
    \vskip 1\p@
  }}
\makeatother

\thispagestyle{empty}
\begin{titlepage}
\begin{center}
{\Large{\bf{Quantum fields in the non-perturbative
regime\newline\\ -- Yang-Mills theory and gravity}}}\newline\\
\vspace{1.5cm}
{\large{\hspace{3cm}Astrid Eichhorn\footnote{aeichhorn@perimeterinstitute.ca}}}\newline\\
\vspace{1.2cm}
\emph{\hspace{2cm}Ku pami\c{e}ci Mireille i Henryka Fr\c{a}ckowiaka.}\newline\\
\vspace{2cm}
\emph{Abstract}
\end{center}
In this thesis we investigate two different sets of physics questions, aiming at a better understanding of the low-energy behaviour of Yang-Mills theories, and the properties connected to confinement, in a first part. In a second part, we consider asymptotically safe quantum gravity, which is a proposal for a UV completion of gravity, based on the existence of an interacting UV fixed point in the Renormalisation Group flow. Both theories are characterised by non-perturbative behaviour, the first in the IR, the second in the UV, thus we apply a functional Renormalisation Group equation which is valid beyond the perturbative regime.
We investigate the ground state of SU(3) Yang-Mills theory, finding the formation of a gluon condensate, which we connect to a model for quark confinement. We further investigate the deconfinement phase transition at finite temperature and shed light on the question what determines the order of the phase transition.
Within quantum gravity, we examine the properties of the Faddeev-Popov ghost sector in a non-perturbative regime, thus extending truncations of Renormalisation Group flows into a new set of directions in theory space.
Finally we establish a connection of quantum gravity to observations of matter, by coupling fermions to gravity. Here we use the existence of light fermions - an observationally well-established fact in our universe - to impose constraints on quantum theories of gravity.

\end{titlepage}

\newpage
\thispagestyle{empty}
\begin{flushleft}
\phantom{x}
\vspace{3cm}
\emph{
This doctoral thesis has been submitted to the Friedrich-Schiller-Universit\"at Jena and defended successfully on September 6th, 2011.\newline\\
The compilation of this thesis is solely due to the
author, however, a large part of the work presented here has
been published in a number of articles and in
collaboration with several authors. 
Chap.~\ref{YM} relies on work in collaboration with Holger Gies and Jan M. Pawlowski \cite{Eichhorn:2010zc}, and a further collaboration also involving Jens Braun \cite{Braun:2010cy}.
Chap.~\ref{AS}
is founded on work in collaboration with Holger Gies \cite{Eichhorn:2010tb,ghostpaper3} and additionally with Michael M. Scherer \cite{Eichhorn:2009ah}. The work presented in chap.~\ref{ferminAS}  was done
in collaboration with Holger Gies \cite{Eichhorn:2011pc}.}
\end{flushleft}

\newpage
\thispagestyle{empty}
\begin{center} 
{\bf Quantum fields in the non-perturbative regime -- Yang-Mills theory and gravity}\\
{\it Summary}
\end{center}
 
In this thesis we study candidates for fundamental quantum field theories, namely non-Abelian gauge theories and asymptotically safe quantum gravity. Whereas the first ones have a strongly-interacting low-energy limit, the second one enters a non-perturbative regime at high energies. Thus, we apply a tool suited to the study of quantum field theories beyond the perturbative regime, namely the Functional Renormalisation Group.

In a first part, we concentrate on the physical properties of non-Abelian gauge theories at low energies.

Focussing on the vacuum properties of the theory, we present
an evaluation of the full effective potential for the field
strength
  invariant $F_{\mu \nu}F^{\mu \nu}$ from non-perturbative gauge  correlation functions and find a non-trivial minimum corresponding
  to the existence of a dimension four gluon condensate in the
  vacuum.
We also relate the infrared asymptotic form of the $\beta$
  function of the running background-gauge coupling to the asymptotic
  behavior of Landau-gauge gluon and ghost propagators and derive an upper bound on their scaling exponents.

We then  consider the theory at finite temperature and study the nature of the confinement phase transition in $d=3+1$
  dimensions in various non-Abelian gauge theories. For
SU(N) with N$=3,\dots,12$
  and $Sp(2)$ we find a first-order phase transition in agreement with
  general expectations.  Moreover our study suggests that the phase
  transition in $E(7)$ Yang-Mills theory also is of first order. Our studies shed light on the question which property of a gauge group determines the order of the phase transition.

In a second part we consider asymptotically safe quantum gravity.  Here, we focus on the Faddeev-Popov ghost sector of the theory, to study its properties in the context of an interacting UV regime. We investigate several truncations, which all lend support to the conjecture that gravity may be asymptotically safe. In a first truncation, we study the ghost anomalous dimension which we find to be negative at the fixed point. This suggests the existence of relevant couplings in the ghost sector.
In an extended truncation, we then discover two fixed points, one of which can be interpreted as an infrared fixed point, thereby allowing the construction of a complete RG-trajectory. Furthermore, the two fixed points differ in the sign of the ghost anomalous dimension, shifting further ghost operators towards relevance or irrelevance, respectively. We further discuss the structure of the ghost sector in the non-perturbative regime and point out that in the vicinity of an interacting fixed point for gravity further ghost couplings will generically be non-zero. We then discuss the implications of relevant operators in the ghost sector and give an explicit example for such an operator, namely a ghost-curvature coupling.

Finally we study the compatibility of quantum gravity with the existence of light fermions.
 We
  specifically address the question as to whether metric fluctuations can
  induce chiral symmetry breaking in a fermionic system. Our results indicate that chiral symmetry is
  left intact even at strong gravitational coupling. In particular, we find
  that asymptotically safe quantum gravity generically admits universes with
  light fermions. Thus our results in this sector also support the asymptotic-safety
  scenario.  We then point out that a study of chiral symmetry breaking through gravitational quantum
  effects is also an important test for other quantum
  gravity scenarios, since a completely broken chiral symmetry at the Planck
  scale would be in severe conflict with the observation of light fermions in
  our universe.  We demonstrate that this elementary observation already
  imposes constraints on a generic UV completion
  of gravity.

\newpage
\setcounter{page}{1}
\thispagestyle{empty}
\tableofcontents

\chapter{Motivation: Challenges in fundamental quantum field theories}
\emph{"In any region of physics where very little is known, one must keep to the experimental basis if one is not to indulge in wild speculation that is almost certain to be wrong. 
I do not wish to condemn speculation altogether. It can be entertaining and may be indirectly useful even if it does turn out to be wrong. One should always keep an open mind receptive to new ideas, so one should not completely oppose speculation, but one must take care not to get too involved in it." (P.A.M. Dirac \cite{blue_book})}\newline\\

Modern high-energy physics is described in terms of quantum field theory, which is a framework determined by the unification of Quantum Mechanics with Special Relativity. The quantisation of a classical theory works, e.g. in the path-integral framework. Within this setting, theories are determined by two properties, namely their field content, and their symmetry properties. Both can, to some extent, be deduced from experimental observation, although the relation between the fields used in the path integral and observable degrees of freedom is not always straightforward. 

In the path integral also field configurations which do not fulfill the classical equations of motion, i.e. which are called "off-shell", contribute to expectation values of operators. All contributing configurations are weighted by a complex phase factor, which is a function of the classical action. Accordingly the solution of the quantum equations of motion for the expectation values of operators can be much more involved than the solution of the classical equations of motion, since a part of the challenge lies in the derivation of the quantum equations of motion.

As observations indicate that all presently known fundamental, i.e. non-bound, matter turns out to be fermionic, the standard model of particle physics is built on theories involving fermionic fields\footnote{If it is found experimentally, a fundamental scalar Higgs boson will of course provide an exception.}. An important class of symmetries is presented by space-time dependent, i.e. local, gauge symmetries. Imposing gauge symmetries on fermionic theories leads to the introduction of bosonic force fields, the gauge bosons. 
Using the Abelian gauge group U(1) as the symmetry group allows to construct Quantum Electrodynamics, which has been tested to extremely high precision, see, e.g. \cite{Hanneke:2008tm}. The framework of quantum field theory itself is rather well-understood and allows to incorporate a wealth of different physical phenomena\footnote{In particular, the framework of quantum field theory does not only allow to describe observations in particle physics, or condensed matter systems, but, within standard cosmological scenarios for inflation even provides for an understanding of the large-scale structure of matter in the universe: Quantum fluctuations in the very early universe form the seeds for later structure-formation processes and therefore allow for an understanding why galaxies like our own are formed.}. In the following, we will state some major challenges of high-energy (particle) physics and discuss, whether it is possible to follow the conservative route to try to incorporate these into this well-tested framework.

Employing non-Abelian symmetry groups such as, e.g. SU(3) results in a very fascinating property: Non-Abelian gauge theories with a limited number of fermions turn out to be asymptotically free, since gluons have an antiscreening effect.
Accordingly perturbative tools allow to access the properties of the theory at high energies, where experimental confirmation from accelerator experiments is possible.
The most prominent example of such a theory is Quantum Chromodynamics (QCD), which describes strong interactions between coloured quarks and gluons. 
Here, the low-energy regime of the theory corresponds to a regime with a large coupling, and therefore shows a number of intriguing physical phenomena: The degrees of freedom of the theory change from quarks and gluons to hadrons, colourless bound states. This property, known as confinement, remains to be explained in terms of a physical mechanism. Different candidates for confining field configurations, typically of topological nature, and distinct criteria for confinement are discussed in the literature. A clear picture has only started to emerge, due to the notorious difficulty of treating a strongly-interacting theory.

Furthermore it remains to be clarified if and how confinement can be linked to the second property determining the appearance of QCD at low energies, namely chiral symmetry breaking. In particular the phase diagram of QCD at finite baryon densities and finite temperature is qualitatively as well as quantitatively only partially under control. The existence of exotic phases, such as a quarkyonic phase \cite{McLerran:2007qj} with restored chiral symmetry and confinement is currently debated. Moreover the existence of a critical endpoint of the chiral as well as the deconfinement phase transition, and the question if the two transition lines lie on top of each other, is a further unresolved question. Answering some of these will also allow us to understand astrophysical observations of neutron stars, as well as the dynamics in the early universe in more detail.

In QCD the main challenge lies in establishing, how these properties of the macroscopic theory emerge from the microscopic physics. It is usually believed that although the problem is hard to solve due to its non-perturbative nature, the framework of quantum field theory can fully account for all physics properties of low-energy QCD. This will also imply, that we then have understood the main origin of our own mass, which is mostly not due to the Higgs mechanism, but arises from the non-perturbative dynamics of QCD. In this sense, one might say that a full understanding of QCD is not a purely academic problem, but directly related to properties that we observe in our everyday world. \newline

A second main challenge in theoretical high-energy physics may even fundamentally change the framework of high-energy physics, namely (local) quantum field theory. It lies in the reconciliation of Quantum Mechanics and General Relativity to a theory of quantum gravity. Unlike QCD, a quantum field theory of gravity cannot be accessed with perturbative methods at high energies, which manifests itself in the well-known perturbative non-renormalisability of General Relativity. 

One can now adopt measures of different "degree of radicalness": One might introduce new degrees of freedom, following the physical idea that the metric is not the (only) fundamental field necessary to describe gravity at high energies. On the technical side these new degrees of freedom cancel the divergences leading to the perturbative non-renormalisability, as is the main idea behind, e.g. supergravity theories. Furthermore one may abandon the requirement of locality. The physical assumption related to this is the existence of a fundamental physical scale, often identified with the Planck scale. Such an idea might be seen in accordance with the development of physics during the last two centuries, where a continuous picture of matter had to give way to discrete atoms, and a continuous notion of energy is given up in many examples in Quantum Mechanics. Similarly a continuous description of space-time may be wrong, and space-time might be fundamentally discrete. This idea is at the heart of proposals such as causal set theory or non-commutative space-times, and might also come out of loop quantum gravity and spinfoams.  

Furthermore one may hypothesise that the symmetry properties of gravity are changed at high energies. In particular (local) Lorentz invariance may either be broken, or deformed at high energies. Such a deformation or breaking of symmetries by quantum gravity effects provides one of the very few possibilities to currently test some properties of quantum gravity experimentally.

Finally some approaches, such as loop quantum gravity, causal and Euclidean dynamical triangulations, as well as the asymptotic-safety scenario suggest that a perturbative approach to gravity is incorrect and genuinely non-perturbative information is crucial to quantise the theory. In analogy to the low-energy regime of QCD, which is expected to describe the physics of the strong interactions correctly, but is characterised by a breakdown of perturbation theory, a quantum field theory of the metric may be a valid description of the physics of quantum gravity, but might not be accessible by perturbative tools at high energies. Hence one also might have the choice to remain in the framework of local quantum field theory without introducing any new degrees of freedom, as is proposed in the asymptotic-safety scenario. This scenario is not excluded by the perturbative non-renormalisability of General Relativity. It simply implies that a quantum field theory of the metric has to be non-perturbatively renormalisable, if it is supposed to make sense as a fundamental, and not just as an effective theory. This scenario seems to be the least radical of the currently available choices, as it stays within the well-tested framework of local quantum field theories. On the other hand it is a rather bold conjecture, that the metric is indeed the fundamental degree of freedom of gravity on all energy scales. However since we do currently not have any experimental hints on what more fundamental gravitational degrees of freedom might be, we may test to what extent a local quantum field theory of the metric is self-consistent and therefore potentially realisable. Of course this does not entail that it is indeed realised in our universe, since nature may have "chosen" a different internally consistent theory.
Ultimately either some of the approaches to quantum gravity will turn out to be inconsistent within themselves, or finally experimental results may shed some light on the question, which of several approaches to quantum gravity is the one favoured by nature. Although the typical scale of quantum gravity, the Planck scale, is not currently experimentally accessible, one should not cast aside the possibility of experimental results on quantum gravity in the near future. In particular cosmology and astrophysics provide settings where even tiny effects may accumulate to a sizable contribution.

As emphasised by Wilczek \cite{Wilczek:1998ma} "whether the next big step will require a sharp break from the principles of quantum field theory or, like the previous ones, a better appreciation of its potentialities, remains to be seen". In this spirit we may try to push the existing framework as far as possible. In one direction, coming from a known microscopic description, this entails that we deduce and understand all observable properties of the macroscopic theory, such as in the example of QCD. In the other direction, it requires us to test whether UV completions of known low-energy theories, such as gravity, can be incorporated into the framework of local quantum field theory. If in particular the second possibility fails, this might require us to completely rethink properties of our theories which we have taken to be fundamental properties of nature, such as, e.g. the assumption of a continuous space-time.

To address such non-perturbative questions adequately we need to evaluate the complete quantum theory, i.e. we need a non-perturbative handle on the generating functional. This may be done within the Functional Renormalisation Group (FRG), which allows us to take into account quantum fluctuations in the path-integral momentum shell by momentum shell. Thereby the functional integral is reformulated into a functional differential equation, which is much easier to handle.

The FRG is a very flexible tool that is applicable to diverse problems, ranging from the BEC-BCS-crossover in ultracold quantum gases \cite{Diehl:2009ma}, to supersymmetric field theories, see, e.g. \cite{Synatschke:2008pv}, the phase diagram of QCD \cite{Braun:2006jd,Braun:2009gm}, the Higgs sector of the Standard model, see, e.g. \cite{Gies:2009hq, Gies:2003dp}, non-commutative quantum field theories \cite{Sfondrini:2010zm} and quantum gravity, see, e.g. \cite{Reuter:1996cp, Reuter:2007rv}. 

In this thesis we will apply the framework of the FRG to QCD, to better understand and derive properties of the macroscopic theory from our microscopic description. In particular we will focus on questions related to confinement at zero and finite temperature.

In the second part of this thesis we will focus on the asymptotic-safety scenario for quantum gravity, testing its internal consistency and its properties in a specific way and also investigating its compatibility with matter.
\newline

This thesis is structured as follows: In chap.~\ref{methodchap} we will introduce the Functional Renormalisation Group, with a particular emphasis on its application to gauge theories. 
We employ the FRG in a study of non-Abelian gauge theories in chap.~\ref{YM},
where we are interested in the physics of the infrared
sector, where the theory is strongly interacting. We investigate the non-perturbative vacuum
structure of Yang-Mills theories, which might contain a gluon condensate. Here we also deduce a bound on
the infrared scaling exponents of gluon and ghost propagators for low
momenta. In a second step we move towards the evaluation of the full phase diagram of QCD and study the deconfinement phase transition in the limit of infinitely heavy quarks.
We determine the critical temperature and the order of the
deconfinement phase transition for diverse gauge groups and
present evidence on the question, what determines the order
of the phase transition.

We then proceed to introduce the asymptotic-safety scenario for quantum gravity in chap.~\ref{AS} and explain how it can be investigated with the help of the FRG on the example of the Einstein-Hilbert term. Here we present a method of evaluating the flow equation in gravity, which avoids making use of heat-kernel techniques. We report on new results concerning the Faddeev-Popov ghost sector of the theory. In particular, we investigate the properties of this sector within a non-perturbative setting, studying the fixed-point structure and the RG flow in several truncations.
In chap.~\ref{ferminAS} we focus on the inclusion of quantised matter into the asymptotic-safety scenario for quantum gravity. In particular we examine if gravity, similar to non-Abelian gauge theories, can break chiral symmetry in a fermionic system and induce fermion masses. Since we observe fermions much lighter than the Planck scale, the compatibility of light fermions with quantum gravity is a crucial test for any quantum theory of gravity. Here we use that the framework of the FRG is also applicable to effective theories, where the UV completion of the theory needs not to be known in order to study the RG flow within a finite range of scales. In the case of quantum gravity this allows us to derive conditions for the existence of light fermions within other UV completions for gravity.

\chapter{The Functional Renormalisation Group}\label{methodchap}
\section{The basic physical idea: Connecting microscopic and macroscopic physics}\label{base_phys}

Physics looks very different on different scales, and effective descriptions of the same system on different scales can be structurally as well as conceptually very different. Consider the example of nuclear forces, which are mediated by pions between neutrons and protons. For a large part of our understanding of this system we do not have to know the microscopic structure, which, according to our current understanding, consists of quarks and gluons. Similarly the nuclear structure is not relevant for the description of physics on atomic scales, and an effective description suffices. 
In particular the effective degrees of freedom as well as the realisation of fundamental symmetries may be altered on different scales, since spontaneous symmetry breaking may occur. 
In such cases the details of the microscopic physics do not play a role for the description of the effective macroscopic dynamics, which can often be parametrised by only a few effective parameters.
The microscopic theory then allows to determine the values of these couplings, and determines the relations between the effective and the microscopic degrees of freedom.

To obtain a fundamental description of nature, we ultimately want to derive the effective theories governing physics on large scales from the microscopic dynamics. We want to establish a connection between the dynamics over a large range of scales, and determine the parameters of effective theories from the microscopic theory. This connection is from small to large scales, which intuitively makes sense: Knowing a microscopic, fundamental theory, we can deduce an effective description on larger scales. In particular, different microscopic theories can lead to the same effective dynamics. In some sense, the information on microscopic details gets "washed out", when we go to an effective description on larger scales.

In certain areas of physics on the other hand we only know the effective, macroscopic dynamics, and do not have any experimental guidance as to the nature of the microscopic, fundamental theory. A quantum theory of gravity is one of the examples. Here we want to establish a connection from large to small scales, and find the microscopic theory underlying the effective description that is currently accessible to experiments.

In both cases, when making the transition from the microscopic to the macroscopic regime and vice-versa, we need a tool that allows us to connect effective descriptions on different scales, and derive macroscopic physics from underlying microscopic descriptions, including the effect of quantum fluctuations on all intermediate scales. In particular we want to access regimes where physics is governed by strong correlations and non-perturbative effects, such as, e.g. in QCD at large, or quantum gravity at small scales. 

Here, we will introduce a tool that is particularly suited to these situations, namely the Functional Renormalisation Group (FRG). 

\section{Coarse graining and the effective average action}\label{Gammak}
Quantum field theories (QFTs) can be defined by a path integral that weighs quantum fluctuations with a complex phase factor $e^{iS}$, where $S$ is the classical (or microscopic) action\footnote{Mathematically, the path integral is challenging, in particular for interacting theories, however it beautifully generalises the quantum mechanical idea that a particle simultaneously "takes all possible paths", weighted by phase factors, instead of just travelling along the classical trajectory. Therefore it presents a very intuitive approach to quantum field theories.}.
The central object in the path-integral approach to a QFT is the generating functional from which all $n$-point correlation functions are calculable, thus allowing to access all observables. In Euclidean space\footnote{The transition to Euclidean space implies that we will focus on the vacuum properties as well as equilibrium physics of the theory. Real-time dynamics are accessible in a Lorentzian setting.}, the generating functional for a scalar field $\varphi$ coupled to a source $J$ is given by 
\begin{equation}
 Z[J]= \int \mathcal{D}_{\Lambda}\varphi\, e^{-S[\varphi]+ J \cdot \varphi}.\label{genfunct}
\end{equation}
Equivalent definitions hold for fermion, vector and tensor
fields, which may also transform non-trivially under
internal, local or global, symmetries. We denote the
appropriate index contractions by a dot, which also includes
an integral over real space, where the dependence of the
fields on space-time coordinates is understood implicitly.
The path-integral measure $\mathcal{D}_{\Lambda}\varphi$ is
understood to be UV-regularised, which may be a highly
non-trivial issue in theories, where no regularisation may
exist that is compatible with the symmetries. Such a theory
is called anomalous, which simply means that quantum effects
break the classical symmetry. We will neglect this in the
following, and simply assume that the path integral is
UV-regularised.

The generating functional for all one-particle irreducible correlation functions, the effective action, is defined by a Legendre transform:
\begin{equation}
 \Gamma[\phi]= \underset{J}{\rm sup} \left(\int J \cdot \phi -\ln Z[J]\right).
\end{equation}
Here the expectation value $\phi = \langle \varphi \rangle$ is evaluated at the supremum $J = J_{\rm sup}$, which automatically ensures the convexity of the effective action.

The quantum equations of motion, which govern the dynamics of expectation values, can be derived from the effective action by functional variation:
\begin{equation}
J = \frac{\delta \Gamma[\phi ]}{\delta \phi}.\label{qeom}
\end{equation}
Ultimately we are interested in solving these in theories such as non-Abelian gauge theories or quantum gravity, to understand the vacuum state of, e.g. QCD or our universe and derive the properties of excitations on top of this state.

The microscopic equations of motion can be vastly different
from the effective, macroscopic equations of motion for the
expectation values of the quantum fields, see \Eqref{qeom}.
These take into account the effect of all quantum
fluctuations, and will therefore generically contain
effective interactions, that are not present in the
 microscopic dynamics. 

The main purpose of the Functional Renormalisation Group (FRG) is to connect the description of physics on different momentum scales, in weakly as well as strongly interacting regimes. It presents a tool that allows to derive effective dynamics from the underlying microscopic dynamics, even in cases where perturbative tools become inapplicable.

The main idea of the FRG states that in order to describe dynamics at a momentum scale $k$ it is not necessary to consider the microscopic interactions at scales greater than the scale $k$. Instead it suffices to consider an effective theory that is constructed from the microscopic theory by integrating out quantum fluctuations at high momenta. This idea implies that the infrared, i.e. low-energy physics, decouples from the ultraviolet, i.e. high-energy physics: High-energy degrees of freedom do not explicitly show up in the theory in the infrared, their effect is only indirect by determining the values of the coupling constants of the effective theory. Of course such a decoupling does not directly hold for massless degrees of freedom, unless phenomena such as confinement or dynamical mass generation occur.

The implementation relies on the Wilsonian idea of performing the path-integral momentum- shell wise \cite{Wilson:1971dh,Wilson:1971bg,Wilson:1973jj}, by introducing a floating infrared(IR)-cutoff $k$, which can be identified with an inverse coarse-graining scale. This is most easily realised in a Euclidean formulation, see \Eqref{genfunct}.

In real space the procedure can best be exemplified by Kadanoff's idea of block spinning \cite{Kadanoff:1966wm}: If one is interested in the low-momentum, i.e. large distance, physics of an Ising spin system, one can imagine to average microscopic spins over a finite region of space. The system is then constituted by the averaged spins. Subsequently one rescales the system, which implies, that now one effectively considers a larger number of degrees of freedom (microscopic spins), when looking at the same size of the sample. The effect of this procedure exemplifies the basic property of this coarse-graining procedure: While the model we started with typically contains only nearest-neighbour interactions, the effective, i.e. coarse-grained spin system after averaging and rescaling contains all possible interactions that are compatible with the symmetries, e.g. also next-to-nearest-neighbour interactions.

This procedure indeed relies crucially on the concept of locality: If an interaction is non-local in the sense that it cannot be rewritten as a finite number of terms in an expansion in powers of derivatives, it implies that one cannot meaningfully average quantum fluctuations over a finite region in space. Typically one considers only local microscopic dynamics in QFTs. 
Non-local interactions should only emerge in the limit where
all quantum fluctuations have been integrated out.
A well-known example is the Polyakov action in two
dimensions, where, e.g. a scalar field is coupled minimally
to gravity. Integrating out the scalar field explicitly
yields a non-local action for gravity
\cite{Polyakov:1981rd}, see also \cite{Codello:2010mj}. The non-locality is in this
sense an emergent phenomenon.

Let us stress that the requirement of locality is also at
the heart of the necessity to renormalise: Since
interactions are local, their expansion in Fourier
components requires the inclusion of components of
arbitrarily high momentum $p$. This leads to divergent
loop-integrals in perturbation theory, where the divergences
are then removed by a regularisation and subsequent
renormalisation procedure. Introducing a physical cutoff
into the theory, which can be interpreted as the scale of
non-locality implies that physical results will depend on
this scale, but it removes the UV-divergences in
perturbative loop integrals.

The simple example of Kadanoff's block spinning procedure may be misleading in a crucial point, as it suggests that the coarse-graining scale may be identified with a length scale. This may not be always correct: Degrees of freedom are usually integrated out according to their eigenvalue of the kinetic operator\footnote{It is also possible to study theories with a trivial kinetic term such as matrix models for 2-dimensional gravity. Here the procedure of integration out quantum fluctuations proceeds in a more abstract space, allowing the continuum limit of such theories to be studied.}. In many theories this corresponds precisely to the simple Laplacian operator and therefore to an inverse length scale. In particular quantum field theories on a flat background with vanishing background fields typically have a kinetic operator which is just given by the momentum squared.
In the case of theories with UV-IR mixing (e.g. field
theories on a non-commutative background) such a separation does not occur. Also considering quantum fluctuations around a non-trivial classical background field will typically result in the spectrum of the kinetic operator depending on this background field.

\begin{figure}[!here]
\begin{minipage}{0.65\linewidth}
 \includegraphics[scale=0.7]{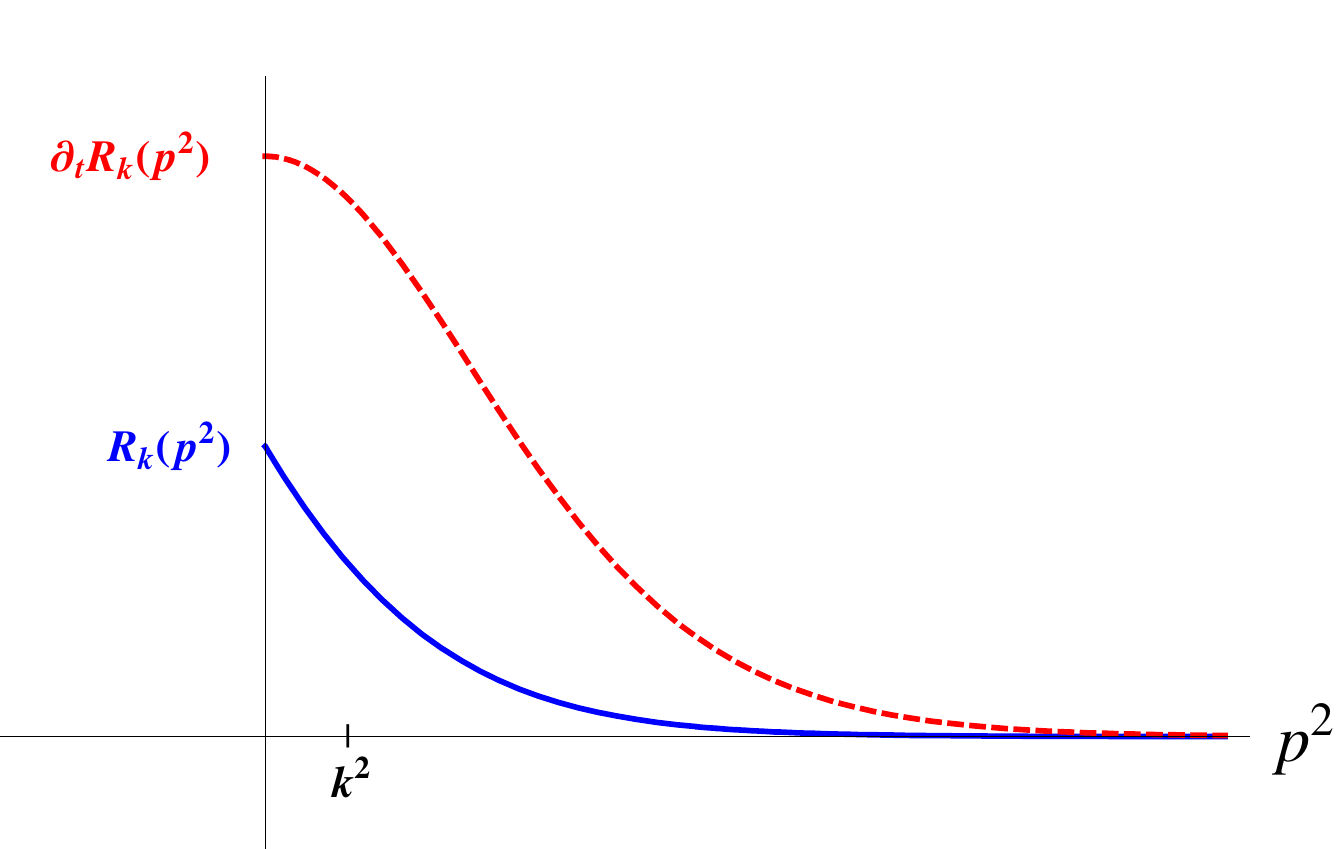}
\end{minipage}
\begin{minipage}{0.3\linewidth}
\caption{We show a regulator of the type $R_k(p^2)= \frac{p^2}{e^{\frac{p^2}{k^2}}-1}$ and its scale-derivative $\partial_t R_k(p^2)$, to exemplify the suppression of low-energy modes. The scale-dependent mass-term vanishes for $p^2 \gg k^2$.}
\label{regplot}
\end{minipage}
\end{figure}

Let us now formalise the above ideas: A momentum-shell-wise integration of quantum fluctuations can be implemented by defining a scale-dependent generating functional
\begin{equation}
 Z_k[J]= \int \mathcal{D}_{\Lambda}\varphi e^{-S[\varphi]+ \int J \varphi -\Delta S_k} \, \, \mbox{ with } \Delta S_k= \frac{1}{2}\int_p \varphi(-p) \cdot R_k(p^2) \cdot \varphi(p).
\end{equation}
Here the infrared regulator function $R_k(p^2)$ with $R_k(p^2)>0$ for $\frac{p^2}{k^2}\rightarrow 0$ ensures that the contribution of quantum fluctuations with momenta below $k^2$ is suppressed by a $k$-dependent mass-like term\footnote{Although we use a notation that suggests that the regulator depends on the momentum, it generically depends on the kinetic operator, which may be the momentum squared in simple cases, but can also be an appropriate covariant differential operator or similar. In such cases the regulator distinguishes quantum fluctuations with respect to their eigenvalues of the kinetic operator. The variable $p^2$ therefore is to be understood as a placeholder for the eigenvalues of the kinetic operator, and $\int_p$ correspondingly can also be a sum over discrete eigenvalues.}. 
As the regulator function is chosen to vanish for $p^2 >k^2$, high-momentum quantum fluctuations are unsuppressed and fully contribute to the path integral, see fig.~\ref{regplot}.

\noindent Furthermore the limit $R_k(p^2) \overset{k \rightarrow \infty}{\longrightarrow} \infty$ ensures that the effective average or flowing action, defined by a modified Legendre transform
\begin{equation}
 \Gamma_k[\phi]= \underset{J}{{\rm sup}} \left(\int J \cdot \phi - \ln Z_{k}[J] \right)- \Delta S_k.
\end{equation}
fulfills $\Gamma_k\overset{k \rightarrow \infty}{\longrightarrow} S$, see, e.g. \cite{Reuter:1996eg}:
The exponential of the flowing action satisfies 
\begin{equation}
e^{-\Gamma_k[\phi]}=\int \mathcal{D}\varphi\, e^{-S[\varphi]+ \int \frac{\delta \Gamma[\phi]}{\delta \phi}\left(\varphi- \phi \right)}e^{-\frac{1}{2}\int \left(\varphi -\phi \right)R_k \left(\varphi -\phi \right)}.
\end{equation}
As in the limit $k \rightarrow \infty$ the regulator suppresses all modes $\sim k^2$, the second exponential is proportional to a delta function $\delta (\varphi -\phi)$.

In the limit $k \rightarrow 0$ we recover the effective action which includes the effect of all quantum fluctuations, since the regulator function vanishes in this limit.

The flowing action defines a family of effective theories, labelled by the scale $k$, which can be used to describe dynamics at the momentum scale $k$ and which interpolate smoothly between the classical action in the ultraviolet and the effective action in the infrared.\\
To evaluate the main contributions to a process that involves external momenta at the scale $k$, a tree level evaluation of $\Gamma_k$ suffices, since external momenta effectively act as a cutoff in loop diagrams\footnote{This is the main rationale, e.g. behind a particular type of RG-improvement: Assuming the validity of the classical equations of motion, the effect of quantum fluctuations in a semi-classical regime can be included by substituting the couplings with their running counterparts. A crucial step in this procedure is the suitable identification of $k$ with a physical scale of the problem. In the context of asymptotically safe quantum gravity many results on the effect of quantum gravity in cosmology and astrophysics can be derived in this way, for a review see \cite{Bonanno:2009nj}.}.

To better understand the effective average action let us turn to theory space, which is the space spanned by the couplings of all operators compatible with the field content and the symmetries of the theory. Clearly this space is typically infinite dimensional, so that we can only depict a subspace.

\begin{figure}[!here]
 \begin{minipage}{0.6\linewidth}
  \includegraphics[scale=0.4]{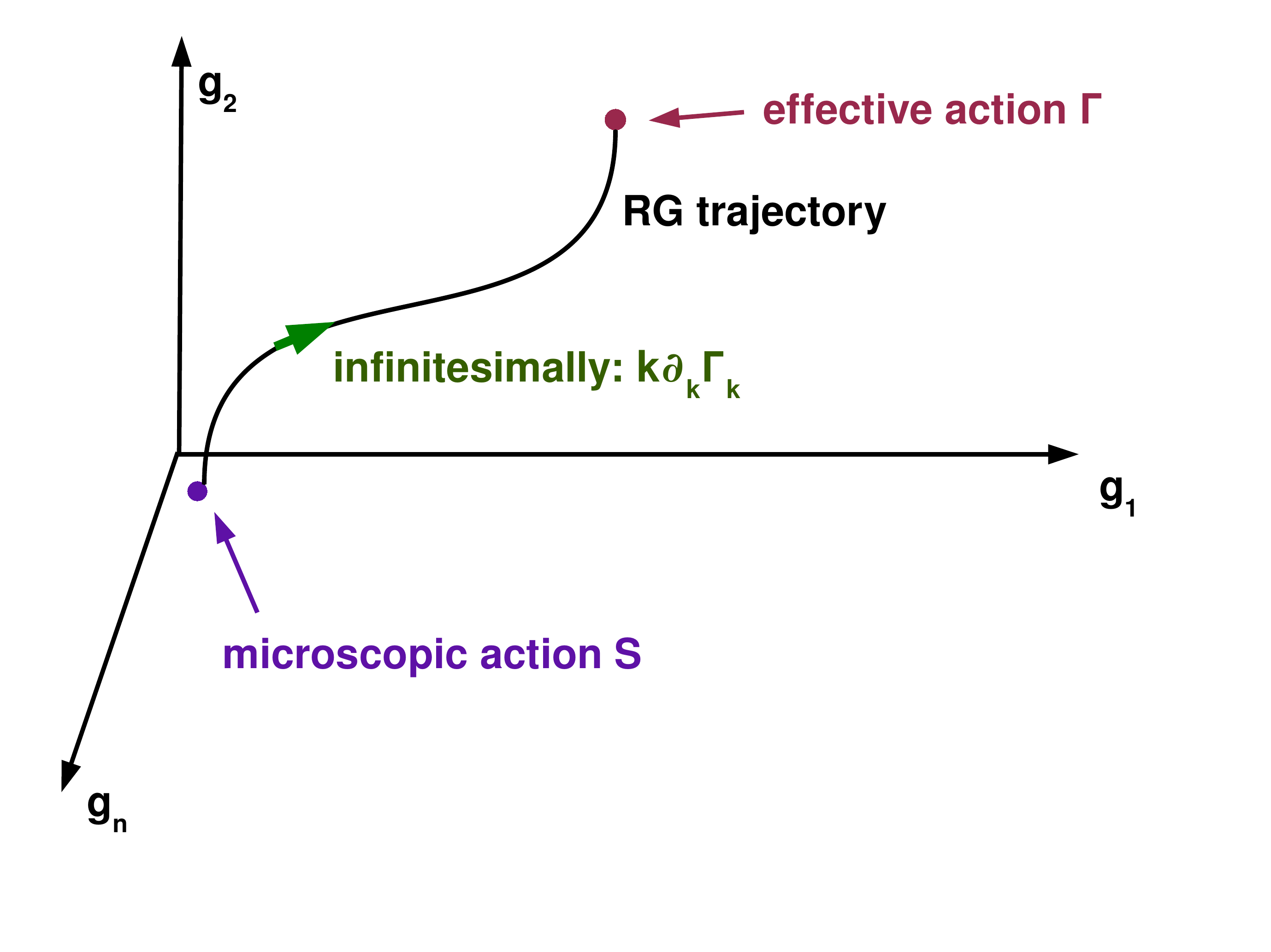}
 \end{minipage}
\begin{minipage}{0.35\linewidth}
 \caption{Integrating out quantum fluctuations results in a flow in theory space, which connects the microscopic action $S_{\rm cl}$ to the full effective action $\Gamma_{k \rightarrow 0}$.\label{flowintheoryspace}}
\end{minipage}
\end{figure}

The effective average action at a scale $k$ is specified by giving the values of all couplings at this scale, defining a point in theory space. Integrating out quantum fluctuations in the momentum shell $\delta k$ then results in a shift of the couplings. For a theory with a known microscopic or classical action we can thus start in the far ultraviolet and integrate out fluctuations all the way down to $k \rightarrow 0$, where we reach the full effective action, cf. fig.~\ref{flowintheoryspace}. 

\noindent The effect of quantum fluctuations hence is to induce a flow in theory space, which connects the classical action in the ultraviolet to the full effective action in the infrared. The tangent vectors to the flow lines are given by the scale derivative of the effective average action, which is governed by the Wetterich equation.

\section{Wetterich equation for gauge theories}\label{Wetteq}

The Wetterich equation \cite{Wetterich:1993yh} is an exact equation for the scale derivative of the effective action, which does not rely on the existence of a small parameter and holds for arbitrary values of the couplings. Reviews can be found in \cite{Berges:2000ew,Bagnuls:2001pr,Polonyi:2001se,Pawlowski:2005xe,Delamotte:2007pf,Rosten:2010vm}. For the specific case of gauge theories, see \cite{Litim:1998nf,Gies:2006wv,Igarashi:2009tj}.

For gauge theories we have a choice between two formulations: One may either construct a gauge-invariant flow equation \cite{Pawlowski:2003sk,Arnone:2005fb,Arnone:2006ie}, or work in a gauge-fixed formulation. The first may be considered to be cleaner conceptually; the second is more adapted to practical calculations.

We therefore proceed to gauge-fix, using the well-known Faddeev-Popov procedure (see, e.g. \cite{Weinberg_book} or \cite{Reuter:1996cp} for the case of gravity), which is a procedure developed in the context of perturbation theory. 
In a gauge-fixed approach we may encounter a serious problem, namely the Gribov problem \cite{Gribov:1977wm,Singer:1978dk}: The perturbative gauge-fixing procedure is not well-defined in some gauges in the non-perturbative regime. One example is the Landau gauge in Yang-Mills theory, where the gauge condition is
\begin{equation}
 \mathcal{F}=\partial_{\mu}A^{\mu}=0,\label{Landaugauge}
\end{equation}
and the corresponding Faddeev-Popov operator is
\begin{equation}
\mathcal{M}_{ab}= -\partial_{\mu}D^{\mu}_{ab},\label{FPLandau}
\end{equation}
where by Latin indices we denote colour indices.

\begin{figure}[!here]
\begin{minipage}{0.65\linewidth}
 \includegraphics[scale=0.4]{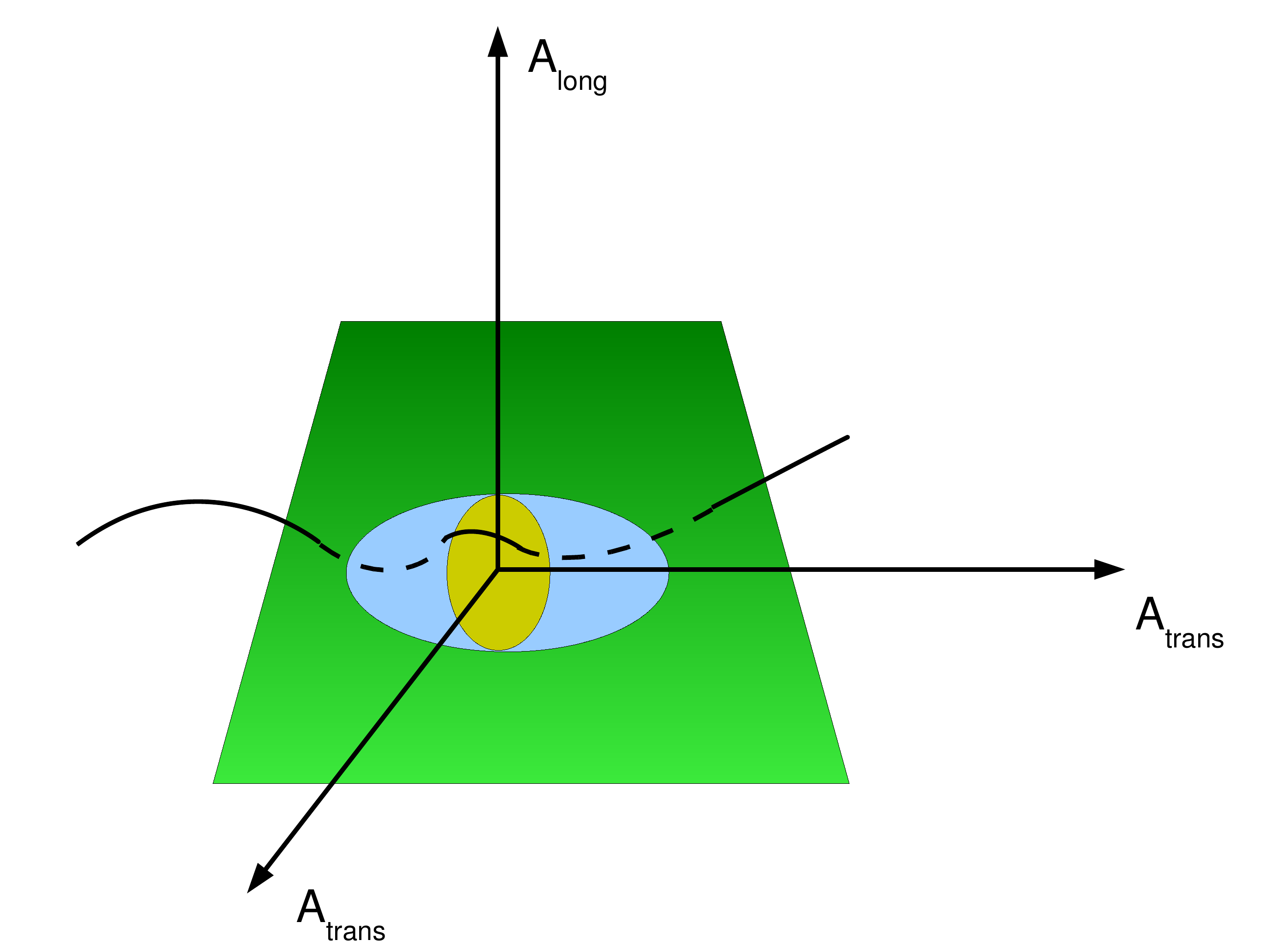}
\end{minipage}
 \begin{minipage}{0.3\linewidth}
 \caption{We show a sketch of the (infinite dimensional) gauge field configuration space, indicating a gauge orbit by the blue line. The gauge-fixing hypersurface is indicated by the green plain, and the first Gribov region and the fundamental modular region are shown in blue and yellow, respectively. Note that these include the origin of gauge-field configuration space and share a common boundary. \label{gribovplot}}
\end{minipage}
\end{figure}

The gauge-fixing condition \Eqref{Landaugauge} is not unique, so each gauge field configuration has several Gribov-copies, which are related by a gauge transformation and nevertheless also fulfill the gauge condition \Eqref{Landaugauge}. Thus a gauge orbit, which corresponds to only one particular physical field configuration, intersects the gauge-fixing hypersurface in gauge field configuration space multiple times, cf. fig.~\ref{gribovplot}. 
Furthermore the Faddeev-Popov operator \Eqref{FPLandau} is not positive definite for large values of the gauge field. This property follows directly from a consideration of gauge copies of a field configuration which also fulfill the gauge condition \Eqref{Landaugauge}. For an infinitesimal gauge transformation, there exists a gauge copy also fulfilling \Eqref{Landaugauge}, if the Faddeev-Popov operator has a zero eigenvalue\footnote{To see this explicitly, consider a gauge transformed configuration $\tilde{A}_{\mu}$ of a configuration $A_{\mu}$ that fulfills the gauge condition, i.e. $\partial_{\mu}A^{\mu}=0$. Requiring that the gauge-transformed field also satisfies the gauge condition, results in $\partial^{\mu}(U^{\dagger}\partial_{\mu}U+ U^{\dagger}A_{\mu}U)=0$, where $U$ is an element of the gauge group. Specialising to infinitesimal transformations $U = 1 + \omega$, and using that $A_{\mu}$ satisfies the gauge condition, we finally get $\partial^2 \omega - \partial_{\mu}\omega A^{\mu}+ A^{\mu}\partial_{\mu}\omega=0$, which we recognise as the Faddeev-Popov-operator acting on $\omega$. Thus the Faddeev-Popov operator has to have a zero eigenvalue.}.
For perturbation theory the problem is non-existent, since $-\partial_{\mu}D^{\mu}_{ab} \rightarrow -\partial^2 \delta^{ab}$ has a positive spectrum for vanishing coupling. 
Both problems imply that the generating functional, on which the flow equation is founded conceptually, is ill-defined non-perturbatively.

\noindent A solution to the Gribov problem is given by a restriction of the domain of integration in the generating functional to the first Gribov region, or even the fundamental modular region, which both have a positive definite Faddeev-Popov operator, and the second of which singles out exactly one representative per gauge orbit, thus uniquely implementing the gauge condition \cite{Zwanziger:2003cf}, cf. fig.~\ref{gribovplot}. It can be shown that the origin of gauge field configuration space is contained in both regions, and both are bounded and convex regions. Interestingly this restriction results in non-trivial boundary conditions for the ghost and gluon propagator in the deep infrared, which can be incorporated in the flow equation. For more details see, e.g. the review articles \cite{Gies:2006wv, Fischer:2006ub}. For details on how the restriction to the first Gribov region is implemented within the flow equation, see sec.~\ref{Landaugaugeprops}.

It is known that in Yang-Mills theory no local and Lorentz covariant gauge exists, which singles out only one representative per gauge orbit. In other words, all these gauges suffer from the Gribov problem \cite{Singer:1978dk}. The non-uniqueness of gauge-fixing ultimately follows from the topology of the gauge group, which is why a unique gauge-fixing can be defined locally, e.g. in perturbation theory, but not globally. In the case of gravity the Gribov problem also exists \cite{Das:1978gk,Esposito:2004zn}, but it has not been studied yet, how it can be solved and what the consequences, e.g. for the metric and the ghost propagator will be in a strongly-interacting regime.

Let us now proceed to state and explain the Wetterich equation in a gauge-fixed setting, keeping in mind that depending on the gauge we might have to deal with the Gribov problem.

The generating functional in a gauge-fixed formulation with source terms for the gauge and ghost fields is then given by
\begin{equation}
 Z_k[J]= \int \mathcal{D}A_{\mu}\, \mathcal{D}c\, \mathcal{D}\bar{c} e^{-S[A]-S_{\rm gf}- S_{\rm gh}+ \int J \cdot A+ \int \bar{\eta}c - \int \bar{c}\eta -\Delta S_k},
\end{equation}
where the classical action $S[A]$ is supplemented by a gauge-fixing term 
\begin{equation}
S_{\rm gf}= \frac{-1}{2\alpha} \int_x \mathcal{F}\cdot \mathcal{F}
\end{equation}
with the gauge-fixing functional $\mathcal{F}$. For the sake of simplicity we suppress whatever indices it might carry. The corresponding Faddeev-Popov ghost term reads
\begin{equation}
S_{\rm gh}= -\int \bar{c} \cdot \mathcal{M} \cdot c, 
\end{equation}
where $\mathcal{M}$ is obtained by deriving the gauge-transformed condition $\mathcal{F}$ with respect to the gauge parameter. Note that the regulator term is present for all quantum fields, i.e. also for the ghosts.

The Wetterich equation can then be derived straightforwardly:
\begin{equation}
\partial_t \Gamma_k[A, \bar{c}, c]= k \partial_k \Gamma_k[A, \bar{c}, c]= \frac{1}{2} {\rm STr} \left[\partial_t R_k \left(\Gamma_k^{(2)}[A,\bar{c},c]+R_k \right)^{-1}\right].
\end{equation}
Here $\Gamma_k^{(2)}$ denotes the second functional derivative of the flowing action with respect to the gauge field $A_{\mu}$ and the ghost and antighost fields $c$ and $\bar{c}$. It is therefore a (not necessarily diagonal) matrix in field space, and also carries Lorentz and internal indices as well as space-time dependence. The supertrace $\rm STr$ implements a trace over all, continuous as well as discrete indices and introduces an additional negative sign for Grassmann valued fields. For minimally coupled fields on a flat background and zero classical background fields, it implies a simple integration over the momentum\footnote{Note that one can also derive the Wetterich equation by assuming that the theory is defined by some generating functional for the $n$-point correlation functions. No path-integral representation needs to be invoked at any point in the derivation here \cite{lecturenotesJan}, which clarifies, why the Wetterich equation is not directly influenced by issues related to the path-integral measure such as anomalies etc. This does of course not preclude the treatment of an anomalous theory within the framework presented here. The boundary conditions required to solve the Wetterich equation can include such effects. Furthermore terms that arise due to an anomaly can be included in the effective average action, and their physical implications can be studied.}.

The flow equation is automatically UV as well as IR finite: The IR finiteness follows by construction. The UV finiteness follows from the scale derivative of the regulator in the numerator, which vanishes for $p^2 \gg k^2$ and is typically peaked around $p^2 \approx k^2$. The trace on the right-hand side of the flow equation receives the main contribution from eigenvalues of the inverse propagator which are comparable to $k$. This implements the idea of performing the functional integral momentum-shell wise.

Structurally, the flow equation, although in spirit based on the path integral, is independent of the question of the path integral being well-defined. It is a functional differential equation, allowing for an analytical as well as numerical treatment also beyond the perturbative regime and in regions where, e.g. numerical simulations of the path integral based on Monte-Carlo techniques break down. 

Note that the flow equation has a one-loop structure, which is technically very favorable, as no overlapping loop integrations, as they do occur, e.g. in other non-perturbative functional equations such as Dyson-Schwinger equations, have to be performed. Nevertheless, the equation is exact and does not miss contributions that are formulated as two-loop or higher terms in other approaches. Using fully dressed vertices and propagators corresponds to a particular type of resummation of diagrams, which accounts for the compatibility of being exact and one-loop. 
In particular, perturbation theory can be reproduced to any order by iteratively applying the Wetterich equation \cite{Papenbrock:1994kf,Litim:2002xm}.

One may also choose to regularise the theory with an
operator insertion that depends on higher powers of the
field. As discussed in \cite{Litim:2002xm}, the fact that
expectation values with more than two fields involve
multi-loop integrals will result in the flow equation not
being of one-loop type. Since this is a highly desirable
property for computational reasons, the regulator insertion
is chosen to be quadratic in the fields.

The flow equation has a diagrammatic representation: Denoting the full propagator by a straight line for gauge bosons and a dashed line for Faddeev-Popov ghosts, it reads:
\begin{figure}[!here]
\begin{minipage}{0.15\linewidth}
\begin{flushright}
$\partial_t \Gamma_k[A]= \frac{1}{2}$
\end{flushright}
\end{minipage}
\begin{minipage}{0.43\linewidth}
\includegraphics[scale=0.25]{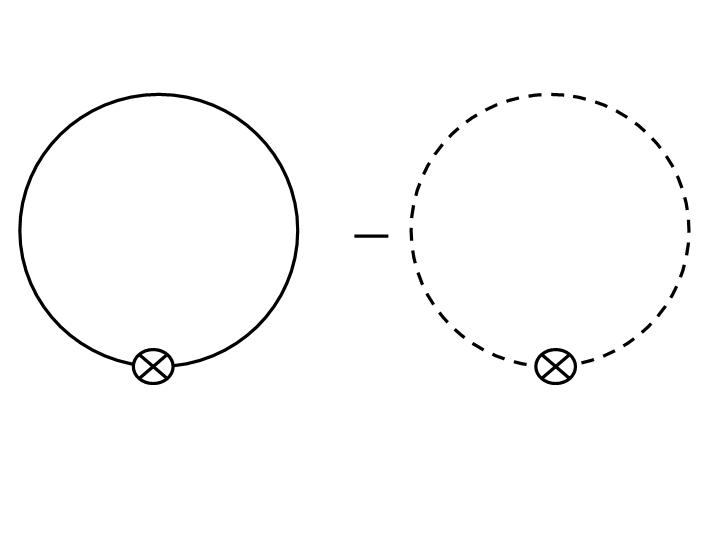}
\end{minipage}
\begin{minipage}{0.37\linewidth}
\caption{Diagrammatic representation of the flow equation: The trace over the full propagator gives a closed circle, with the regulator insertion $\partial_t R_k$ denoted by a crossed circle.}
\end{minipage}
\end{figure} 

\noindent This diagrammatic representation, reminiscent of Feynman diagrams, emphasises again that the flow equation does not contain any functional integrals.

Typical applications of the flow equation will be theories, which show a transition from weak to strong interactions over a range of scales, which prohibits the use of perturbation theory. Often such a transition is accompanied by a change in the effective degrees of freedom (e.g. in QCD from quarks and gluons to hadrons, or in cold atoms in the BEC-BCS crossover), and by a spontaneous breakdown of symmetries (such as chiral symmetry in QCD). Here the huge advantage of this approach is that it also works in cases, where we do not a priori know the effective degrees of freedom, or the realisation of a symmetry. The functional RG comes with a toolbox that allows to "ask" the theory, which degrees of freedom are relevant, and what is the status of fundamental symmetries. The first is implemented simply by checking, which degrees of freedom give the dominant contribution to physics at a scale $k$. Phenomena such as, e.g. hadronisation in QCD are accounted for by including effective boson fields through a (scale-dependent) bosonisation \cite{Gies:2001nw,Gies:2002hq}. 
The spontaneous breaking of global symmetries is accessible through the evaluation of the full effective potential, which determines the vacuum expectation value of the field.

\subsection{Fundamental theories from $\beta$ functions and fixed points}\label{funtheories}
Expanding the effective average action in the infinite sum
\begin{equation}
 \Gamma_k = \sum_i g_{i}(k) \mathcal{O}_i
\end{equation}
of operators $\mathcal{O}_i$ multiplied by running couplings, the Wetterich equation can be rewritten as an infinite tower of coupled differential equations.
The scale dependence of the couplings is captured in the $\beta$ functions, which are defined by
\begin{equation}
 \beta_i = \partial_t\, g_i(k).
\end{equation}
$\beta$ functions thus form a vector field in theory space, the "flow", which yields an RG-trajectory upon integration.

Of special interest are fixed points in theory space, where $\beta_i=0\, \, \forall i$, hence the theory is scale-independent. Here we are interested in the $\beta$-functions of the dimensionless couplings $\tilde{g}_i = k^{-n} g_i$, where $n$ is the canonical dimension of the coupling. 
Using dimensionless couplings ensures that we have a true scale-independence of the effective average action at a fixed point. If, e.g. the dimensionful couplings tend to a constant, this implies, that we have kept a scale in the theory. We are interested in discovering truly scale-free theories, thus we should work with dimensionless couplings. 

We are particularly interested in fixed points of essential couplings, as these cannot be set to unity by a redefinition of the fields. Examples for inessential couplings for which no fixed-point condition holds (and the $\beta$ functions of which are algebraic functions of the other couplings only) are usually the wave-function renormalisation factors; in the case of the metric this is actually more subtle, see \cite{Percacci:2004sb}.

Fixed points allow us to take the UV ($ k \rightarrow \infty $) limit in such a way as to avoid divergences in couplings and thus also in measurable quantities. If a $\beta$ function has a (UV-attractive) fixed point, then the couplings approach their fixed-point values when $k \rightarrow \infty$. \\
Therefore UV fixed points are interesting as they allow to define a UV completion of an effective theory. Here we would like to clarify one issue, as the statement that the FRG can be used to search for UV completions may be confusing at first sight. This is, since the "natural" direction of the flow is from the UV to the IR, where high-momentum degrees of freedom are integrated out. However FRG equations such as the Wetterich equation can also be used explicitly to discover a possible UV completion of an effective theory. Technically a necessary condition for this to work is the fact that the classical action does not enter the Wetterich equation. Instead of specifying a classical action, we determine a theory space. Then the Wetterich equation determines the $\beta$ functions in this theory space, which may admit fixed points. Such fixed points can then be used to construct a UV completion. The Wetterich equation therefore is a tool that allows to \emph{predict} the classical action, given a field content and symmetries. \newline
In this case one may wonder, how the RG flow can actually be used "backwards", since one actually loses microscopic information when using the RG flow from the UV to the IR (i.e. in the natural direction). Due to universality many different kinds of UV completions can result in the same effective theory in the infrared. 
If the values of all running couplings where known to arbitrary precision at some IR scale, this would determine a unique RG trajectory, the UV limit of which could be investigated. If this trajectory ran into a FP, this would define a possible UV completion, however not necessarily a unique one, since a different microscopic theory might show similar behaviour in the IR. 
In particular in cases where the UV degrees of freedom are actually different from the effective IR degrees of freedom, the theory space built from the IR degrees of freedom is not the correct one to search for a UV completion. From a "bottom-up view" there is no possibility to decide whether the degrees of freedom change at some very high scale. This is precisely due to universality: Totally different UV completions may all have the same effective low-momentum description. Using the FRG to search for UV completions therefore only tests whether there is a consistent \emph{possibility} to find a UV completion for an effective theory in the \emph{same} theory space.\\

Ultimately having established the existence of the fixed point one then uses the flow in the "natural" direction to integrate out quantum fluctuations to get to the IR and investigate whether the low-momentum regime agrees with expectations from effective theories such as General Relativity, or the Standard Model.\newline

We may then distinguish two types of fixed points: The Gau\ss{}ian fixed point (GFP) is defined by $\beta_i =0$ with fixed-point values $g_{i\, \ast} =0 \, \, \forall i$. At a GFP all interactions vanish, and only the kinetic terms of a theory remain. In its vicinity, physical observables can then be calculated by perturbative tools in an expansion in small couplings.\footnote{Note however that some observables may depend on the coupling non-perturbatively even for small coupling, i.e. a resummation of the perturbative expansion may be necessary to recover the correct behaviour. That is to say, the small-coupling expansion and a perturbative expansion are not necessarily the same thing.
As an example, consider the small-coupling expansion of the free energy of the quark-gluon plasma, which contains genuinely non-perturbative coefficients at $\mathcal{O}(g^6)$, see \cite{Arnold:2007pg} and references therein.}

The most prominent example of a Gau\ss{}ian fixed point is given by non-Abelian gauge theories with a limited number of fermions in the fundamental representation. The Gau\ss{}ian fixed point is UV stable, as can be seen from the negative coefficient of the one-loop $\beta$-function of such theories. The theory then exhibits highly non-trivial IR behaviour, due to the running of the relevant coupling.

A less well-studied case is given by a non-Gau\ss{}ian fixed point (NGFP), where $\beta_i =0$ at $g_{i\, \ast}\neq 0$ (for at least one $i$). This defines a theory with residual (and possibly strong) interactions at the fixed point. Perturbative calculations become highly challenging here and typically cannot be implemented straightforwardly. Nevertheless defining the UV completion with a NGFP yields a fundamental theory, as does also the use of a GFP.

The classification of fixed points works by the number of attractive directions and the values of the critical exponents, which are universal (i.e. regularisation-scheme independent) numbers that parametrise the flow in the vicinity of the fixed point. Many universality classes are well-known from thermodynamics, where they describe the dependence of observables on external parameters such as the temperature in the vicinity of a second-order phase transition. Fixed points can actually be linked to second order phase transitions, as there the correlation length diverges which implies that the theory becomes scale-free at the phase transition. In other words, fluctuations on all scales are important for the dynamics of the theory. A scale-free theory in turn is one that lives at a fixed point.

\begin{figure}[!here]
 \begin{minipage}{0.55\linewidth}
  \includegraphics[scale=0.4]{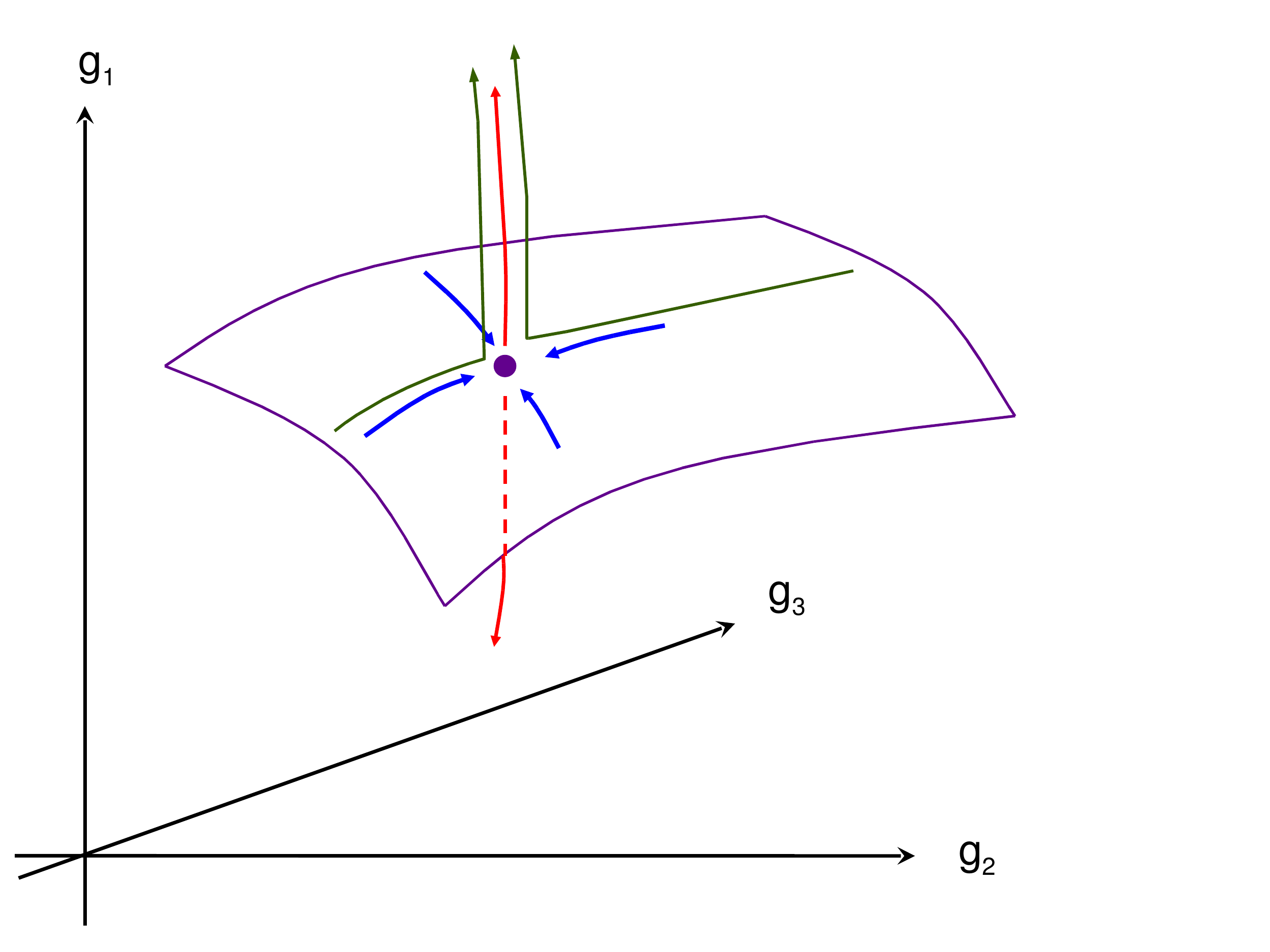}
 \end{minipage}
\begin{minipage}{0.4\linewidth}
\caption{Sketch of the flow towards the ultraviolet in a three-dimensional subspace of theory space: The critical surface and the NGFP are indicated in purple, relevant directions are blue, irrelevant directions red. Trajectories that lie slightly off the critical surface (green) are attracted by the NGFP, but then flow away from it due to the irrelevant couplings.\label{theoryspacesketch}}
\end{minipage}
\end{figure}

Let us introduce the critical exponents by considering the linearised flow around the fixed point:
\begin{eqnarray}
\beta_{g_i}&=&\partial_t g_i=  B_{ij}\left(g_j(k)-g_j^{\ast} \right)+ \mathcal{O}(g_j(k)-g_j^{\ast})^2,  \, \, \mbox{ where }\label{linflow1}\\
B_{ij}&=& \frac{\partial \beta_{g_i}}{\partial g_j}\Big|_{g^{\ast}} \label{generalStabm}
\end{eqnarray}
is the stability matrix.
A solution to \Eqref{linflow1} is given by:
\begin{eqnarray}
g_i (k)= g_i^{\ast}+ \sum_n C_n V_i^n \left( \frac{k}{k_0} \right)^{-\theta_n}.
\end{eqnarray}
Herein $\{\theta\}= -{\mathrm{spect}}(B_{ij})$ are the eigenvalues of the stability matrix (including an additional negative sign) and $V^n$ are the (right) eigenvectors of $B_{ij}$.
The scale $k_0$ is a reference scale and the $C_n$ are constants of integration.
The behaviour of the couplings $g_i(k)$ clearly depends on the eigenvalues $\theta_n$, see fig.~\ref{theoryspacesketch}: In order for $g_i(k)$ to hit the fixed point in the ultraviolet, the constants $C_n$ have to be set to zero for those $n$ where $\theta_n<0$. 
Directions with $\theta_n<0$ are called irrelevant directions. They do not contain any free parameter. In cases where the stability matrix has zero eigenvalues, the behaviour of these marginally (ir) relevant directions is determined by the next order in the linearised flow. If the zero persists to all orders such a direction is truly marginal.
If we have set all $C_n=0$ for $\theta_n<0$, we are on the critical surface. This implies that the $\theta_n>0$ which belong to relevant directions, will ensure that we are attracted into the NGFP towards the ultraviolet. This happens irrespective of the value of the $C_n$ of the relevant directions, which implies that these $C_n$ correspond to free parameters. Note that typically the operators entering the effective action do not simply correspond to (ir)relevant directions at a NGFP; non-trivial superpositions of these typically do.\\

\noindent The issue of predictivity is related to the flow towards the IR:
UV attractive directions are of course IR repulsive, therefore the IR observable value of the coupling is not determined by the fixed point, and has to be fixed by an experiment. This feature leads to the name "relevant" coupling, and it is linked to a free parameter, as the constant of integration remains unfixed.
The IR values of irrelevant couplings are predictable from the values of the relevant couplings. For examples of this in the context of a NGFP see, e.g. \cite{Gies:2009hq, Harst:2011zx}. In order to approach the NGFP in the UV the initial conditions in the IR have to lie exactly on a trajectory that ends up within the critical surface. A slight shift away from the critical surface suffices that, at possibly very large $k$, the flow is driven away from the NGFP along a repulsive direction. Since the couplings thus have to agree with values in the critical surface to arbitrary precision, the requirement to hit the NGFP might, loosely speaking, be understood as a certain type of fine-tuning problem \footnote{Note that here we mean something quite different from the usual fine-tuning problem, which pertains to a hierarchy of scales and implies that values of couplings have to be tuned very precisely at high energies, in order to allow for dimensionful couplings to be of $\mathcal{O}(1)$ at low energies.}.

The search for a UV completion is a very interesting issue in the case of gravity. On the other hand the use of the flow equation is also highly useful in the context of Yang-Mills theories, where due to asymptotic freedom perturbative calculations break down in the infrared. In both cases we are interested in applying the Wetterich equation to a gauge theory, so we have to understand the relation between symmetries and the FRG.

\subsection{Symmetries in the Functional Renormalisation Group}\label{symmetries}
In the case of gauge theories the introduction of a cutoff is a rather subtle issue: A simple momentum cutoff clearly breaks the gauge invariance by cutting off modes that are gauge equivalent to modes that are integrated out. Another way of saying this is that the cutoff corresponds to a mass-like term, which is clearly incompatible with gauge invariance, as we are not in the Higgs-phase of the theory, and the mass is not induced by a non-trivial vacuum expectation value (VEV)\footnote{And of course due to Elitzur's theorem \cite{Elitzur:1975im} a non-zero VEV in a gauge theory is only observable after having fixed a gauge.}. Thus, a cutoff term will appear in the Ward-identities, which we briefly discuss here.

A crucial aspect of gauge theories is the consideration of the remnants of gauge symmetry in a gauge-fixed formulation. Using translation invariance of the path integral in field space, one may easily derive the Ward-identities:
\begin{equation}
\mathcal{G}\Gamma_k =- \mathcal{G} \Delta S_k + \langle \mathcal{G} \left(S_{\rm gf}+ S_{\rm gh}+ \Delta S_k \right)\rangle.\label{Ward}
\end{equation}
Herein $\mathcal{G}$ is the symmetry generator of the symmetry under consideration. 

Accordingly the regulator term simply adds an additional contribution to the Ward identity, which at $k \rightarrow 0$ reduces to the standard Ward identity. From \Eqref{Ward} it is also clear that one may in principle choose to regularise a theory with a symmetry using a regulator which breaks that symmetry. Then the only term in \Eqref{Ward} results from the regulator. Such a construction should however be avoided if possible since a symmetry-breaking regulator implies that the flow will take place in the larger theory space which is subject to the remnant and not the physical symmetry.

It is possible to show that the modified Ward-identity holds under the flow, if it holds at an initial scale, see, e.g. \cite{Gies:2006wv}. For practical purposes however an exact solution of the flow equation is impossible to find, see sec.~\ref{truncation}. Thus the Ward identity will typically be violated.

\subsection{Gauge theories: Background field method}\label{backgroundfieldmethod}

 The background field method allows to construct an
effective average action that is gauge-invariant in the
limit $k \rightarrow 0$ \cite{Abbott:1980hw}. This is
accomplished by gauge-fixing with respect to an auxiliary
background field $\bar{A}_{\mu}$ (or $\bar{g}_{\mu \nu}$ in
the case of gravity). 

To this end the physical field is split into a background field and a fluctuation field
\begin{equation}
A_{\mu}= \bar{A}_{\mu}+ a_{\mu} \phantom{blablabla} g_{\mu \nu}= \bar{g}_{\mu \nu}+ h_{\mu \nu}.
\end{equation}
Note that for the metric this entails that the inverse metric will have an expansion in powers of the fluctuation field $h_{\mu \nu}$ with terms of arbitrary high order, since  $g_{\mu \nu}g^{\nu \kappa}= \delta_{\mu}^{\kappa}$ holds. This property will (in part) be responsible for a larger number of different interaction vertices that can be constructed in gravity from very basic truncations. In particular the ubiquitous metric determinant $\sqrt{g}$ in the volume factor generates couplings to arbitrary powers of the fluctuation metric. Therefore in the case of gravity, every truncation involving at least a cosmological term $\lambda \int d^dx \sqrt{g} $ can be expanded to arbitrary order in fluctuation-$n$-point functions. This is very different from Yang-Mills theory, and one of the reasons why terms of a similar structure (e.g. minimally coupled fermions), may give rise to very different flows in Yang-Mills theory and gravity, see chap.~\ref{ferminAS} for details.

In the case of gravity this split has the additional advantage that one can use the background metric to construct a background-covariant Laplacian with respect to which one can classify fluctuation modes into "high-momentum" and "low-momentum" modes, for details see chap.~\ref{AS}.

One may choose the background field to fulfill, e.g. the quantum equations of motion, but this is not strictly necessary. The background field need not be understood as a physical background, around which small quantum fluctuations are considered. In particular in the case of gravity it is crucial, that the background field method does not imply that $h_{\mu \nu}$ is a small fluctuation around a flat (or possibly cosmological) background. Let us emphasise that the background field method is to be understood primarily as a technical tool, and for its use does neither require the background field to be the true physical expectation value of the gauge field, nor assume that the fluctuations are restricted in amplitude.

We now gauge-fix the fluctuation field with respect to the background field, by generalising covariant gauge conditions without a background (resp. a trivial, i.e. flat one in the case of gravity).
The gauge-fixing condition is, e.g. given by
\begin{equation}
\mathcal{F}=\bar{D}_{\mu}a^{\mu}=0 , \label{YMbackgroundgauge}
\end{equation}
for the case of non-Abelian gauge theories. In the case of
gravity, the background field gauge condition will typically
contain several terms, which is a simple consequence of the
metric being a tensor instead of a vector. Accordingly the
gauge comes with two parameters in gravity, as the different
terms in the gauge condition may have different weights.
Explicitly it is given by
\begin{equation}
\mathcal{F}_{\mu}=\bar{D}_{\mu}h^{\mu}_{\nu}-\frac{1+\rho}{d}\bar{D}_{\mu}h^{\nu}_{\nu},
\end{equation}
in $d$ dimensions.
Hence the gauge-fixing term fixes the fluctuation fields with respect to the background fields. In both cases it contains a transversality condition of the fluctuation field with respect to the background. In the case of gravity the second term is a condition to be fulfilled by the trace of the fluctuation field.

The corresponding Faddeev-Popov operator reads 
\begin{equation}
\mathcal{M}^{ac}= -\bar{D}_{\mu}^{ab}D^{\mu \, bc}
\end{equation}
for non-Abelian gauge theories. In gravity we have
\begin{equation}
\mathcal{M}^{\mu}_{\, \nu}=\bar{D}^{\rho}\bar{g}^{\mu \kappa}g_{\kappa \nu}D_{\rho}+ \bar{D}^{\rho}\bar{g}^{\mu \kappa}g_{\rho \nu}D_{\kappa}
- \frac{1}{2}(1+\rho)\bar{D}^{\mu}\bar{g}^{\rho \sigma}g_{\rho \nu}D_{\sigma}.
\end{equation}
Again the fact that the metric is a tensor entails a more complicated structure and also demands the Grassmannian ghost fields to transform as vectors.
The crucial step in the background field method is the introduction of an auxiliary background gauge transformation (for details see, e.g. \cite{Gies:2006wv, Reuter:1996cp,Reuter:2007rv}). Let us stress that this is purely auxiliary and does not acquire a physical meaning at this stage.
The above gauge-fixing term is then invariant under the sum of the auxiliary and the physical gauge transformation.
We therefore conclude that the effective action in the background field formalism is gauge-invariant if we identify the background field with the full gauge field.

These considerations can be directly transmitted to the effective average action $\Gamma_k[a, \bar{A}]$, for details see, e.g. \cite{Reuter:1993kw,Reuter:1996ub,Pawlowski:2005xe,Gies:2006wv,Reuter:1996cp,Reuter:2007rv}. Note however that setting $A_{\mu}=\bar{A}_{\mu}$ ($g_{\mu \nu}=\bar{g}_{\mu \nu}$) before the evaluation of the full RG flow is incorrect, as the flow in the extended theory space contains operators that vanish in this limit. This does however not imply that they cannot contribute to the flow of operators that will be gauge invariant under the identification of background field and gauge field. Therefore the price to pay for the construction of a gauge invariant effective action in the limit $k \rightarrow 0$ is the dependence of the flow on two gauge fields/ metrics. Note that of course the modified WTI's (derived from acting with the \emph{physical} gauge transformations on the effective average action) still have to be fulfilled and impose non-trivial symmetry constraints.

Within the background field formalism, the inverse propagator is given by $\Gamma_k^{(2,0)}[a,\bar{A}]$, hence we need the second functional derivative with respect to the fluctuation field. This quantity is not accessible from the flow of $\Gamma_k[0, \bar{A}]$. Hence the flow equation for the effective action of the background field (i.e. at zero fluctuation field) is not closed  \cite{Pawlowski:2005xe,Pawlowski:2001df,Pawlowski:RG2002,Litim:2002xm,Litim:2002hj,Litim:2002ce}.
  \begin{equation}
\partial_t\Gamma_k[0,\bar{A}]=\frac{1}{2} {\rm STr} \left(\Gamma_k^{(2,0)}
[0,\bar{A}]+R_k\right)^{-1} \partial_t R_k\,.
\label{eq:backflow0}
\end{equation}
Several routes are now open: One may set $A_{\mu}= \bar{A}_{\mu}$ ($g_{\mu \nu}= \bar{g}_{\mu \nu}$) after deriving the inverse propagator, i.e. setting $\Gamma_k^{(2,0)}[0,\bar{A}]= \Gamma_k^{(2)}[\bar{A}]$. What one typically does here is to neglect all operators in theory space which depend on the background field apart from the gauge-fixing term and the ghost term. One then evaluates the second functional derivative with respect to the gauge field, and then proceeds to identify the gauge field and the background field. 
This neglects all differences between the flow of background
quantities and quantities that are constructed from the
physical gauge field, or from a combination of background
and physical fields. In particular, one neglects the
back-coupling of terms depending on both fields into the
flow, except the gauge-fixing and ghost terms. For examples,
see \cite{Reuter:1997gx, Gies:2002af} and
\cite{Reuter:1996cp} for the case of gravity.

One may also work in a truncation where terms that depend on both the full gauge field and the background field are taken into account. This strategy has recently been applied to the case of quantum gravity \cite{Manrique:2010am,Manrique:2010mq}. Working in this extended theory space naturally implies a larger number of terms that couple into the flow and usually necessitates a higher level of technical sophistication in order to distinguish the flow of different operators.

The last possibility is to use a relation between the background field gauge and covariant gauges like the Landau gauge: As is clear from the gauge condition \Eqref{YMbackgroundgauge}, the former is related to the latter for vanishing background field. This allows to reconstruct the inverse propagator in the presence of a background field from the inverse Landau gauge propagators. The idea has first been applied in \cite{Braun:2007bx} and will be used in this thesis to investigate aspects of confinement in Yang-Mills theories, see chap.~\ref{YM}.

\subsection{The necessity to truncate}\label{truncation}

The Wetterich equation is an exact one-loop equation, however for practical computations it usually yields only approximate results, due to the following reason: As discussed above (see sec.~\ref{base_phys}), the flow equation generates a vector field in theory space. In general this vector field has non-vanishing components in all (infinitely many) directions in theory space. Therefore only a treatment including all these directions leads to exact results. This however is in general impossible, as the Wetterich equation constitutes an infinite tower of coupled differential equations. As an example, consider the flow equation for an $n$-point vertex, which implies taking the $n^{\rm th}$ functional derivative of the right-hand-side of the Wetterich equation with respect to the field: Using that (schematically) 
\begin{equation}
\frac{\delta}{\delta \phi}\left( \Gamma_k^{(2)}\right)^{-1} = \left( \Gamma_k^{(2)}\right)^{-1}\Gamma_k^{(3)}\left( \Gamma_k^{(2)}\right)^{-1}, 
\end{equation}
it is easy to see that $\partial_t \Gamma_k^{(n)}$ will then be related to $\Gamma_k^{(n+1)}$ and $\Gamma_k^{(n+2)}$. To deal with this infinite tower of equations is in general impossible\footnote{Note, e.g. the following important exception: In the deep-IR limit where all momenta go to zero, a scaling ansatz for the vertices in Yang-Mills theory presents a consistent solution to the complete tower of equations \cite{Fischer:2006vf, Fischer:2009tn}. }. Therefore it is necessary to truncate the theory space by simply making an ansatz for an (infinite) subset of couplings (generically one simply sets these to zero). Then one can proceed to derive the flow equation in this truncation, which in general is different from the flow equation projected onto the truncation. 

The strength of the approach lies in the fact that truncations may be chosen following very different guiding principles: A truncation adapted to the perturbative regime neglects operators generated by higher powers of the coupling. In such a truncation it is possible to recover perturbation theory to any order \cite{Papenbrock:1994kf,Litim:2002xm}.

On the other hand, e.g. a derivative expansion sorts operators by the number of derivatives they contain and is therefore intrinsically non-perturbative. It is therefore a well-adapted tool to study theories in the non-perturbative regime. It is also possible to work in a vertex expansion, which is "orthogonal" to the derivative expansion in the sense, that an infinite number of different operators from the derivative expansion can in principle contribute to one operator in the vertex expansion (since, e.g. the three-point function can contain an arbitrary function of the momenta). Within gauge theories a derivative expansion allows for a more straightforward incorporation of the gauge symmetry, as gauge symmetric operators can, in Yang-Mills theories as well as gravity, easily be sorted by the number of derivatives they contain. On the other hand an $n$-point function $\Gamma_k^{(n)}$ is a gauge-covariant object in gauge theories, as is clearly visible, e.g. in the case of the inverse propagator.\\
One should however note that the term "expansion" used here simply refers to an organisation scheme for the infinitely many operators, and is not to be confused with an expansion, where higher-order terms are in some sense smaller than the ones from lower orders\footnote{An exception to this is given, e.g. by the search for relevant couplings at a non-Gau\ss{}ian fixed point, where, as explained in sec.~\ref{AS_intro}, only the first few canonically irrelevant couplings in a derivative expansion may be expected to turn into relevant ones.}.

In the non-perturbative regime a control over the error of the calculation is highly desirable. Here we have to state that in the FRG framework, as also in most other non-perturbative methods, this is in general highly challenging, and not always possible. As it is very intricate to devise a small parameter to control the expansion, the size of neglected terms is not easily accessible from a truncation\footnote{As an example, consider a scalar theory in a derivative expansion, where the size of the anomalous dimension $\eta$ can be used to estimate if a derivative expansion works well in this case.}.\\
One may however use two different methods to judge the quality of a truncation, apart from the obvious possibility to compare results obtained with the flow equation with results from other methods (or ideally experimental results). The regulator dependence vanishes in the limit $k \rightarrow 0$ in the untruncated theory space. As soon as one works within a truncation, a residual regulator dependence remains. Studying the extent of the regulator dependence allows to give an estimate for the truncation error; however this is a rather crude estimate and one cannot prove that the distance of the result to the true physical result may not be larger than this estimate.
Here, optimisation techniques have been devised, that allow to construct a regulator that minimises the truncation error \cite{Litim:2001up}. 

At a NGFP the quantities that are universal (i.e. mainly critical exponents) acquire a regulator dependence within truncations. Again variations of the regulator function allow to estimate the quality of the truncation.

To check the importance of terms outside a truncation is obviously possible by studying the stability of the result under an enlargement of the truncation. Of course this only allows to clearly determine the effect of previously neglected operators, but will not yield a reliable estimate of all terms outside the truncation. In the worst scenario a result may be highly stable under several successive enlargements of the truncation, from which one may be tempted to conclude that the result is already very close to the exact one. However an operator outside the largest truncation studied can still spoil this picture completely. \\
Therefore such studies of the stability of results should always be taken with a grain of salt. The reliability of a truncation can only be determined rigorously by a comparison with the exact result, which may not be known in many interesting cases. 

Here it is very useful if complementary methods exist, as is
the case in many field theories that can be simulated
applying lattice discretisation and numerical Monte Carlo
studies of the path integral. Assuming, that these methods
have a good control over their respective error sources
(e.g. finite-size effects and discretisation artifacts),
they can be used to check results obtained with the FRG.

In some settings the necessity to truncate may also be turned into an advantage, as it may allow for a clear investigation of physical effects: As the effect of removing a certain operator from a truncation can be studied, the origin of a physical effect can be identified and understood within the flow equation. Here, truncations can be used to study the mechanisms behind physical phenomena, and understand, e.g. which operators are responsible for phenomena such as spontaneous symmetry breaking (for an example see chap.~\ref{ferminAS}). In particular in the strongly-interacting regime, where potentially many couplings are non-zero, it is important to understand the mechanisms behind physical properties of the theory. Changing the truncation used to study a physical question helps to appreciate which operators are important in a specific question.

\chapter{Aspects of confinement from the Functional Renormalisation Group}\label{YM}

\section{Motivation: Strongly interacting physics in the low-energy limit}
QCD is a paradigmatic example for the application of the ideas of the Renormalisation Group. Its UV behaviour is governed by a Gau\ss{}ian fixed point with one marginally relevant direction, i.e. the theory is asymptotically free and shows a non-trivial RG running towards the IR. In the low-energy regime, it becomes strongly-interacting. The microscopic degrees of freedom, quarks and gluons, are then unobservable, instead effective degrees of freedom emerge in the form of colourless massive bound states, the hadrons, which constitute the observable degrees of freedom at low energies. This regime is then characterised by two central properties, namely confinement and spontaneous chiral symmetry breaking. As a result the low-energy bound states, the hadrons,
become massive. It is important to emphasise that this mass
is not related to the Higgs
mechanism, which only provides the small current quark masses. The hadron mass is mainly a consequence of the dynamics of QCD.

This plethora of interesting physics results from the fact
that the theory does not sit exactly \emph{on} the GFP, but
arbitrarily close to it in the far UV. Since it has one
relevant direction, a non-trivial RG-flow occurs, which
leaves the scale-free regime in the vicinity of the fixed
point. As the coupling is dimensionless in four dimensions
and explicit gauge boson mass terms violate gauge invariance, the theory
does not possess any inherent classical scale. Quantum fluctuations
thus generate a non-trivial scale, $\Lambda_{\rm QCD}$. The
theory becomes strongly interacting at this scale, and
exhibits confinement as well as chiral symmetry breaking. 

At finite temperature, the QCD phase diagram then shows at
least two phases, the confined phase with broken chiral
symmetry and the quark-gluon plasma phase with restored
chiral symmetry. There has to be a crossover or a phase
transition, the deconfinement phase transition, in between.
Also the chiral transition occurs at a similar temperature
(at least for vanishing chemical potential), but its connection
to the deconfinement transition is not fully clear. Further
phases, such as a quarkyonic phase with confinement but
restored chiral symmetry might occur in the QCD phase
diagram at high chemical potential. At very high chemical
potential, even colour superconductivity might be found. This region of the phase diagram remains to be explored from first principles, and the properties of phases as well as the phase boundaries remain to be understood from the microscopic theory.

As phase transitions and crossovers are typically
 linked to a change in the degrees of freedom of the system and the formation of bound states,
they cannot be described by perturbation theory to any
finite order. Even at zero temperature, the properties of the macroscopic theory, such as its ground state, are determined by non-perturbative physics. Hence non-perturbative tools that allow to investigate a
transition between different phases of the theory, and do not break down in the non-perturbative regime are
required.
The FRG can account for a change in the effective degrees of freedom, and can describe the transition between a strongly and a weakly interacting regime. In particular, in the case of QCD, functional methods can incorporate quarks at a finite chemical potential and realistically
small masses, and therefore have access to the
complete phase diagram of QCD \cite{Braun:2009gm,Haas:2010bw,Fischer:2011mz,Alkofer:2011zp}.

Let us stress that functional methods rely on a gauge-fixed
 formulation, which may yield technical
simplifications, as one may adapt the choice of gauge to the problem under investigation\footnote{As an example, we will see that in covariant gauges, such as Landau gauge, both in Yang-Mills theory and gravity certain modes of the gauge-field or metric, resp. drop out of all diagrams with external vertices.}, but is of course more challenging from a
conceptual point of view: As all physical quantities, as
well as dynamical processes are by definition
gauge-invariant, a gauge-fixed calculation may "blur" the
physical picture. For example, a physical mechanism such as
confinement might manifest itself
very differently in different gauges, and a straightforward
relation of results from different gauges might not be
possible. 
Hence one goal of functional methods should be to connect results in different gauges and finally account for the \emph{physical}, i.e. gauge-independent mechanisms of confinement and chiral symmetry breaking, as well as the dynamics of the quark-gluon plasma and of hadronisation processes and so on.

A challenging point of current functional methods
 is their dependence on truncations. It is therefore non-trivial to gain qualitative as well as quantitative control over the non-perturbative regime of a theory. Here
a method that works without truncations, such as lattice gauge
theory, is a very useful counterpart of a functional RG
(or DSE) calculations, as it allows to check the quantitative precision of continuum functional calculations, assuming that the systematic as well as numerical and statistical errors of the lattice calculation are under control. Having found agreement in a certain area of parameter space, one can use functional methods to explore regions which are unaccessible to lattice calculations, for either conceptual reasons, as, e.g. calculations at finite baryon density, or computational power, as, e.g. the investigation of large gauge groups, or chiral fermions. Besides, functional methods are a very useful tool to gain a deeper understanding of the properties of the theory. Here a seeming disadvantage, the necessity to truncate, can be turned into an advantage, as it is possible to cleanly distinguish which operators are necessary to induce a certain physical phenomenon. Clearly lattice simulations and functional methods are
 complementary and could be used concertedly to obtain a full understanding of QCD in the non-perturbative
regime\footnote{Both lattice gauge theory as well as
functional methods work in the Euclidean path integral
formalism, which is appropriate for systems in
thermodynamic equilibrium. To realistically describe
processes in real time, further steps are necessary, as
emphasised in \cite{Alkofer:2010ue}. As the vacuum state and
thermodynamic equilibrium present only very special cases of
the more general formalism admitting non-equilibrium and
real-time dynamics, many interesting physics questions
relating to the equilibration process in the
quark-gluon-plasma, or processes in the early universe are
not easily or not at all accessible from the Euclidean
formulation.}.

A crucial first step in the attempt to fully understand
 QCD is the understanding of the pure gauge sector of the
theory, which one obtains from QCD by taking the limit of
infinite quark masses and thus suppressing quark
fluctuations in the path integral. Then static quarks can still
be employed as fundamental colour sources. As confinement is a property induced by the
gauge sector of the theory, it may be investigated within
this reduced setting. 

In this chapter we will first focus on
 the zero-temperature limit of the theory. Here the ground
state of the theory is an interesting question, which may be linked to the confining properties of the theory.
We then go one step further and study the deconfinement phase transition at finite temperature. 
Here, we will also shed light on the question what determines the order of the phase transition.

\section{Gluon condensate in Yang-Mills theory}
\subsection{The ground state of Yang-Mills theory and confinement}\label{groundstate}
The ground
state of Yang-Mills theory cannot be inferred from
perturbative considerations \cite{Pagels:1978dd}, i.e. the perturbative vacuum is presumably unstable in Yang-Mills theory. This is suggested by the
fact that due to asymptotic freedom large values of the expectation value of the
field strength $\langle F^2 \rangle$ correspond to the perturbative domain,
whereas the domain of low field strengths is controlled by
the strongly-interacting regime\footnote{Here we should clarify the following subtlety: Of course a \emph{small} value of the fluctuation field $a_{\mu}$ corresponds to the regime where perturbation theory is applicable. However a \emph{large} classical background field strength corresponds to a high mass scale and therefore to the perturbative regime.}. Hence, as emphasised in
\cite{Pagels:1978dd}, the vacuum
configuration of Yang-Mills theory will not be deducible from
perturbative calculations.
Notably, in a strongly-interacting regime there is no
reason to suspect that the solution to the classical
equations of motion will also solve the quantum
equations of motion. In more physical terms, quantum fluctuations break the classical scale invariance and can lead to non-vanishing, dimensionful expectation values, whereas these vanish perturbatively.
Hence the Yang-Mills vacuum state is characterised by the values of gauge-invariant
quantities such as $\langle F^2 \rangle$, $\langle (F
\widetilde{F})^2 \rangle$, etc.\footnote{For an overview of our
notation, see app.~\ref{notation_YM}.}.
This suggests a non-trivial vacuum structure
of Yang-Mills theories.
In particular the well-known trace anomaly, i.e. the impossibility 
to quantise the theory in a way that respects
scale invariance, implies $\langle F^2 \rangle \neq 0$. This is also known as dimensional
transmutation, i.e. the generation of a mass-scale by
the quantisation of a classically scale
invariant theory.

An explicit calculation of the one-loop effective 
action by Savvidy \cite{Savvidy:1977as, Matinyan:1976mp}
provides evidence for the instability of the perturbative vacuum to the
formation of a non-vanishing constant colourmagnetic field configuration.
The one-loop potential for such a field strength
shows a non-trivial minimum, which implies $\langle
F^2 \rangle \neq 0$. This may be interpreted as the
condensation of gluons in the vacuum. However this calculation
suffers from two problems: The
chosen background field configuration is
unstable, as indicated by the existence of a tachyonic mode in the
spectrum of the two-point function parametrising small
fluctuations around this background \cite{Nielsen:1978rm}. As this tachyonic mode is a long-range phenomenon, the introduction of a spatial inhomogeneity and a
configuration of finite-sized "patches" of colourmagnetic
background field, also known as the "Kopenhagen vacuum"
\cite{Nielsen:1978rm,Nielsen:1978tr,Ambjorn:1978ff,Ambjorn:1980ms}, renders the configuration stable. A
second problem which calls for a calculation of the effective
action beyond one loop is the fact that the one-loop running
coupling diverges when the effective
potential traverses zero, coming from large field strengths. Accordingly the interesting region
of field strengths, where Savvidy's calculation implies the
formation of a gluon condensate, lies beyond its domain of
validity. 
 
Note that the existence of a gluon condensate $\langle F^2 \rangle \neq 0$ is physically highly attractive, since
it might be linked to confinement. This
connection exists within a model for the effective
potential, which retains the classical form of the effective
potential, but evaluates the one-loop logarithmically running coupling at a
scale set by the field strength:
\begin{equation}
W_{eff}(F^2)= \frac{1}{4 g^2(F^2)}F^2 \sim F^2 \ln F^2.
\end{equation}
(Note that we redefined the gauge field here to scale out the coupling.)
Within this model the vacuum acts as a non-linear dielectric medium, where the dielectric constant exhibits a field-dependence, and develops a zero for small field strengths.
This leading-log model \cite{Pagels:1978dd,Adler:1981as,Adler:1982pj,Adler:1982rk,Lehmann:1983bq,Dittrich:1996cd,Gies:1996ki}, as well
as more sophisticated dielectric confinement models \cite{Fishbane:1986rg,Chanfray:1990fi,Hosek:1989up} incorporate the non-trivial minimum indicated by Savvidy's
calculation, and furthermore have interesting physical
consequences: Considering a static quark-antiquark pair on
this background leads to a linearly rising potential between
the two sources \cite{Pagels:1978dd,Adler:1982rk}. In particular,
the string tension parametrising the potential $V =
\sqrt{\sigma} r$, can be related to the non-trivial minimum for the gauge group SU(3):
\begin{equation}
\sqrt{\sigma} = \left(\frac{1}{3}F^2\Bigl|_{\rm min} \right)^{\frac{1}{4}}.
\end{equation}
Since, in the case of non-dynamical quarks a linearly rising potential can
be used as a criterion for confinement,
the non-trivial minimum results in confinement of static
fundamental colour sources. Consequently we apply a criterion for confinement that
cannot be maintained in full QCD with dynamical quarks. There the potential shows a linearly rising part and then flattens due to string-breaking, where the energy stored in the string suffices for the creation of a quark-antiquark pair and a subsequent formation of two mesonic bound states. (For a further discussion of how criteria for confinement can be devised in full QCD, see, e.g. \cite{Greensite:2009zz}.)

We will evaluate the effective potential from the functional RG here. Accordingly we have to integrate the Wetterich equation for a non-vanishing background field (cf. \Eqref{eq:backflow0}), which requires information on the fluctuation field propagators. Previous work \cite{Reuter:1994zn,Reuter:1997gx} relied on the approximation of setting $A_{\mu}= \bar{A}_{\mu}$ after evaluating the inverse propagator, see also sec.~\ref{backgroundfieldmethod}.

\subsection{Propagators in the background field gauge}\label{Landaugaugeprops}
\subsubsection{The relation between the background field gauge and the Landau gauge}
Here we follow a route to construct the required fluctuation field propagators, that has been put forward in \cite{Braun:2007bx}, where it was successfully applied to connect the properties of the Landau gauge propagators to quark confinement. As we will be interested in projecting the flow equation onto a pure background-field effective potential with vanishing fluctuation field, we only need the fluctuation field propagators evaluated at vanishing fluctuation field.
The spectrum of fluctuations on a non-trivial background can then be reconstructed from the Landau gauge propagators. Here we use that the background field gauge condition reduces to
 the Landau gauge condition in the limit of vanishing
background field:
\begin{equation}
\bar{D}_{\mu}(A_{\mu}- \bar{A}_{\mu})=\bar{D}_{\mu}a_{\mu}=0 \overset{\bar{A}_{\mu} \rightarrow 0}{\longrightarrow} \partial_{\mu}A_{\mu} =0.
\end{equation}
Thus correlation functions in the
 background field gauge must reduce to Landau gauge
correlation functions in the same limit. The background
field formalism requires the effective action to be invariant under simultaneous background and gauge
transformations. Thus the $n$-point correlation functions have to transform as tensors under background field transformations.
 This
allows to - at least partly -  reconstruct background field correlation functions
from Landau gauge correlation functions. In the case of the
two-point function, this correspondence reads explicitly:
\begin{equation}
\Gamma_k^{(2)}[a=0, A] =\Gamma_{k\, \rm Landau}^{(2)}[0,A](-\mathcal{D}^2) + f_{\mu \nu}(D) F^{\mu \nu}.\label{reconstructedinvprop}
\end{equation}
Herein $\mathcal{D}^2$ is a Laplace-type operator that
 reduces to the Laplacian in the limit of vanishing background field. For notational simplicity we drop the bar on the background quantities, as we only work with the background field and the fluctuation field from now on. The function $f_{\mu
\nu}$ that multiplies the background field strength has to
be non-singular in the limit $A_\mu \rightarrow
0$, such that $\Gamma_k^{(2)}[0,0]= \Gamma_{k\, {\rm Landau}}^{(2)}$, but is unrestricted otherwise. Unless the background field configuration has vanishing field
strength, higher-order correlation
functions in the Landau gauge are necessary for the reconstruction of the additional terms.

\subsubsection{Landau gauge propagators}
The inverse ghost and gluon propagators can be parametrised
as
\begin{eqnarray}\label{gluon} 
(\Gamma_{k,A}^{(2,0)})_{\mu \nu}^{ab}[0,0](p^2)&=&p^2 Z_A(p^2) \mathbf{P}_{T\, \mu \nu} (p) \delta^{ab} + p^2 \frac{Z_{\text{L}}(p^2)}{\alpha} \mathbf{P}_{L\, \mu \nu}(p) \delta^{ab} \,,
\end{eqnarray} 
for the gluon\footnote{Here we have used the
projection operators onto transversal and longitudinal
directions: $\mathbf{P}_{T\, \mu \nu}(p)=\delta_{\mu \nu}-
\frac{p_{\mu}p_{\nu}}{p^2} $ and $\mathbf{P}_{L\, \mu \nu}(p)=
\delta_{\mu \nu}- \mathcal{P}_{T\, \mu \nu}(p)$.} with the gauge parameter $\alpha$ and 
\begin{eqnarray}\label{ghost} 
(\Gamma_{k,C}^{(2,0)})^{ab}[0,0](p^2)=p^2 Z_C(p^2)\delta^{ab}\,
\end{eqnarray} 
for the ghost. The complete effect of the quantum 
fluctuations, in particular the non-perturbative physics, is
encoded in these wave-function renormalisation factors\footnote{Note that the dressing functions $G$ and $Z$, conventionally used in DSE approaches, are defined to be the inverse of the wave-function renormalisations $Z_{C, A}$, respectively.}.
 
For the longitudinal wave-function renormalisation, we have
$Z_{\text{L}}=1+\mathcal{O}(\alpha)$.  Hence it drops 
out of all diagrams beyond one
loop in Landau gauge $\alpha \rightarrow 0$.

\begin{figure}[!here]
\begin{minipage}{0.45 \linewidth}
\includegraphics[scale=0.22]{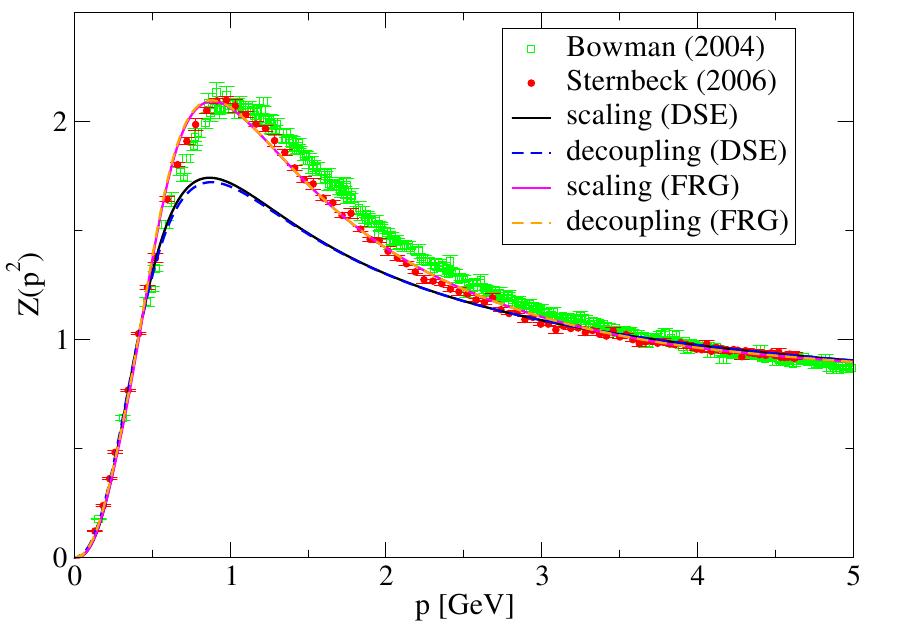}
\end{minipage}
\begin{minipage}{0.45 \linewidth}
\includegraphics[scale=0.22]{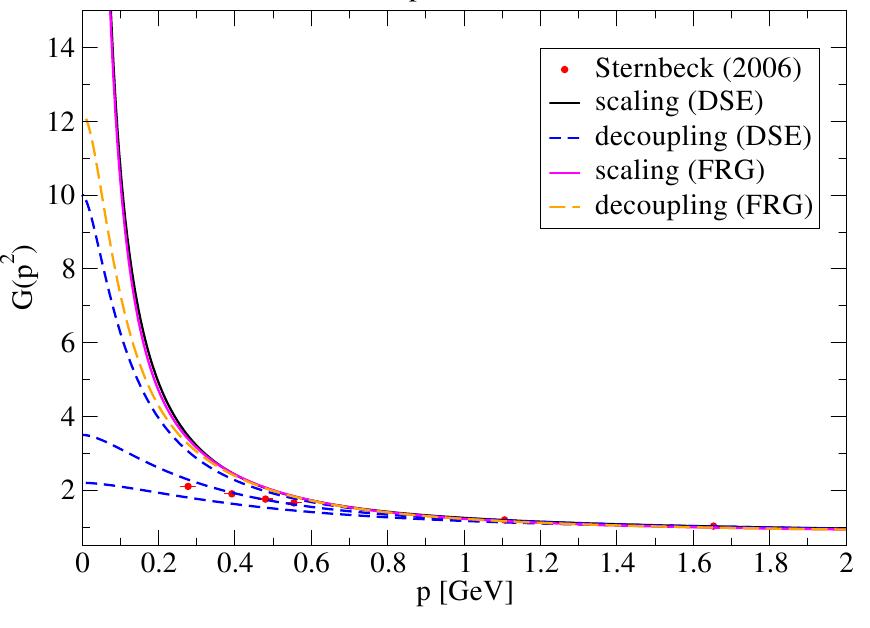}
\end{minipage}
\caption{The inverse wave-function renormalisation 
factors (i.e. dressing functions) for the gluon (left panel) and the ghost (right panel) propagator \cite{Fischer:2008uz} carry the non-trivial
momentum dependence of the full propagator. Results from lattice calculations and the FRG only deviate in the deep
IR regime. The deviation of the DSE calculations to the FRG and lattice results in the mid-momentum regime arises from a truncation of the full DSEs.\label{ghost_gluon_prop}}
\end{figure}

Following extensive work during the past decade, FRG calculations \cite{Ellwanger:1995qf,Ellwanger:1996wy,Bergerhoff:1997cv,Kato:2004ry,Pawlowski:2003hq,Pawlowski:2004ip, Fischer:2004uk,Fischer:2008uz,JMP_to_publish}, DSE calculations
\cite{vonSmekal:1997is,vonSmekal:1997vx,Lerche:2002ep,
Fischer:2002hn,Fischer:2006vf,
Fischer:2008uz,Fischer:2009tn}, and lattice calculations
\cite{Bonnet:2000kw,Gattnar:2004bf,CucchieriMendes,
SternbeckSmekal,Bogolubsky} are in agreement in the
dynamically important
momentum regime, and only deviate in the far infrared (see
fig.~\ref{ghost_gluon_prop}).

\noindent In the deep infrared, the wave-function
renormalisations
$Z_{A,C}$ exhibit a leading
momentum behaviour
\begin{eqnarray}\label{ir}
Z_A(p^2\to 0)\simeq (p^2)^{\kappa_A}\,,\quad
Z_C(p^2\to 0)\simeq (p^2)^{\kappa_c}.
\end{eqnarray} 
Here, functional methods and lattice results differ: In accordance with the fact that one expects a NGFP for the Landau gauge coupling in the deep infrared
\cite{vonSmekal:1997is,Lerche:2002ep,Fischer:2006vf},
functional methods observe a scaling solution
\cite{Fischer:2002hna,Lerche:2002ep}, first found in
\cite{vonSmekal:1997is,vonSmekal:1997vx}
  with
\begin{eqnarray}
\kappa_A =-2 \kappa_c \label{sumrule}\\
\kappa_c \approx 0.595.\label{kappaval}
\end{eqnarray}
The sum rule \Eqref{sumrule} follows from
 the existence of a scaling solution: As observed in
\cite{Fischer:2006vf,Fischer:2009tn}, a scaling ansatz that
simultaneously solves the tower of functional RG as well as
Dyson-Schwinger equations implies that the ghost-gluon
vertex remains bare. This property, also known as a
non-renormalisation theorem, was deduced in \cite{Taylor:1971ff} from the transversality of the gluon propagator in Landau gauge. The
value for $\kappa_c$ is regulator-dependent, with
\Eqref{kappaval} being the value for the optimised regulator
\cite{Pawlowski:2003hq}, which agrees with the originally found value \cite{Lerche:2002ep}. Indeed the critical exponent $\kappa_c$ parametrises the scaling behaviour of all $n$-point ghost and gluon correlation functions for the scaling solution \cite{Alkofer:2004it,Huber:2007kc}.
For the scaling solution, the Kugo-Ojima criterion $\kappa_c > 0$ \cite{Kugo:1979gm, Kugo:1995km} and the Gribov-Zwanziger condition $\kappa_c>\frac{1}{2}$ \cite{Gribov:1977wm,Zwanziger:1993dh} are satisfied. The first follows from the definition of the physical Hilbert space with the help of a well-defined global BRST charge. Hence colour confinement is realised for the scaling solution. The second is related to a solution of the Gribov problem by restricting the domain of integration in the path integral to the first Gribov region, or the fundamental modular region, where each gauge orbit has a unique representative and the Faddeev-Popov operator is positive semi-definite. Here one may wonder how boundary 
conditions on the path integral, such as the restriction to the first Gribov region, are implemented correctly within the FRG, since the form of the differential equations for the $n$-point correlation functions is not altered
by restricting the path integral to the first Gribov (or
even the fundamental modular) region. However solving the
equations with the non-trivial boundary condition on the
ghost propagator then implements a solution to the Gribov
problem within this setting. The Gribov-Zwanziger scenario
is also related to confinement, since it states that
configurations close to the Gribov horizon dominate the IR,
and thus are responsible for confinement\footnote{A direct
relation to an almost linearly rising quark potential can be
established in Coulomb gauge, see, e.g. \cite{Fischer:2006ub}
and references therein.}.

The scaling solution \Eqref{kappaval} and \Eqref{sumrule} implies that the gluon propagator vanishes in the deep infrared, which means that a simple picture of confinement due to a diverging gluon propagator cannot be sustained\footnote{In fact the quark-antiquark-gluon vertex diverges strongly enough to ensure confinement \cite{Alkofer:2007zb}.}. Instead the ghosts become dynamically enhanced, as their propagator is even more divergent than a perturbative propagator. This dominance of an unphysical sector of the theory is naively rather unexpected and implies that ghosts (at least in Landau gauge) can be crucial to determine the physical properties of the theory.

On the other hand, lattice calculations, as well as some
 DSE studies \cite{Aguilar:2008xm,Boucaud:2008ky,Dudal:2008sp} indicate
\begin{eqnarray}
\kappa_{A\, \rm lat}& =& -1 \phantom{xxxx}\kappa_{c\, \rm lat}= 0, \label{decoupling}
\end{eqnarray} 
which has become known under the name decoupling solution,
 since \Eqref{decoupling} implies an infrared finite,
and therefore massive gluon propagator. This entails the decoupling of gluons in the deep infrared. As the gluon propagator is positivity-violating
for the decoupling solution  \cite{Fischer:2008uz,Cucchieri:2004mf}, gluons are not
part of the asymptotic state space, although the Kugo-Ojima criterion is not satisfied.

Finite-size effects \cite{Fischer:2007mc} cannot be made 
responsible for this difference \cite{Cucchieri:2007rg}. As argued in
\cite{Maas:2009se}, the non-perturbative incompleteness of Landau gauge
allows for a further gauge fixing condition in the non-perturbative regime, which might resolve the Gribov ambiguity and is related
to the infrared value of the ghost propagator.
Hence the deviating results on the deep-IR behaviour from
functional methods and lattice might be understood as arising
from the implementation of two different gauge conditions. 

For a part of the investigations in this and the following sec.~\ref{deconfinement}, these differences will play no role, as we observe a separation of scales between the deep IR and the scale of the deconfinement phase transition as well as the gluon condensate. As an important conclusion our results are independent of the IR asymptotics.

Let us emphasise that the definition of a global BRST charge is necessary for the construction of a physical Hilbert space within the framework of Kugo and Ojima.
This singles out the scaling solution in contradistinction
to the decoupling solution \cite{Fischer:2008yv}, as only
the former satisfies the criterion that well-definiteness of
the global BRST charge imposes on the ghost propagator.

As the propagators that we will employ encode colour confinement due to the Kugo-Ojima/
Gribov-Zwanziger scenario, we will establish a
connection between this scenario and a model showing confinement of static quarks, the
leading-log model.\\
Finding such connections between seemingly unrelated confinement scenarios will hopefully ultimately allow to arrive at a fully consistent and gauge-independent understanding of the confinement mechanism. Here we only perform a first step in such a direction, as we link correlation functions respecting a confinement criterion in one particular gauge to a classical model for the Yang-Mills ground state that shows confinement of static quarks.

\subsubsection{Propagators on a self-dual background}
As we are interested in the
effective potential $W_k(F^2)=
\frac{\Gamma_k(F^2)}{\Omega}$, where $\Omega$ denotes the
space-time volume, the reconstruction of the
background gauge propagators requires knowledge on the function $f_{\mu \nu}(D)$ in \Eqref{reconstructedinvprop}.
This function is related to higher order correlation functions in Landau gauge. Here, we perform a "minimal" reconstruction, by generalising $p^2\to \mathcal{D}_{\text
  T}, \mathcal{D}_{\text L}, \mathcal{D}_{\text{gh}}$.
\begin{eqnarray}
\mbox{for the transversal gluon:}&{}& \mathcal{D}_{\,T\, \mu \nu}=\left(D^2\delta_{\mu \nu} + 2 ig F_{\mu \nu} \right) \\
\mbox{for the longitudinal gluon:}&{}& \mathcal{D}_{L\, \mu \nu}= D_{\mu}D_{\nu}\\
\mbox{for the ghost:}&{}& \mathcal{D}_{\rm gh}= D^2\label{Laplops}.
\end{eqnarray}
We include the spin-one coupling of the transversal gluon fluctuation field to the background and use the  covariant derivative with respect to the background field, which is given by
  $D_{\mu}^{ab}= \partial_{\mu}\delta^{ab}+ g f^{abc}A_{\mu}^c$. This choice is motivated by a correct perturbative limit.

\subsection{Self-dual field configuration}
We now choose a background field configuration that 
allows to project onto the effective potential $W(F^2)$.
Hence a covariantly constant field strength with
$D_{\mu} F^{\kappa \lambda}=0$ suffices. As the spectrum of
the above Laplace-type operators \Eqref{Laplops}, or at least the
heat-kernel trace for these operators has to be known, we
have a limited choice in the possible background field
configurations. To avoid problems with tachyonic modes that
indicate the instability of a background \cite{Nielsen:1978rm,Gies:1999jt} we project onto the only known stable
covariantly constant background, which is selfdual, hence
\begin{equation}
 \widetilde{F}_{\mu \nu}= \frac{1}{2}\epsilon_{\mu \nu}^{\, \, \, \, \kappa \lambda}F_{\kappa \lambda}= F_{\mu \nu}.
\end{equation}
This implies that we have to give up on the uniqueness of the projection, as the desired potential $W(F^2)$ is indistinguishable from functions depending also on the dual background field strength. In the possible case that terms such as $(F \widetilde{F})^2$ etc. are negligible in comparison to $(F^2)^2$-terms\footnote{Note that due to parity conservation only even powers of $(F \widetilde{F})$ can be non-zero.}, the non-uniqueness will not affect our result much.

Note that, as we are interested in a non-trivial 
vacuum expectation value, the background field has a direct
physical meaning here, as we will choose a stable field
configuration that gives $\langle F^2 \rangle \neq 0$. In this case, one may think of the background
field as the physical solution to the quantum equations of
motion, around which the vacuum expectation value of quantum
fluctuations is zero. This should enhance the stability of the flow, since we expand around a physical point \cite{Litim:2002hj}. 
We have to remark however, that our background is only locally a candidate for the true ground state. For a more realistic configuration, see, e.g. \cite{Galilo:2010fn} and references therein.

The gauge configuration yielding such a field strength can then be chosen to be
\begin{equation}
A_\mu^a = - \frac{1}{2} F_{\mu\nu} x_\nu n^a, \mbox{ with } n^a=\text{const.}, \phantom{xx}
n^2=1. \label{selfdualgaugefieldconf}
\end{equation}
We then set $ F_{\mu \nu}=0$ apart from $F_{01}=F_{23}\equiv f=\text{const.}$; all other non-zero components follow from the antisymmetry of the field strength tensor.
This field configuration, fluctuations around it and its
stability properties have first been analysed in
\cite{Minkowski:1981ma,Leutwyler:1980ma,
Leutwyler:1980ev}. Due to the enhanced symmetry properties connected to the self-duality, zero-modes, so-called chromons, exist. These carry important information (e.g. a main contribution to the one-loop Yang-Mills $\beta$ function), and have to be regularised with care, since the standard choice $R_k \sim \Gamma_k^{(2)}$ is zero on the zero-mode subspace.

Our truncation reads:
\begin{eqnarray}
\Gamma_k[a,A]&=& \int d^4x \Bigl[ \frac{1}{2} a_{\text T \mu}^{a}
  \big(\Gamma_{k,\text T}^{(2,0)}(\mathcal{D}_{\,\text{T}})\big)_{\mu\nu}^{ab}\,
  a_{\text T \nu}^{b}+ \frac{1}{2}a_{\text L \mu}^a{}
  \big(\Gamma_{k,\text L}^{(2,0)}(-D^2)\big)_{\mu\nu}^{ab}\,
  a_{\text L \nu}^b  \nonumber\\
&{}& + \bar c^a
  \big(\Gamma^{(2,0)}_{k,\text{gh}}(-D^2) \big)^{ab} c^b \Bigr],
\end{eqnarray}
where $a_{ \text T \mu}^a$, $a_{ \text L \mu}^a$ and $\bar{c}^a$ and $c^a$ denote the transversal and longitudinal gluonic and ghost and antighost fluctuations, respectively. The inverse propagators depend on the Laplace-type operators introduced above, see \Eqref{Laplops}, which satisfy \cite{Flory:1983dx}
 \begin{eqnarray}
{\rm spec} \left({\mathcal{D}_{\,T}}\right)&=& 2 g f_l (n+m+2) \mbox{ with }n,m \in \mathbb{N} \mbox{ and with multiplicity 2 in 4 dimensions,} \nonumber\\\
&{}& 2g f_l (n+m) \mbox{ with }n,m \in \mathbb{N} \mbox{  with multiplicity 2 in 4 dimensions.}\label{spectransD}\\
{\rm spec}\left({-D^2}\right)&=&2 g f_l( n + m +1) \mbox{ with } n,m \in \mathbb{N}\label{specD}
\end{eqnarray}
with a degeneracy factor $\frac{f^2}{(2 \pi)^2}$. Herein $f_l = \vert
\nu_l \vert f$ and $\nu_l$ is given by
\begin{equation}
\nu_l= \text{spec}\{ (T^a n^a)^{bc} | n^2=1 \}, \label{eq:nul}
\end{equation}
with the generators in the adjoint representation $(T^a)^{bc}$ and therefore depends on the direction of the unit 
vector $n^a$.

As a regulator we choose a cutoff of the following type
\begin{equation}
R_k= \Gamma_k^{(2,0)}(k^2)\, r(y), \quad y= \frac{\mathcal{D}}{k^2},
\end{equation}
where $\mathcal{D}\rightarrow \mathcal{D}_{\text
  T}, \mathcal{D}_{\text L}, \mathcal{D}_{\text{gh}}$.
For $r(y)= e^{-y}$ the regulator meets all standard requirements\footnote{$\underset{\frac{x}{k^2}\rightarrow 0}{lim}R_k(x)= \Gamma_k^{(2)}(k^2)= Z_k\, k^2$. In particular, using the IR and the UV limit of the propagators, we find $\underset{\frac{k^2}{x}\rightarrow 0}{lim}R_k(x)\rightarrow 0$, and $\underset{k^2\rightarrow \Lambda^2 \rightarrow \infty}{lim}R_k(x)\rightarrow \infty$, since $\Gamma_k^{(2)}(k^2)\sim k^2$ for $k^2 \rightarrow \Lambda^2$.}.
This choice will allow us to establish an explicit connection between the deep-IR asymptotics of the Yang-Mills $\beta$ function and the critical exponents $\kappa_{A,c}$ in the following. 

For the regularisation of the zero modes, we cannot set $\mathcal{D}=\mathcal{D}_{\,\text{T}}$ in the argument of the regulator function, since then
$y=\mathcal{D}/k^2 \to 0$ (because $\mathcal{D}_{\,\text{T}}=0$ on the zero-mode
subspace). In particular, this would not properly account for the decoupling
of the zero mode, once it acquires a field-dependent mass, which already happens perturbatively \cite{Leutwyler:1980ma}. Instead, we choose
$\mathcal{D}=-D^2$ as the argument of the regulator on the zero-mode subspace which
makes the regulator satisfy all standard requirements. On the zero-mode
subspace, we have $\mathcal{D}=-D^2\to 2f_l$, cf. \Eqref{specD} for
$n=m=0$.

In a first step, we do not evaluate the full effective potential, but instead project the flow onto the running background field coupling.

\subsection{An upper bound on the critical exponents $\kappa_{A,c}$}

The $\beta$ function of the running background field 
coupling can be extracted from the flow of the effective
potential by projecting onto the first field-dependent term
in the Taylor-expansion of the effective potential:
\begin{equation}
\Gamma_k\Big|_{F^2}= \frac{1}{4\, g^2}\int d^4x \, F_{\mu \nu}F^{\mu \nu}.
\end{equation}
Thus we can evaluate the $\beta$ function of the background running coupling by projection onto the first field-dependent term in a Taylor expansion of the effective potential in powers of the background field strength. For our specific background, this corresponds to a projection onto the term $\sim f^2$:
\begin{equation}
\beta_{g^2} := \partial_t g^2 = -g^4\, \partial_t \frac{1}{g^2} = -\frac{g^4}{\Omega}\, \partial_t
\Gamma_k[f]\Big|_{f^2}.\label{beta}
\end{equation}
Note that unlike the complete effective potential 
this first term is not affected by the ambiguity arising
from the use of a self-dual background field configuration:
Operators containing an uneven power of $\widetilde{F}_{\mu \nu}$ would break parity conservation. Since our regulator also respects this symmetry, the flow does not leave the symmetric theory space, and hence the couplings of the terms
$(F_{\mu \nu}\tilde{F}^{\mu \nu})^{(2n+1)}$ (with
$n = 0,1,...$) are strictly zero. 

We are now interested in the deep-IR asymptotic form of the $\beta$ function, which requires a parameterisation of the regularised asymptotic form of the inverse propagators:
\begin{equation}
(\Gamma_{k,A}^{(2,0)})_{\mu \nu}^{ab}(p^2) =\Gamma^{(2)}_{k,\,A}(p^2)  \mathbf{P}_{T\, \mu \nu} (p) \delta^{ab} , \quad
(\Gamma_{k,c}^{(2,0)})^{ab}(p^2)=\Gamma^{(2)}_{k,\,c}(p^2)\delta^{ab},
\end{equation}
where the scalar functions $\Gamma^{(2)}_{k,\,A/c}(p^2)$ are given by \cite{JMP_to_publish}, 
\begin{eqnarray}
\Gamma_{k,\,A}^{(2)}(p^2) &=& \gamma_{A} \frac{(p^2 + c_A
  k^2)^{1+\kappa_A}}{(\Lambda_{\text {QCD}}^2)^{\kappa_A}} \phantom{xxx}\Gamma_{k,\,c}^{{(2)}}(p^2) = \gamma_{c} \frac{p^2(p^2 + c_c k^2)^{\kappa_c}}{(\Lambda_{\text QCD}^2)^{\kappa_c}}. \label{IRprops}
\end{eqnarray}
The dimensionless quantities $\gamma_{A,c}$ account
for the difference of scales between $\Lambda_{\text{{QCD}}}$ and the onset of the asymptotic regime. The regulator dependence of the 2-point functions manifests itself in the constants $c_{A,c} = \mathcal O(1)$. In the absence of any IR regularisation, i.e. $k\to0$
  or $c_{A,c}\to 0$, \Eqref{IRprops} reduces to the standard form \Eqref{gluon}, \Eqref{ghost} and \Eqref{ir}.

Since the degeneracy factor in the trace already carries a factor of 
$\frac{f^2}{(2 \pi)^2}$, all $f$ dependence outside the operator trace can
already be ignored due to the projection in \Eqref{beta}, since we are only interested in the term $\mathcal{O} (f^2)$. Note that the trace
  over the Laplace-type spectra and the projection onto the Taylor coefficient of the $f^2$ term do not
  commute, so that we get
\begin{eqnarray}
\partial_t \Gamma_k[A]\Big|_{f^2}
&=& \Biggl[\Biggl(\frac{3}{2} \frac{\partial_t \Gamma_{k,
    A}^{(2)}(k^2)}{\Gamma_{k, A}^{(2)}(k^2)} \left(\tfrac{\Gamma_{k,
      A}^{(2)}(0)}{\Gamma_{k,A}^{(2)}(k^2)}+1 \right)^{-1}
- \frac{\partial_t \Gamma_{k,c}^{(2)}(k^2)}{\Gamma_{k, c}^{(2)}(k^2)} 
\left(\tfrac{\Gamma_{k, c}^{(2)}(0)}{\Gamma_{k,c}^{(2)}(k^2)}+1 \right)^{-1}+1 \Biggr) \tr\,e^{-\frac{-D^2}{k^2}}\nonumber\\
&{}& +  \frac{\partial_t \Gamma_{k, A}^{(2)}(k^2)}{\Gamma_{k, A}^{(2)}(k^2)}\left(\tfrac{\Gamma_{k, A}^{(2)}(0)}{\Gamma_{k,A}^{(2)}(k^2)}+1 \right)^{-1}\tr\, e^{-\frac{2f_l}{k^2}}\Biggr] \Bigg|_{f^2}
. \label{eq:betaflow}
\end{eqnarray}
To obtain the $\beta$ function, we extract the coefficient of the
expansion of the heat-kernel trace over coordinate and colour space at
second order in $f$,
\begin{eqnarray}
\Tr_{x\text{c}} e^{- \left(\frac{-D^2}{k^2} \right)} &=&
\frac{\Omega}{(4\pi)^2}\,  \sum_{l=1}^{{\rm{N}}_c^2 -1} \,
\frac{f_l^2}{\sinh^2 \left( \frac{f_l}{k^2}
  \right)}=\frac{\Omega}{(4 \pi)^2}\sum_l\left(1-\frac{f_l^2}{3} +\mathcal{O}(f^4)\right).
\end{eqnarray}
The IR form of the propagators \Eqref{IRprops} then yields the 
$\beta$ function in the asymptotic regime:
\begin{equation}
\beta_{g^2} = - \frac{{\rm N_c}g^4}{(4 \pi)^2}
\Biggl(
-\frac{(1+\kappa_A)}{c_{\kappa}+1}+\frac{2}{3}(1+\kappa_c)-\frac{1}{3}+8
\frac{1+\kappa_A}{c_{\kappa}+1}\Biggr),
\label{eq:betaIR} 
\end{equation}
where we have used that $\sum_l^{{\rm N_c}^2 -1} \nu_l^2
={\rm N_c}$. We introduced the regulator dependent constant
\begin{eqnarray}
c_{\kappa} = \frac{\Gamma_{k\, A}^{(2)}(0)}{\Gamma_{k\, A}^{(2)}(k^2)}=
 \left(\frac{c_A}{1+c_A} \right)^{1+\kappa_A}. 
\end{eqnarray}
In the limit $c_{\kappa} \rightarrow 0$ and $\kappa_{A/c}\rightarrow 0$ we recover the perturbative one-loop form of the $\beta$ function.

\begin{figure}[!here]
\begin{minipage}{0.38\linewidth}
\includegraphics[scale=0.6]{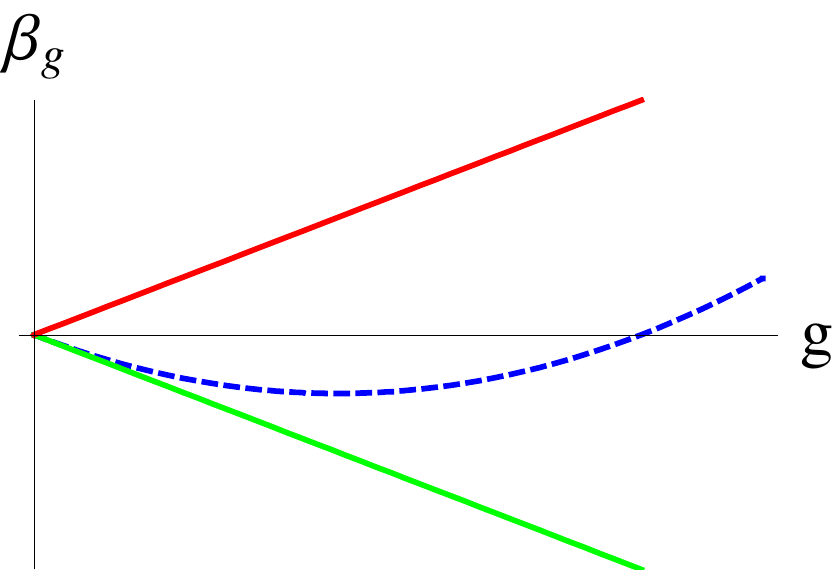}
\end{minipage}
\begin{minipage}{0.57\linewidth}
\caption{Sketch of the background coupling $\beta$ function: The form displaying asymptotic freedom as well as an interacting IR fixed point is plotted in blue (dashed curve). The red line shows a $\beta$ function with an IR-attractive GFP, whereas the green line exemplifies a $\beta$ function where the sign of the first coefficient is compatible with an interacting IR regime.\label{betaplot}}
\end{minipage}
\end{figure}

The seemingly trivial observations that asymptotic freedom requires the Gau\ss{}ian fixed point to be UV-attractive (i.e. IR-repulsive), and that for the realisation of physical properties such as confinement and chiral symmetry breaking, it is necessary that the theory is strongly-interacting in the IR, now allow us to deduce requirements on $\kappa_{A,c}$. A strongly-interacting IR regime necessarily requires that the GFP is infrared repulsive, see fig.~\ref{betaplot}. Thus we have:
\begin{equation}
\frac{\partial_{g^2} \beta_{IR}}{g^2} < 0 \mbox{ for } g^2 \rightarrow 0.
\end{equation}

\noindent In addition, this condition is also suggested by consistency between the Landau
gauge and the Landau-deWitt gauge. Conjecturing that the running couplings are linked on all scales, the existence of an IR fixed
point for the Landau gauge coupling should imply a fixed
point for the background coupling.  In the Landau gauge an
interacting IR fixed point was shown to exist
\cite{vonSmekal:1997is,Lerche:2002ep,Fischer:2006vf}, where
the running coupling was fixed
at the ghost-gluon vertex.  The operators that actually
induce the fixed
point for the background running coupling are beyond our truncation here, and have been included in \cite{Gies:2002af}. However
we can still infer a necessary condition in order for the
NGFP to be infrared attractive: The
$\beta$ function has to have positive slope there, otherwise
the fixed point would be infrared repulsive. Thus the GFP must be infrared repulsive (see fig.~\ref{betaplot}).

Hence the requirement of a strongly-interacting infrared regime as well as the consistency requirement with the Landau gauge result in the same condition on the $\beta$ function.
Therefore we have established the following criterion 
for the critical exponents:
\begin{equation}
 -\frac{(1+\kappa_A)}{c_{\kappa}+1}+\frac{2}{3}(1+\kappa_c)-\frac{1}{3}+8 \frac{1+\kappa_A}{c_{\kappa}+1} > 0.\label{kappacrit}
\end{equation}
As $c_{\kappa}$ is a regulator-dependent quantity, 
we consider the most restrictive bound from the inequality
\Eqref{kappacrit} in the following. This ensures that, irrespective of the
choice of regulator, the corresponding background coupling $\beta$ function always has an IR repulsive GFP.

Using the scaling relation $\kappa_A = -2 \kappa_c$ 
and the limit $c_A \rightarrow \infty$ we obtain the
bound
\begin{equation}
 \kappa_c < \frac{23}{38}.\label{kappa_crit}
\end{equation}

\begin{figure}[!here]
\begin{minipage}{0.32\linewidth}
\includegraphics[scale=0.4]{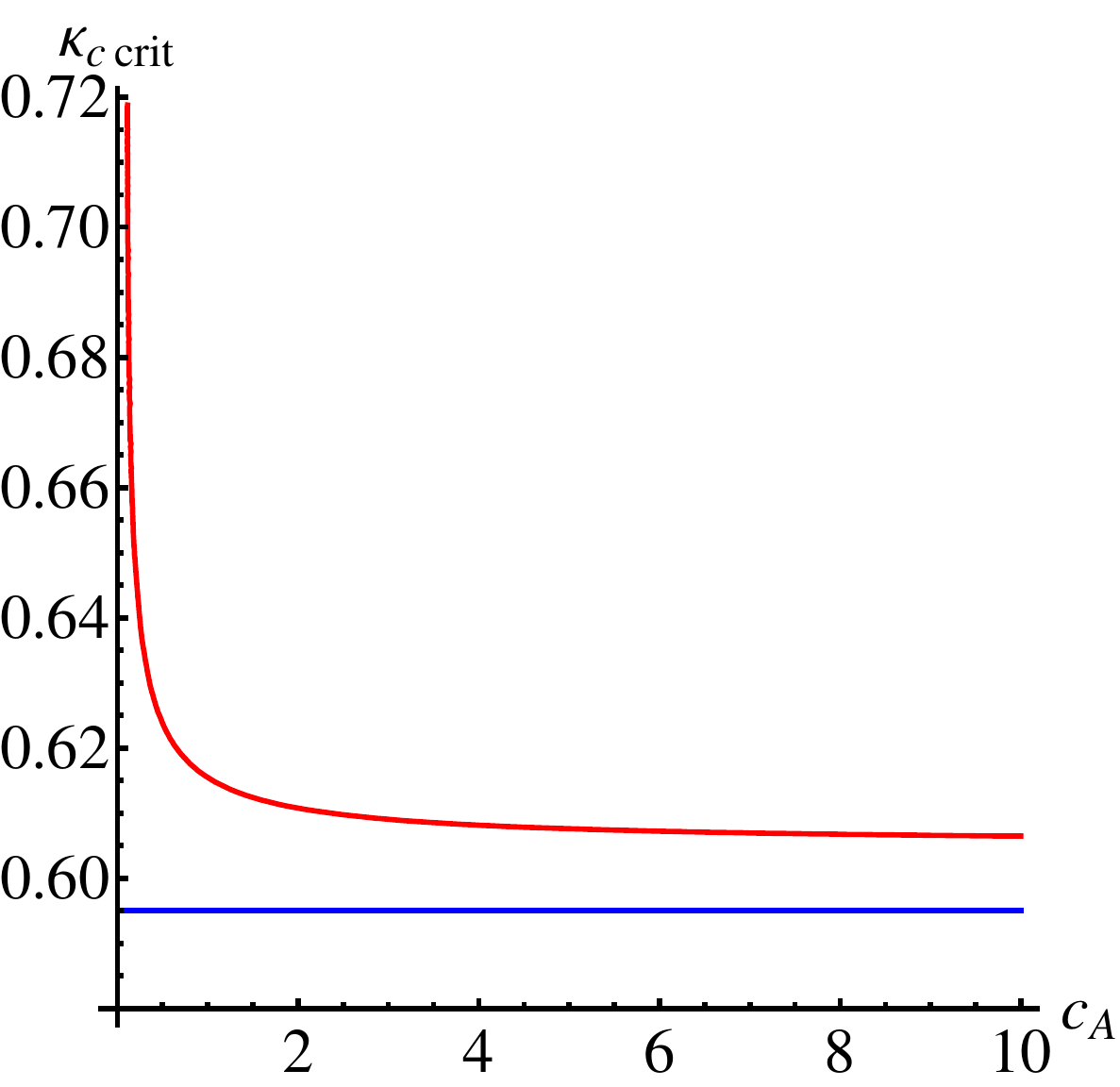}
\end{minipage}
\begin{minipage}{0.32\linewidth}
\includegraphics[scale=0.4]{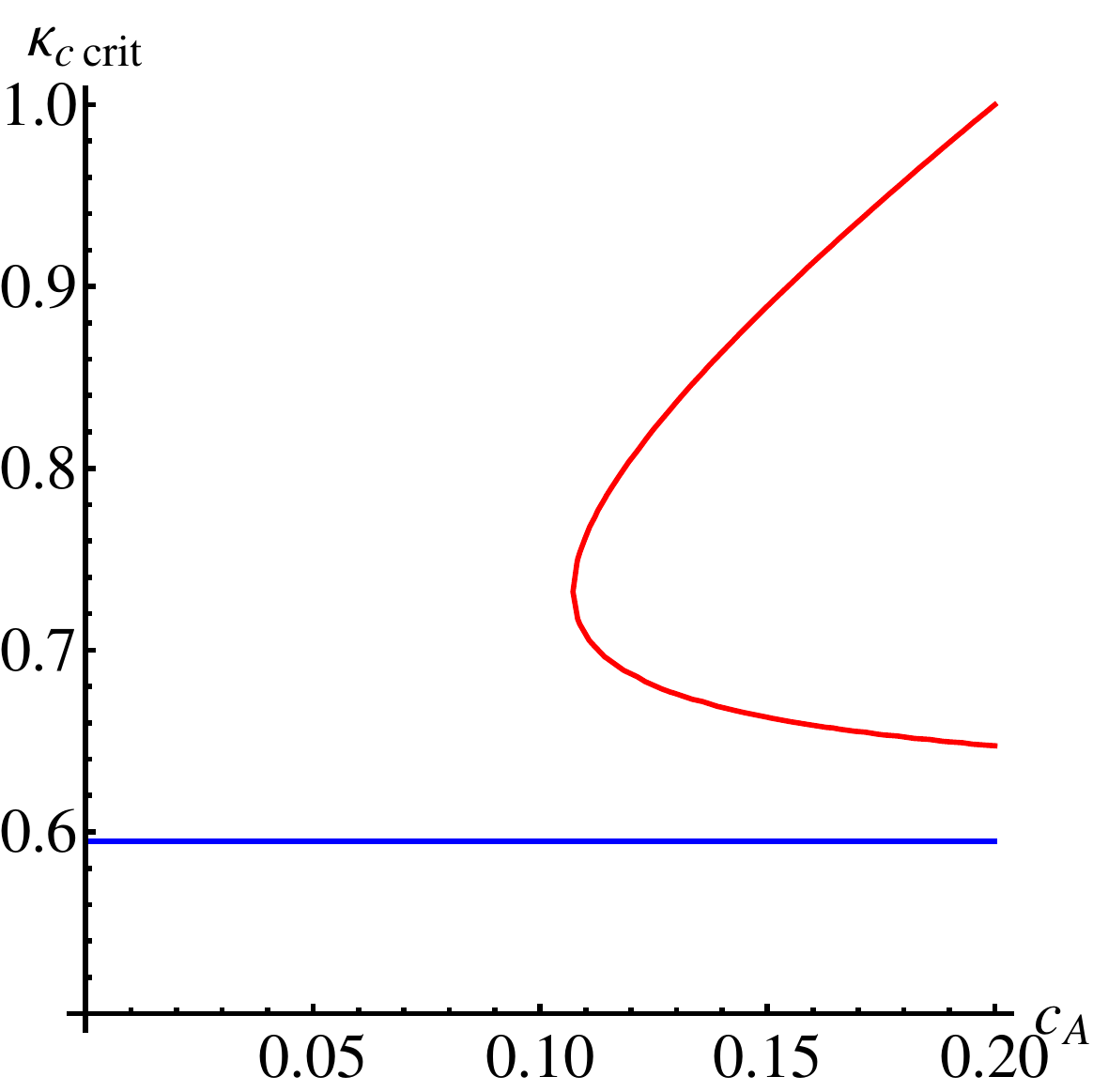}
\end{minipage}
\begin{minipage}{0.3\linewidth}
\caption{$\kappa_{c\, \rm crit}$ as a function of $c_A$. In the left panel the red curve represents an upper bound for $\kappa_c$. For $0.1 \lesssim c_A \lesssim 0.2$ the bound bifurcates (right panel), such that all values of $\kappa_c$ to the left of the red curve are allowed, i.e. in addition to the upper bound $\kappa_{c\, \rm crit}$ a lower bound $\kappa_{c\, \rm crit,2}$ exists, and only the region between these two is excluded. The blue line is $\kappa_c= 0.595$.\label{kappacritplot}}
\end{minipage}
\end{figure}

The maximum upper limit for $\kappa_{c, \rm crit}$ is reached for $c_A
\approx 0.1073$, where $\kappa_{c, \rm crit}\simeq 0.72767$. For values of $c_A\lesssim 0.2$, the inequality \Eqref{kappacrit} also holds if
$\kappa_{c}> \kappa_{c, {\rm crit}\, 2}$, see fig.~\ref{kappacritplot}, as the nonlinear inequality
  \Eqref{kappacrit} bifurcates. We find that allowed values for $\kappa_c$ lie to
the left of the {red/upper} curve (see right panel of
fig.~\ref{kappacritplot}). For $0\leq c_A\lesssim 0.1073$, only the bound
  $\kappa_c<1$ from unitarity \cite{Lerche:2002ep} remains. However, let us
  stress that the limit $c_A\to0$ corresponds to a highly asymmetric
  regularisation as the contribution from transverse gluons to the $\beta$
  function is removed in this limit (note that $c_\kappa\to \infty$ for
  $c_A\to 0$ and $\kappa_A<-1$).  For this case, the Gau\ss{}ian fixed point
is naturally IR repulsive for all values of $\kappa_c > \frac{1}{2}$.

Note that the decoupling solution also satisfies our requirement \Eqref{kappacrit}.

\noindent We may read this new upper bound as a criterion for 
confinement and chiral symmetry breaking: QCD is not
confining in the perturbative regime. A necessary condition
for confinement is a strongly-interacting regime in the
infrared. As our criterion ensures the infrared instability
of the GFP, it implies the existence of a strongly-interacting regime and therefore is a
necessary condition for confinement. Moreover chiral
symmetry breaking can be related to the fact that the gauge
coupling exceeds a critical value
\cite{Braun:2005uj,Braun:2006jd}. Having an IR attractive GFP for the coupling would accordingly not drive chiral
symmetry breaking. Let us emphasise, that this requirement can be fulfilled for $\beta$ functions which also admit an infrared NGFP, as well as for $\beta$ functions which correspond to a diverging coupling in the infrared, i.e. both the blue and the green curve in fig.~\ref{betaplot} are compatible with our criterion. 

Together with the quark confinement criterion 
$\kappa_c > \frac{1}{4}$ \cite{Braun:2007bx}, see also sec.~\ref{deconfphasetrans}, and the Kugo-Ojima/ Gribov-Zwanziger criterion
$\kappa_c > \frac{1}{2}$ this defines a rather narrow window
for the critical exponents. Results for the critical exponent from functional methods lie in the range $\kappa_c\in [0.539\,,\, 0.595]$ (see, e.g. \cite{Fischer:2004uk}) and therefore fall
right into this window. Hence we may conclude that the
asymptotic regime of the ghost and gluon propagator in
Landau gauge fulfills all known criteria for confinement.

\subsection{Effective potential: Gluon condensation in Yang-Mills theory}

We now focus on the full effective potential,
thereby extending previous calculations \cite{Reuter:1994zn,Reuter:1997gx}. The latter did not use the
reconstructed fluctuation-field propagators, but instead
employed the standard approximation of setting $A_{\mu}= \bar{A}_{\mu}$ after the evaluation of the inverse propagator. All explicit background-field dependence beyond the gauge-fixing term was neglected here.
Besides the effective potential was evaluated in a
polynomial truncation. As in our work we are motivated by models of the form $W(F^2)\sim F^2 \ln
F^2$ which have a non-terminating Taylor-series, a polynomial truncation would not suffice.
  
  As a simple parameterisation of the full correlation
  functions, we still use the asymptotic form displayed in \Eqref{IRprops} which
  we amend with $k$-dependent critical exponents $\kappa_A(k)$ and
  $\kappa_c(k)$ in accordance with the propagators in \cite{JMP_to_publish}. A suitable
  interpolation between $\kappa_{A,c}(k\to\infty)\to 0$ and the corresponding
  IR values $\kappa_{A,c}(k\to0)\to \kappa_{A,c}$ can
  parametrise the full momentum dependence of the
correlation functions.

After taking the trace over Lorentz indices (for details see app. \ref{selfdualspec}) the flow equation for the background potential in the selfdual background can
 be written as
\begin{eqnarray}
\partial_t \Gamma_k[A]&=& \frac{3}{2} \tr \,\partial_t R_{k, A}(-D^2)\Bigl(\Gamma_{k, A}^{(2)}+R_{k, A}(-D^2) \Bigr)^{-1}
- \tr \,\partial_t R_{k, c}(-D^2)\Bigl(\Gamma_{k,c}^{(2)}+R_{k, c}(-D^2) \Bigr)^{-1} \nonumber\\
&+&\frac{1}{2} \tr\, \partial_t R_{k, L}(-D^2) \left(\Gamma_{k, L}^{(2)}+ R_{k, L}(-D^2) \right)^{-1}\nonumber\\
&+& \frac{1}{2} \tr\, P_0 \partial_t R_{k, A}(-D^2)\left(\Gamma_{k, A}^{(2)}+R_{k, A}(-D^2) \right)^{-1}\label{eq:Gint}
, 
\end{eqnarray}
where $P_0$ denotes the projector onto the zero-mode subspace. 

This allows us to determine the effective potential by integrating the
flow equation:
\begin{eqnarray}
W_k({F^2})&=& -\frac{1}{\Omega}\int_0^{k_{\rm UV}}
\frac{dk}{k} 
\frac{1}{2}{\rm STr} \partial_t R_k \left(\Gamma_k^{(2)}+R_k
\right)^{-1} + W_{k_{\rm UV}}({F^2})\label{intflow},
\end{eqnarray}
where $k_{\rm UV}$ is an initial ultraviolet scale, which we choose $k_{\rm UV}= 10 \, \rm GeV$, as this lies well within the perturbative regime. Then the effective
action at the UV-scale is given by the form
\begin{equation}
\Gamma_{\rm UV} = \frac{1}{4\, g^2(k_{\rm UV})}F^2 \Omega.
\end{equation}
Here we require the value of the running coupling at the UV-scale 
$k_{\rm UV}$, which can be self-consistently
determined from our input, namely the Landau gauge propagators.
The running coupling 
\begin{equation}
\frac{g^2(p)}{4 \pi}=\alpha (p^2)= \alpha(\mu^2) \left(Z_A^{\frac{1}{2}}(p^2) Z_c(p^2)\right)^{-1},
\end{equation}
defined from the ghost-antighost-gluon vertex approaches a fixed point in the infrared \cite{Alkofer:2004it} with the renormalisation point $\mu^2$ (see, e.g. \cite{Fischer:2008yv}).
Using the IR fixed-point value $\alpha(0) \approx 2.9$ \cite{Lerche:2002ep} then allows to deduce the 
value of the running coupling at any other scale, requiring
as input only the Landau gauge propagators.
Thereby we
arrive at $\frac{1}{4 g^2}\vert_{k=k_{\rm UV}}\approx \frac{1}{16 \pi \,0.2294}$.

Here we have used that the maximum of the gluon dressing function
  $1/Z_A(p^2)$ defines a scale, such that our YM scales can be related to that used in lattice computations.  The
  normalisation is such that the related string tension $\sigma$ (computed on
  the lattice) is given by $\sqrt{\sigma}=440$ MeV.

We can then integrate the effective potential 
from the UV scale down to $k=0$.

\begin{figure}[!here]
\begin{minipage}{0.58\linewidth}
 \includegraphics[scale=0.27]{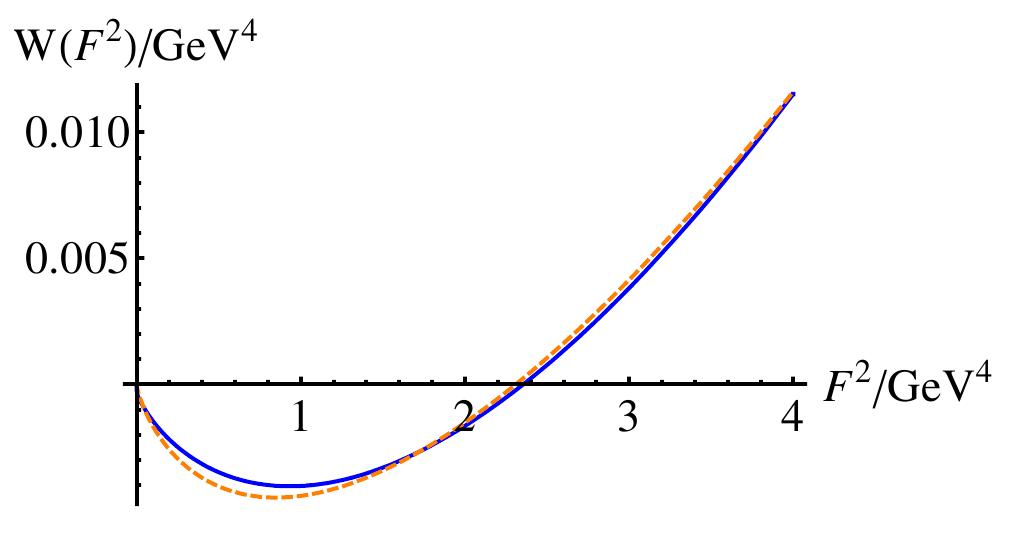}
\label{fit_pot}
\end{minipage}
\begin{minipage}{0.37\linewidth}
\caption{The effective potential for SU(3) as a function of $F^2$
  (thick blue line) shows a non-trivial minimum. The functional form can be approximated by the one-loop inspired fit to the numerical
  data of the form $a F^2 \, \ln b F^2$ (orange dashed line).}
\end{minipage}
\end{figure}

\noindent For SU(3) we find a non-trivial minimum at 
\begin{equation}
  F^2  \approx  0.93 \,{\rm GeV}^4. \label{eq:Fqvalue}
\end{equation}
We can check the dependence on the direction in colour space, since  the colour eigenvalues
$\vert \nu_l \vert$ enter on the right-hand side of the flow equation. Choosing these along one of the two directions
of the Cartan subalgebra for SU(${\rm N}_c=3$) yields an uncertainty of $\sim 10 \%$ in the
value of the minimum.

The result $\langle F^2 \rangle/4\pi
  \simeq0.074 \text{GeV}^4$ is well compatible with recent
phenomenological estimates, $\langle F^2 \rangle/4\pi \simeq0.068(13)
  \text{GeV}^4$ from spectral sum rules \cite{Narison:2009vy} (note
  that our field definition differs from that of \cite{Narison:2009vy} by a
  rescaling with the coupling). 
This good agreement might either indicate that corrections
due to $\widetilde{F}F$-terms, which are included in our estimate
for $W(F^2)$ are rather small. In addition terms that we
neglected in the reconstruction of the propagators could also be subleading. On the other hand an
accidental cancellation of these systematic errors cannot
be excluded. One can check for the second effect with the help of a colourmagnetic background field configuration, for which the selfduality condition does not hold. 
As the spectrum of fluctuations around such a background contains a tachyonic mode, it is necessary to evaluate the Taylor-coefficients of the effective potential separately, as then the tachyonic mode cannot contribute. A comparison to our result would then allow to estimate the contribution from terms $\sim F \widetilde{F}$.

Note that the main effect in the build-up of the condensate is due to the propagators in
the mid-momentum regime and above. The deep-IR asymptotics
do not play a decisive role. This does not automatically follow from the separation of scales between the condensate scale and the deep IR, 
 but it can be checked
explicitly in this example.
 We again emphasise that accordingly we would observe a similar value for the gluon condensate using the decoupling solution for the propagators.

Interestingly, the functional form of the full effective potential can qualitatively be well approximated by a parameterisation of the form
\begin{equation}
 W(F^2)= a F^2 \,\ln (b F^2), \label{eq:logfit}
\end{equation}
which is inspired by the corresponding one-loop results
\cite{Savvidy:1977as} with two fit parameters $a$
and $b$, cf. dashed line in fig.~\ref{fit_pot}). The fit yields $a=0.00528$ and
$b=0.433 \,\text{GeV}^{-4}$.

This implies, that, although the one-loop calculation itself
is not reliable, its prediction for the functional form of
the effective potential is recovered here. This lends
non-trivial support to the leading-log model and in
particular allows us to deduce a value for the string
tension between a static quark-antiquark pair from our
calculation. We find
\begin{equation}
\sigma^{1/2} \simeq 747 \,\text{MeV}
\end{equation}
to be compared to the correct value $\sigma^{1/2} \simeq 440\,\text{MeV}$. We should stress that the string-tension
  $\sigma$ appears in two very different meanings in our calculation: It
    first occurs as an input scale for fixing the initial condition for the flow
    equation, i.e. it fixes the absolute scale of the propagators. We then
  derive the \emph{physical} string tension in a nontrivial way from
  the minimum of the effective potential via the leading-log
    model. This output is linked to a mechanism of confinement, and has therefore
  acquired a physical meaning beyond pure scale fixing.

The discrepancy of our result for $\sqrt{\sigma}$ to the correct value has several sources: The mapping between our full effective potential and a potential of the type $F^2 \ln F^2$ implies that we will not have quantitative precision, since the agreement between our form of the effective potential and the leading-log effective potential is only at the qualitative level. 
Further, we do not distinguish between the condensate $\langle F^2 \rangle$, and condensates involving the dual field strength. Assuming that the second type also exists, our value is to be read as an upper bound, and hence, upon resolving this ambiguity, our value for the string-tension will also be lowered.

\subsection{Outlook: Gluon condensation at finite temperature and in full QCD}
We have determined the effective potential for $F^2$ from generalised Landau gauge ghost and gluon propagators. The effective potential shows a non-trivial minimum at $ F^2 /4\pi
  \simeq0.074 \,\text{GeV}^4$, indicating the condensation of gluons in the vacuum. We have established a connection between a \emph{model} for the YM vacuum that shows confinement of static colour charges, and the propagators in Landau gauge, which satisfy the Kugo-Ojima confinement criterion and the Gribov-Zwanziger scenario. We have thus performed a  first step in connecting two specific pictures of confinement.

As several systematic errors enter our study, it will be interesting to examine their effect to get a quantitatively precise result in future computations. The first error arises from neglecting further terms $\sim F$ in the reconstruction of the propagators. Secondly, our choice of a self-dual background introduces a further source of error, which can be investigated by checking the agreement between the first few Taylor coefficients of the effective potential for the self-dual and a second type of background, e.g. a colourmagnetic one. We emphasise that even within our approximations the difference between our result and values obtained from phenomenological estimates is less than ten percent.

As the formation of a gluon condensate can be linked to
 a linearly rising potential for a static quark-antiquark
pair, it would be interesting to study the effective
potential at finite temperature to observe the "melting" of the condensate.
Thus the onset of quark confinement as determined by the
Polyakov loop (see sec.~\ref{deconfinement}), and the formation of a non-trivial minimum of
the effective potential for $F^2$ might be linked. It will
be interesting to connect the critical temperature for
both cases.

Further it is straightforwardly possible to use gluon  and ghost
propagators that include the effects of quark fluctuations.
Thus our calculation can be extended to a setting within full QCD at finite temperature, which allows to establish a connection to the QCD phase diagram.

Furthermore, as explained above, the Yang-Mills vacuum
 is not characterised by the invariant $\langle F^2 \rangle$
alone. Using similar techniques as explained here, however
on a more elaborate background, or a combination of several
background configurations, would allow for a determination
of other invariants.

Finally let us add that it is of course also an interesting question to access not only the vacuum state but also the complete spectrum of the theory, i.e. the glueball spectrum in the case of Yang-Mills theory. Since glueballs are expected to become manifest in the spectrum of appropriate operators in momentum space \cite{Dudal:2010cd}, a calculation along similar lines presented in this chapter should allow to access the masses of the lowest glueball states.

\section{Deconfinement phase transition in Yang-Mills theories}\label{deconfinement}

The current time is an exciting time for research in QCD,
since, e.g. the Large Hadron Collider at CERN allows to study
phases of QCD, that have only very recently become
accessible experimentally. In particular, a study of the
deconfinement and chiral transition or crossover at finite
temperature and the properties of the quark-gluon plasma can
directly be connected to experimental results, which
unfortunately has become a rare experience in some areas of
high-energy physics. 

Here we will aim at an understanding of the deconfinement phase transition from the FRG. Again we deform QCD to contain infinitely massive fundamental colour sources, i.e. static quarks, only. This already allows to study the deconfinement phase transition, and circumvents conceptual issues such as a proper definition of confinement within full QCD. Let us remark that the methods presented here can also be employed successfully in the context of full QCD with dynamical quarks \cite{Braun:2009gm,Haas:2010bw}.

\subsection{Order parameter for the deconfinement phase transition}
Here we apply the framework of equilibrium
finite-temperature field theory, where the partition function
can be written as a path integral with a compactified
Euclidean time direction of extent $\beta = \frac{1}{T}$.
Then space-time has the topology $\mathbbm{R}^3\, x\, S^1$.
Bose symmetry implies that gauge bosons have to satisfy
periodic boundary conditions in the "time" direction\footnote{Note that ghost
fields also satisfy periodic boundary conditions, although
they anticommute. This follows from the Faddeev-Popov
procedure, where ghost fields are simply used to
exponentiate the Faddeev-Popov determinant, which inherits
the periodic boundary conditions from the gauge field.}.
Accordingly the zeroth component of the momentum is replaced
by the discrete Matsubara frequency $p_0 \rightarrow 2 \pi T n$
with $n \in \mathbbm{Z}$.

To study the confinement of quarks, let us consider
the wordline of a static quark, where spatial fluctuations can be neglected due to the large mass of the quark and
\begin{equation}
\partial_{0} \psi(x^0, \vec{x})= i g A_0 \psi(x^0, \vec{x})
\end{equation}
holds. Using the path-ordering operator $\mathcal{P}$, the solution reads
\begin{equation}
\psi(\beta,\vec{x}) = \mathcal{P} e^{i g \int_0^{\beta}dx^0\, A_0} \psi(0,\vec{x}).
\end{equation}
Taking the trace in the fundamental representation gives the Polyakov loop \cite{Polyakov:1978vu,Susskind:1979up} 
\begin{equation}
 L[A_0] = \frac{1}{{\rm N}_c} {\rm tr}\, \mathcal{P}\, \exp \left( i g
  \int\limits_0^\beta dx^0\, A_0(x^0,\vec{x}) \right), \label{Polyakov}
\end{equation}
which is
invariant under periodic gauge transformations, i.e. gauge
transformations $U(x^0, \vec{x})$ where
$U(x^0+\beta,\vec{x})= U(x^0, \vec{x})$. 
It constitutes an order parameter for the deconfinement phase transition as
\begin{equation}
\vert \langle L[A_0] \rangle \vert^2 \sim e^{-\beta F_{q\bar{q}}},
\end{equation}
where $F_{q\bar{q}}$ is the free energy of a static quark-antiquark pair at large separation. Since the free energy increases with distance in the confined phase, we conclude $\langle L[A_0]\rangle =0$, whereas in the deconfined phase the free energy goes to a constant at large separations and hence $\langle L[A_0]\rangle \neq 0$. 
The value of the Polyakov loop therefore allows to distinguish the two different phases. Indeed in pure Yang-Mills theory these are separated by a phase transition which is marked by the breaking of center symmetry in the ground state. The Polyakov loop serves as an order parameter for this phase transition.

At finite temperature global center symmetry is a
 symmetry of the theory (for a review
see, e.g. \cite{Holland:2000uj}): 
Under a topologically non-trivial gauge transformation with
 $U(x_0,\vec{x})= z U(x_0+\beta, \vec{x})$, where $z$ is an
element of the center\footnote{Recall that the center of
a gauge group is given by all group elements which commute
with the rest of the group. In the case of SU(N), e.g. the
center is given by $z\in \mathbf{Z}_{\rm N}$, i.e. $z =
\mathbf{1} e^{2 \pi i \frac{n}{N}}$ with $n \in
\{1,...,N\}$.}, the action remains invariant, yet the
Polyakov loop transforms non-trivially. Due to the trace, the
net effect of a periodic gauge transformation is none, but
for the topologically non-trivial gauge transformations we find
\begin{equation}
L[A_0] \rightarrow z L[A_0].
\end{equation}
The global center symmetry of the theory therefore is
 broken if the Polyakov loop acquires a non-zero vacuum
expectation value, as is the case in the deconfined phase. 
Note that the
high-temperature phase is the one with the broken symmetry,
whereas the symmetry is restored at low temperature, in contrast to a more standard setting.

Since fields that transform in the fundamental
representation, such as quarks, explicitly break the center
symmetry, the center symmetry does not survive the transition from Yang-Mills theory to full QCD. Then the Polyakov loop is not a
good order parameter any more. Similar considerations apply
to other order parameters that rely on detecting center
symmetry breaking, such as the dual condensate \cite{Gattringer:2006ci,Synatschke:2007bz,Fischer:2009wc,Fischer:2009gk,Braun:2009gm}.

Clearly the Polyakov loop is a non-local quantity as it
 involves an integral over the time direction. It exemplifies a connection between non-perturbative quantities related to confinement and non-locality,
which is emphasised, e.g. in \cite{Alkofer:2010ue}: In a theory with confinement, one should expect non-local objects to carry the physical information on confinement.

The expectation value of the Polyakov loop is a
 quantity that is hard to access from the knowledge of
gauge-field correlation functions, as the expansion of the
exponential contains correlation functions of $A_0$
to any order. Therefore a related quantity has been
established as an order parameter for center symmetry
breaking, namely $L[\langle A_0 \rangle]$
\cite{Braun:2007bx, Marhauser:2008fz}. It can be
accessed from the knowledge of two-point correlation
functions of gluons and ghosts alone and does not require
the knowledge of any higher-order correlation functions. 
$L[\langle A_0 \rangle]$ satisfies\footnote{To see why $L[\langle A_0 \rangle]$ is an order parameter, let us follow \cite{Marhauser:2008fz}, and consider the Polyakov gauge, where $\partial_0 A_0 =0$ and 
$A_0^a = n^a \mathbf{A}_0$ with $n^2 =1$ is rotated into the
Cartan subalgebra. (The Cartan subalgebra is the largest
subset of commuting generators.) Here we parametrise the
direction in colour space by the unit vector $n^a$, which
has non-vanishing components in directions corresponding to
the Cartan generators, only. Then the time-integral in the
Polyakov loop becomes a multiplication with
$\frac{1}{\beta}$. Hence the Polyakov loop, e.g. for SU(2) is given
by
\begin{equation}
 L[A_0]_{\rm SU(2)}= \cos\left[\frac{1}{2}g \beta \mathbf{A}_0\right],
\end{equation}
where the factor $\frac{1}{2}$ arises as the eigenvalues
 of the Cartan generator in the fundamental representation
are $\pm \frac{1}{2}$ in SU(2).
Then $L[\langle A_0 \rangle]$ is an upper bound for 
$\langle L[A_0]\rangle$ in the region $\frac{1}{2} g \beta
\mathbf{A}_0 < \frac{\pi}{2}$, since there the Polyakov loop is positive. This region suffices to study
the deconfinement phase transition, as negative values of
the Polyakov loop arise from positive values by
a center transformation, and larger values are physically
equivalent due to the periodicity of the effective potential, which we will explain below.
(A similar argument can be made for SU(N) with $\rm N>2$.)
As explained in \cite{Marhauser:2008fz}, 
$L[\langle A_0 \rangle]$ is exactly zero in the confined
phase, as one can deduce directly from $\langle L[A_0]\rangle =0$.}
\begin{eqnarray}
L[\langle A_0 \rangle] &\geq& \langle L[A_0]\rangle\,\,
\mbox{ in the deconfined phase}\\
L[\langle A_0 \rangle] &=&0= \langle L[A_0]\rangle\, \, \mbox{ in the confined phase}
\end{eqnarray}
and therefore $L[\langle A_0 \rangle]$ and accordingly also $\langle A_0 \rangle$ constitute valid order parameters.

To evaluate $\langle A_0 \rangle$, we can use the Wetterich
 equation to determine the minimum of the effective action
$\Gamma[A_0]$ as a function of the temperature. Since the only input of the Wetterich equation are the full propagators, the order parameter $\langle A_0 \rangle$ can be evaluated from 2-point functions only, instead of all $n$-point functions. 

As we are interested in a constant field configuration,
 the effective action is simply given by a volume factor
times the effective potential: $
\Gamma[A_0] = \Omega V[A_0]$.

\subsection{Fluctuation field propagators at finite temperature}
To evaluate the effective potential, we again apply the background field method (see \ref{backgroundfieldmethod}), where we specialise to a background with non-vanishing static $A_0$. Accordingly this background has a vanishing field strength, which simplifies the construction of the fluctuations propagators. At finite temperature we have
\begin{equation}
\Gamma_k^{(2,0)}[0, A]= \Gamma_{k\, \rm Landau}^{(2)}[0,A](-D^2)+ (L-\rm terms).\label{Polyakovprops}
\end{equation}
For the case of non-zero
temperature, the inverse propagator in the background field
gauge may also depend on the Polyakov loop ("$L$-terms"), as this is a
further invariant. It is related to the second derivative of the effective potential $V_k^{(2)}$, see \cite{Pawlowski:2005xe,Pawlowski:2001df}. We will neglect this term in our calculation, and comment on this below.

At non-zero temperature, the heat bath determines a preferred frame and the gluon propagator can be
 decomposed into a part longitudinal and transversal to the heat bath. The former acquires a thermal mass, and moreover the dressing functions can differ from their zero-temperature versions over the whole momentum range.
We neglect these effects in our calculation and take only
the temperature-dependence which arises from the Matsubara
frequencies into account. Such a strategy has been shown to give quantitative insight into the
finite-temperature phase structure for scalar theories \cite{Braun:2009si}.
Studies of the finite-temperature
propagators suggest
that this approximation works well for 
Matsubara frequencies $2 \pi T n$ with $|n|\gtrsim 2,3$, see \cite{Gruter:2004bb,Fischer:2010fx}.

We now introduce the variable $\phi^a$ by
\begin{equation}
\beta g \langle \mathsf{A}_0^a\rangle= 2\pi\!\!\!\!\!\!\! \sum_{T^a \in
  \rm{Cartan}} \!\!\!\!\!\!T^a\phi^a
= 2\pi\!\!\!\!\!\!\! \sum_{T^a \in \rm{Cartan}}\!\!\!\!\!\! T^a n^a |\phi|,\,\, n^2=1 \label{eq:HermColM}.
\end{equation}
Then the spectrum of the background covariant Laplacian becomes
\begin{equation}
\text{spec}\{-D^2\} =\vec{p}^{\,2} + (2\pi T)^2 (n - |\phi|
\nu_l)^2\,,\label{spec}
\end{equation}
where $n\in \mathbbm{Z}$, and $(T^a)^{bc}=-i f^{abc}$
denotes the generators
of the adjoint representation of the gauge group under
consideration and $\nu_l$ is defined as in \Eqref{eq:nul}. The number of these eigenvalues is of course
equal to the dimension
 of the adjoint representation of the gauge 
group $d_{\rm adj}$. For each non-vanishing eigenvalue $\nu_l$ there
exists an
eigenvalue $-\nu_l$. 
This implies the symmetry $\phi
\rightarrow -\phi$ in the effective potential. Inspection of \Eqref{spec} indeed reveals that $\phi$
is a periodic variable.
 
Here we have used that a constant field configuration
 in colour space can always be rotated into the Cartan
subalgebra. 
Therefore the
effective potential will be defined on a $d_C$-dimensional
space, where $d_C$ is the dimension of the Cartan
subalgebra. 
In a first step, we will use only the perturbative form of the propagators to investigate the corresponding one-loop effective potential. This will allow to deduce a confinement criterion for the scaling exponents of the propagators. Further we will see that, as expected, the perturbative propagators do not carry the information on confinement, and the knowledge on the full momentum dependence of the non-perturbative propagators is necessary.

\subsection{Perturbative potential: Deconfined phase}
\begin{figure}[!here]
\begin{minipage}{0.45\linewidth}
\begin{flushright}
 \includegraphics[scale=0.4]{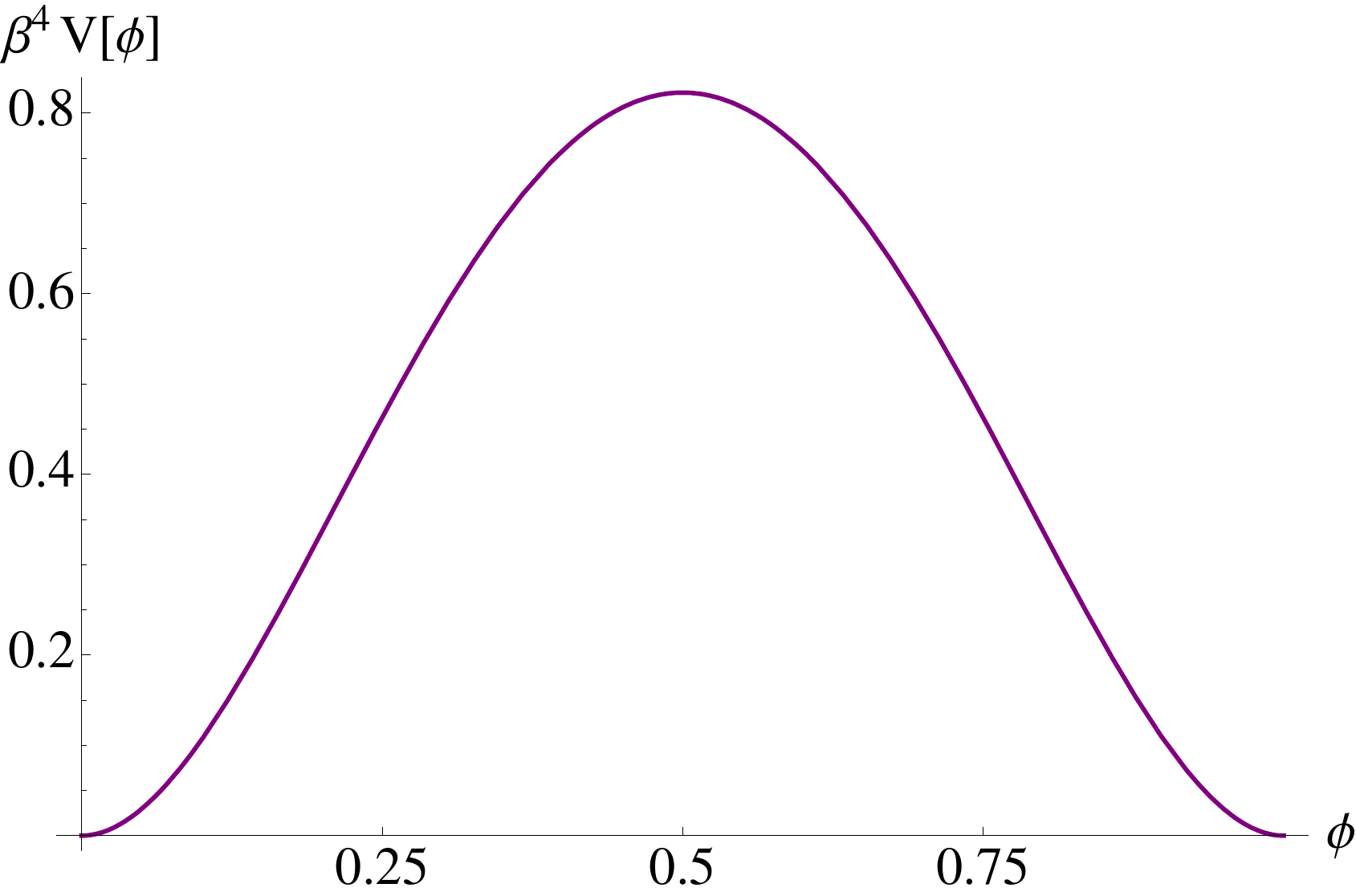}
\end{flushright}
\end{minipage}
\begin{minipage}{0.5\linewidth}
 \caption{One-loop effective potential for SU(2) (to the left) and SU(3) (lower panels), 
normalised to $V[\phi =0]=0$. For SU(3) we depict the potential in the two-dimensional Cartan subalgebra, where dark shades indicate smaller values. We then specialise to a cut through this plane at $\phi_8=0$, along which one of the maxima can be found. 
\label{pertpots}}
\end{minipage}\newline\\
\begin{minipage}{0.48\linewidth}
 \includegraphics[scale=0.7]{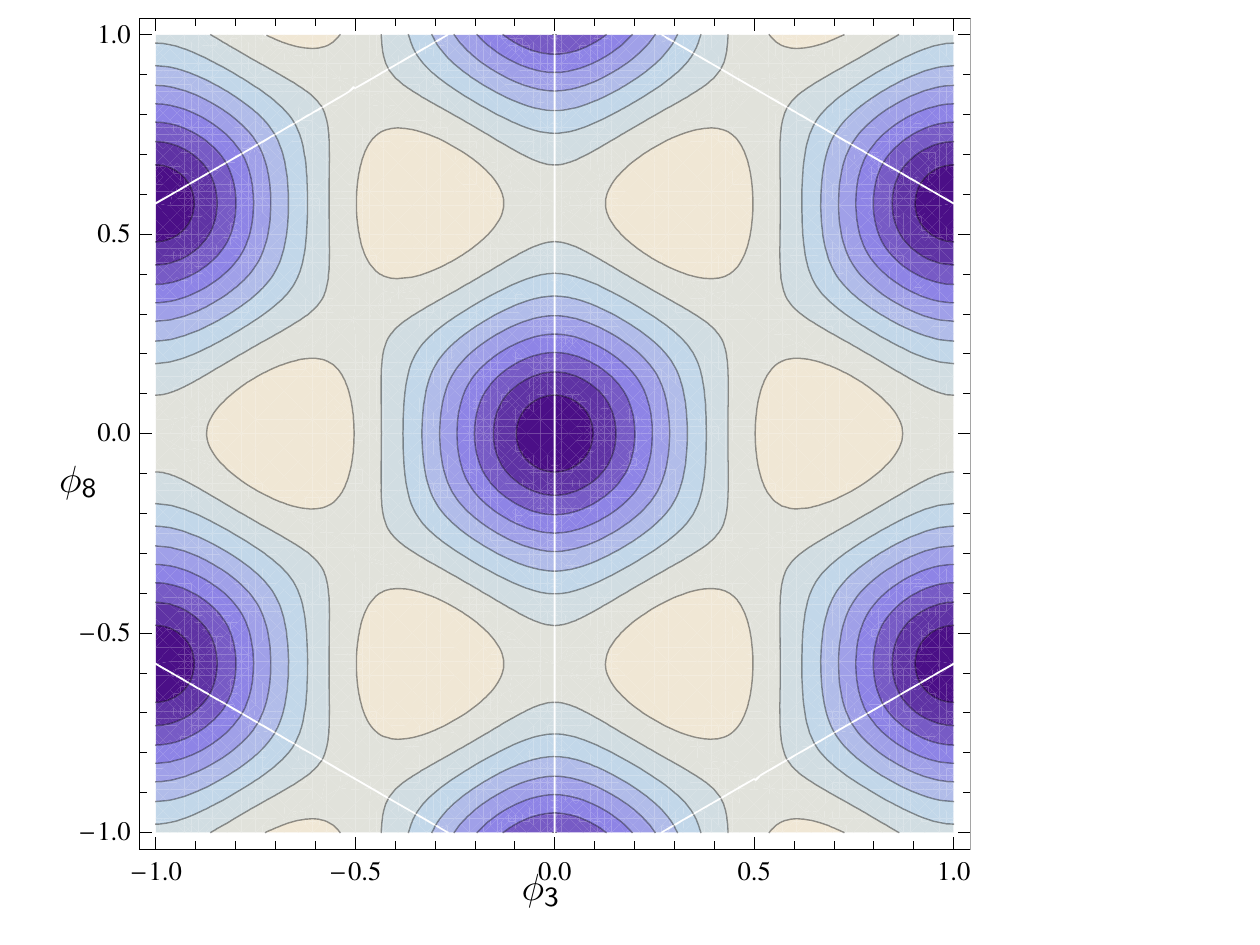}
\end{minipage}
\begin{minipage}{0.05\linewidth}
$\longrightarrow$
 \end{minipage}
\begin{minipage}{0.43\linewidth}
 \includegraphics[scale=0.8]{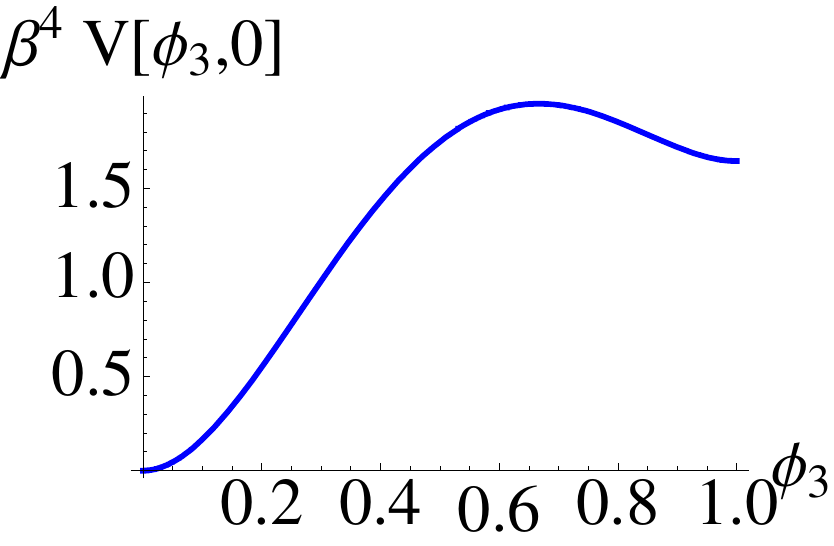}
\end{minipage}
\end{figure}

We now use the perturbative form of the propagators, where $\Gamma_{k\, pert}^{(2)}(p^2)= p^2 \rightarrow -D^2$.
We can easily trace over the spectrum of $-D^2$ to
 get the one-loop effective action:
\begin{equation}
\Gamma_{\rm 1-loop}= \frac{1}{2}{\rm Str} \ln S^{(2)}= \frac{1}{2}\left((d-1)+1-2 \right) \tr \ln (-D^2),\label{1loop}
\end{equation}
where the terms are due to $(d-1)$ transversal and one timelike mode 
(after tracing over Lorentz indices), one longitudinal mode and
two ghosts. The perturbative effective potential for
the Polyakov loop field $\phi$ is clearly an example where
the cancellation between non-physical (timelike and
longitudinal) gluon modes and ghosts works directly, and
only a contribution from the transversal gluon modes
remains. 
The resulting potential is the well-known Weiss potential 
\cite{Weiss:1980rj}. For the cases of SU(2) and SU(3) it is
shown in fig.~\ref{pertpots}.

\noindent The confining values are $\phi=\frac{1}{2}$ for SU(2)
 and $\phi_3 = \frac{2}{3}$ for SU(3), as can be checked by
a direct inspection of the Polyakov loop, using the Cartan
generators in the fundamental representation. The coordinates of the maximum of the one-loop potential coincide with exactly these points
in each of the two cases.
Accordingly the perturbative propagators do not encode
quark confinement, as one would expect. Therefore we will use the non-perturbative propagators, obtained from a generalisation of the fully momentum-dependent Landau gauge propagators in the following.

\subsection{Non-perturbative potential: Deconfinement phase transition}\label{deconfphasetrans}
The flow equation can be rewritten as
\begin{equation}
\partial_t \Gamma_k = \frac{1}{2} {\rm STr} \partial_t 
\ln \left(\Gamma_k^{(2)}+R_k \right)- \frac{1}{2} {\rm STr}(
\partial_t \Gamma_k^{(2)})\left(\Gamma_k^{(2)}+R_k
\right)^{-1},
\end{equation}
where $\partial_t$ in the first term acts on both $\Gamma_k^{(2)}$ as well as $R_k$.
Integrating the flow over $k$ from $\Lambda$ to zero then yields the 
following expression for the effective action:
\begin{equation}
\Gamma_{k \rightarrow 0}= \frac{1}{2}{\rm STr} 
\ln \left(\Gamma_{k \rightarrow 0}^{(2)} \right)+
\int_{0}^{\Lambda} \frac{dk}{k}\frac{1}{2} {\rm STr}
\partial_{t} \Gamma_{k}^{(2)}\left(\Gamma_{k}^{(2)}+R_{k}
\right)^{-1}+\Gamma_{\Lambda} - \frac{1}{2}{\rm STr} \ln
\left(\Gamma_{\Lambda}^{(2)} +R_{\Lambda}\right).\label{Polyakovpotflow}
\end{equation}
Here we observe a clear analogy to the one-loop 
effective action, cf. \Eqref{1loop}. In this more general setting the
full propagator, including the effect of all quantum
fluctuations, enters. Clearly this term is not regulator dependent. The last two terms constitute initial conditions.
The term $\sim \partial_t
\Gamma_k^{(2)}$, which can be considered as an RG-improvement term,
turns out to be subdominant in this setting. This
automatically implies that the main contribution to the
effective potential is not regulator dependent.

A confinement criterion has been put forward in 
\cite{Braun:2007bx}, see also \cite{Braun:2010cy}, relating the infrared behaviour of the
ghost and the gluon to a confining Polyakov-loop potential.
It relies on the fact that the effective potential at a
temperature scale $T$ is mainly driven by momentum modes
with $p^2 \sim (2 \pi\, T)^2$. Accordingly only the deep-IR form of the
propagators plays a role for the effective potential at $T
\rightarrow 0$. Neglecting the RG-improvement term, which actually turns out to be subleading, the
effective potential can then be rewritten as the one-loop
effective potential with a factor that depends on the
critical exponents:
\begin{equation}
 V_{\text{IR}}(\phi)
= \left\{ 1+ \frac{(d-1)\kappa_A-2\kappa_C}{d-2} \right\}
V_{\rm 1-loop}(\phi). \label{eq:WeissIR}
\end{equation}
Demanding a confining potential in the
infrared thereby restricts the critical exponents.
Interestingly, the critical exponents of both the scaling as well as the decoupling solution satisfy this restriction. Hence confinement of (static) quarks can
thereby be inferred from the pure knowledge of gauge two-point
correlation functions.
 
Before we study the transition between these two phases, the deconfining and the confining one, let us discuss the order of this transition, and how it can be related to the underlying gauge group.

\subsection{What determines the order of the phase
transition?}
To understand the properties of QCD, let us consider an enlarged parameter space, where, e.g. the gauge group is not fixed. Thus, we consider, among others, the rank of the group as an "external" parameter.
Such a deformation of QCD allows to understand, in which way certain physical properties of the theory depend on a specific parameter. In
particular, we would like to understand what determines the
order of the phase transition. The order of the phase
transition is, e.g. important for the dynamics of our universe, since the physical properties accompanying a first order phase transition, such as the coexistence of both phases at the transition, are vastly different from the properties of a second order phase transition. Besides, physics beyond the standard model might also rely on QCD-like theories, such as the technicolour proposal for the Higgs sector, see, e.g. \cite{Sannino:2009za}. Here, the gauge group might be different, and therefore studying Yang-Mills theory in a more general setting than for SU($\rm N_c=3$) might also be of interest.

The universality class of the deconfinement phase
transition in the case of a second order phase transition is conjectured to be linked to the center of the gauge group \cite{Svetitsky:1982gs},
since integrating out all degrees of freedom in the path integral
of a $(d+1)$-dimensional Yang-Mills theory except the Polyakov loop results in a center-symmetric
 scalar field theory in $d$
dimensions. The dimensional reduction comes about, as the
Polyakov loop already contains an integral over the
Euclidean time direction. In the case of a second order phase transition the
complicated microscopic dynamics of this theory is "washed
out" due to the diverging correlation length, and the
universality class of the theory is determined by the
dimensionality, the field content and the symmetry.  This
explains why SU(2) in 4 dimensions falls into
the 3-dimensional Ising universality class \cite{Engels:1989fz,Engels:1992fs,Marhauser:2008fz}.

Interestingly some gauge theories show non-universal phase transitions despite the availability of
universality classes, as, e.g. in the case of the symplectic
groups Sp(N) having center $\mathbf{Z}_2$
\cite{Holland:2003kg}. Here the dynamics of the theory
seems to prevent the occurrence of second order phase
transitions. A physical
argument to explain this observation relates to the dynamical degrees of freedom in the two phases \cite{Holland:2003kg, Pepe:2006er}: A large
mismatch in the number of dynamical degrees of freedom on
both sides of the phase transition makes a smooth (i.e.
second order) transition between both regimes unlikely.
The degrees of freedom on the deconfined side of
the
phase transition are free gluons, the number of which is
given by the number of generators of the gauge
group. On the other side of the phase transition, glueballs are the dynamical degrees of freedom. Their
number is essentially independent of the size of the gauge
group, since they are colourless objects.
Accordingly for large gauge groups the number of dynamical degrees of freedom changes considerably.
This is difficult to reconcile with a
smooth change in the physical quantities of the theory, such
as order parameters. Therefore we are led to expect that
large gauge groups will not show universal behaviour, but
will instead have a first order deconfinement phase
transition.

We will investigate this conjecture for different gauge
groups, confirming studies based on lattice gauge theory, as
well as extending these to very large gauge groups which
are currently unaccessible to lattice gauge theory due to
computational resources. Here we are in the situation to check results from the FRG against results from a method which is usually quantitatively more precise. After having ascertained that our results reproduce lattice results for small gauge groups, we can use the FRG to go beyond the regime that is accessible to lattice gauge theory, and study the above conjecture for several large gauge groups.

\begin{figure}[!here]
\begin{minipage}{0.55\linewidth}
\includegraphics[scale=0.65]{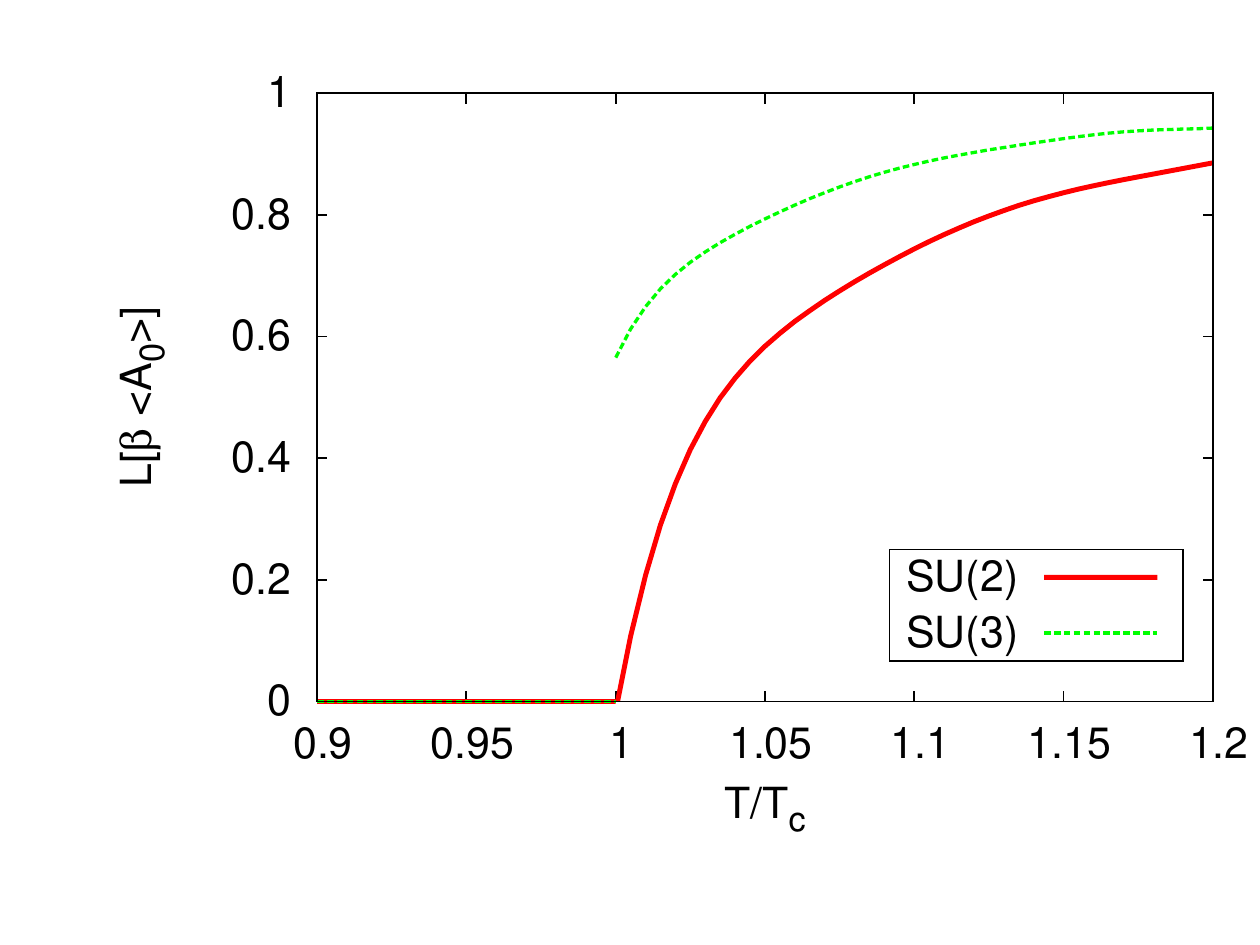}
\end{minipage}
\begin{minipage}{0.4\linewidth}
\caption{The order parameter for the deconfinement phase transition for SU(2) (red, full line) and SU(3) (green, dashed line) as a function of $T/T_c$ shows a smooth, second order phase transition for SU(2), and a discontinuous, first order phase transition for SU(3), see also \cite{Braun:2007bx}.\label{Su2Su3PLoop}}
\end{minipage}
\end{figure}

We find a second order phase transition for SU(2) (cf. fig.~\ref{Su2Su3PLoop}), which corresponds to a continuous change in the order parameter, 
and a first order phase transition for SU(3)-SU(12), which
agrees with findings of lattice gauge theory, which are available up to SU(8), see \cite{Teper:2008yi,Panero:2009tv} for reviews. 

\noindent Before we explain these
results in more detail we consider the mechanism of second
and first order phase transitions in our setting.

\subsection{Mechanism of second and first order phase
transitions}

\begin{figure}[!here]
\begin{minipage}{0.55\linewidth}
  \includegraphics[scale=0.45]{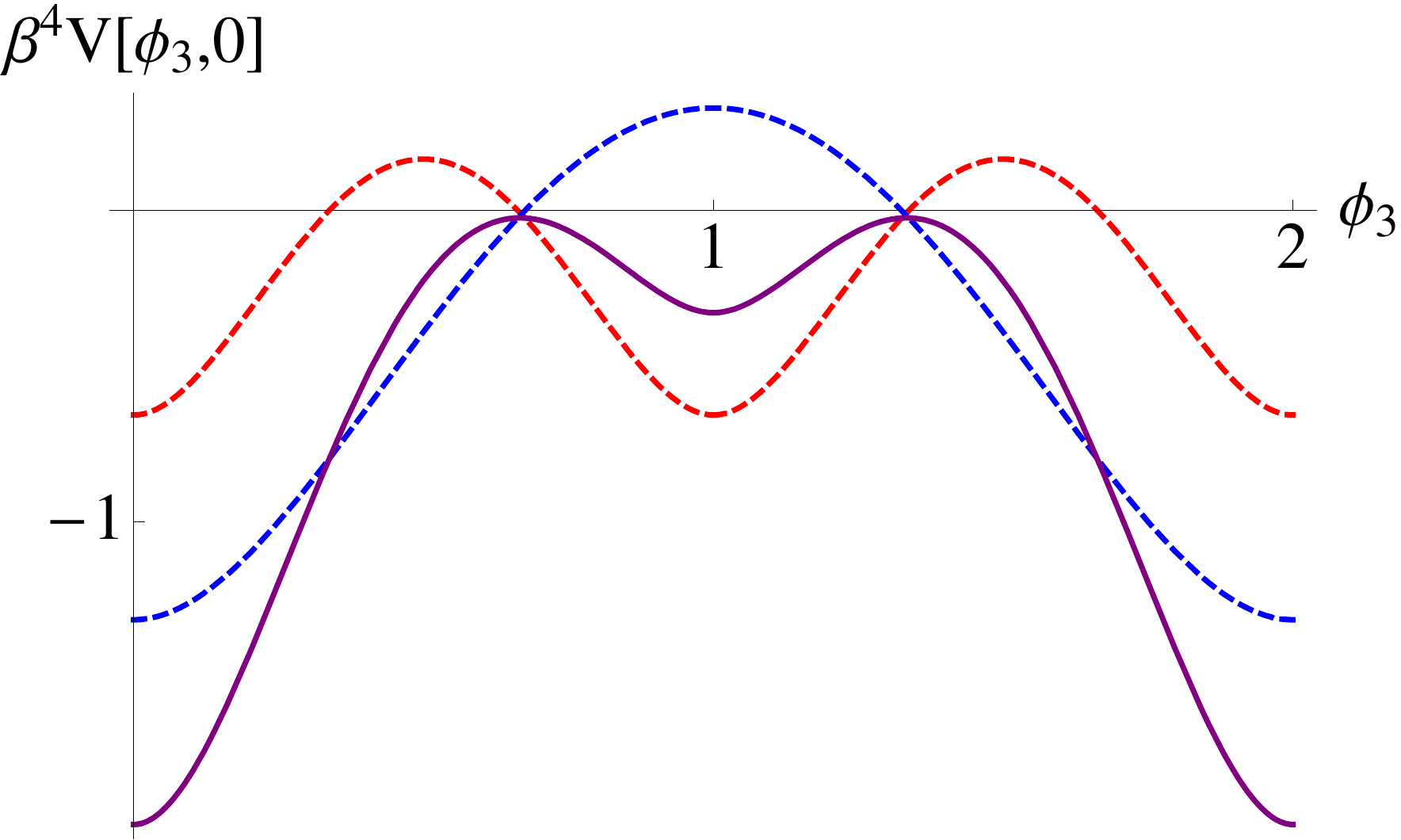}
\end{minipage}
\begin{minipage}{0.4\linewidth}
\caption{Perturbative effective potential for SU(3) in
$\phi_3$-direction: 
The two dashed lines are the contributions of the
eigenvalues $\nu=\pm\frac{1}{2}$ (blue) and $\nu = \pm 1$
(red), respectively. The resulting potential is plotted in purple and
clearly shows a shifted maximum in comparison to the SU(2)
potentials as well as local minima.\label{SU3superpositionpot}}
\end{minipage}
\end{figure}

Here it is crucial that the effective potential of a general gauge group obeys
\begin{equation}
V(\phi^a) = \sum_l \frac{1}{2}V_{\rm SU(2)}(\nu_l \vert \phi \vert), \label{potsum}
\end{equation}
where the eigenvalues $\nu_l$ depend on the specified direction $n^a$ in the Cartan subalgebra. Note that since the Cartan subalgebra of SU(N) is (N-1) dimensional the 
effective potential is a function of an (N-1) dimensional
variable.
The superposition \Eqref{potsum} leads to shifted maxima in 
comparison to an SU(2) potential and most importantly to the
formation of local minima (cf. fig.~\ref{SU3superpositionpot}), which  can induce a first order phase transition.

\begin{figure}[!here]
\begin{minipage}{0.55\linewidth}
 \includegraphics[scale=0.75]{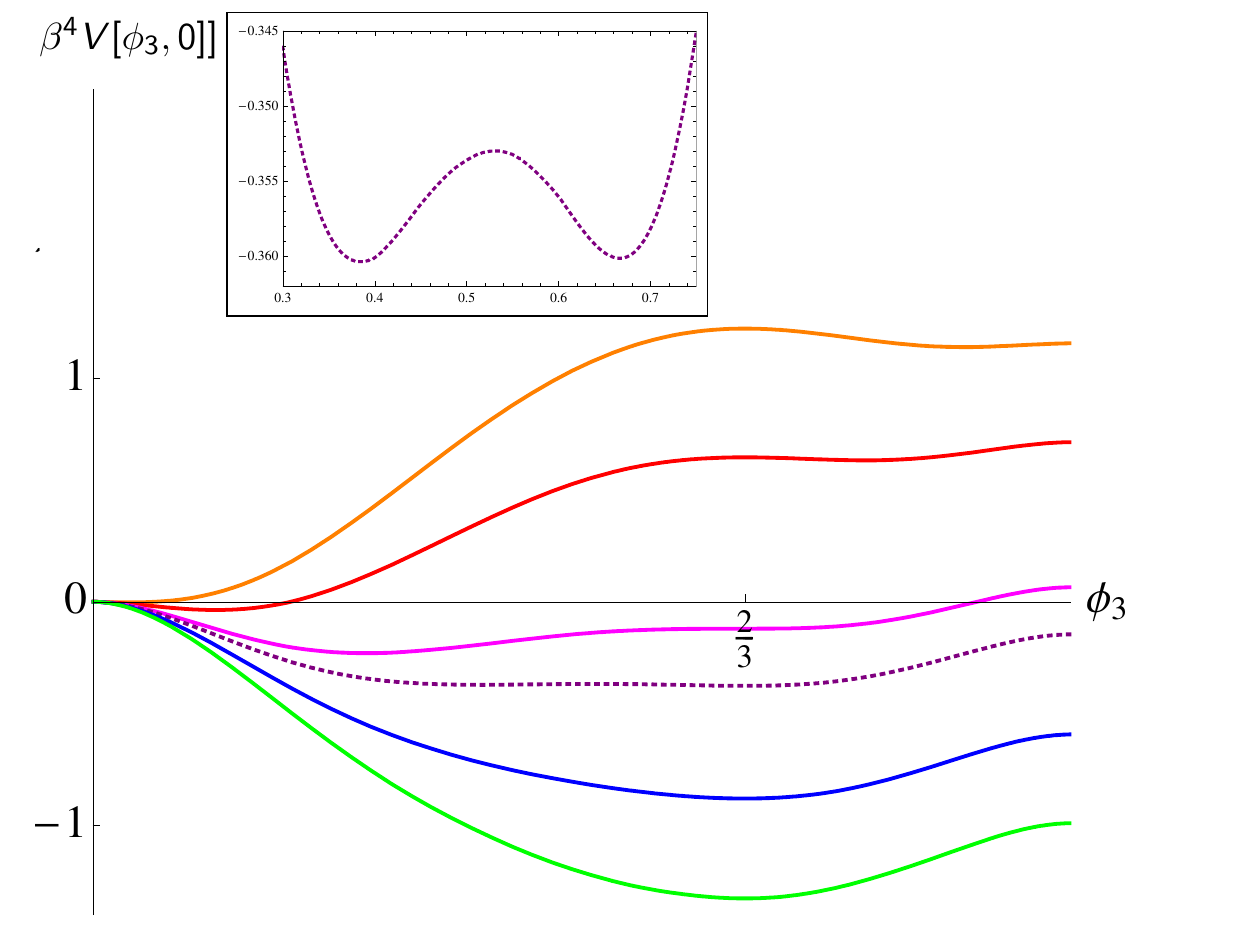}
\end{minipage}
\begin{minipage}{0.4\linewidth}
\caption{Temperature-dependent effective potential for 
SU(3) in $\phi_3$ direction. Different temperatures are
indicated by different colours, starting with yellow at $T>T_c$ and ending with green at $T<T_c$. The local minimum deepens
with decreasing temperature, until at $T= T_c$ (dashed line, see also the inlay)
the minima are degenerate and $\phi_3$ jumps to $\phi_3 = \frac{2}{3}$, which is the
confining minimum.\label{su3temppot}}
\end{minipage}
\end{figure}

\noindent
In SU(2) the global minimum starts to move
from $\phi=0$ to larger values, until it reaches the value
$\phi=\frac{1}{2}$ at the critical temperature. This results in a second order phase
transition for SU(2). As soon as two SU(2) potentials with
different periodicities are superposed, in addition to the
global minimum local minima develop. This obviously allows
for a first order phase transition, if the local minimum
reaches the same depth as the global minimum for
$\phi_{min,\, loc}^a\neq \phi_{min,\, glob}^a$. 
The temperature-dependent effective potential for SU(3) (see fig.~\ref{su3temppot}) clearly exemplifies this
behaviour. We only show the potential in the
$\phi_3$-direction, as indeed the global minimum always lies
at $\phi_8=0$ (or center-symmetric points). Note, however,
that such a simplification is in general not possible: For
an effective potential that is spanned by several directions
we have to follow the global minimum in the full space, as
it generically does not follow a straight line in this
space.

\noindent Here we have to note how the approximations that we used 
in our setting could alter this behaviour:
\begin{itemize}
 \item We can only rewrite the complete potential as a 
sum over SU(2) potentials as we use the same propagators for
all gauge groups. However the differences between the
propagators for different SU(N) are only
$\mathcal{O}\left(\frac{1}{N}\right)$ and therefore
will not lead to strong modifications\footnote{Even the propagators for SU(2) and SU(3) have been shown to agree within errors in lattice calculations \cite{Cucchieri:2007zm}.}. Indeed the full
potential will then still be a superposition of SU(2)
potentials plus a small correction, which should not alter
our findings.
\item Including the $V^{(2)}$ term (cf. \Eqref{Polyakovprops}) leads to a change of 
the potential that depends on the curvature of the full $d_C$-dimensional
potential, not of the single one-dimensional SU(2) potentials. Therefore
this term cannot be written as a superposition of SU(2)
terms. Such a term is important for the study of second order phase transitions and the calculation of critical exponents, since a second order phase transition is signalled by
long-range correlations. For a study taking this term into account in SU(2), where it allows to recover the correct critical exponents, see \cite{Marhauser:2008fz}. Thus we expect this term to be
sub-leading for first order phase transitions. Of course, the fluctuations corresponding to the $V^{(2)}$ term could weaken a first order phase transition, possibly resulting in a second order phase transition.
\end{itemize}

The approach described above for SU(2) and SU(3) can then be
straightforwardly generalised to arbitrary gauge groups. In our approach the calculation of the order parameter 
$L[\langle A_0 \rangle]$ only requires the knowledge of the
eigenvalues of the Cartan generators in the fundamental and
in the adjoint representation. A single numerical evaluation
of the effective potential for SU(2) then suffices to
determine the critical temperature and the order of the
phase transition for any other gauge group.

\subsection{Results: Finite-temperature deconfinement phase transition}
For the numerical evaluation of the full effective potential we employ an optimised regulator of the form
 \cite{Pawlowski:2005xe} for the calculation of the RG improvement term $\sim \partial_t\Gamma_k^{(2)}$ in \Eqref{Polyakovpotflow}:
\begin{eqnarray}
R_{\rm opt}=(\Gamma_0^{(2)}(k^2)-
\Gamma_k^{(2)}(p^2))\theta(\Gamma_0^{(2)}(k^2)-\Gamma_0^{(2)}(p^2)).
\label{eq:optreg}
\end{eqnarray}
We again stress that the leading contribution to the effective potential is regulator-independent (cf. \eqref{Polyakovpotflow}), and only the RG-improvement term $\sim \partial_t \Gamma_k^{(2)}$ depends on our choice of regularisation scheme.

\subsubsection{SU(N)}
We investigate the groups SU(N) for ${\rm N= 2,...,12}$. 
 Whereas for the case of SU(3) the global
minimum of the effective potential only moves in the
direction associated with the $T_3$ generator, we do not observe a
straight line on which the minimum of the potential moves
for the higher gauge groups, ${\rm N} > 4$. Therefore we investigate the
minimum of the effective potential on the complete (N-1)
dimensional Cartan subalgebra.

We find a first order 
phase transition for all SU(N) with ${\rm N}\geq 3$. This is in
agreement with lattice results for SU(3) to SU(8)
\cite{Teper:2008yi} and shows that the first order phase
transition persists beyond the already investigated regime. The critical temperatures are listed in the
following table, where the numerical uncertainties are approximately $2\, \rm MeV$:
\begin{center}
\begin{tabular}[!here]{l|c|c|c|c|c|c|c|c|c|c|c|c}
 $\rm{N}_c$&2&3&4&5&6&7&8&9&10&11&12\\
\hline
$T_c/ \rm MeV$ & 265&291&
292&295&295&295&295&295&295&295&295.
\end{tabular}
\end{center}

\begin{figure}[!here]
\begin{minipage}{0.55\linewidth}
\includegraphics[scale=0.7]{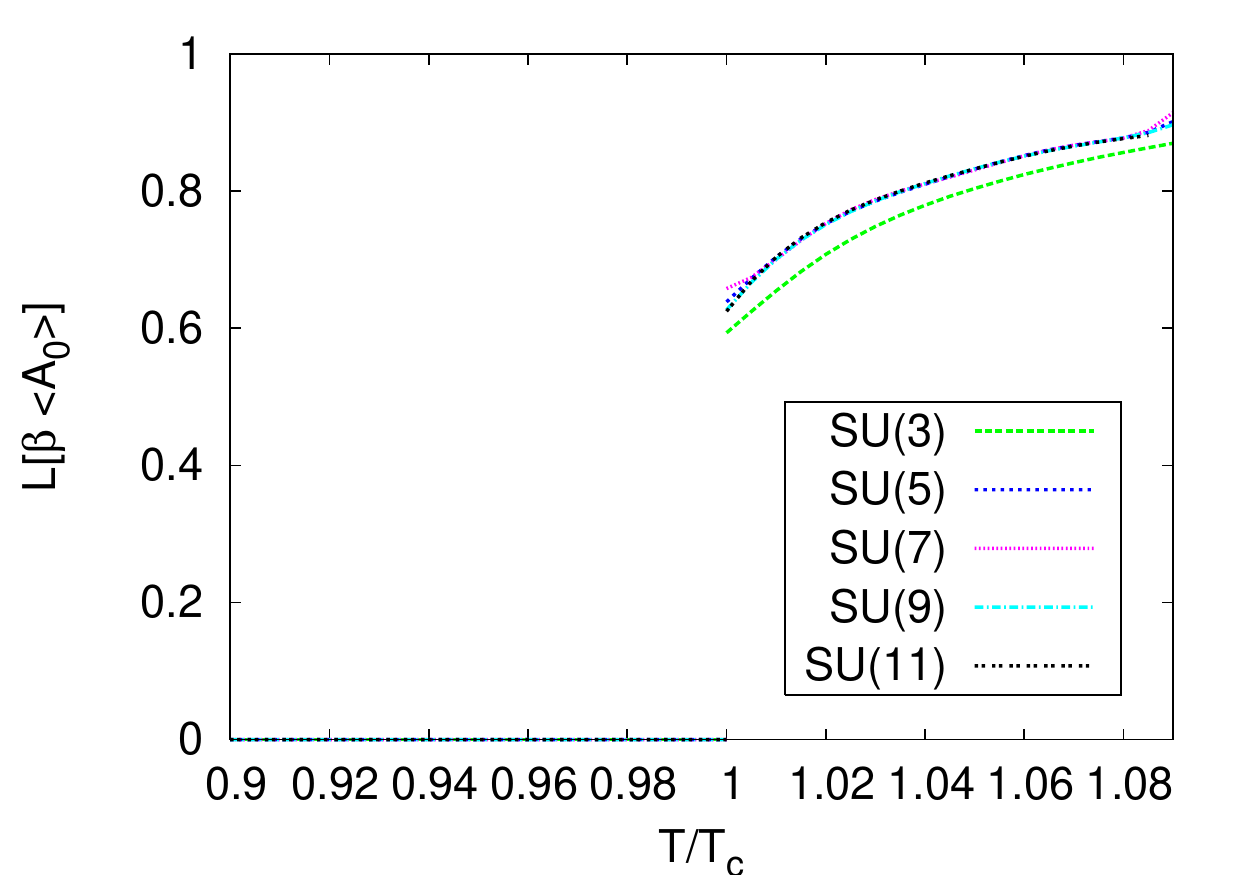}
\end{minipage}
\begin{minipage}{0.4\linewidth}
\caption{The order parameter for SU(N) for $N=3,5,7,9,11$ shows a first order phase transition for these gauge groups as a function of $T/T_c$. The functions $L[\langle A_0 \rangle](T)$ are highly similar for $N>5$, where also the critical temperatures $T_c$ agree. }
\end{minipage}
\end{figure}

\noindent In particular, the critical temperatures normalised to the string tension yield a dimensionless quantity that can be compared to lattice calculations. Here we have, e.g.
\begin{eqnarray}
 \frac{T_{c\, \rm SU(3)}}{\sqrt{\sigma}}&\approx& 0.66.
\end{eqnarray}
Lattice results find a similar value, see, e.g. \cite{Teper:2008yi}.

 We observe that the height of the jump as well as the
critical temperature become independent of N for ${\rm
N}>5$, in accordance with lattice results \cite{Panero:2009tv}.

Let us emphasise, that the order of the phase transition as well as the critical temperature are independent of the deep-IR asymptotics of the propagators. We again find, as for our value of the gluon condensate, that the scaling as well as the decoupling solution both yield the same result.

\subsubsection{Sp(2)}
For our choice of conventions regarding the symplectic
group Sp(2), see app.~\ref{SP2}.
The dimension of the group is $(2\rm N+1)N$, the center
$\mathbf{Z}_2$ and the rank is 
$\rm N$. Sp(N) gauge groups hence are very useful to investigate
what determines the order of the gauge group, as the Ising
universality class is always available for a second order
phase transition in 4 dimensions. On the other hand the
growing size of the gauge group will trigger a first order
phase transition for a critical size of the gauge group. 
Since Sp(1)= SU(2) has a second order phase transition 
the next group to investigate is Sp(2).
 
\begin{figure}[!here]
\begin{minipage}{0.55\linewidth}
\includegraphics[scale=0.3]{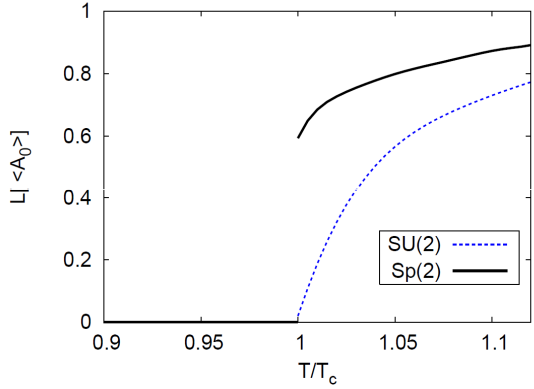}
\end{minipage}
\begin{minipage}{0.4\linewidth}
 \caption{The order parameter for Sp(2) in comparison to SU(2) shows a first order phase transition, although both gauge groups have the center $\mathbbm{Z}_2$. The number of generators of Sp(2) is 10, and therefore comparable to SU(3) with 8 generators.\label{Sp2plot}}
\end{minipage}
\end{figure}

\noindent In accordance with
lattice results \cite{Pepe:2006er,Holland:2003kg} we find a first order phase
transition here (cf. fig.~\ref{Sp2plot}), which presumably is induced by the comparably large number of generators. 
 
\subsubsection{E(7)}
E(7) is one of the exceptional groups. It has 133 
generators and therefore a huge number of free gluons in the deconfined phase.
The center of E(7) is again $\mathbf{Z}_2$, hence the 
Ising universality class is available for a second order
phase transition. However, according to the conjecture
advanced in \cite{Holland:2003kg, Pepe:2006er}, the large number of generators should
induce a strong first order phase transition, unless the
glueball spectrum of E(7) is for some reason very different
from other gauge groups. Evaluating the deconfinement order
parameter for E(7) in the seven-dimensional Cartan subalgebra, we indeed observe a first order phase
transition. 
\begin{figure}[!here]
\begin{minipage}{0.55\linewidth}
\includegraphics[scale=0.3]{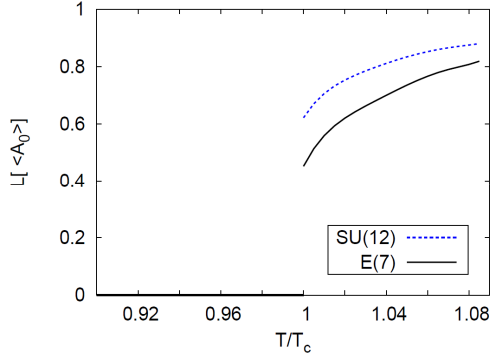}
\end{minipage}
\begin{minipage}{0.4\linewidth}
\caption{The order parameter for E(7) shows a first order phase transition. Comparing to a gauge group with a similar number of generators, namely SU(12), the jump in the order parameter is smaller, which implies a weaker first order phase transition.}
 \end{minipage}
\end{figure}

\noindent Note that we cannot definitely exclude a
second order phase transition, which might possibly be induced by the missing term $\sim V^{(2)}$.
 
\begin{figure}[!here]
 \begin{minipage}{0.6\linewidth}
  \includegraphics[scale=0.28]{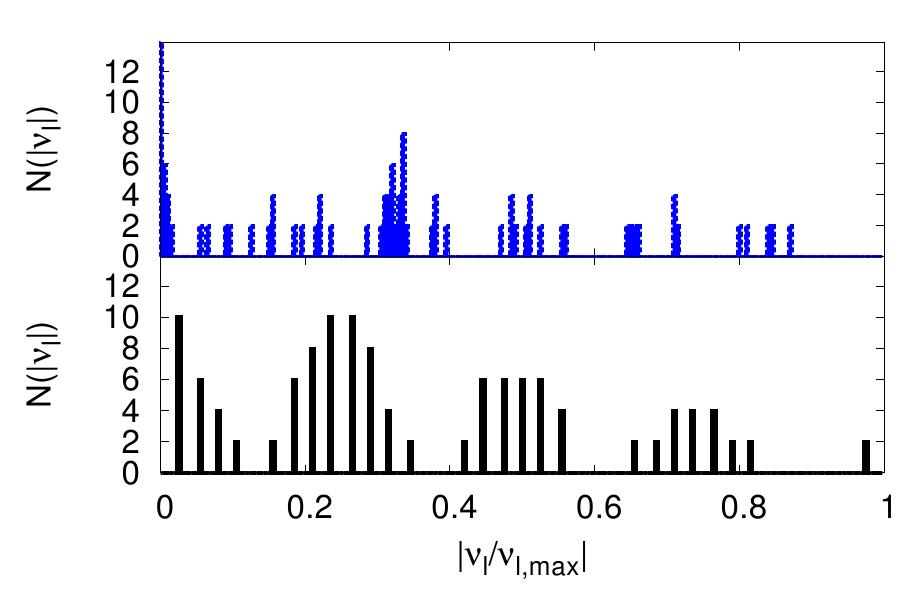}
 \end{minipage}
\begin{minipage}{0.35\linewidth}
 \caption{Eigenvalue distribution $N(|\nu_l|)$ of the spectrum as a function of the (normalized) eigenvalue $|\nu_l/\nu_{l\, \rm max}|$ for $E(7)$ 
(lower panel) and $SU(12)$ (upper panel) at the ground state of the potential 
for $T \rightarrow T_{\rm c}^{+}$. The case of $E(7)$ shows a dominance of eigenvalues with $|\nu_l/\nu_{l\, \rm max}|=0.25;\, 0.5;\, 0.75$, whereas in the case of $SU(12)$ many eigenvalues occur with a similar weight, and no clear dominance can be seen.
\label{interference}}
\end{minipage}
\end{figure}

Surprisingly a comparison with similar-sized SU(N) gauge 
groups reveals, that the height in the jump of the
Polyakov loop is smaller for E(7). As the Polyakov loop is
not an RG invariant, the height itself
does not necessarily carry physical meaning, however in our
calculation it is possible to link it to the properties of
the eigenvalue spectrum $\{\nu_l\}_{l=1}^{d_{\rm adj}}$: As
the effective potential is a sum over SU(2)-effective
potentials, depending on the eigenvalues $\nu_l$ even a very
large gauge group can still have a second order phase
transition. This happens, if at the phase transition
temperature, the value of the field $\phi^a$ in the Cartan
subalgebra is such that $\vert \phi \vert \nu_l$ is equal for all $l$.
One can now imagine two scenarios which can be termed 
"destructive" or "constructive interference" of SU(2)
potentials: Remember that due to equation \Eqref{potsum} the complete
effective potential is given by a sum over SU(2) potentials
with different periodicities. The periodicities are
determined by the eigenvalues of the Cartan generators in
the adjoint representation. For many eigenvalues of the same
value, the complete potential will be dominated by an SU(2)
potential of the corresponding periodicity. As the SU(2)
potential induces a second order phase transition, a
dominance of one eigenvalue-size implies a weak first order,
or in the extreme case of all eigenvalues being of the same
size, even a second order phase transition. This mechanism
works irrespective of the size of the gauge group, and can be understood as a constructive interference of SU(2) potentials. In the
other case, where the complete potential is a sum over many
similar-weighted SU(2) potentials with different
periodicities, we observe a destructive interference of
SU(2) potentials: The complete potential will have competing
local minima, such that a (strong) first order phase
transition will be induced. The two opposite cases are
indeed observed in E(7) vs. SU(N) (N>3), cf. fig.~\ref{interference}.

\subsection{Outlook: Phase transition in 2+1 dimensions and thermodynamics in the deconfined phase}
In this section we have presented results on a first step 
in the evaluation of the QCD phase diagram from first principles, namely the
determination of the critical temperature and the order of
the deconfinement phase transition in the limit of
infinitely heavy quarks. 
Since the standard order parameter, the Polyakov loop $\langle L[A_0] \rangle$ requires knowledge on 
all $n$-point correlation functions of the gauge field, its
determination is challenging. Using a related order
parameter, $L[\langle A_0 \rangle]$, circumvents this
difficulty. 
Within the flow-equation setup we evaluate 
$\langle A_0 \rangle$ by numerically determining the minimum
of the effective potential. Accordingly only the gluon and the ghost propagator enter the determination of the phase boundary in this setting.

In our approximation the complete potential is a superposition of SU(2) potentials with different periodicities. The group structure of the gauge group enters here, since these are given by the  eigenvalues of the Cartan generators in
the adjoint representation.  
Hence we can easily evaluate our
order parameter for quark confinement for different
gauge groups, after performing the numerical
evaluation of the temperature-dependent effective
potential for SU(2). Thus we can go beyond gauge groups studied in lattice simulations and consider SU(N) for 2$\leq$ N $\leq$12, Sp(2) and E(7). This allows
to investigate which property of Lie groups determines the order of the phase
transition. Our findings, showing first order phase
transitions for all investigated gauge groups except SU(2),
indicate that the size of the gauge group, and not the
center of the group, is the decisive quantity. This supports the conjecture that a large mismatch in the dynamical degrees of freedom, resulting from a large gauge group, induces a first order phase transition. We further relate the approach to the phase transition to the structure of the gauge group, by establishing a picture of destructive and constructive interference for the order-parameter potential.

In (2+1) dimensions, there are less transversal degrees 
of freedom. Thus, for a given gauge group,
the mismatch in the dynamical degrees of freedom is reduced
when passing from (3+1) to (2+1) dimensions. Accordingly one might expect more groups to show second order phase transitions, which is indeed supported by lattice calculations: SU(3) and Sp(2) have second order phase transitions in (2+1) dimensions, whereas the order of the
phase transition in SU(4) is still under debate. 

To implement the order parameter $L[\langle A_0\rangle]$ 
in (2+1) dimensions, several changes are necessary in
comparison to the (3+1) dimensional case: The dimensional
factors in the momentum integration in the effective
potential clearly change. Furthermore the Landau
gauge propagators themselves are dimension-dependent.
Implementing (2+1) dimensional propagators\footnote{We are
grateful to Daniel Spielmann who provided lattice data on
SU(2) propagators in (2+1) dimensions.} in a (2+1)
dimensional potential however fails to reproduce the second
order phase transition expected in SU(3). This is due to the back-coupling of the potential $V^{(2)}$
(cf. \Eqref{Polyakovprops}), which
is neglected in these studies. It turns out to be essential in the
vicinity of a second order phase transition, as there long-range correlations are important. In
contradistinction to SU(2), where the group structure,
namely a single pair of eigenvalues, induces a second
order phase transition in (2+1) and (3+1) dimensions, the
$V^{(2)}$-term is crucial for larger groups. Hence it should
be included in studies of (2+1) dimensional Yang-Mills theories for SU(3) and SU(4).

Center-free gauge groups are another interesting field to
study. Lattice studies suggest that, e.g. for G(2)
the Polyakov loop exhibits a jump \cite{Pepe:2006er,Wellegehausen:2009rq},
although no symmetry exists to be broken at this
transition. Furthermore G(2) is of particular interest, as
its generators fall into two categories: A subset is given
by the SU(3) generators. Introducing a Higgs mechanism for
the other gluons allows to interpolate between the
center-free group G(2) and SU(3). Furthermore G(2) mimics
QCD with dynamical quarks in the following sense: The charge of a
static quark can be screened by gluons in G(2), such that
string-breaking can occur even without dynamical quarks. In our study we also observe a fast change in the Polyakov loop for G(2) at the critical temperature. However, using SU(3) propagators in the (3+1) dimensional 
G(2)-potential yields an order parameter $L[\langle A_0
\rangle]$ that is slightly negative in the vicinity of $T_c$. As this
quantity is an upper bound for the expectation value of the
Polyakov loop, it would follow that also $\langle L[A_0]
\rangle <0$. This is in contrast to results on the lattice,
where for $T \leq T_c$, $\langle L[A_0]\rangle$ assumes a
small, non-zero positive value.
A first improvement consists in using G(2) Landau gauge propagators. 
As these are available in (2+1) dimensions, we implement the
evaluation of the effective potential for G(2) there\footnote{We are grateful to Axel Maas, who provided lattice
data on G(2) Landau gauge propagators.}. This does not
ameliorate our findings. The deviation in our result from the expectation can be traced back to the observation that in the vicinity of $T_c$ there are several local minima, where the value of the potential is very close to the value at the global minimum. Therefore our study of G(2) requires a very high level of quantitative precision, such that also subleading terms such as the $V^{(2)}$ term are crucial here. We conclude that the
$V^{(2)}$ term should be included here in order to recover an order parameter that is positive everywhere in agreement with the expectation.

Since it is notoriously difficult to do calculations in full QCD in the infrared, many models, such as the Polyakov-Quark-Meson model have been devised to study the QCD phase diagram. A crucial input here is the effective potential for the Polyakov loop. This can be deduced from our order-parameter potential and can be used as an input for model calculations, see, e.g. \cite{Braun:2011fw}. 

We can furthermore derive thermodynamic properties from our setting, since we can directly calculate the free energy 
$F$ which is related to the partition function by $F = - T\,
\ln Z= T \Gamma_{min}$. This will allow to determine quantities such as the pressure as a function of the temperature. Here, the full temperature-dependence of the propagators presumably plays a crucial role.

\chapter{Asymptotically safe quantum gravity}\label{AS}
\section{Asymptotic safety: A UV completion for gravity?}
\subsection{The problem with quantum gravity}

In this chapter we will study a particular scenario for the
quantisation of gravity. The unification of the two
"pillars" of modern theoretical physics, namely General
Relativity, and quantum theory, to a theory that
incorporates quantum effects in gravity, naively pictured as
fluctuations of the space-time geometry itself, is a
long-standing problem.\\
In view of the successful application of perturbative
quantisation techniques to the standard model of particle
physics a first attempt at quantising gravity might be along the
same lines. However one can easily see why this is bound to
fail:
In a (local) gravity theory built from
diffeomorphism-invariant quantities like $R$ (for our
conventions regarding gravity see
app.~\ref{gravvariations}), contractions of $R_{\mu \nu}$
and $R_{\mu \nu \rho \sigma}$, the propagator will be $\sim
\frac{1}{(p^2)^n}$, where $n$ is the power of curvature
(e.g. $R^n, (R_{\mu \nu}R^{\mu \nu})^{n/2}$etc.). The
vertices will then be proportional to $(p^2)^n$,
irrespective of the number of fields that are attached to
them\footnote{Here we neglect the tensor structure of the propagator and assume that we quantise linearised metric fluctuations around a flat background, so that we can make the transition to Fourier space.}. It therefore follows that the superficial degree of
divergence in a gravity theory is given by
\begin{equation}
\mathcal{D}= d L +2 n V -2 n I,
\end{equation}
in $d$ dimensions, where of course $L$ is the number of
loops, $V$ the number of vertices and $I$ the number of
internal lines. Using the topological relation $I = V+L-1$,
we arrive at
\begin{equation}
\mathcal{D}= (d-2n)L +2 n. \label{degofdiv}
\end{equation}
Accordingly, the theory is presumably
perturbatively renormalisable  if it is built from
objects of the form $R^{d/2}$ in $d$ dimensions. This
suggests that Einstein gravity is renormalisable in 2
dimensions. Indeed explicit calculations in 4 dimensions
show the necessity to include further counterterms at
two-loop order (or even at one-loop order if matter is
included). Perturbative renormalisability of Einstein
gravity in 4 dimensions therefore fails
\cite{'tHooft:1974bx,Deser:1974zza,Deser:1974zzc,
Goroff:1985th,vandeVen:1991gw}. 

In the language of fixed points of the RG flow the canonical
dimensionality of the Newton coupling and the cosmological
constant implies that the Gau\ss{}ian fixed point is UV
repulsive in the first direction. It follows that, although
the observable values of the couplings in the IR are rather
small, perturbative quantisation schemes fail. For some
reason the universe seems to sit on a trajectory that passes
very close to the GFP in the IR. Yet it is not
possible to construct a trajectory that terminates in the
GFP in the ultraviolet. Consequently the perturbative
quantisation fails. 

One might then conclude that a quantum theory of gravity
cannot successfully be constructed as a theory of metric
fluctuations within the framework of local QFT, and indeed
many candidate theories for quantum gravity start from
different assumptions, e.g. by including additional degrees of
freedom, departing from the assumption of locality, or
describing the quantum excitations of space-time in an
altogether different language than standard QFT.\\
Since virtually no experimental guidance is currently
available in the realm of quantum gravity, there is a large
number of open questions concerning the physical as well as
mathematical aspects of the quantisation procedure and its
consequences in gravity, see, e.g. \cite{Sorkin:1997gi}:\\
Starting with a quantisation procedure, one has a choice between the canonical framework, and the path-integral quantisation. The former is mainly being followed in Loop Quantum Gravity (for reviews see, e.g. \cite{Rovelli:2011eq, Oriti:2009zz}), and allows for the construction of a kinematical Hilbert space. Defining the dynamics of the theory then becomes a major challenge. 

Here we will focus on the path-integral framework, or
sum-over-histories approach,  where one is interested
in evaluating the following schematic expression
\begin{equation}
Z=\int_{\rm geometries} \, e^{i\,S}, \label{statementofintent}
\end{equation}
that is some kind of weighted sum over geometries, but is
at this point of course no well-defined expression.
At this stage, we face a physical choice:
Geometries may be smooth Riemannian geometries, or discrete
ones (see, e.g. the causal set programme
\cite{Bombelli:1987aa,Sorkin:2003bx,Dowker:2005tz}, as well as spin foams, see, e.g. \cite{Rovelli:2011eq, Oriti:2009zz}). Indeed
the Planck length $L_{\rm Pl} \approx 1.6 \, 10^{-35}m$
may play the role of a fundamental physical cutoff, such
that no smaller lengths can exist, thus automatically regularising quantum gravity as well as quantum field theories\footnote{Such a fundamental physical cutoff scale might then also help to resolve the singularities in cosmological as well as black-hole solutions of Einstein's equations. Within a setting where no such cutoff scale exists, singularity resolution might also be possible, but is more challenging, see, e.g. \cite{Bonanno:2000ep}.}. Indications for such a scenario are found, e.g. in Loop Quantum Gravity, where the area as well as the volume operator are gapped in the kinematical Hilbert space, see, e.g. \cite{Rovelli:2011eq}. 

Apart from introducing discretisation as a property of
physics, one may also consider it just as a technical device
in order to regularise the expression
\Eqref{statementofintent}. In this case it is necessary to
take a continuum limit, very similar to lattice gauge
theory, in order to arrive at physical results. This
approach is being followed in matrix models for two-dimensional gravity, see, e.g.
\cite{Ginsparg:1991bi,Ambjorn:1994yv}, as well as Euclidean
and Causal Dynamical Triangulations \cite{Ambjorn:1998xu},
see also the review article \cite{Ambjorn:2010rx}.

Furthermore it is unclear whether the fluctuations that are
being summed over in \Eqref{statementofintent} should also
include fluctuations in topology, and if the number of
space-time dimensions and the signature of the metric should
be fixed by hand or emerge dynamically. An example where
apart from the signature all these parameters are allowed to
fluctuate is constituted by causal set theory
\cite{Bombelli:1987aa,Sorkin:2003bx,Dowker:2005tz}.
Naturally an approach which allows "quantum space-time" to
depart so radically from classical space-time is severely
challenged when it comes to the recovery of a semi-classical
limit, see also \cite{Oriti:2009zz}.

Many of these aspects can be subsumed in the physical question: What are the degrees of freedom of quantum gravity?

Here no experimental guidance is available to answer that
question at very high energies\footnote{Let us emphasise
that proposals to study quantum gravity by means of analogue
models, where, e.g. effective horizons for sound waves can
be studied in laboratory experiments see, e.g.
\cite{Schutzhold:2008zzb} only allow to access
\emph{kinematical} aspects of the theory. For instance, the
effect of an underlying discreteness of space-time on
Hakwing radiation can be probed in these models. Such
analogue systems do however not provide a doorway to access
questions on the \emph{dynamics} of quantum gravity.}.
Therefore many possible answers, ranging from excitations of
a string to discrete elements of a causal set have been
given. Being guided by examples such as QCD, where the
effective low-energy degrees of freedom (hadrons) are rather
different from the high-energy degrees of freedom (quarks
and gluons), one may argue that the metric might only be an
effective degree of freedom. This assumption is being tested
within the asymptotic-safety scenario, as it opens a
possibility to construct a local, continuum quantum field theory of the
metric within the path-integral framework. A failure of this
scenario accordingly would suggest that the metric is not
the fundamental degree of freedom of quantum gravity.

If one chooses to stay within the framework of local quantum
field theory, \Eqref{degofdiv} suggests one possible way to
build a perturbatively renormalisable theory of gravity in
four dimensions, by including higher-derivative terms up to
four powers in momentum, i.e. two powers in the curvature.
As shown in \cite{Stelle:1976gc,Julve:1978xn,Fradkin:1981iu,
Fradkin:1981hx,Avramidi:1985ki}, this theory is
perturbatively renormalisable, as it is asymptotically free
in the higher-derivative couplings. It however has a
different problem, namely unitarity. The $1/ p^4$ propagator
can be rewritten as a sum of two $1/p^2$ propagators, one of
which has a mass and a negative residue at the pole, i.e. it
is a ghost and therefore presumably spoils the unitarity of
the theory in a perturbative framework, see
\cite{Stelle:1977ry}. 

To summarise the results from applying perturbation theory to the quantisation of gravity, we find a non-renormalisable theory at lowest derivative order and a non-unitary theory at higher derivatives, thus we conclude that the construction of a quantum field theory of the metric might necessitate a non-perturbative framework.\\
We now make a physical assumption about quantum gravity,
namely that its dynamics will be governed by a NGFP instead
of a GFP in the far UV, which implies that residual
interactions dominate the high-energy behaviour. Of
course the validity of this claim must be supported by
explicit calculations. Within this scenario the perturbative
non-renormalisability of a QFT with the Einstein-Hilbert
action as the fundamental action is interpreted simply as a
breakdown of perturbation theory. In analogy to QCD, where
the existence of a strongly-coupled regime in the IR is
simply inaccessible to perturbation theory, but nevertheless
a valid description of physics, a non-perturbatively
constructed QFT of the metric could be valid, and requires
the existence of a NGFP.

The approach to quantise gravity as a local quantum field
theory within the path-integral framework is very
conservative, as it avoids the assumption of new physical
degrees of freedom\footnote{Note that the degrees of freedom are not
completely specified, once the field is specified, since the
number of initial conditions needed to solve the equations
of motion depends on the kinetic operator of the theory. In
this sense higher-derivative gravity has more degrees of
freedom than Einstein-Hilbert gravity. This choice is
actually a priori open in the asymptotic-safety scenario,
since it is not clear which operators are part of the
classical action.}, or the
introduction of a wholly new framework. Without being
prejudiced about the question if this route to the
quantisation is the one that "nature chose", it is still
highly interesting to examine where the framework of local
quantum field theories and the path integral really breaks
down, and if it cannot be pushed to incorporate the
quantisation of gravity with the metric as the fundamental
field\footnote{Note that in the
strongly-interacting regime it might happen that within the
path integral it is possible to map the metric degrees of
freedom onto different degrees of freedom, in which it is
considerably more natural to describe gravity. However the
basic variable in the path integral would still be the
metric, and no new degrees of freedom would be introduced
here, one might merely realise that a change of variables in
the path integral is a more convenient description.}.

\subsection{The asymptotic-safety scenario}\label{AS_intro}

As stressed by Weinberg, the concept of perturbative renormalisability is very restrictive, and we may overlook valid quantum field theories by making perturbative renormalisability a central requirement. Indeed "non-renormalisable theories are just as renormalisable as renormalisable theories, as long as we include all possible terms in the Lagrangian" \cite{Weinberg_book}. As, disregarding the issue of anomalies, counterterms being required for renormalisation are always subject to the symmetry constraints of the theory, we immediately see that including all terms allowed by symmetry will provide an appropriate term for each possible divergence. As each term comes with its own coupling constant, it might at a first glance seem to be difficult to arrive at a predictive theory in such a way.

Nevertheless perturbatively non-renormalisable theories may
indeed be non-perturbatively renormalisable and
parametrised by just a finite number of free parameters.

Asymptotic safety, as proposed in \cite{Weinberg:1980gg},
see also
\cite{Weinberg:1976xy,Weinberg:2009ca,Weinberg:2009bg,
Weinberg:2009wa}, is a generalisation of the concept of
asymptotic freedom, where a theory is consistent up to
arbitrarily high momenta as it sits on a trajectory that is
connected to the Gau\ss{}ian fixed point. Similarly a
trajectory that runs into a non-Gau\ss{}ian fixed point can be used to define a
theory that is valid up to arbitrarily high momenta (see
sec.~\ref{funtheories})\footnote{Note, however, the
following subtlety: A theory may be defined exactly at the
NGFP, in which case one may take the UV-cutoff $\Lambda
\rightarrow \infty$. Then the infrared physics is also
described by the fixed-point values of the couplings. If the
theory is supposed to have a non-trivial scale-dependence,
we have to start arbitrarily close to the fixed point in the
UV and then flow towards the region of theory space which is
compatible with the values of the couplings that we infer
from experiments at low energies. This implies however, that
we have to define the theory at some, arbitrary high, but
still finite cutoff $\Lambda$ with coupling values which are
very close to the fixed-point values. Although for practical
purposes this does not play a role, as $\Lambda$ may indeed
be arbitrarily high, and the flow may stay in the vicinity
of the fixed point for many orders of magnitude in the
momenta, this still leaves open the question of what
determines the initial values of the couplings at
$\Lambda$.}. 
Such a theory, where the UV-cutoff can be removed without running into a divergence is called a fundamental theory. In particular, theories which are fundamental and also non-trivial in the infrared, provide building blocks for a complete and consistent description of high-energy physics.

The question of whether such a theory is predictive is then analogous to a theory that is defined at a Gau\ss{}ian fixed point in the UV: If there is a finite number of UV-relevant operators, this theory is predictive. Of course predictivity requires the couplings to be precisely tuned: The theory has to sit \emph{exactly} on the critical surface in order for it to hit the NGFP in the UV. One may well wonder why "nature" would "choose" to place a theory exactly on the critical surface, and expect an even more microscopic theory to account for this. On the other hand the fine tuning is not more severe than for perturbatively renormalisable theories, where all couplings with canonical negative mass dimension are set to zero. The principle that one invokes here, perturbative renormalisability, is not a directly physical requirement.

Hence a theory is asymptotically safe if it has a
 NGFP with a finite number of
relevant directions\footnote{Note that it may actually be
possible, by a clever change of variables, to formulate the
theory in such a way, that it admits a Gau\ss{}ian fixed
point in the new formulation. Then the asymptotic-safety
scenario, in changed variables, could be investigated by the
same techniques as, e.g. Yang-Mills theories at high
energies.}. 
One can find an argument why only a finite number of
relevant directions should be expected at a NGFP, if,
indeed, it exists: The canonical dimension of the couplings
of all diffeomorphism invariant operators at higher order in
momenta than the Einstein-Hilbert term is zero or negative
and decreases further with the number of momenta. Therefore
the contribution to the critical exponent resulting from
quantum fluctuations would have to grow arbitrarily large in
order to outweigh the canonical dimension and turn
infinitely many power-counting irrelevant into relevant
operators. 

The asymptotic-safety scenario is of course not restricted to the case of
gravity. In particular, the Higgs sector of the Standard
Model may also be asymptotically safe \cite{Gies:2009hq,
Gies:2009sv, Scherer:2009wu}, as may, e.g. also be the
non-linear sigma model, see, e.g.
\cite{Percacci:2009dt}, or the
Gross-Neveu model in 3 dimensions \cite{Braun:2010tt}. In
the last case the interpretation of the NGFP is mainly in
terms of quantum phase transitions. Furthermore SU(N) gauge
theories can be asymptotically safe in 5 dimensions
\cite{Gies:2003ic}.

Let us remark on the scenario and its relation to other frameworks for quantum gravity:
\begin{itemize}
\item Note the crucial difference between the asymptotic-safety scenario and the framework of effective theories, that can be applied to General Relativity straightforwardly. Here one does not make any assumption about the UV completion of gravity, but uses that in the low-energy regime one can calculate the effect of quantum fluctuations in gravity to any desired precision with just a finite number of free parameters \cite{Burgess:2003jk}, assuming a decoupling between high and low energies, such that the low-energy dynamics is independent of the UV completion (see, e.g. \cite{Donoghue:1993eb} for a determination of the quantum corrections to the Newtonian gravitational potential). 

The asymptotic-safety scenario goes beyond such a framework
in that it provides for a UV completion of gravity. This
framework will then allow to test if the decoupling
assumption of effective theories indeed holds.\newline
Nevertheless the $\beta$ functions for the gravitational
couplings evaluated with the help of the Wetterich equation
are valid generally: One may either be interested in the
asymptotic-safety scenario, where fixed-point conditions for
the $\beta$ functions will be of particular interest. On the
other hand one may assume that an effective description of
quantum gravity in terms of metric fluctuations will hold
below some scale $k_0$ (e.g. the Planck scale), irrespective
of what the UV completion is. Such an assumption can also be
valid if the UV completion is formulated in a totally
different framework than local QFT, and, e.g. assumes a
physical discreteness of space-time at small scales. The
microscopic theory, whatever it may be, and how its degrees
of freedom look like, then determines the initial condition
at $k_0$ in the theory space that we are working in. Below
$k_0$ our parameterisation holds, such that it is possible
to follow the flow further towards the IR and study the
compatibility of the initial conditions with experimental
observations of gravity as well as matter. In such a way it
is possible to restrict possible UV completions of gravity,
as some of them may determine initial conditions that will
lead to a clash with observations at lower scales. (Of
course the non-trivial step here is to determine the initial
condition from the microscopic theory.) This allows to study certain aspects of other UV completions for gravity besides the asymptotic-safety scenario in our framework.

As experimentally, in the IR, it is not possible to distinguish a trajectory starting close to the NGFP on the critical surface from one that is just slightly off the critical surface (as the irrelevant couplings are drawn towards the critical surface towards the IR), it is difficult to distinguish between an asymptotically safe trajectory and one that arises from a different UV completion and has initial conditions such that it sits on a trajectory slightly off the critical surface (see also \cite{Percacci:2010af}). Thus for a large range of scales the RG flow for other microscopic theories of gravity may be very similar to the RG flow within the asymptotic-safety scenario.

\item Formulating the theory in terms of the metric, or in terms of the vielbein and the spin-connection (which allows for the introduction of non-zero torsion), are inequivalent choices in the quantum theory. Recent results suggest that both formulations may be asymptotically safe \cite{Daum:2010qt}.

\item Since a theory defined at a NGFP is scale-free, it can
be analysed by conformal-field-theory techniques. In
particular, it may be conceivable, that a duality
relation similar to the  AdS-CFT conjecture also applies
to the, as yet unknown, fixed-point theory. Thus it might
be possible to understand the scenario, or aspects of it,
with the help of perturbative techniques applied to a dual theory.

\item There might be a further possibility how asymptotically
safe quantum gravity can be linked to other UV completions
for gravity: Fixed points in theory space describe
scale invariant theories. Physically, scale
invariance may be realised in second-order phase
transitions, where the correlation length diverges, and
therefore the system becomes scale-free. Hence the fixed
point underlying the asymptotic-safety scenario might
potentially be interpreted as being related to a phase
transition. Here, a connection to other approaches to
quantum gravity may be found: Several candidate theories to
quantum gravity rely on a discretisation of space-time, and
introduce some kind of fundamental building blocks. These do
often not have a direct geometrical interpretation, as for
instance the fields in group field theory, which live on a
group manifold, see, e.g.
\cite{Freidel:2005qe,Oriti:2006se}. One may then postulate
that at very high energies, these microscopic degrees of
freedom exist in some kind of "pre-geometric" phase. A
second order phase transition may then be related to a kind
of "condensation" mechanism of these fundamental degrees of
freedom \cite{Oriti:2007qd}. The "condensed" phase on the other side of the
phase transition might then be describable by a quantum
field theory of the metric, and geometric notions would make
sense in this phase. Such a scenario would receive support
if the critical exponents found at the NGFP in
asymptotically safe quantum gravity would agree with
critical exponents evaluated in the discrete theory. Of
course these ideas are highly speculative, but show that
seemingly very different approaches to quantum gravity may
in some sense be different sides of the same picture. 

Let us nevertheless emphasise that, while the recovery of a
regime where space-time can be described in terms of a differentiable
manifold with a dynamical metric is a crucial challenge for
the discrete approaches to quantum gravity, the
asymptotic-safety scenario as it stands can also be taken as
complete. Here one assumes that the metric carries the
fundamental degrees of freedom of gravity up to
arbitrarily high energies, and the UV behaviour of the
theory is simply governed by the fixed point. The question,
whether this scenario, or the scenario with a phase
transition to a different (possibly non-geometric) phase is
realised is ultimately an experimental question.
\end{itemize}

Let us briefly discuss the implications of the
fixed-point requirement in gravity. Here, the Newton 
coupling $G_N$ is not dimensionless in four dimensions.
Accordingly, the $\beta$ function for the dimensionless
Newton coupling $G$ will take the form
\begin{equation}
 \beta_{G}= 2 G + G \eta_N,
\end{equation}
 where the factor $2 G$ reflects the dimensionality, and $\eta_N$ is a function of all couplings of the theory, which is non-zero as an effect of quantum fluctuations. The fixed point condition for a NGFP clearly requires
\begin{equation}
 \eta_N= -2.
\end{equation}
The non-trivial fixed point hence emerges due to a balancing between dimensional and quantum scaling.

The dimensionful coupling at high momenta, in the vicinity of the fixed point, can then be rewritten with the fixed-point value $G_{\ast}$ as
\begin{equation}
 G_N \sim \frac{G_{\ast}}{k^2}.
\end{equation}
Taking $k^2 \rightarrow \infty$ results in the dimensionful
gravitational coupling going to zero. This effect is due to
the dimensionality, and should be contrasted with the case
of Yang-Mills theory, where the coupling goes to zero in
the UV as it is dimensionless and attracted by a
Gau\ss{}ian fixed point. The fixed-point requirement
$\eta_N=-2$ can then be interpreted as a sign of a dynamical
reduction to 2 dimensions in the vicinity of the fixed
point, see, e.g. \cite{Lauscher:2005qz,Lauscher:2005xz}.

The next important test that such a candidate for a
fundamental theory has to pass is of course the connection
to observations, since a theory may be completely consistent
in itself, but simply not realised in nature. The minimal
requirement here is that within the experimentally
accessible range of energies the couplings coincide with
their values inferred from measurements. In the case of gravity this implies
that there has to be a regime where $G_N =\rm const$ and also the dimensionful cosmological constant is non-zero:
$\bar{\lambda} = \rm const$. For the dimensionless couplings this
translates into $G \sim \frac{1}{k^2}$ and $\lambda
\sim k^2$ and hence $G \sim \frac{1}{\lambda}$. This
feature is indeed observed in some trajectories passing close
to the Gau\ss{}ian fixed point in the Einstein-Hilbert
truncation, cf. \cite{Reuter:2004nx}, see also sec.~\ref{EHtrunc}.

\subsection{The search for asymptotic safety with the Wetterich equation}

First studies of asymptotic safety in gravity involved a
perturbative expansion in $2+\epsilon$ dimensions \cite{Christensen:1978sc,Gastmans:1977ad},
and a $\frac{1}{N}$ expansion in matter fields
\cite{Smolin:1981rm}, where the large number of matter
fields allows to discard metric fluctuations. These
calculations constitute evidence for the realisation of the
asymptotic-safety scenario. 
In a symmetry-reduced setting, where metric fluctuations are severely restricted due to symmetry, a NGFP can also be found \cite{Niedermaier:2003fz,Niedermaier:2009zz,Niedermaier:2010zz}. Further, numerical simulations of the gravitational path integral also find support for the scenario, see, e.g. \cite{Hamber:2009zz,Ambjorn:1998xu,Ambjorn:2010rx}.

The Wetterich equation is a natural tool to search for a
NGFP as a UV completion of gravity, as it allows to search
the theory space for fixed points also in the
non-perturbative regime. Following the pioneering work of M.
Reuter \cite{Reuter:1996cp}, extensive work on this topic
has been done
\cite{Dou:1997fg,Souma:1999at,Lauscher:2001ya,
Lauscher:2001rz,Reuter:2001ag,Lauscher:2001cq,
Lauscher:2002sq,Reuter:2002kd,Percacci:2002ie,
Percacci:2003jz,Reuter:2003yb,Litim:2003vp,Percacci:2004sb,
Bonanno:2004sy,Lauscher:2005xz,Reuter:2005bb,Percacci:2005wu, Fischer:2006fz,Litim:2006dx,Fischer:2006at,Codello:2006in,
Reuter:2006qh, Reuter:2006zq,Codello:2007bd,Machado:2007ea,
Reuter:2008wj,Reuter:2008qx,Codello:2008vh,Manrique:2008zw,
Benedetti:2009rx,
Reuter:2009kq,
Machado:2009ph,Manrique:2009tj,Benedetti:2009gn,
Benedetti:2009iq,Narain:2009qa,
Eichhorn:2009ah,Eichhorn:2010tb,Groh:2010ta,
Manrique:2010am,Percacci:2010yk,Daum:2010qt,Manrique:2011jc}. Reviews can be found in
\cite{Lauscher:2007zz,Reuter:2007rv,Percacci:2007sz,
Niedermaier:2006ns,Litim:2008tt,Niedermaier:2006wt,
Litim:2011cp}. For work on the phenomenological
applications in astrophysics and cosmology, see, e.g.
\cite{Bonanno:2009nj} and references therein.

Technically, the use of the flow equation in gravity relies
on the background field method \cite{Abbott:1980hw}, which is also extensively used in Yang-Mills theory, since it allows to define a gauge-invariant effective action.
In gravity the use of this method is mandatory, since a metric is necessary
to define a scale with respect to which quantum fluctuations
are suppressed. Said in
different words, the notion of coarse graining is one which
presupposes the existence of some external scale, which
allows to distinguish coarse-grained from fine-grained
geometries. Solving the quantum equations of motion
defines the physical expectation value of the metric which
in turn can be used to define a notion of lengths. In the
case of the effective average action such a construction would
presuppose that one solves the quantum equations of motion
before one even has integrated out all quantum fluctuations
to arrive at the effective action. Clearly for technical reasons some external
notion of length scales is therefore necessary, with respect to which
the coarse-graining procedure can be defined. Accordingly the
background field method is applied which allows not only to
define a gauge-invariant effective action, but also provides
a background metric with respect to which one can define
low- and high-momentum modes. This entails that a
straightforward implementation of the requirement of
background independence is impossible. 
However one may work on a fixed background, using it only as
a technical tool, and making sure that none of the physical results
depend on the choice of background. This approach
circumvents some of the problems of strictly
background-independent approaches\footnote{Challenges
arise when no
background geometry at all is admitted, and gravity has to
be rephrased as a purely relational theory. 
Most of these approaches are then formulated in terms of
discrete variables, where the discreteness is either
fundamental, or some kind of continuum limit needs to be
taken. In either case, the recovery of a regime where small
fluctuations around a fixed, smooth background geometry
parametrised by a Lorentzian metric describe the dynamics
appropriately, becomes a highly non-trivial task.}, and may
be termed "background-covariant". The application of the
Wetterich equation to investigate the asymptotic-safety
scenario is indeed such an approach.
Here we have a more subtle realisation of background
independence, in the sense that the background is just a
technical necessity. If physical quantities of the resulting
quantum gravity scenario turn out to be independent of the
choice of background, this theory should also be considered
background independent. One may then argue that the
background dependence is a spurious property of our
formulation of the theory, but not one of its physical
properties.

In particular the use of the background method does not
imply that only small, linearised metric perturbations
around a fixed physical background are considered. The
background is simply a technical tool and can be chosen
arbitrarily. Furthermore the fluctuations around it are not
restricted to be small, hence the use of the background
field method does not presuppose that gravity can be
quantised perturbatively, with linearised excitations.

The use of the flow equation in the search for a NGFP is
possible since the microscopic action does not enter the
Wetterich equation. It only serves as an initial condition
when the flow is integrated to $k \rightarrow 0$. Therefore
in circumstances when the flow equation is used in the
direction of $k \rightarrow \infty$, to search for an
ultraviolet completion of the theory, the fixed-point action
$\Gamma_{k \rightarrow \infty}$ is a genuine prediction of
the theory. This distinguishes the construction of a quantum
theory of gravity with the help of the asymptotic-safety
scenario in the framework of the Wetterich equation from
other approaches to quantum gravity, where the microscopic
action is one of the axioms of the theory.

As detailed in sec.~\ref{truncation} theory space has to
be truncated for practical calculations. In the case of
gravity, theory space is spanned by directions which are
defined by diffeomorphism invariant quantities such as $\int
d^d x\sqrt{g}\, R^n$,$\int d^d x\sqrt{g}\, R_{\mu \nu}R^{\mu
\nu}$ etc. After gauge fixing operators depending on the
background metric as well as the full metric enlarge the
theory space considerably. Accompanying the gauge fixing
also operators involving Faddeev-Popov ghosts have to be
taken into account.
Terms that depend on background field quantities only do not
couple into the flow, as $\Gamma_k^{(2,0)}[h, \bar{g}]=0$ on
this subspace, however their flow is being driven by the
other operators.

The truncations that have been investigated until now
include extensive studies of the Einstein-Hilbert truncation
\cite{Reuter:1996cp,Dou:1997fg,Souma:1999at,Reuter:2000nt,
Lauscher:2001ya,Lauscher:2001cq,Reuter:2001ag}, truncations
involving higher terms in the curvature scalar $R$
\cite{Lauscher:2002sq}, in particular polynomial truncations
up to $R^8$
\cite{Codello:2007bd,Codello:2006in,Codello:2008vh,Machado:2007ea} and
truncations involving the square of the Weyl tensor
\cite{Benedetti:2009rx,Benedetti:2009iq}. The inclusion of
matter has also been studied  \cite{Percacci:2002ie,
Percacci:2003jz,Zanusso:2009bs, Narain:2009gb, Daum:2009dn, Daum:2010bc,Vacca:2010mj,
Folkerts:2011jz, Harst:2011zx}; for more details see
chap.~\ref{ferminAS}. 
 First truncations distinguishing between the background metric $\bar{g}_{\mu \nu}$ and the full metric $g_{\mu \nu}$ have included the Einstein-Hilbert term for the metric $g_{\mu \nu}$ as well as a background metric Einstein-Hilbert sector, and also bimetric terms \cite{Manrique:2010am,Manrique:2010mq}. 
 
In total these studies provide rather convincing evidence
for the possible realisation of the asymptotic-safety
scenario in quantum gravity: All truncations investigated so
far show a NGFP. Although the fixed-point values of the
couplings vary, the fixed-point value of Newton's coupling
is always found to be positive. This is a crucial
requirement, as a negative value of Newton's coupling would
lead to an unstable theory. On the other hand the
microscopic value of the cosmological constant is not
restricted by observations and might be either positive or
negative; it is only the IR value that is restricted by observations. Furthermore the number of relevant directions has
been found to be $\leq 3$ in most of these studies. In
particular in polynomial truncations involving the curvature
scalar $R$, quantum fluctuations change the
power-counting marginal operator $R^2$ into a relevant
operator, but all power-counting irrelevant operators $R^n$
with $n \geq 3$ stay irrelevant. This is in agreement with
the expectation that quantum fluctuations only turn a finite
number of power-counting irrelevant operators into relevant
ones.

In the case of the inclusion of higher-derivative terms, a
challenge is posed to the asymptotic-safety scenario by
unitarity. From the point of view of perturbation theory,
higher-derivative terms directly lead to a massive
propagator with a negative residue at the pole, i.e. a
ghost (not to be confused with Faddeev-Popov ghosts). This leads to a violation of unitarity. Within a
non-perturbative setting, this might be evaded, since the
mass of the negative-norm state is a function of the
running couplings, hence it is scale-dependent itself. The
mass at a momentum scale $p^2$ may therefore be such that
the full propagator $\left(\Gamma_k^{(2)}(p^2)\right)^{-1}$
evaluated at $p^2$ does not have a pole. Depending on its
scale dependence, the pole may, e.g. be shifted to arbitrary
high energies (see comments in
\cite{Lauscher:2002sq,Niedermaier:2006wt,
Percacci:2007sz,Benedetti:2009rx}). Let us stress that a
complete answer to this problem can only be given after a
full determination of the fixed point and the trajectories
emanating from it.  

We now turn to a first calculation within the FRG framework for quantum gravity, namely the study of the Einstein-Hilbert truncation. This truncation has been studied very extensively \cite{Reuter:1996cp,Dou:1997fg,Souma:1999at,Reuter:2000nt,Lauscher:2001ya,Lauscher:2001cq,Reuter:2001ag}. Here, we add further to the evidence for the existence of the NGFP, as we study a further combination of gauge parameters and regularisation scheme. We also introduce a different computational scheme which does not rely on heat-kernel methods.

\section{Einstein-Hilbert truncation}\label{EHtrunc}
\subsection{Method: The Einstein-Hilbert truncation on a maximally symmetric background}
The Einstein-Hilbert truncation is defined by
\begin{equation}
 \Gamma_k = \Gamma_{k \,\rm EH}+ \Gamma_{k\,\rm gf}+ \Gamma_{k\, \rm gh},
\end{equation}
where 
\begin{eqnarray}
\Gamma_{k\,\mathrm{EH}}&=& 2 \kappa^2 Z_{\text{N}} (k)\int 
d^4 x \sqrt{g}(-R+ 2 \bar{\lambda}(k))\label{GEH},\\
\Gamma_{k\,\mathrm{gf}}&=& \frac{Z_{\text{N}}(k)}{2\alpha}\int d^4 x
\sqrt{\bar g}\, \bar{g}^{\mu \nu}F_{\mu}[\bar{g}, h]F_{\nu}[\bar{g},h]\label{Ggf},
\end{eqnarray}
with
\begin{equation}
 F_{\mu}[\bar{g}, h]= \sqrt{2} \kappa \left(\bar{D}^{\nu}h_{\mu
   \nu}-\frac{1+\rho}{d}\bar{D}_{\mu}h^{\nu}{}_{\nu}
\right). \label{gaugefixing}
\end{equation}
We will focus on $d=4$ in the following. The gauge
fixing depends on two parameters $\alpha$ and $\rho$, which
in principle should be treated as running couplings, too.
Harmonic gauge (also known as deDonder gauge) is realised
for $\alpha=1$, whereas $\alpha =0$ corresponds to the
Landau-deWitt gauge and constitutes a fixed point of the RG
flow \cite{Ellwanger:1995qf,Litim:1998qi}\footnote{For
gravity, the same argument suggesting that $\alpha=0$ is a fixed point
applies as for Yang-Mills theory, see
\cite{Lauscher:2001ya}: Landau-deWitt gauge $\alpha =0$
implements the gauge condition $F_{\mu}=0$ exactly, i.e. by
a delta-function in the path integral. Therefore all modes
that are integrated out always exactly satisfy the gauge
condition, and hence $\alpha=0$ is stable under a change in
the infrared cutoff $k$.}.
In the above, $\kappa= (32 \pi G_{\text{N}})^{-\frac{1}{2}}$
is related to the
bare Newton constant $G_{\text{N}}$. The ghost term with a
wave-function renormalisation $Z_c$ is given by
\begin{equation}
 \Gamma_{k\, \rm gh}= -\sqrt{2}Z_c\, \int d^4x \sqrt{\bar{g}}\,\bar{c}_{\mu} 
\Bigl(\bar{D}^{\rho}\bar{g}^{\mu \kappa}g_{\kappa \nu}D_{\rho}+ \bar{D}^{\rho}\bar{g}^{\mu \kappa}g_{\rho \nu}D_{\kappa}
- 2\frac{1+\rho}{d}\bar{D}^{\mu}\bar{g}^{\rho \sigma}g_{\rho
\nu}D_{\sigma} \Bigr)c^{\nu}.\label{eq:Ggh}
\end{equation}
In this first truncation any non-trivial running in the ghost and gauge-fixing sector is neglected. We therefore set $Z_c=1$ and let $\alpha =\const$ and $\rho = \const$.

Evaluating the right-hand side of the flow equation we have two different strategies at our disposal: Either the spectrum and the eigenfunctions of $\Gamma_k^{(2)}$ are known, then the full propagator can be constructed explicitly in the basis provided by the eigenfunctions. The trace is then implemented straightforwardly as a summation/integration over the discrete/continuous spectrum, including possible trace measure factors corresponding to the density of states with one eigenvalue. If this information is not available, but the heat-kernel trace of the Laplace-type operator $\mathcal{D}$ occurring in $\Gamma_k^{(2)}$, i.e. the trace over  ${\rm exp}(- \mathcal{D} s)$, is known, one can use a Laplace representation of the right-hand side of the flow equation. In this case one rewrites
\begin{equation}
 {\rm tr}\,f(-\mathcal{D}) = \int_0^{\infty}  ds\,
\tilde{f}(s)\,{\rm tr }\,e^{-(-\mathcal{D})s},
\end{equation}
where $\tilde{f}(s)$ is the inverse Laplace transform of $f(\mathcal{D})$. Note that here the spectrum of the differential operator is not required, the reduced information contained in the heat-kernel trace suffices.

In earlier studies the method employing a Laplace transform
and the heat-kernel trace was used (see, e.g. \cite{Reuter:1996cp, Reuter:2007rv, Lauscher:2001ya, Codello:2007bd}). Considering
the flow equation with external fields, such as ghost fields
or fermion fields, is technically rather involved then, and
has only been considered recently
\cite{Groh:2010ta,Benedetti:2010nr}. However here knowledge on off-diagonal heat-kernel traces is necessary, and the calculations can become technically challenging.

Therefore, to project
onto the running of operators containing, e.g. ghosts or
matter fields, a
different technique is useful. Although truncations
involving no running in the ghost sector can be investigated
by both methods, we will introduce the method on the
example of the Einstein-Hilbert truncation, and then use it
for a study of the ghost sector \cite{Eichhorn:2009ah,Eichhorn:2010tb,ghostpaper3} (see also \cite{thesis_M_Scherer}).

In the case of the Einstein-Hilbert truncation a background which allows to distinguish $\partial_t Z_N(k)$ and $\partial_t \bar{\lambda}(k)$ is sufficient. As we require knowledge on the spectrum of $-\bar{D}^2$, we choose a background fulfilling both requirements. Here, it is convenient to choose a maximally symmetric background, where due to symmetry reasons
\begin{equation}
 \bar{R}_{\rho \sigma \mu \nu}= \frac{\bar{R}}{d(d-1)}(\bar{g}_{\rho \mu}\bar{g}_{\sigma \nu}- \bar{g}_{\rho \nu}\bar{g}_{\sigma \mu}),
\end{equation}
where the dimension-dependent factor is fixed by requiring $\bar{g}^{\mu \nu}\bar{R}^{\kappa}_{\,\,\mu \kappa \nu}=\bar{R}$.

The basic classification is then as follows: One distinguishes spaces with $\bar{R}=0$ and positive or negative $\bar{R}$, respectively. For Euclidean signature $\bar{R}=0$ corresponds to $d$-dimensional flat space. Spaces characterised by $\bar{R}<0$ are hyperboloids and $\bar{R}>0$ spheres, which we will concentrate on. Introducing the curvature radius $r$, the following relations hold:
\begin{eqnarray}
 \bar{R}&=& \frac{d(d-1)}{r^2}, \quad \quad \bar{R}_{\mu \nu}=\bar{g}_{\mu \nu}\frac{\bar{R}}{d}, \quad \quad \int d^d x
\sqrt{\bar{g}}= \frac{\Gamma(d/2)}{\Gamma(d)}(4 \pi r^2)^{\frac{d}{2}}.
\label{eq:dspheres}
\end{eqnarray}
To bring the inverse propagator into a form where the only
dependence on the covariant derivative is through the
appearance of a covariant Laplacian $-\bar{D}^2$ we employ a
York decomposition \cite{York:1973ia} of the fluctuation
metric, following \cite{Lauscher:2001ya}:
\begin{equation}
 h_{\mu \nu}= h^{TT}_{\mu \nu}+\bar{D}_{\mu}v_{\nu}+ \bar{D}_{\nu}v_{\mu}+ \bar{D}_{\mu} \bar{D}_{\nu}\sigma - \frac{1}{d}\bar{D}^2 \bar{g}_{\mu \nu}\sigma + \frac{1}{d}\bar{g}_{\mu \nu}h.\label{York}
\end{equation}
Here the following transversality and tracelessness conditions hold:
\begin{eqnarray}
\bar{D}^{\mu}h_{\mu \nu}^{TT}&=&0,\quad \quad \bar{D}^{\mu}v_{\mu}=0 ,\quad \quad \bar{g}^{\mu \nu}h_{\mu \nu}^{TT}=0. \label{transv}
\end{eqnarray}
Similarly, we decompose the ghost and antighost according to
\begin{equation}
 c^{\mu}= c^{\mu \, T}+ \bar{D}^{\mu} c^L \, \, \mbox{ and }  \bar{c}^{\mu}= \bar{c}^{\mu \, T}+ \bar{D}^{\mu} \bar{c}^L, \label{ghostYork}
\end{equation}
with the transversal component $\bar{D}^{\mu}c_{\mu}^T=0$.

This decomposition allows to bring $\Gamma_k^{(2)}$ into a form which only depends on the covariant
Laplacian $\bar{D}^2$ and the curvature tensor and its
contractions. Due to the transversality conditions all
uncontracted derivatives $\bar{D}^{\mu}$ can be traded for a
covariant Laplacian and a curvature tensor, using that for a
vector $V^{\mu}$
\begin{equation}
[\bar{D}_{\mu},\bar{D}_{\nu}]V^{\kappa} = R^{\kappa}_{\,\, \lambda \mu \nu} V^{\lambda}.
\end{equation}
The change of variables  in the form of the York decomposition introduces a Jacobian into the path integral which can be cancelled by the further rescaling (see, e.g. \cite{Lauscher:2001ya})
\begin{eqnarray}
\sigma &\rightarrow& \left(\sqrt{\left(\bar{D}^2\right)^2
+\frac{d}{d-1}\bar{D}_{\mu}\bar{R}^{\mu
\nu}\bar{D}_{\nu}}\right)^{-1}\sigma, \phantom{xxx}
v_{\mu} \rightarrow \left(\sqrt{-\bar{D}^2 - \widebar{\rm Ric}}\right)^{-1}v^{\mu}\\
c_L &\rightarrow& \left(\sqrt{-\bar{D}^2}\right)^{-1}c_L,
\phantom{xxx}
\bar{c}_L \rightarrow \left(\sqrt{-\bar{D}^2}\right)^{-1}\bar{c}_L,\label{rescaling_jaco}
\end{eqnarray}
where $\left(\widebar{\rm Ric}\,v\right)^{\mu}= \bar{R}^{\mu \nu}v_{\nu}$.

The second functional derivative of the Einstein-Hilbert
action on a spherical background takes its well-known form
(for explicit representations, see app.~\ref{gravvariations}). The propagator matrix is
non-diagonal, except in the case $\rho = \alpha$ for $d=4$,
which will be our choice of gauge condition in the
following. We then observe that the vector and scalar
inverse propagators are both $\sim \alpha$. Therefore these
components drop out of all diagrams with at least one
external vertex for $\rho = \alpha =0$, if we choose the
scalar part of the regulator function $R_k =
\Gamma_k^{(2)}r(y)$, where $y
=\frac{\bar{\mathcal{D}}}{k^2}$ with some appropriate
Laplace-type operator $\bar{\mathcal{D}}$. Let us illustrate
this by noting that, schematically, the derivative with
respect to the external field leads to a form $\partial_t
R_k \left(\Gamma_k^{(2)}(1+r(y)) \right)^{-2}$ which is
$\sim \alpha$.
Accordingly these modes will drop out of the right-hand-side
of the flow equation as soon as a derivative with respect to
an external field is performed. Interestingly no choice of
$\rho$ and $\alpha$ allows to remove the contribution of
the trace mode completely. Running couplings in gravity are
therefore driven by transverse traceless tensor fluctuations
(being related to the spin-2-properties of the metric), as
well as trace fluctuations, and of course ghosts.

We then proceed to specify a choice of regulator which will
allow us to evaluate the trace on the right-hand side of the
flow equation analytically. We use a spectrally and
RG-adjusted regulator \cite{Gies:2002af,Litim:2002xm} with
an exponential shape function 
\begin{equation}
R_k= \Gamma_k^{(2)} \frac{1}{e^{\frac{\Gamma_k^{(2)}}{Z_{\rm mode} k^2}}-1},\label{expreg}
\end{equation}
where $Z_{\rm mode}$ depends on the mode of the York
decomposition such that the effective cutoff scale is the
same for all modes. 
Here, one includes a negative sign for the trace mode, the
kinetic term of which has a negative sign\footnote{In the
Euclidean functional integral with the action being the
Einstein-Hilbert action this leads to a stability problem,
as a strongly fluctuating conformal mode can make the factor
$e^{-S_{\rm EH}}$ arbitrarily large \cite{Hawking:1976ja}.
A negative inverse propagator can lead to stability problems
in the flow equation, if we choose the regulator to be
positive, as then the regularised propagator is
schematically given by $\left(-p^2 +k^2\right)^{-1}$, which
clearly may lead to divergences. The correct choice of
$Z_{\rm mode}$ for the conformal mode therefore involves a
negative sign, leading to the schematic regularised
propagator $-\left(p^2 +k^2\right)^{-1}$, which does not
introduce any spurious divergences into the flow equation
\cite{Reuter:1996cp}.}. 

The complete regulator function carries an appropriate
tensor structure for each sector, since
\begin{eqnarray}
 R_{k\, \mu \nu \kappa \lambda}^{TT}& \sim & \Gamma_{k\,
\mu \nu \kappa \lambda}^{(2)\, TT}
\end{eqnarray}
and similar for the other modes.

Here, we employ a spectrally adjusted cutoff, i.e. a cutoff where the regulator function has the full inverse propagator $\Gamma_k^{(2)}$ as its argument, instead of the Laplacian, only. This has the property that the momentum shells that are integrated out are adapted to the scale-dependent two-point function, see \cite{Gies:2002af}. Thus the flow adjusts to the change of the two-point function, and can thereby be expected to be closer to the projection of the trajectory in the untruncated theory space. 
Technically it amounts to the right-hand-side of the flow equation depending on the couplings and the beta functions, instead of the couplings only. This might be viewed as effectively resumming a larger class of diagrams.

We can then use the eigenvalues and degeneracy factors of
$-\bar{D}^2$ acting on scalar, vector and tensor hyperspherical
harmonics, which form a basis in the respective space on a
spherical background
\cite{Rubin:1984tc,Rubin:1983be}, for details turn to app.~\ref{hypersphericals}.

The propagators can then be constructed explicitly in a
basis of the appropriate hyperspherical harmonics, which we
exemplify for the transverse traceless tensor mode:
As the hyperspherical harmonics $T^{lm}_{\mu \nu}(x)$ form a
basis for tensor functions on the $d$ sphere, we
can expand the Green's function as follows:
\begin{equation}
 G(x-x')_{\mu \nu \rho \sigma}= \sum_{l=2}^{\infty}\sum_{m=1}^{D_l(d,2)}
a_{lm}\,T^{lm}_{\mu \nu}(x)T^{lm}_{\rho \sigma}(x')\label{greenh},
\end{equation}
where there is a $D_l(d,2)$-fold degeneracy of
the hyperspherical harmonics for fixed $l$ but different $m$
\cite{Rubin:1984tc, Rubin:1983be} where
\begin{equation}
D_l(d,2)=
\frac{(d+1)(d-2)(l+d)(l-1)(2l+d-1)(l+d-3)!}{2(d-1)!(l+1)!}.
\end{equation}
Now we can invert the two-point function to arrive at the regularised
$k$-dependent propagator, 
\begin{equation}
\left(\Gamma_{TT}^{(2)}+R_{k,TT}\right)_{\mu \nu \kappa \lambda} 
G_{\ \ \ \rho \sigma }^{\kappa \lambda}
= \frac{1}{\sqrt{\bar{g}}}\delta^d(x-x')\frac{1}{2}\left(\bar{g}_{\mu \rho}\bar{g}_{\sigma \nu}+\bar{g}_{\mu \sigma}\bar{g}_{\nu \rho}\right).
\label{hGreen}
\end{equation}
We will need several properties of the hyperspherical
harmonics:
They fulfil a completeness and an
orthogonality relation, and are eigenfunctions of the covariant Laplacian
$\bar{D}^2$:
\begin{eqnarray}
 \frac{\delta^d(x-x')}{\sqrt{\bar g}}\frac{1}{2}\left(\bar{g}_{\mu \rho}\bar{g}_{\nu \sigma}+\bar{g}_{\mu \sigma}\bar{g}_{\nu \rho}\right)&=&
 \sum_{l=2}^{\infty}\sum_{m=1}^{D_l(d,2)}T_{\mu \nu}^{lm}(x)T_{\rho
   \sigma}^{lm}(x'),\qquad \label{completenessa}\\
\delta^{lk}\delta^{mn}&=&\int d^d x \, \sqrt{\bar{g}}\, \frac{1}{2}\left(\bar{g}^{\mu
  \rho}\bar{g}^{\nu \sigma}+\bar{g}^{\mu \sigma}\bar{g}^{\nu \rho}\right)T^{lm}_{\mu \nu}(x)T^{kn}_{\rho \sigma}(x), \label{tortho11}\nonumber\\
 -\bar{D}^2 T_{\mu \nu}^{lm}(x)&=& \Lambda_{l,2}(d)T_{\mu \nu}^{lm}(x).\label{tsh}
\end{eqnarray}
The eigenvalues of the Laplacian are given by
\begin{equation}
\Lambda_{l,2}(d)=\frac{l(l+d-1)-2}{r^2}.\label{lapleigenvalues}
\end{equation}
We insert our expression
\Eqref{greenh} into \Eqref{hGreen}, and use the eigenvalue equation
\Eqref{tsh}. As the regulator is some function of $-\bar{D}^2$, it turns into
the same function of $\Lambda_{l, 2}(d)$ in the
hyperspherical-harmonics basis, schematically
$f(-\bar{D^2})\rightarrow f(\Lambda_{l,2}(d))$.

Applying the completeness relation allows to rewrite the right-hand side of
the definition of the Green's function \Eqref{hGreen}. By a comparison of
coefficients with respect to the hyperspherical-harmonics basis, we obtain
\begin{eqnarray}
 a_{lm}&=&\bigg( \kappa^2
 Z_{\text{N}}\Bigl(\frac{d(d-3)+4}{r^2}-2\bar{\lambda}
 +\Lambda_{l,2}(d)\Bigr)+R_{k,l}\bigg)^{-1}\,,\label{glmh}
\end{eqnarray}
for $l \geq 2$. 
Inserting our expression for the propagator on the
right-hand side of the flow equation we have to take the
appropriate traces, which involve an integration over the
hyperspherical harmonics, where we again invoke
\Eqref{completenessa}. Finally we end up with a summation
over the discrete label $l$. Here we have to take care of
the fact that some modes do not contribute to the trace, see
\cite{Lauscher:2001ya}.

The discrete sum on the right-hand side of the flow equation can then be brought into an integral form with the help of the Euler-MacLaurin formula, which reads
\begin{equation}
 \sum_{l=0}^{\infty}f(l)= \int_0^{\infty}dl\, f(l) + \frac{1}{2}f(0)+ \frac{1}{2}f(l)+\sum_{k=1}^{\infty}(-1)^{k+1}\frac{B_{k+1}}{(k+1)!}\left(f^{(k)}(l)-f^{(k)}(0)\right)+ \rm{const}.
\end{equation}
Here $B_k$ are the Bernoulli numbers.
If the infinite sum over derivatives of the function
contributed in our case, this rewriting would not be of much
use. Using the Euler-MacLaurin representation for the
right-hand side of the Wetterich equation with the
propagator given by \Eqref{greenh}, \Eqref{glmh} and
\Eqref{lapleigenvalues} and the regulator \Eqref{expreg}, a
detailed inspection reveals that none of the terms in the
sum over derivatives can contribute to the flow in the
Einstein-Hilbert truncation. This is due to the following
reason: Terms at large $l$ are suppressed due to the
UV cutoff presented by the scale derivative of the regulator
function. At small $l$, the derivatives with respect to $l$
cannot contribute to a projection on the running of Newton's
coupling and the cosmological term\footnote{In fact at higher orders
in the curvature a finite number of these terms will give a
non-vanishing contribution.}.

We finally project the right-hand side of the flow equation
on $\partial_t Z_{N}(k)$ and $\partial_t \lambda(k)$ by
expanding it in powers of the curvature radius $r^2$. As we
are interested in evaluating the fixed point structure, we
introduce dimensionless renormalised couplings $G$ and
$\lambda$ which are
related to the bare quantities by
\begin{eqnarray}
G&=&\frac{G_{\text{N}}}{Z_{\text{N}} k^{2-d}}
=  \frac{1}{32\,  \pi\,  \kappa^2\,  Z_{\text{N}}\, k^{2-d}},\nonumber\\
\lambda&=& \bar{\lambda} k^{-2}, \quad \Rightarrow \partial_t \lambda = -2
\lambda + k^{-2}\partial_t \bar{\lambda}.
\end{eqnarray}
The resulting $\beta$ functions are given in app.~\ref{betafunctionsghost} in the limit $\eta_c \rightarrow 0$.

They admit a Gau\ss{}ian as well as a non-Gau\ss{}ian fixed point, the coordinates of which are given by \cite{Eichhorn:2010tb}
\begin{eqnarray}
G_{\ast}= 0.2701 &{}&\phantom{xxxxxx}\lambda_{\ast}=0.3785 \mbox{ in Landau deWitt gauge, where } \rho = \alpha =0,\nonumber\\
G_{\ast}=0.1812&{}&\phantom{xxxxxx} \lambda_{\ast}=0.4807\mbox{ in deDonder gauge, where } \rho = \alpha =1.
\end{eqnarray}
Let us consider the product $G_{\ast}\lambda_{\ast}$, which
is a universal quantity, since it is dimensionless even for
the dimensionful couplings, and is invariant under constant
rescalings of the metric. The values $G_{\ast}\lambda_{\ast}
\approx 0.102$  in Landau deWitt gauge and
$G_{\ast}\lambda_{\ast} \approx 0.087$ in deDonder gauge are
close to values observed in other schemes, see, e.g.
\cite{Lauscher:2001ya,Codello:2008vh}. 

In the convention adopted here the critical exponents of the Gau\ss{}ian fixed point are simply given by the canonical dimensions, which implies that the direction corresponding to the Newton coupling is irrelevant and therefore UV repulsive. At the NGFP we find a non-trivial mixing between the two directions, and two relevant critical exponents given by
\begin{eqnarray}
\theta_{1,2}&=&2.10 \pm i 1.69 \quad \mbox{ in Landau deWitt gauge }\nonumber\\
 \theta_{1,2}&=& 1.41 \pm i 1.67 \quad \mbox{ in deDonder gauge.}
\end{eqnarray}
These lie within the range ${\rm Re}[\theta_{1/2}]\in
\{1.1,2.3\}$  that has been found with other regularisation
schemes and gauges \cite{Litim:2008tt}. The imaginary part
is slightly lower than the typical range ${\rm
Im}[\theta_{1/2}]\in \{2.4,7.0\}$ \cite{Litim:2008tt}. Note that its existence
leads to the spiralling into the FP. The imaginary part may
seem a bit unusual, in particular from the view of condensed
matter, where fixed points signal a phase transition and
critical exponents show how physical quantities scale in the
vicinity of the second order phase transition. Depending on
the quantity under consideration a physical interpretation
of an imaginary part may be possible, and indeed however complex
critical exponents are also encountered in condensed matter physics. Interestingly they are connected to a discrete form of scale invariance there \cite{Sornette} \footnote{If a similar implication can be inferred from the complex pair of critical exponents in the case of gravity remains to be investigated.}.
As emphasised in \cite{Kadanoff:2011aj},
only the real part decides about the relevance. Complex
critical exponents are also familiar from classical
mechanics, where trajectories in phase space can be
attracted to fixed points. In this case an imaginary part is
rather common and again implies that the trajectories spiral
into the fixed point.

\begin{figure}[!here]
\begin{minipage}{0.45\linewidth}
 \includegraphics[scale=0.6]{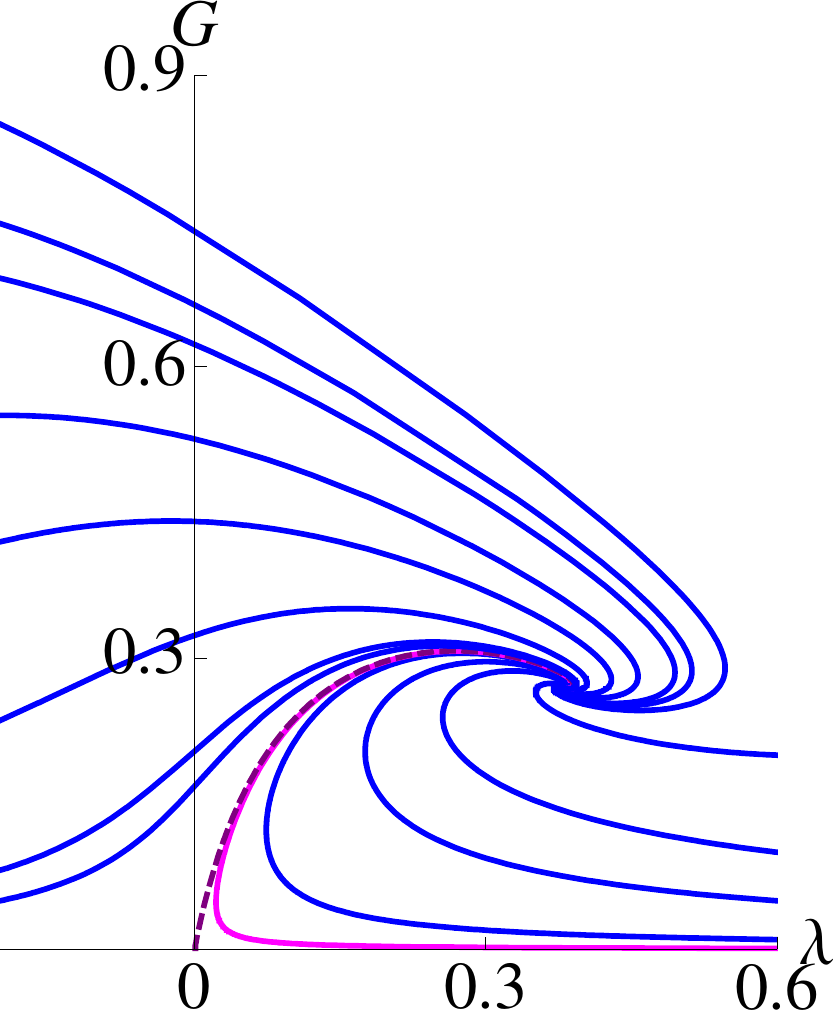}
\end{minipage}
\begin{minipage}{0.5\linewidth}
\caption{Flow of the dimensionless couplings $G$ and
$\lambda$ in the Einstein-Hilbert truncation. Flowing
towards the infrared, the trajectories emanate from the NGFP
at $G_{\ast}\approx 0.2701,\, \lambda_{\ast}\approx 0.3785$.
The separatrix, which connects the non-Gau\ss{}ian and the
Gau\ss{}ian fixed point, is plotted by a purple dashed line.
An asymptotically safe trajectory which passes very close to
the Gau\ss{}ian fixed point is shown in magenta as an
example of a trajectory that has a classical regime existing
over a large range of scales.\label{EHtruncationplot}}
\end{minipage}
\end{figure}

The phase portrait of the Einstein-Hilbert truncation is presented in fig.~\ref{EHtruncationplot}.

\noindent Note that this truncation does not seem to allow
an infrared limit $k \rightarrow 0$, since the flow cannot
be extended to $k \rightarrow 0$, when using a regulator
that is not proportional to the full two-point function but
only to the covariant Laplacian (multiplied by some
appropriate tensor structure to account for transversality
etc). Then this feature is clearly visible as trajectories
terminate at $\lambda = \frac{1}{2}$ due to a singularity in
the $\beta$ functions. With a spectrally adjusted cutoff one
can shift this point such that the trajectories can be
extended in a singularity-free fashion for arbitrary
$\lambda$, see \cite{Codello:2008vh}.\\
An enlargement of the truncation is presumably mandatory to
search for possible infrared fixed points governing, e.g.
cosmology.

\subsection{The TT-approximation}
Having found a fixed point in a variety of different
truncations, one may wonder, what physical mechanism is
responsible for the generation of the fixed point. This
question can be investigated by altering structural aspects
of the set-up and checking for the existence of the fixed
point. If one induced a vanishing of the fixed point by a
definite change in the set-up this would imply that this
particular aspect is possibly behind the physical mechanism
for the fixed-point generation. To be more definite, one may
ask, e.g. 
\begin{itemize}
\item Is there a fixed point when only the conformal mode is allowed to fluctuate? (If not, then the spin-2 dynamics is crucial for the generation of the fixed point.)
\item Is there a fixed point in both signatures, Lorentzian, as well as Euclidean, i.e., is the causal structure crucial to get a well-behaved UV limit?
\item Is there a fixed point in higher dimensions $d > 4$?
\item Is there a fixed point when using the vielbein and the connection as variables, instead of the metric?
\item Which of the metric modes carries the physical information on the fixed point?
\end{itemize}
The first four questions have been addressed in \cite{Reuter:2008qx,Reuter:2008wj,Machado:2009ph,Manrique:2011jc,Litim:2003vp, Fischer:2006fz,Daum:2010qt}. The first one can be answered in the affirmative in the Einstein-Hilbert truncation \cite{Reuter:2008qx,Reuter:2008wj}, but for extended truncations the existence of a fixed point may depend on the presence of suitable matter fields \cite{Machado:2009ph}. Astonishingly, the change in signature does not result in a qualitative change in fixed point properties \cite{Manrique:2011jc} in the Einstein-Hilbert truncation.  The fixed point also exists for any number of dimensions $d \geq 4$ \cite{Litim:2003vp, Fischer:2006fz} in the Einstein-Hilbert truncation, and for the Holst action \cite{Daum:2010qt}.

As follows from \Eqref{delta2Gammagf}, only the transverse traceless mode is unaffected by a change of the gauge parameters. One might therefore expect, that this mode, which is related to the spin-2-dynamics of the metric field, carries physical information on the properties of the theory. 

Here we will briefly focus on the corresponding question, to
what degree it is possible to simplify calculations by
suppressing the contributions from the other modes in
the path integral completely. Accordingly we only let
$h_{\mu \nu}^{TT}$ fluctuate, and only take the resulting
loop into account.

Within this so-called TT-approximation, we observe that the NGFP exists in the Einstein-Hilbert truncation and that its universal properties are preserved under the approximation, i.e. we again find two critical exponents with a positive real part ($\theta_1 = 2.808$, $\theta_2 = 1.057$ \cite{Eichhorn:2010tb}).

We therefore conclude that at least within the Einstein-Hilbert truncation, the transverse traceless fluctuations already carry the dynamics necessary to generate the fixed point with the same universal properties as in the full calculation.

If this feature persists beyond the simple Einstein-Hilbert
truncation, it allows for a simplification of calculations,
as we only have to take into account transverse traceless
fluctuations. Of course for quantitative precision it is
crucial to include all fluctuations modes. However for a
first investigation of the relevance of a new coupling in an
extended truncation, the TT-approximation may suffice. We
will apply this line of reasoning in sec.~\ref{ghostcurv}.

\section{Ghost sector of asymptotically safe quantum gravity}\label{ghosts}

\subsection{Why investigate the ghost sector? - Ghost scenarios in a non-perturbative regime}

A subspace of theory space that has been included in a very minimal fashion in studies of asymptotically safe quantum gravity is the ghost sector, where by ghosts we mean Faddeev-Popov ghosts, and not ghosts as implied by non-unitarity. Let us first motivate why an extension of truncations in the ghost sector is of interest, in particular in a non-perturbative regime, before we introduce specific truncations in the next sections.

Ghosts are not part of the asymptotic state space of the theory and as such are unphysical. One may therefore be tempted to conclude that their inclusion in the Wetterich equation is not mandatory. However this is clearly not the case: In a gauge-fixed formulation, ghost fluctuations generically drive running couplings as do metric fluctuations (which are of course also not all physical). Indeed the diagrammatic representation of the Wetterich equation directly shows that the ghost loop enters with a negative sign. Therefore the effect of metric fluctuations may be counterbalanced or even outweighed by the effect of ghost fluctuations. As the \emph{full} ghost propagator enters the Wetterich equation, not only the ghost wave-function renormalisation, but also many further ghost-antighost couplings will directly appear in the $\beta$ functions of couplings from the metric sector, such as the Newton coupling or the cosmological constant. Due to the one-loop nature of the flow equation, operators containing more than one ghost-antighost pair cannot couple directly into the flow of these couplings. Still an infinite number of operators coupling a ghost-antighost-pair to metric invariants such as $R^n$, $(R_{\mu \nu}R^{\mu \nu})^n$ etc. exist.

Therefore the evaluation of running couplings in the ghost sector is an important step towards getting a more complete picture of theory space, and investigating the existence of a NGFP as well as the properties of trajectories emanating from it.

By applying the background field gauge we have enlarged the theory space by a huge number of directions. Assuming the existence of a NGFP we can distinguish four basic scenarios with respect to ghost couplings:
\begin{itemize}
 \item In a simple scenario, all ghost couplings beyond
those in the Faddeev-Popov operator are zero at the NGFP,
which we refer to as $(\rm NGFP_{metric}-GFP_{ghost})$, and
correspond to irrelevant directions. Then the NGFP is
compatible with a simple form of gauge-fixing, which should
not necessarily be expected, since at a NGFP we are in the
non-perturbative regime. In many gauges, a
perturbative gauge-fixing might then be insufficient due to the
Gribov problem.
Furthermore all relevant directions, which
correspond to free parameters to be fixed by an experiment,
do not contain any ghost operators, but the metric only.
Therefore it is -- in principle -- possible to infer their values from measurements.
\item In a not so simple scenario, theory space again shows
a $(\rm NGFP_{metric}-GFP_{ghost})$ structure, but relevant
directions correspond to ghost directions, or non-trivial
superpositions of ghost and metric operators. Again the
compatibility of the NGFP with a simple Faddeev-Popov gauge
fixing remains, but now we have to consider relevant
directions which contain ghost fields. Here it is necessary
to remind ourselves that within a gauge-fixed formulation,
the (modified) Ward-identities will relate different
operators and restrict the flow to a hypersurface in theory
space. Thus it might be possible to relate such ghost
couplings to measurable quantities through a solution of
the Ward-identities.
\item In a third scenario, further couplings beyond those in
the Faddeev-Popov operator have interacting fixed points,
and correspond to irrelevant directions. Here the
perturbative gauge-fixing turns out to be insufficient in a
non-perturbative setting, and the flow automatically
generates further couplings. One may speculate that this
effect may also solve the Gribov problem, and lead to a
unique gauge-fixing. 
\item In a last scenario, we have both an interacting fixed point in the ghost sector, and relevant directions containing ghost operators. Clearly our comments on the non-perturbative extension of gauge fixing and relevant ghost directions also apply here.
\end{itemize}
The first, simple scenario seems to be the least challenging one at a first glance. Note, however, that the first two scenarios will in many gauges require a solution of the Gribov problem. In fact, as we will explain below (see sec.~\ref{ghostNGFPs}) one should generically expect an interacting fixed point to be induced in the ghost sector by a non-vanishing fixed-point value for the Newton coupling, already starting from the Einstein-Hilbert truncation with a simple Faddeev-Popov operator.

Besides one might generically expect relevant operators to
exist in the ghost sector: As power-counting marginal
operators have been found to be relevant at the NGFP in the
metric sector, see, e.g. \cite{Codello:2008vh}, one may
expect that quantum fluctuations induce anomalous scaling of
a similar size also in the ghost sector. Then adding
ghost-curvature couplings such as $\bar{c}^{\mu}R^n c_{\mu}$
with, e.g. $n=1,2$ to a truncation might result in further
positive critical exponents.

The next issue is then the interpretation of relevant directions which contain ghost directions. Trajectories emanating from fixed points of the flow equation are also restricted to lie on a hypersurface satisfying the modified Ward-identities. This may indeed reveal that different directions in theory space are not independent. Thus, two relevant couplings may turn out to be related by a Ward-identity and hence correspond to one free parameter only. Therefore, within a given truncation, the number of relevant operators counted without solving the modified Ward-identities is an upper limit on the total number of free parameters within this truncation.

An important clue about the effect of quantum fluctuations on the relevance of a coupling at a NGFP is given by the anomalous dimension of the corresponding field: Considering an operator $\mathcal{O}^n$, which is independent of the ghosts (e.g. a Laplace-type operator, or an operator built from the metric such as a curvature tensor), we can construct a term 
\begin{equation}
 g_n \bar{c}^{n/2} \cdot \mathcal{O}^n  \cdot c^{n/2}
\end{equation}
for the ghost and antighost fields. Then the renormalised dimensionless coupling
$\tilde{g}_n$ is defined to be
\begin{equation}
 \tilde{g}_n = \frac{g_n}{Z_c^{n}}k^{-d_n},
\end{equation}
where $Z_c$ is the ghost wave-function renormalisation and $d_n$ is the canonical dimension of $g_n$. The $\beta$ function of this coupling will then be of the form
\begin{equation}
 \beta_{g_n}= \left(n \eta_c-d_n\right) \tilde{g}_n+... \label{relevance},
\end{equation}
where further terms of course depend  on the explicit coupling and truncation under consideration. This implies that the anomalous dimension 
\begin{equation}
\eta_c = -\partial_t \ln Z_c
\end{equation}
 will contribute to the critical exponent. In particular, a negative anomalous dimension enhances the probability that such couplings turn out to be relevant. 

Accordingly we will study the ghost anomalous dimension in sec.~\ref{ghostanomdim} to get a first indication if relevant directions are to be expected in the ghost sector. 

Since the (ir)relevance of a coupling naturally depends not only on the anomalous dimension, but also on further terms induced by quantum fluctuations, we will consider explicit examples for ghost couplings that may become relevant in sec.~\ref{exttrunc} and \ref{ghostcurv}.

A further motivation for a study of the ghost sector results from Yang-Mills theory: As explained in sec.~\ref{Landaugaugeprops}, ghosts become dynamically enhanced in the non-perturbative regime in some gauges. Since within the asymptotic-safety scenario gravity enters a non-perturbative regime at high energies, we may argue by analogy that a similar mechanism may also be realised in gravity. In the case of Yang-Mills theory this enhancement occurs in the deep-IR, confining regime of the theory. What the physical consequences of ghost enhancement in gravity might be remains to be investigated. Further the scaling exponents of all $n$-point correlation functions turn out to be dependent on the ghost scaling exponent in Yang-Mills theory. The knowledge of this exponent therefore allows to deduce the momentum-dependence of all $n$-point correlation functions in the asymptotic regime, which is an important step towards a full understanding of the dynamics in this regime. 
As a first step towards the investigation of these questions in gravity we study the ghost wave-function renormalisation.

\subsection{Ghost anomalous dimension}\label{ghostanomdim}

We will now study the following truncation of the
effective action in four dimensions:
\begin{equation}
 \Gamma_k = \Gamma_{k \,\rm EH}+ \Gamma_{k\,\rm gf}+ \Gamma_{k\, \rm gh},
\end{equation}
where the ghost term is now endowed with a non-trivial wave-function renormalisation $Z_c (k)$. The other terms are given by the expressions in \Eqref{GEH}, \Eqref{Ggf} and \Eqref{eq:Ggh}.
This truncation has been studied in \cite{Eichhorn:2010tb} and \cite{Groh:2010ta} in different gauges and with different choices for the regulator. An extension to a higher number of dimensions is also straightforward, and has been performed in \cite{Groh:2010ta}\footnote{Since current evidence suggests that the asymptotic-safety scenario for gravity might indeed be realised in four dimensions, such an extension, although not excluded, is not necessary for the consistency of the scenario. Naturally, as detailed in \cite{Litim:2007iu} the existence of further, compactified dimensions at the TeV-scale would be very interesting since it would allow for experimental insight into the realisation of asymptotic safety at current accelerator experiments.}.

Let us first comment on the role of the ghost wave-function renormalisation among the couplings in theory space: It belongs to the inessential couplings which implies that the kinetic term can always be redefined to have a unit normalisation by a redefinition of the fields. This implies that we will not have a fixed-point condition for the $Z_c$, instead the anomalous dimension $\eta_c$
will enter the $\beta$ functions of other running couplings
as detailed above, see \Eqref{relevance}. In the set of
coupled differential equations for the running couplings it
will then be possible to re-express $\eta_c$ as a function
of the essential couplings and their $\beta$ functions. This
expression can be reinserted into the other $\beta$
functions, thus reducing the system of coupled differential
equations for the fixed-point search by one. Having found a
fixed point we then determine the value of $\eta_c$ at
the fixed point from the algebraic expression for $\eta_c$.

The back-coupling of $\eta_c$ into the flow in the
Einstein-Hilbert sector can be evaluated along exactly the
same lines as in the previous section. Using a spectrally
adjusted regulator implies that $\eta_c$ will appear on the
right-hand side of the Wetterich equation.

In order to extract the anomalous dimension of the ghost, we project the flow
equation onto the running of the ghost wave-function
renormalisation. We use a
decomposition of $\Gamma^{(2)}_k+R_k=\mathcal{P}_k+\mathcal{F}_k$ into an inverse
propagator matrix $\mathcal{P}_k=\Gamma_k^{(2)}[\bar{c}=0=c]+R_k$, including the
regulator but no external ghost fields, and a fluctuation matrix
$\mathcal{F}_k=\Gamma_k^{(2)}[\bar{c},c]-\Gamma_k^{(2)}[\bar{c}=0=c]$
containing external ghost fields. The components of $\mathcal{F}_k$ are either
linear or bilinear in the ghost fields, as higher orders do not occur in our
truncation. We may now expand the right-hand side of the flow equation as
follows:
\begin{equation}
 \partial_t \Gamma_k= \frac{1}{2}{\rm STr} \{
 [\Gamma_k^{(2)}+R_k]^{-1}(\partial_t R_k)\}\label{eq:flowexp}= \frac{1}{2} {\rm STr} \tilde{\partial}_t\ln \mathcal{P}_k
+\frac{1}{2}\sum_{n=1}^{\infty}\frac{(-1)^{n-1}}{n} {\rm
  STr} \tilde{\partial}_t(\mathcal{P}_k^{-1}\mathcal{F}_k)^n,
\end{equation}
where the derivative $\tilde{\partial}_t$ by definition acts only on the $k$ dependence of the regulator\footnote{Using a spectrally adjusted regulator $R_k \sim \Gamma_k^{(2)}r(y)$ will then also produce terms $\sim \partial_t \Gamma_k^{(2)}$ on the right-hand side of the flow equation.}, $\tilde{\partial}_t=
 \partial_t R_k\cdot\frac{\delta}{\delta R_k}$. As each factor of $\mathcal {F}$ contains a
coupling to external fields, this expansion simply corresponds to an expansion
in the number of vertices that are coupled to external fields.

\begin{figure}[!here]
 \begin{minipage}{0.2\linewidth}
  \includegraphics[scale=0.08]{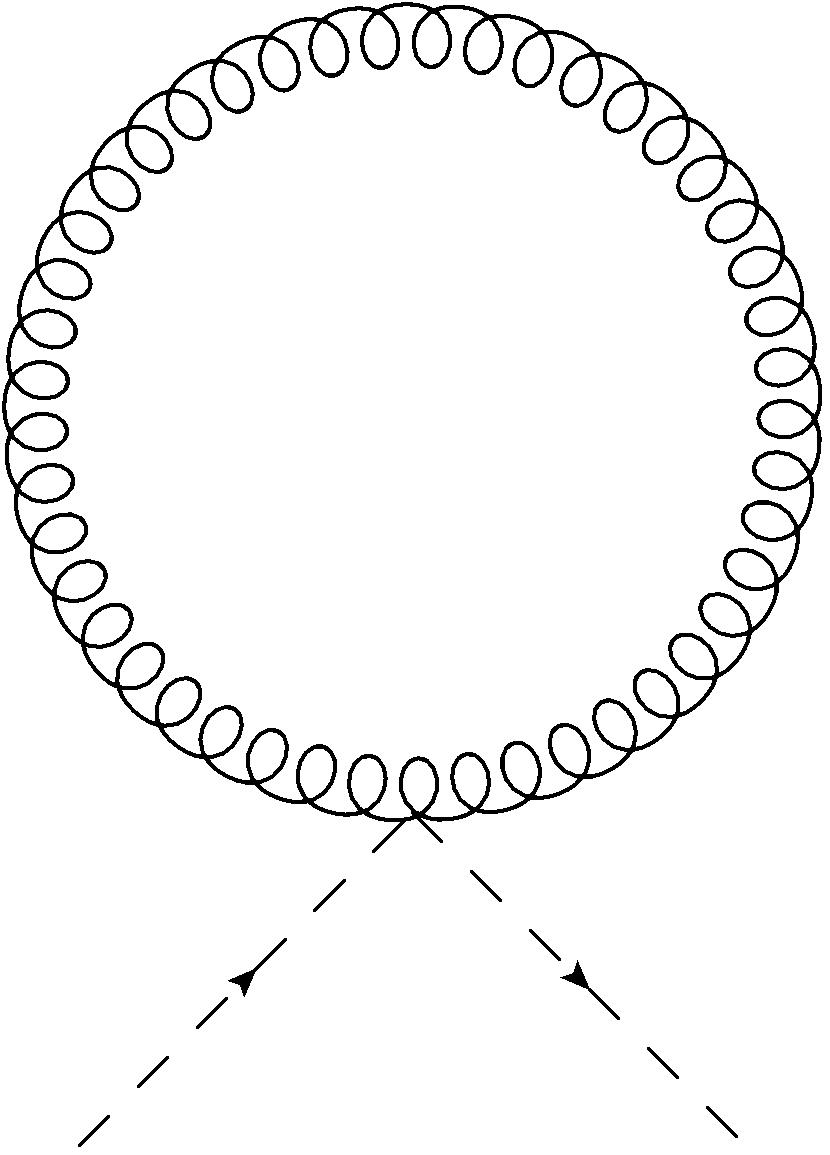}
 \end{minipage}
\begin{minipage}{0.3\linewidth}
  \includegraphics[scale=0.08]{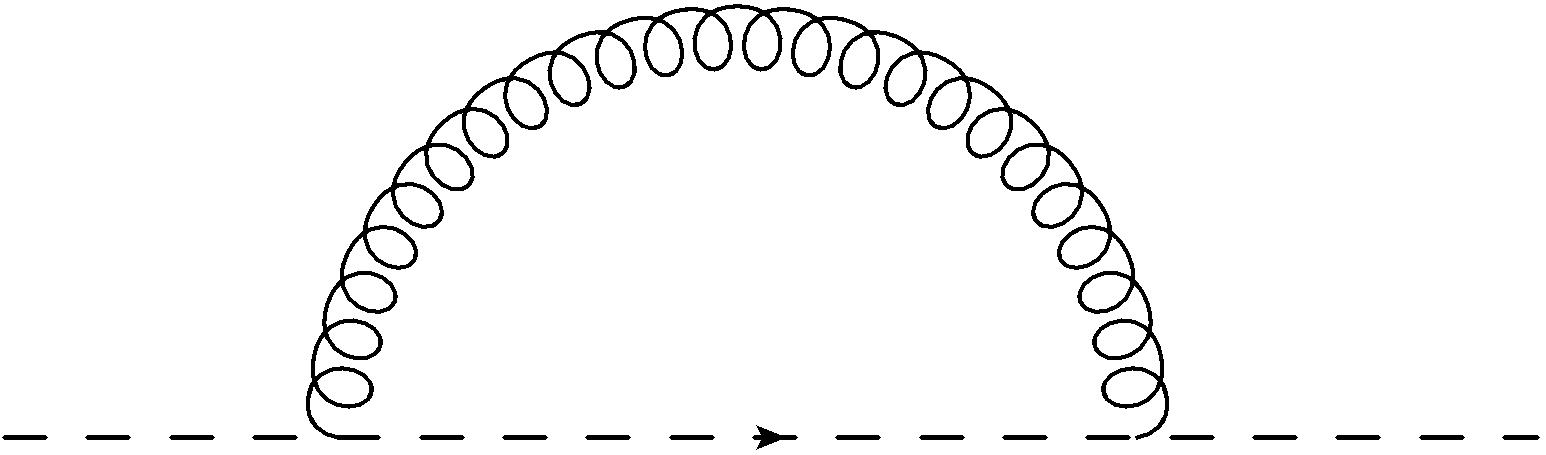}
 \end{minipage}
\begin{minipage}{0.42\linewidth}
\caption{Diagrams contributing to $\partial_t Z_c(k)$: The dashed line denotes the ghost propagator and external ghost fields, the curly line denotes the metric. Regulator insertions are produced by the $\tilde{\partial}_t$-derivative acting on the propagators.\label{ghostanomdimflowdiags}}
 \end{minipage}
\end{figure}

To bilinear order in the external ghost and antighost, we may directly neglect
all contributions beyond $(\mathcal{P}^{-1}\mathcal{F})^2$.
Diagrammatically, the remaining terms correspond to a
tadpole and a self-energy diagram, see
fig.~\ref{ghostanomdimflowdiags}.

\noindent Irrespective of our choice of gauge in the ghost
sector, i.e. for all values
of the gauge parameter $\rho$, we find that the tadpole does not contribute to
the ghost anomalous dimension as the corresponding vertex is zero; this is
shown in app.~\ref{tadpole}.

We now choose a flat Euclidean background as we are interested in projecting onto a term that is non-vanishing also for vanishing curvature, cf. \Eqref{eq:Ggh}. Note that $\beta$ functions might be dependent on the topology of the background: As observed in \cite{Reuter:2008qx} the infrared part of the $\beta$ functions may be differ when evaluated on backgrounds of different topology. This feature also distinguishes $\beta$ functions at finite temperature from those evaluated at zero temperature, since within the Matsubara formalism the only change between the two settings is one in the topology, where the $x^0$ directions is compactified on a circle at finite temperature.
However, we expect that the UV part of the $\beta$
function should be insensitive to topological changes, as
also observed in \cite{Reuter:2008qx}. Thus it is
consistent to use topologically distinct backgrounds to
evaluate different $\beta$ functions in the same truncation,
as long as one is interested in UV fixed points. Here we
will use this observation to evaluate the $\beta$ functions of the
cosmological constant and the Newton coupling on a maximally
symmetric background of non-vanishing positive curvature as
before (see sec.~\ref{EHtrunc}), whereas we will choose
the technically simpler evaluation of the ghost
wave-function renormalisation on a flat background. As this
truncation is presumably insufficient to capture all
infrared effects (as indicated by the divergence at
$\lambda= \frac{1}{2}$ in many regularisation schemes), the
possible incompatibility of the two backgrounds in this
regime is of no importance here. 

We again apply a York decomposition as in \Eqref{York} and \Eqref{ghostYork}, where now the background covariant derivative reduces to the partial derivative, and we can directly work in Fourier space. The vertices contributing to the above diagrams are given in app.~\ref{etacvertices}.

As the propagator matrix in the metric sector is diagonal
only for $\rho = \alpha$, we work for this choice of gauge
parameters. We neglect any running in these parameters.%

On a flat background, we project the flow equation in Fourier space onto
the running wave-function renormalisation by
\begin{eqnarray}
 \eta_c&=& -\frac{1}{\sqrt{2}Z_c} 
\frac{1}{4}\delta^{\alpha \gamma}\frac{\partial}{\partial \tilde{p}^2}
\int \frac{d^4 \tilde{q}}{(2\pi)^4}
\left(\frac{\overset{\rightarrow}{\delta}}{\delta \bar{c}^{\alpha}(\tilde{p})} 
\partial_t \Gamma_k
\frac{\overset{\leftarrow}{\delta}}{\delta c^{\gamma}(\tilde{q})}\right)\Big|_{c, \bar{c}=0}.\label{projection}
\end{eqnarray}
Our conventions for the functional Grassmannian derivatives are such that
\begin{equation}
 \frac{\overset{\rightarrow}{\delta}}{\delta \bar{c}^{\alpha}(\tilde{p})}
 \int \frac{d^4p}{(2 \pi)^4}\bar{c}^{\mu}(p)M_{\mu \nu}(p)c^{\nu}(p) 
\frac{\overset{\leftarrow}{\delta}}{\delta c^{\gamma}(\tilde{q})}
= \delta(\tilde{p},\tilde{q})M_{\alpha
  \gamma}(\tilde{p}), 
\end{equation}
where $\delta(p,q)=(2\pi)^4 \delta^{(4)}(p-q)$.

Using a general regulator of the form
\begin{equation}
 R_{k\, \rm mode}= \Gamma_k^{(2)} r\left(\frac{\Gamma_k^{(2)}}{Z_{\rm mode}k^2} \right),
\end{equation}
we arrive at the following flow equation:
\begin{eqnarray}
\eta_c &=&\sqrt{2} Z_c\tilde{\partial}_{t}  \int \frac{dp^2}{16 \pi^2}p^2\,  \Biggl[ \frac{5(\alpha-7)}{18 (\alpha-3) \bar{\kappa}^2Z_{\text{N}} \left(p^2-2\bar{\lambda}(k)\right)\left(1+r\left(\frac{p^2-2 \bar{\lambda}(k)}{k^2} \right) \right) \sqrt{2}Z_c \left(1+r\left(\frac{p^2}{k^2} \right) \right)}\nonumber\\
&{}&+\frac{ \alpha }{\bar{\kappa}^2Z_{\text{N}}\left(p^2-2\bar{\lambda}(k)\right)\left(1+r\left(\frac{p^2-2 \bar{\lambda}(k)}{k^2} \right) \right)\sqrt{2}Z_c \left(1+r\left(\frac{p^2}{k^2} \right) \right)}\left(\frac{1}{3}-\frac{1}{4}\frac{p^2}{k^2}\frac{r'\left(\frac{p^2}{k^2}\right)}{1+r\left(\frac{p^2}{k^2} \right)} \right)
\nonumber\\
&{}&-  \frac{\alpha}{18 (\alpha-3)\bar{\kappa}^2Z_{\text{N}}\left((\alpha-3)p^2+4 \alpha\bar{\lambda}(k)\right)\left(1+r\left(\frac{p^2+\frac{4 \alpha \bar{\lambda}(k)}{\alpha-3}}{k^2} \right) \right)\sqrt{2}Z_c \left(1+r\left(\frac{p^2}{k^2} \right) \right)}\nonumber\\
&{}&\cdot \left(2 (\alpha-7)+3 \frac{p^2}{k^2}\frac{r'\left(\frac{p^2}{k^2}\right)}{1+r\left(\frac{p^2}{k^2} \right)}\right)\nonumber\\
&{}&+ \frac{ 1}{6 (\alpha-3)\bar{\kappa}^2Z_{\text{N}}\left((\alpha-3)p^2+4 \bar{\lambda}(k)\right)\left(1+r\left(\frac{p^2+\frac{4 \bar{\lambda}(k)}{\alpha-3}}{k^2} \right) \right)\sqrt{2}Z_c \left(1+r\left(\frac{p^2}{k^2} \right) \right)}\nonumber\\
&{}&\cdot \left((3 \alpha-4)-\alpha\frac{p^2}{k^2}\frac{r'\left(\frac{p^2}{k^2}\right)}{1+r\left(\frac{p^2}{k^2} \right)} \right)
\Biggr].\label{eq:13}
\end{eqnarray}
Here the first term is the
transverse traceless contribution. The second term is due to the transverse
vector mode, and the last terms result from the two scalar modes,
respectively. The terms $ \sim r'$ are due to the external momentum $\tilde{p}$ flowing through the internal ghost line and being acted upon by the $\partial_{\tilde{p}^2}$-derivative in \Eqref{projection}. 
The Landau-deWitt gauge $\alpha =0$ clearly plays a distinguished role
as only the transverse traceless and the trace mode
propagate. This favours the Landau-deWitt gauge from a
computational point of view. As it is moreover a fixed point of the
Renormalisation Group flow
\cite{Ellwanger:1995qf,Litim:1998qi}, we consider
the Landau-deWitt gauge as our preferred gauge choice.

We observe that for $\rho =\alpha  \rightarrow 3$ the inverse ghost as well as the inverse scalar propagator develop a zero mode, see app.~\ref{etacvertices}. Accordingly they are not invertible, which simply implies that the gauge fixing is not complete for this choice. Since we will mainly be interested in $\alpha =0$ and $\alpha=1$, our results are not affected.

The expressions for $\beta_{\lambda}$ and $\beta_{G}$  and
explicit representations for $\eta_c$ are given in app.~\ref{betafunctionsghost}, where we specialise to a
spectrally adjusted regulator with exponential shape
function as in \Eqref{expreg} such that the regularisation
scheme in the Einstein-Hilbert and the ghost sector is
consistent. For technical details on the evaluation of the
$\tilde{\partial}_t$-derivative see app.~\ref{tilde_partial}.

We can perform a numerical fixed-point search of these equations, where we insert the algebraic expression $\eta_{c}(G, \lambda, \eta_N, \partial_t \lambda)$ into the expressions for $\eta_N$ and $\partial_t \lambda$ and then demand the fixed-point conditions $\partial_t \lambda =0$ and $\eta_N =-2$.

The table \ref{table_FPvalues} lists numerical results for the NGFP in the deDonder gauge ($\alpha=1$) and the Landau-deWitt gauge ($\alpha =0$).

\begin{table}[!here]
\begin{tabular}{c|c|c|c|c|c}
 gauge&$\qquad G_{\ast}$&$\lambda_{\ast}$&$G_{\ast}\lambda_{\ast}$&$\theta_{1,2}$&$\eta_c$\\
\hline
$\alpha=0$ with $\eta_c =0$ &\, 0.270068 & 0.378519&0.102226&2.10152 $\pm i$1.68512&0\\
$\alpha=0$ with $\eta_c \neq0$ &\, 0.28706 & 0.316824 & 0.0909475&2.03409$\pm i$ 1.49895&-0.778944\\
$\alpha=1$ with $\eta_c=0$&\, 0.181179 &\, 0.480729 &0.0870979&1.40864 $\pm i$ 1.6715&0\\
$\alpha=1$ with $\eta_c \neq0$&\, 0.207738 &0.348335&0.0723625&1.38757$\pm i$1.283&-1.31245
\end{tabular}
\caption{Fixed-point values for the dimensionless Newton coupling $G$,
  cosmological constant $\lambda$, associated critical exponents
  $\theta_{1,2}$ and the ghost anomalous dimension $\eta_c$ for different
  gauges and approximations.} 
\label{table_FPvalues}
\end{table}

The first important conclusion is that this enlarged truncation adds to the evidence for the existence of the NGFP. We also note that the inclusion of this further coupling does not lead to large modifications in the values. In particular, the critical exponents and the universal product $G_{\ast}\lambda_{\ast}$ are rather stable. 

Interestingly, the gauge dependence of the result is reduced
in some quantities, as, e.g. in the real part of the
critical exponents.
We interpret this as a sign of a stabilisation of the flow under the inclusion of a non-trivial ghost sector, as the gauge dependence of universal quantities such as critical exponents is slightly decreased.

We observe that the inclusion of the non-trivial ghost
anomalous dimension results in a reduced fixed-point value
for the cosmological constant. Interestingly this implies a
slightly reduced backcoupling of metric fluctuations into
the flow of other couplings. This is due to the fact, that
the metric propagator, schematically
$\frac{1}{p^2-2\lambda}$, is enhanced for positive
$\lambda$, and suppressed for negative $\lambda$. A smaller
value of $\lambda$ therefore reduces the effect of metric
fluctuations; for a further discussion of this, see
sec.~\ref{results}.

In our enlarged truncation we may also consider the complete flow in the $(G, \lambda)$-plane, for a comparison with the Einstein-Hilbert truncation where $\eta_c=0$. Since we observe a high degree of similarity between the RG flow in both truncations, we interpret this as a further non-trivial confirmation of the asymptotic-safety scenario.

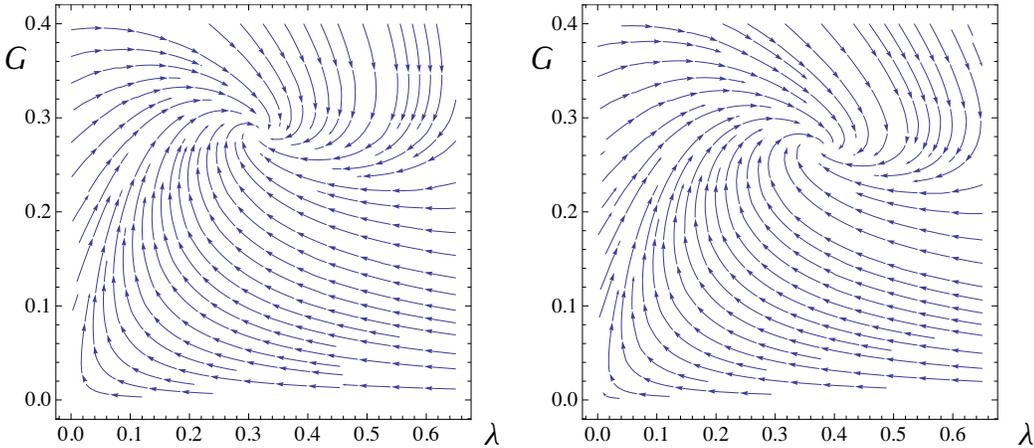
\begin{figure}[!here]
 \setlength{\unitlength}{1cm}
 \begin{picture}(18,6)
\thicklines
\put(0.3,0){\includegraphics[scale=0.7]{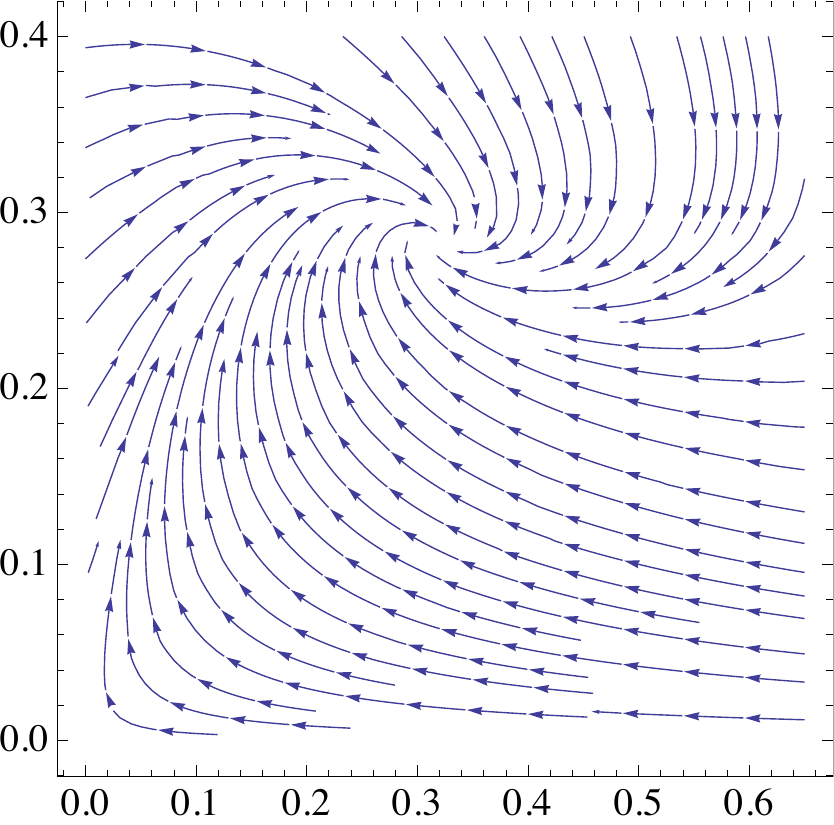}}
\put(6.4,0){$\lambda$}
\put(0,5){$G$}
\put(7.3,0){\includegraphics[scale=0.7]{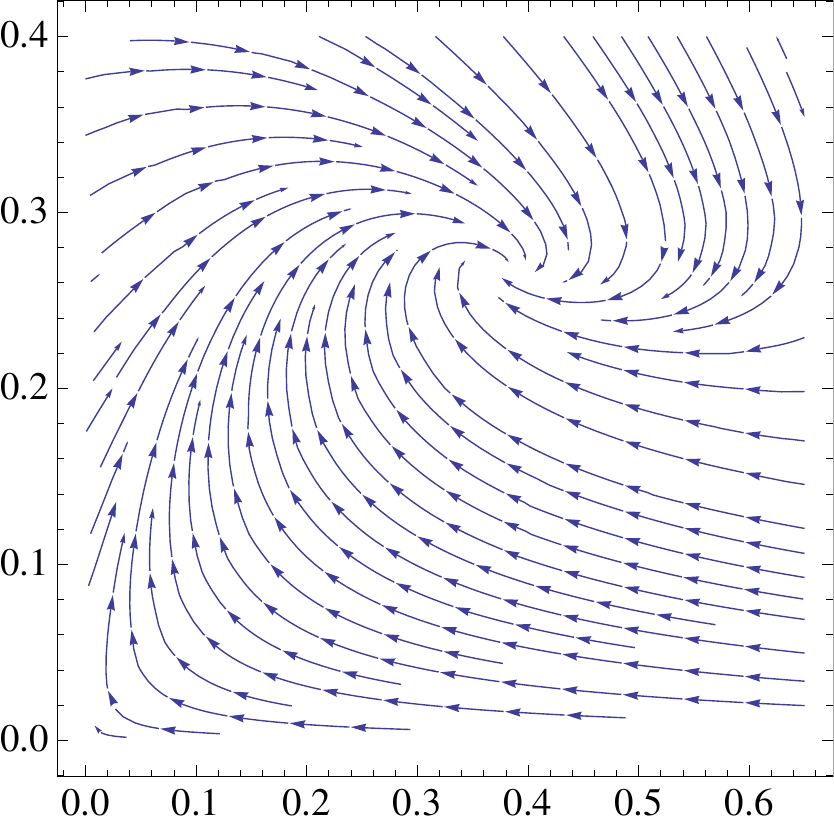}}
\put(13.5,0){$\lambda$}
\put(7,5){$G$}
\end{picture}
\caption{Since we are interested in the approach to the NGFP towards the UV, we plot the flow towards the UV. In the left panel, we plot the RG flow 
 in the
Einstein-Hilbert truncation with a non-trivial ghost
anomalous dimension $\eta_c$, and with
$\eta_c=0$ in the right panel. Apart from a slight shift in
the fixed-point position, the agreement between the flow in
this extended truncation and the simpler Einstein-Hilbert
truncation is very high. Both directions are UV-attractive in both truncations.\label{EHtruncationplotwithetac}
}
\end{figure}

\noindent We thus have asserted that this first step in
enlarging truncations in a new class of directions in theory
space, namely those containing couplings to Faddeev-Popov
ghosts, supports former findings of a NGFP at positive
$G_{\ast}$ and with two relevant directions in the
Einstein-Hilbert sector. We now turn to interpret the
implications of our value for $\eta_c$.

In our calculation we observe that $\eta_c < 0$ in both gauges. In particular, the value in Landau-deWitt gauge can be interpreted as an upper bound (within our truncation) under a variation of the gauge, since for $\alpha >0$ the contribution of the scalar and the vector mode have to be taken into account. These contribute with a negative sign.

The value of the anomalous dimension determines the momentum dependence of the propagator in the vicinity of the fixed point:
As $\partial_t Z_c = -\eta_c Z_c \label{etadef}$
 with $\eta_c = \rm const$ at the fixed point, \Eqref{etadef} can be integrated to yield $Z_c(p^2) \sim (p^2)^{-\frac{\eta}{2}}$ and accordingly
\begin{equation}
 \Gamma^{(2)}(p^2) \sim (p^2)^{1-\frac{\eta}{2}}.
\end{equation}
where we have gone to flat space for simplicity. Fourier
transforming this expression for $\eta \neq 0$ yields real
space propagators that scale logarithmically for $\eta=-2$,
i.e. like a 2-dimensional propagator. Since $\eta_N =-2$,
this is interpreted as one signal of a dynamical dimensional
reduction to two dimensions in the vicinity of the fixed
point \cite{Lauscher:2005qz,Lauscher:2005xz}. This might
provide for a link to other approaches to quantum gravity,
where indications for a similar effect have been found
\cite{Ambjorn:2005db,Modesto:2009kq,Carlip:2009km}.  If the
effective physical manifold indeed goes through such a
dimensional reduction, by being, e.g. fractal or
fractal-like, one might expect that any field propagating on
this background should reflect the effective
two-dimensionality in the behaviour of the propagator. Our
result $\eta_c \neq -2$ can then be interpreted in several
ways:\\
First, crucial operators may be missing from the truncation that will result in $\eta_c \approx-2$.

As a second possibility, one might also expect that only physical fields (i.e. fields from which observables can be constructed) have to show the logarithmic falloff of the propagator in real space. Since ghosts are not part of the physical Hilbert space, their anomalous dimension would remain unrestricted in this case.

The last option is that $\eta_N =-2$ implies that fluctuations of the metric interact so strongly that the metric propagator changes to a logarithmic falloff, but this need not indicate that the same effect applies to other fields coupled to gravity.

The fact that $\eta_c \neq -2$ might also be puzzling at first sight for the following reason: As ghosts are introduced into the theory to cancel non-physical components of the metric, one might naively expect that in order for this to work also non-perturbatively, the scaling exponents of the propagators have to agree. In fact this need not be the case if vertices also acquire a non-trivial scaling, as then ghost diagrams can still cancel the effect of non-physical metric modes. Besides, the value $\eta_N=-2$ holds for the background metric, or within a single-metric approximation, whereas the anomalous dimension of the fluctuations metric may be different.

Since our result is compatible with $\eta_c > \eta_N$ it indicates that the ghost-propagator is less suppressed at large momenta than the metric propagator. We may interpret this as a sign for a possible ghost dominance at the fixed point, where the propagation of ghosts is dynamically enhanced in comparison to the metric. This is reminiscent of Yang-Mills theory in the IR, where the ghost propagator is strongly dynamically enhanced, whereas the gluon propagator is suppressed (see sec.~\ref{Landaugaugeprops} for details). The physical consequences of a dynamic ghost enhancement in gravity remain to be investigated. 
Note, however, that for a full understanding of these issues
it is necessary to depart from the standard single-metric
approximation and evaluate the anomalous dimension of the
fluctuation metric which is not restricted to be $\eta_{\rm
fluc}=-2$, see, e.g. \cite{Codello_thesis}.

An important conclusion can be drawn from the negative sign
of the anomalous dimension: As explained above (see
\Eqref{relevance}), the value of the anomalous dimension is
decisive to determine the relevance of a coupling. Naturally
the critical exponent of a coupling depends on the specific
coupling under consideration and one has to evaluate its
$\beta$ function fully to answer the question of relevance.
However a negative anomalous dimension may shift critical
exponents such that, e.g. power-counting marginal operators
become relevant, since then the corresponding
contribution to the critical exponent is positive. Therefore
$\eta_c <0$ implies that couplings in the ghost sector may
turn out to be relevant. 

We now have to understand the implications of this observation: As we have
already explained, the effective average action in a gauge
theory has to satisfy two equations, namely the Wetterich
equation as well as the modified Ward identity (see sec.~\ref{symmetries}). Therefore determining the number of
relevant operators from the linearised flow around the NGFP
does not determine the number of free parameters of the
theory: A subset of relevant couplings may indeed be related
by the (modified) Ward identities, and therefore correspond
only to a reduced number of free parameters. All operators
containing ghost fields are subject to the Ward identities.
This implies that relevant operators found in the ghost
sector do presumably not directly correspond to measurable quantities.
However, the solution of the (modified) Ward identities is
highly non-trivial. One may speculate that the ghost
sector may offer the opportunity to find relevant operators
related to metric operators that are technically much more
challenging to include in a truncation. Investigating the
question of relevant operators in the ghost sector therefore
corresponds to accessing the critical surface from a
completely different direction in theory space, and can be
used to find an upper bound on the number of free parameters
of the theory. \newline\\

A simple check of effects of terms neglected in this
truncation is given by the following idea: As, e.g.
four-ghost operators cannot couple directly into the flow of
the Einstein-Hilbert term due to the one-loop structure of
the Wetterich equation, an inclusion of such terms into our truncation would only
alter the value of $\eta_c$. Thus we examine the existence
and the properties of the fixed point under variations of
$\eta_c$. This will indicate if higher-order ghost terms can
indirectly lead to a destabilisation of the fixed point in
the Einstein-Hilbert sector. Of course it is still necessary
to check whether the $\beta$ functions of such ghost
couplings themselves also admit a fixed point. 

\begin{figure}[!here]
\begin{minipage}{0.73\linewidth}
\includegraphics[scale=0.24]{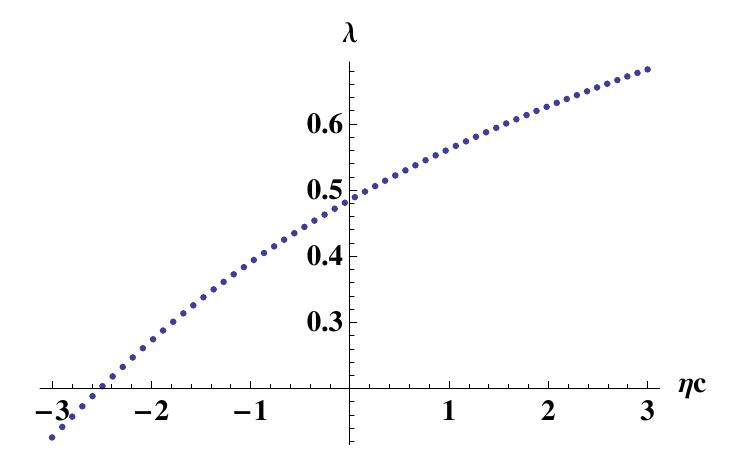}
\includegraphics[scale=0.24]{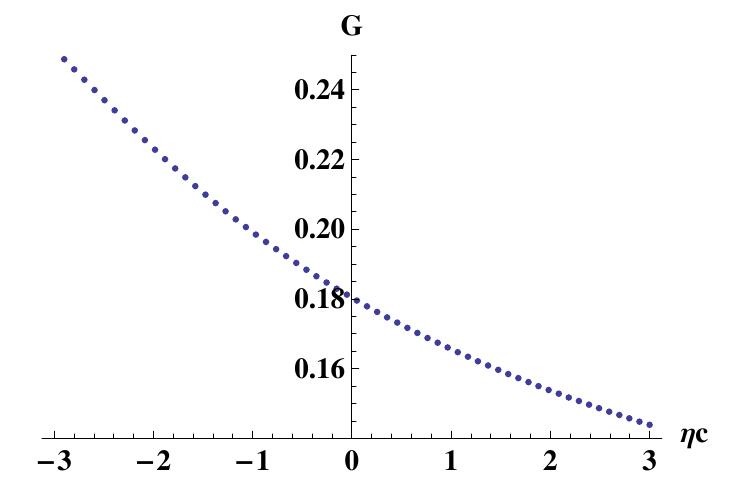}
\end{minipage}
\begin{minipage}{0.24\linewidth}
\caption{For the deDonder gauge $\rho = 
\alpha =1$ we show the fixed point values $G_{\ast}$ and $\lambda_{\ast}$ as a function of $\eta_c$, that we treat as a free parameter here. The self-consistently determined value is $\eta_c\approx -1.31$.\label{etaplots1}}
\end{minipage}
\end{figure}

\begin{figure}[!here]
\begin{minipage}{0.55\linewidth}
\includegraphics[scale=0.24]{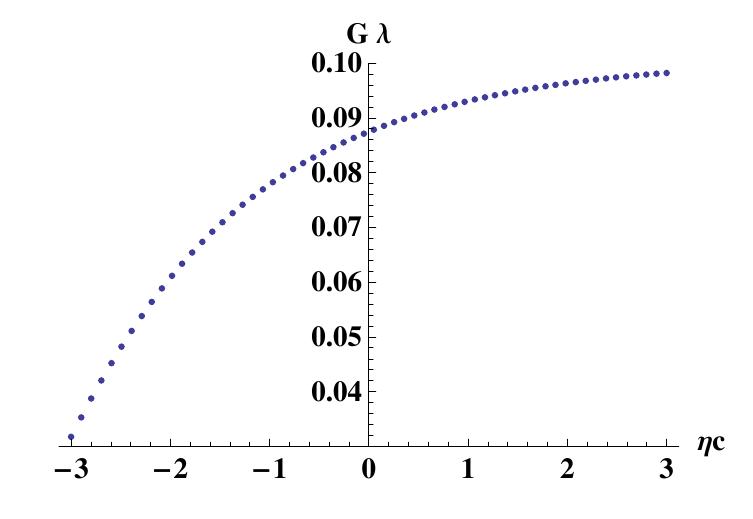}
\end{minipage}
\begin{minipage}{0.4\linewidth}
\caption{We show the universal product $G_{\ast}\lambda_{\ast}$ of the fixed-point values in the Einstein-Hilbert sector as a function of $\eta_c$, again for the deDonder gauge.\label{etaplots2}}
\end{minipage}
\end{figure}

\begin{figure}[!here]
\includegraphics[scale=0.24]{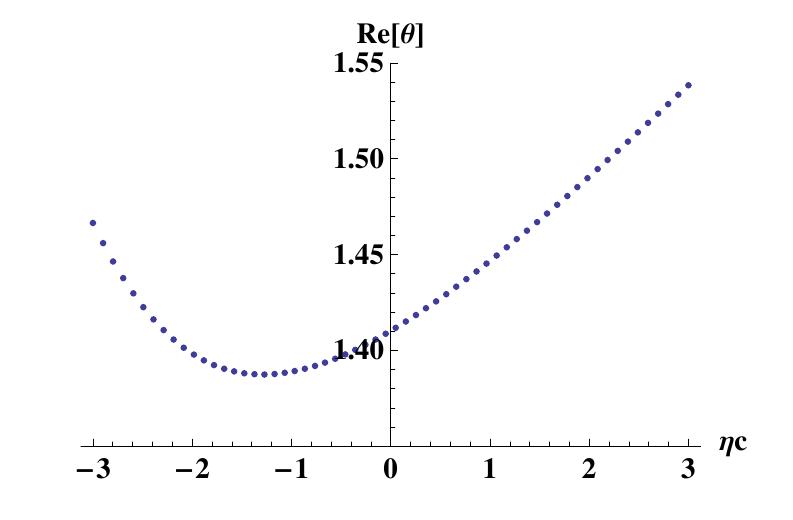}
\includegraphics[scale=0.24]{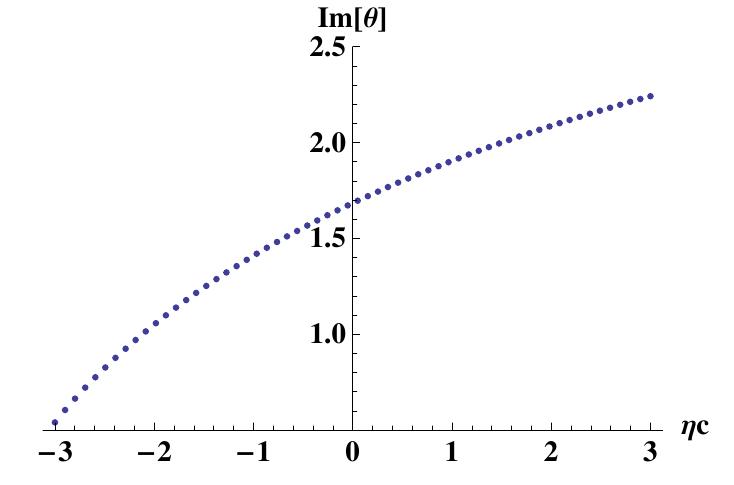}
\caption{We show the real and the imaginary part of the critical exponents in the Einstein-Hilbert sector as a function of $\eta_c$ for the deDonder gauge.\label{etaplots3}}
\end{figure}

We find that
in the Einstein-Hilbert sector the fixed point exists for
values $\eta_c$ between (-3,3), see figs.~\ref{etaplots1}, \ref{etaplots2}, \ref{etaplots3}, which should cover the
result for $\eta_c$ in full theory space. Furthermore it
lies at $G_{\ast}>0$ for all these values and accordingly is
physically admissible, see fig.~\ref{etaplots1}.\\
 
\noindent In particular, we observe that the real part of the critical exponent is not affected much by variations of $\eta_c$, which implies a high degree of stability under enlargements of the truncation in this specific direction. Interestingly the imaginary part of the critical exponent can be reduced considerably, which is in accordance with the observation, that it can depend strongly on the regularisation and gauge-fixing scheme (${\rm Im}(\theta_{1/2})\in (2.4,7)$ as found in other schemes \cite{Litim:2008tt}).

The fixed-point value of the cosmological constant depends
strongly on $\eta_c$. Therefore further couplings in the
ghost sector can have an important indirect effect on the flow:
Driving $\lambda_{\ast} \rightarrow 0$ (or even negative
values) suppresses metric fluctuations in comparison to
positive $\lambda$. Such a suppression can strongly change
the RG flow in the vicinity of the fixed point and alter the
fixed point properties. In particular, since no analogous
suppression mechanism works for the ghost sector, ghost
fluctuations might thus contribute to the flow with a
higher relative weight in the vicinity of the fixed point.\\

As already stressed in sec.~\ref{AS_intro}, the
calculation of $\beta$ functions for gravity from the
Wetterich equation also holds if the UV completion of
gravity is not given by the asymptotic-safety scenario.
Then, our calculations hold below some scale $k_0$
(presumably below the Planck scale), for which an effective
description in terms of metric fluctuations is possible.
Here we may assume, that quantum effects are already
smaller, and consider a perturbative setting where we assume $\eta_N \approx 0$. To
leading order in $G$ (and for vanishing $\lambda$), we then
have that
\begin{equation}
\eta_c=-\frac{\left(5 \alpha ^3-26 \alpha ^2+5 \alpha +124\right) G}{4 \pi  (\alpha
  -3)^2} + \mathcal{O}(G^2).
\label{eq:PT}
\end{equation}
We observe that $\eta_c$ is negative for all admissible gauge
parameters $\alpha\geq0$. The incompleteness of the gauge-fixing at $\alpha =3$ is again reflected in the divergence at this point.
 
The Landau-deWitt-gauge limit is
$\eta_c=-\frac{31}{9\pi} G \simeq -1.0964 G$. Of course, this result is non-universal, i.e.
scheme dependent in four dimensions, as the power-counting RG critical
dimension is $d=2$.

\subsection{Extension of the truncation - first steps beyond Faddeev-Popov gauge fixing}\label{exttrunc}
As motivated, the non-perturbative regime may require to go beyond a simple form of Faddeev-Popov gauge fixing in order to solve the Gribov problem. As a very first step in such a direction we study an extended truncation that distinguishes between the running of different tensor structures.

Any $n$-point function can be decomposed into
scalar dressing functions according to
\begin{equation}
 \Gamma_{k\,
\mu_1,...\mu_m}^{(n)}(p_1,...,p_n)=\sum_{i}a_{k\,i}(p_1,...,
p_n)T^i_{ \mu_1 , ...\mu_m}(p_1,...,p_n) ,
\end{equation}
where we have gone to flat space for simplicity. On a
generic curved background a generalisation holds.
The $a_{k\,i}(p_1,...,
p_n)$ are running scalar dressing functions and the $T^i_{
\mu_1 , ...\mu_n}(p_1,...,p_n)$ denote the different tensor
structures that can be constructed. In general there is no reason
to assume that the different $a_{k\, i}$ satisfy the same
flow equation. Hence our truncation in the ghost sector
with just a single wave-function renormalisation
corresponds to a simplifying assumption. Here we will
generalise the above truncation to 
\begin{equation}
 \Gamma_{k \, \rm gh}= - \sqrt{2} Z_c(k) \int d^4 x \sqrt{\bar{g}}\bar{c}_{\mu} \Bigl(\bar{D}^{\rho}\bar{g}^{\mu \kappa}\gamma_{\kappa \nu}D_{\rho}+\delta Z_c(k) \Bigl(\bar{D}^{\rho}\bar{g}^{\mu \kappa}\gamma_{\rho \nu}D_{\kappa}-\tfrac{1+\rho}{2}\bar{D}^{\mu}\bar{g}^{\rho \sigma}\gamma_{\rho \nu}D_{\sigma} \Bigr) \Bigr)c^{\nu},\label{trunc}
\end{equation}
where we have introduced an additional running coupling $\delta Z_c(k)$ \cite{ghostpaper3}, which was set to $\delta Z_c=1$ in former truncations.
This truncation is motivated by observing that within a single-metric approximation, $g_{\mu \nu}= \bar{g}_{\mu \nu}$ there are four power-counting marginal operators that involve two powers of the ghost field:
\begin{equation}
 \bar{c}^{\mu}D^2 g_{\mu \nu}c^{\nu}\quad \bar{c}^{\mu}[D_{\mu},D_{\nu}]c^{\nu}\quad \bar{c}^{\mu}\{D_{\mu}, D_{\nu}\}c^{\nu} \quad \bar{c}^{\mu}R g_{\mu \nu}c^{\nu}.
\end{equation}
The first three of these naturally appear in the Faddeev-Popov determinant. In former calculations \cite{Eichhorn:2010tb,Groh:2010ta} no distinction has been made between these three operators. Their running couplings have been treated as one single wave-function renormalisation for the ghost field. Here we report on the first step in a more detailed distinction between these three different tensor structures. Our truncation distinguishes the term $D^2 g_{\mu \nu}$ from the two terms $[D_{\mu},D_{\nu}]$ and $\{D_{\mu},D_{\nu}\}$.

On a flat background we can distinguish the two couplings $Z_c$ and $\delta Z_c$ after a York decomposition of the ghost into a transversal and a longitudinal component. The propagators and vertices which arise from this truncation and the projection rules onto the running couplings are given in app.~\ref{vertices_deltaZc}.

\begin{figure}[!here]
 \begin{minipage}{0.22\linewidth}
  \includegraphics[scale=0.07]{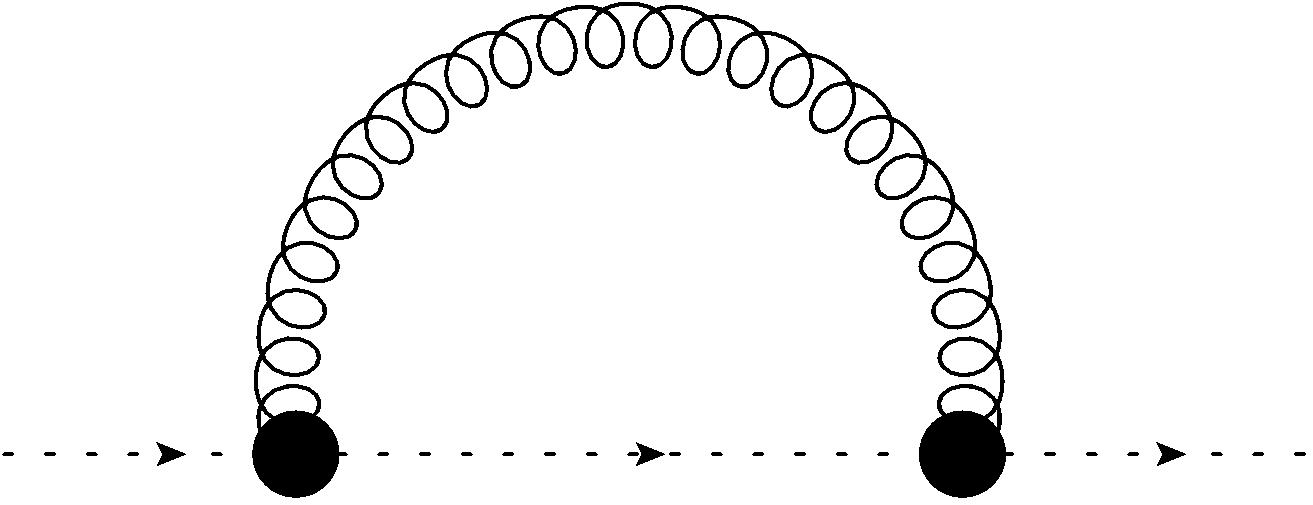}\newline\\
  \includegraphics[scale=0.07]{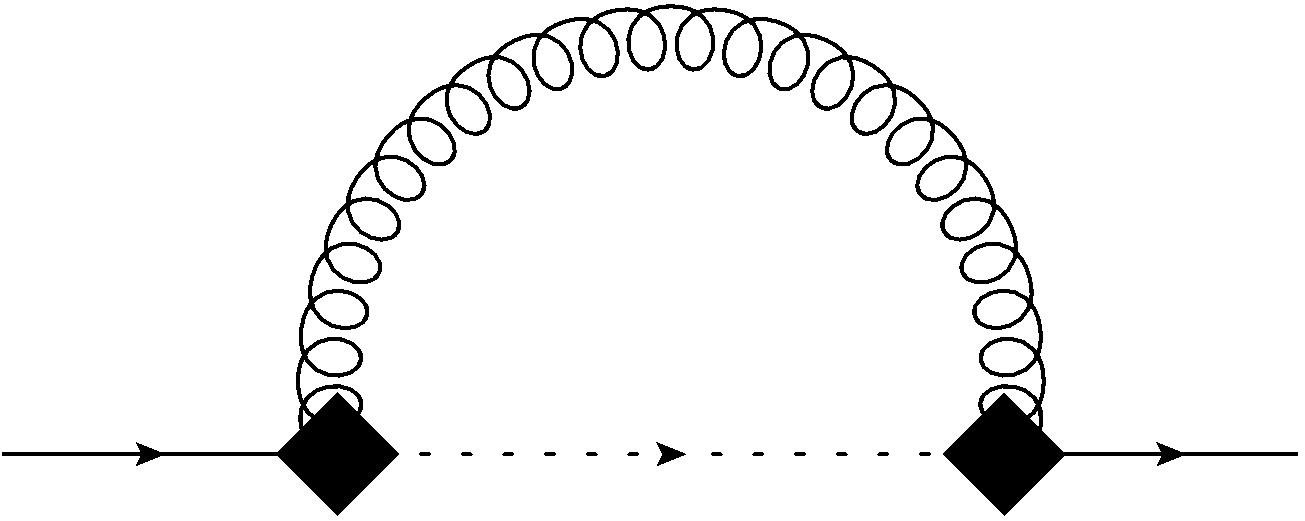}
 \end{minipage}
\begin{minipage}{0.22\linewidth}
    \includegraphics[scale=0.07]{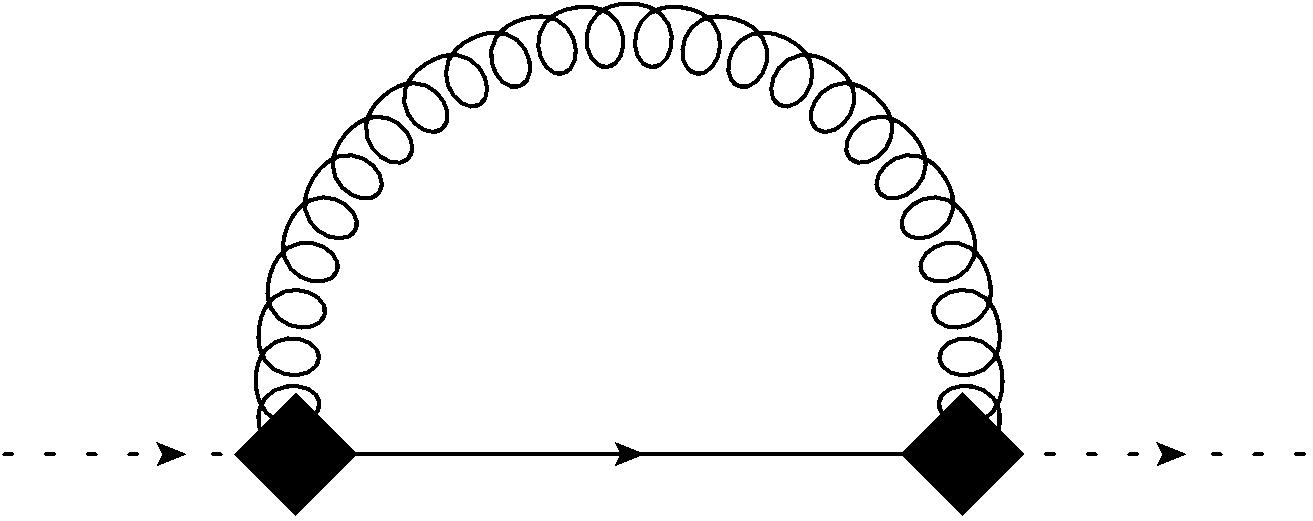}\newline\\
  \includegraphics[scale=0.07]{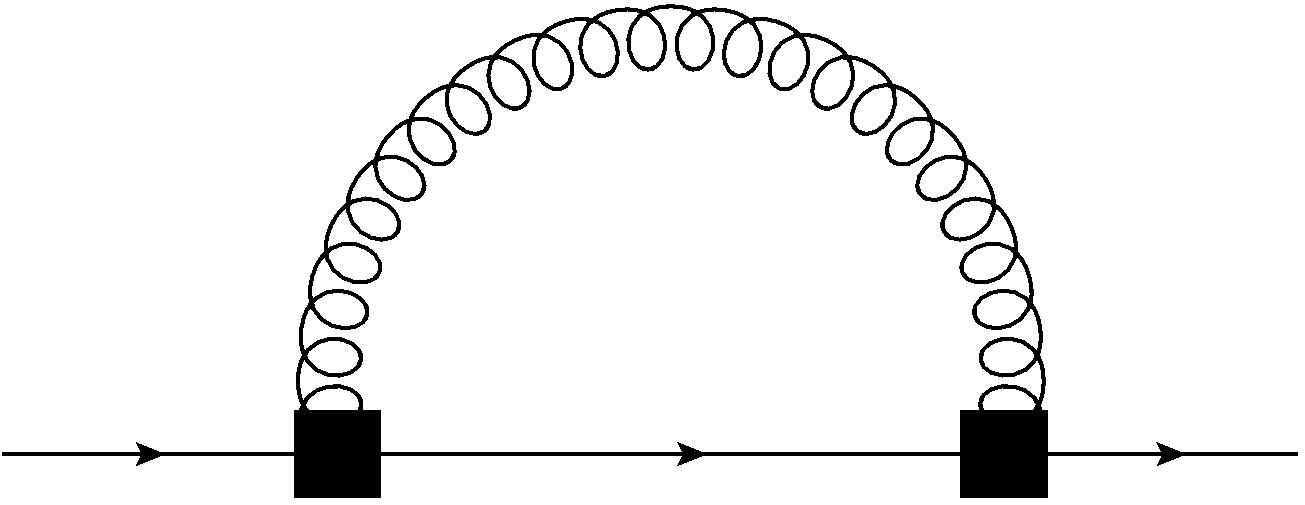}
 \end{minipage}
\begin{minipage}{0.22\linewidth}
    \includegraphics[scale=0.07]{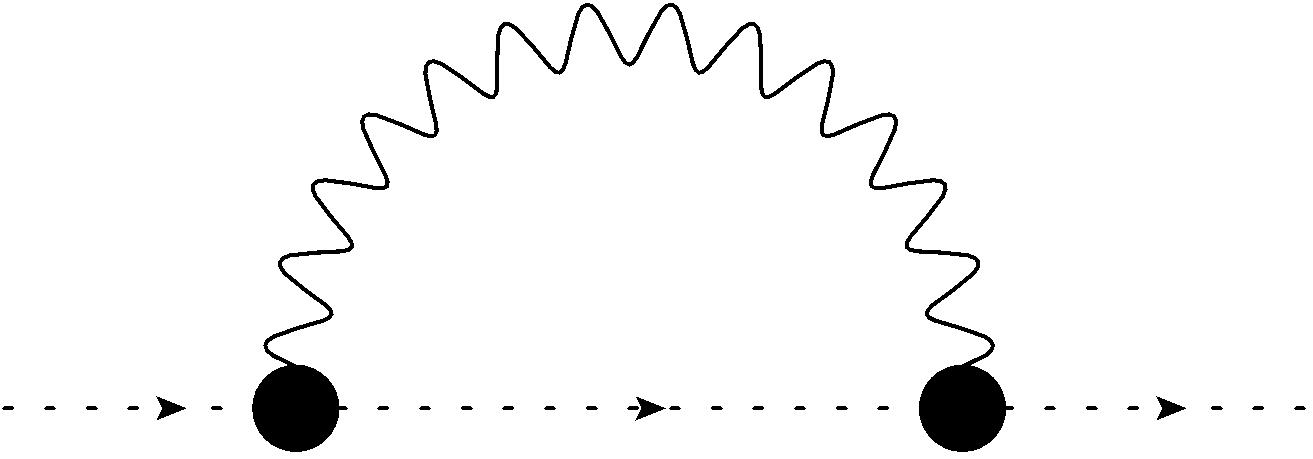}\newline\\
  \includegraphics[scale=0.07]{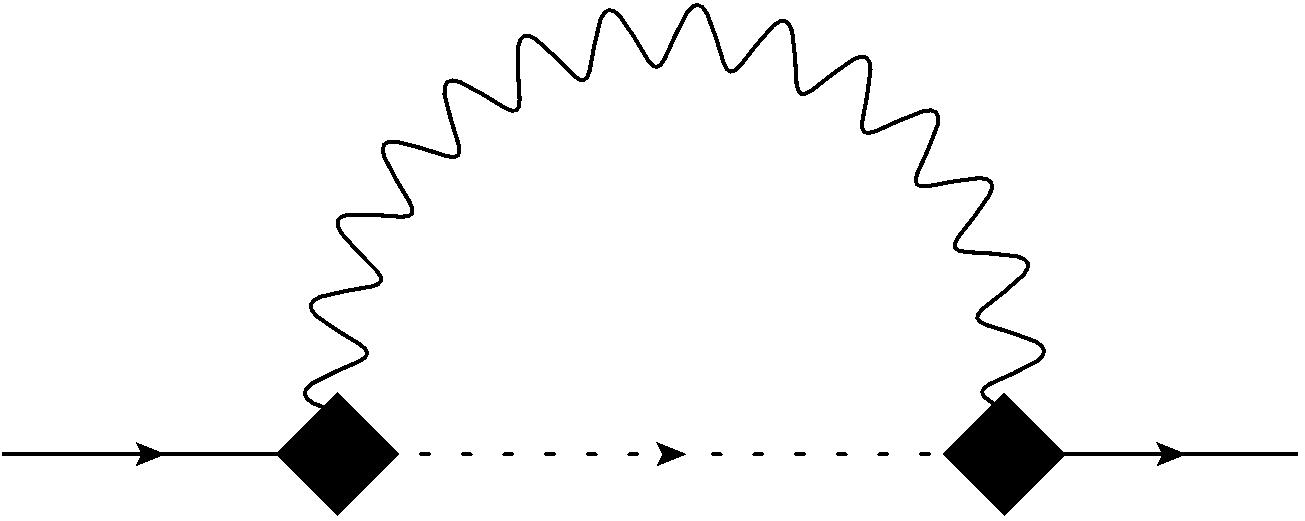}
 \end{minipage}
\begin{minipage}{0.22\linewidth}
    \includegraphics[scale=0.07]{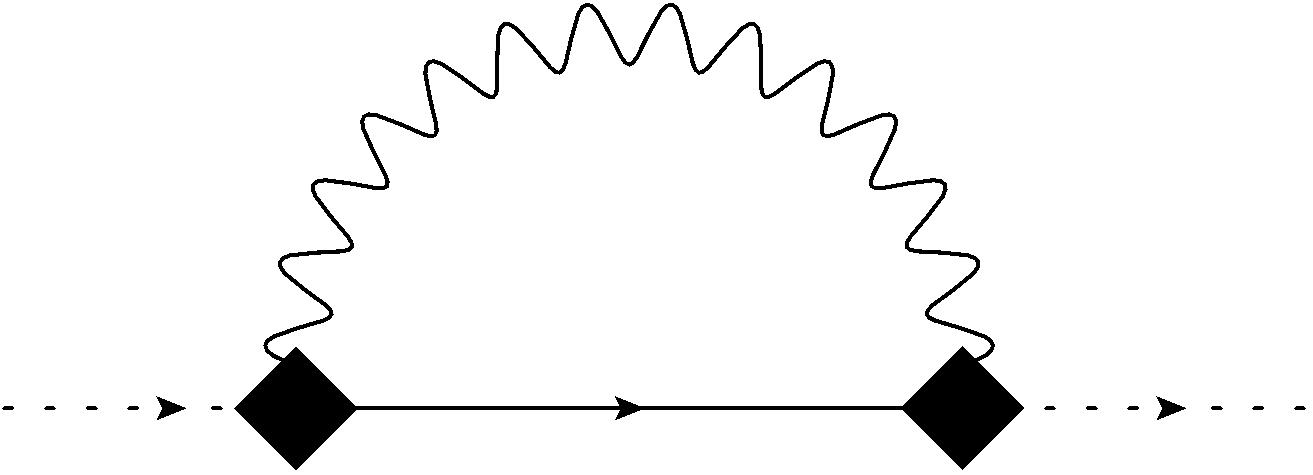}\newline\\
  \includegraphics[scale=0.07]{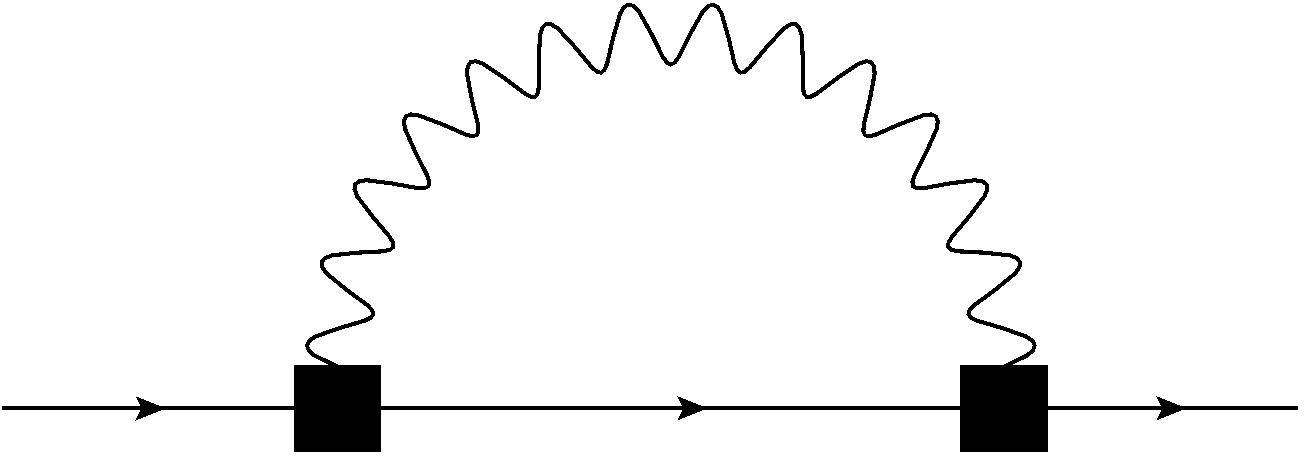}
 \end{minipage}

\caption{Self-energy diagrams contributing to the flow of $\eta_c$ and $\partial_t \delta Z_c$ in the Landau-deWitt gauge (for the projection onto these two couplings see \Eqref{projetacext} and \Eqref{projdeltaZ}). Here, the wavy line denotes the trace mode and the curly line the transverse traceless mode. The dashed line denotes the longitudinal ghosts/antighosts and the full line the transverse ghosts/antighosts. Circles denote vertices coupling external longitudinal (anti) ghosts to internal metric and longitudinal ghost propagators. Squares couple transverse external (anti)ghosts to internal metric and transverse ghost propagators. Finally diamonds denote vertices that mix transverse and longitudinal ghosts, which are also all non-zero.
\label{diagsexttrunc}}
\end{figure}

Diagrammatically, the flow of these couplings is driven by a
self-energy diagram as in fig.~\ref{ghostanomdimflowdiags},
the difference being that we apply a York-decomposition for
the fluctuation and the background ghosts. We thereby get a
self-energy diagram with external transversal ghost and
antighost lines, and a second self-energy diagram with
external longitudinal ghost and antighost lines. Both of
these exist for the trace and the transverse traceless mode,
and for internal longitudinal and transversal ghosts, cf.
fig.~\ref{diagsexttrunc}. 

\noindent Using the same techniques as outlined in the above sections, we determine the coupled system of $\eta_N, \partial_t \lambda, \eta_c $ and $\partial_t \delta Z_c$, which are given in app.~\ref{deltaZc}. The large number of diagrams and the more complex structure of the vertices arising from the additional coupling ultimately accounts for the rather lengthy representation of $\beta$ functions in a specific regularisation scheme, where we again choose a spectrally and RG adjusted regulator with exponential shape function.

Note that $\delta Z_c$ is not an inessential coupling and therefore has to satisfy a fixed-point requirement. This can be seen directly from the $\beta$ function for $\delta Z_c$, which depends on all couplings in our truncation, including $\delta Z_c$. In contrast, the equation for the inessential coupling $Z_c$ only depends on all other couplings and $\beta$ functions and can therefore be eliminated from the set of differential equations determining the fixed points.

Interestingly, we observe two fixed points, the properties of which are given in table \ref{FP_deltaZc} for Landau-deWitt gauge ($\alpha =0$):

\begin{table}[!here]
\begin{center}
\begin{tabular}{c|c|c|c|c|c}
 $G_{\ast}$& $\lambda_{\ast}$& $\eta_c$& $\delta Z_{c\, \ast}$& $\theta_{1/2}$& $\theta_3$\\
\hline
0.287&0.317&-0.780&1&2.034 $ \pm$ i 1.499&0.646\\
0.262&0.372&0.241&1.562&2.130 $ \pm$ i 1.634&-0.826
\end{tabular}
\caption{We observe two NGFPs with differing universality properties. The values are given for the Landau-deWitt gauge $\alpha =0$.} 
\label{FP_deltaZc}
\end{center}
\end{table}

\noindent Let us comment on the fixed-point properties: At the first fixed point $\delta Z_{c\ast}=1$ holds\footnote{This exact value arises from a global factor of $(\delta Z_c -1)$ in the fixed-point equation for this coupling (which is not readily visible from \Eqref{deltaZflowloooong}). In  the Landau-deWitt gauge $\rho =0$, this factor appears in the vertices, cf. \Eqref{transvertices} and \Eqref{confvertices}. As in a further extension of the truncation the form of the vertices will typically change, the value $\delta Z_{c\ast}=1$ will presumably not hold beyond our truncation. Note further that it will also be gauge-dependent, since for $\rho \neq 0$ the corresponding factor changes to $(2 \rho\, \delta Z_c -1+\delta Z_c )$ in some of the vertices.}, which implies that its properties are the same as in the truncation where we set $\delta Z_c=1$ by hand.

We observe numerically, that the new direction, corresponding to $\delta Z_c$ is not mixed with the directions in the Einstein-Hilbert sector, in other words it constitutes a separate eigendirection of the fixed point. As it is relevant, this implies that we will observe a non-trivial flow of $\delta Z_c$ towards the infrared. Therefore the Faddeev-Popov operator will change in the flow towards the infrared. This is a first manifestation of the assumption that we should not expect that a simple Faddeev-Popov gauge fixing holds in the non-perturbative regime. 

Let us speculate on the possible implications of this observation: In the strongly-interacting regime in Yang-Mills theory, the dominant field configurations in gauge-field configuration space lie close to the first Gribov horizon, which is defined by the first zero eigenvalue of the Faddeev-Popov operator. Then, the dominant field configurations change, depending on the scale, as the theory shows a transition from the perturbative, weakly coupled, into the strongly coupled regime. In this transition, the dominant field configurations change from perturbative configurations to configurations in the proximity of the Gribov horizon, which carry the physical information on confinement.
If we now assume, that in the strongly coupled regime in
gravity the dominant field configurations also lie close to
the first Gribov horizon (and if we assume that the first
Gribov region is also bounded\footnote{This may not be the
case; for an example where the Gribov region is unbounded
in some directions in configuration space in Yang-Mills
theory, consider the maximal Abelian gauge
\cite{Capri:2010an}.}), then gravity
shows an additional feature in contrast to Yang-Mills
theory: As the Faddeev-Popov operator changes, depending on
the scale, this implies that the boundary of the first
Gribov region changes with scale. Thus, the dominant field
configurations in gravity would not only differ between the
weakly coupled and the strongly coupled regime, but in
addition change, depending on the scale, \emph{within} the
 non-perturbative regime.

 The second fixed point shows a non-trivial value of $\delta Z_c$, which changes the ghost anomalous dimension to a positive value. This implies that this fixed point shifts further ghost operators towards irrelevance. Furthermore the additional critical exponent at this fixed point is irrelevant, too. Since, after imposing the Ward identities, each relevant direction corresponds to a free parameter, this second fixed point might correspond to a UV completion for gravity with a greater predictive power, since less free parameters remain to be fixed by experiment.

In view of the scenarios that we introduced for the ghost sector, this second fixed point also offers the possibility to more directly identify relevant couplings with measurable free parameters and presumably circumvents the difficulties that accompany the interpretation of relevant directions in the ghost sector.

If this structure persists beyond our simple truncation, the ghost sector might offer a choice between two UV fixed points: One of these potentially shows a larger number of positive critical exponents, whereas the other is characterised by a larger number of irrelevant directions. Furthermore adding new directions in the ghost sector presumably only increases the number of irrelevant directions at this second fixed point, since the ghost anomalous dimension is positive there. 

\begin{figure}[!here]
\begin{minipage}{0.6\linewidth}
 \includegraphics[scale=1]{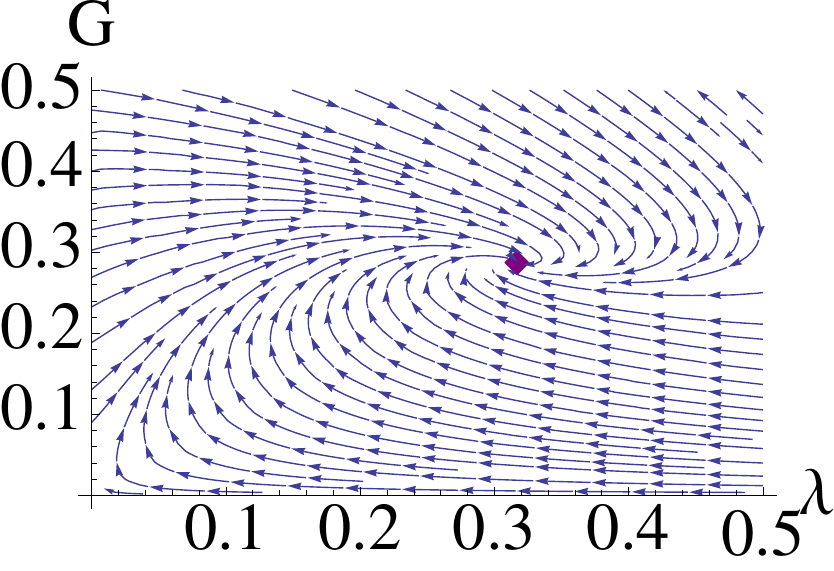}\\
 \includegraphics[scale=1]{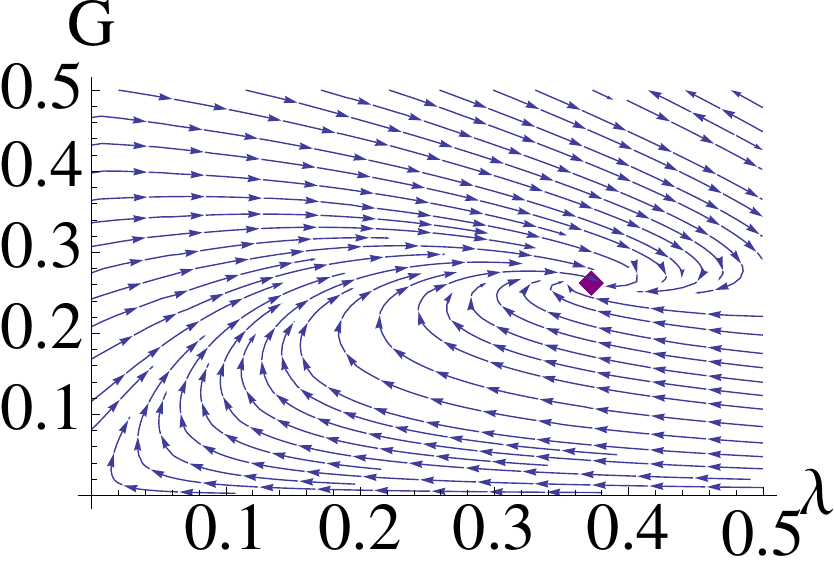}
\end{minipage}
\begin{minipage}{0.35\linewidth}
\caption{The flow towards the ultraviolet in the $(G, \lambda)$-plane for $\delta Z_c =1$ (upper panel) shows the well-known spiral into the NGFP, and also trajectories passing close to the Gau\ss{}ian fixed point, where a long classical regime can be supported. The flow at  $\delta Z_c =1.562$ (lower panel) shows a very high degree of similarity. The corresponding fixed-point values are indicated by a purple square.\label{EHdeltaZflow}}
\end{minipage}
\end{figure}

\begin{figure}[!here]
\begin{minipage}{0.6\linewidth}
\setlength{\unitlength}{1cm}
\begin{picture}(5,5)
\put(0,3){\includegraphics[scale=0.9]{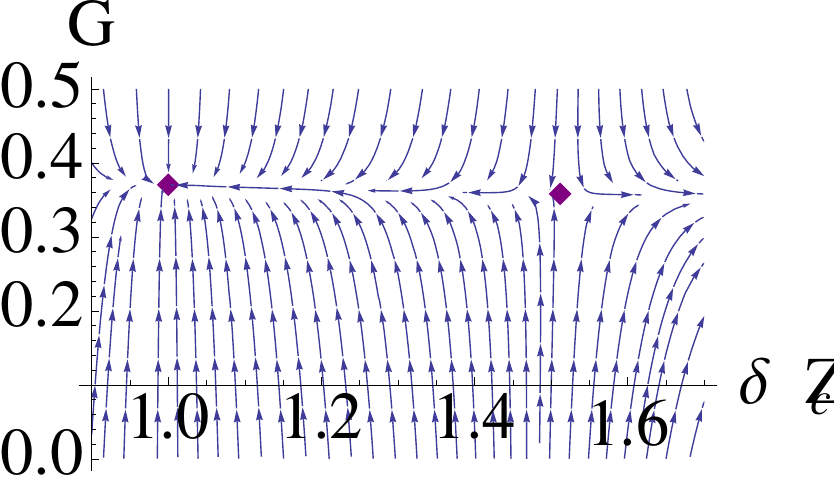}}
\put(7,3.8){\color{white}\circle*{3}}
\put(7.5,3.8){\color{white}\circle*{3}}
\put(7,4){\color{white}\circle*{3}}
\put(7.5,4){\color{white}\circle*{3}}
\put(6.8,3.8){\large $\delta Z_c$}
\put(0,-2){\includegraphics[scale=0.9]{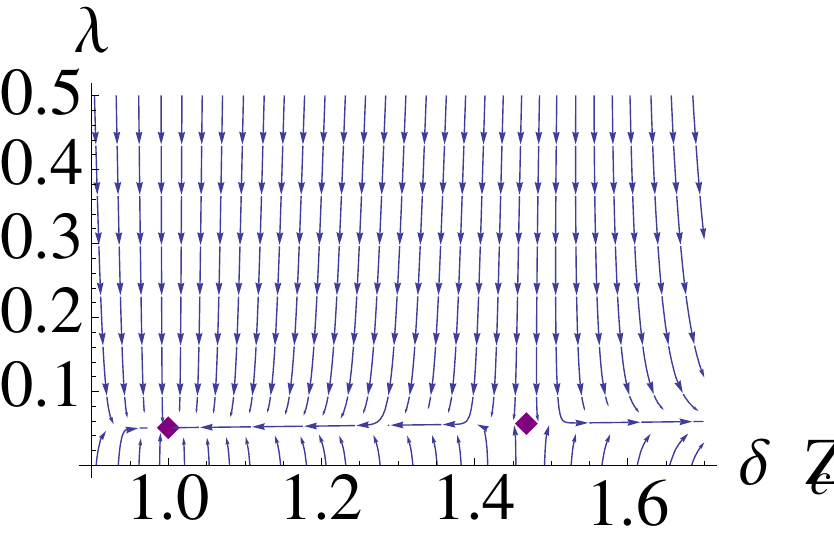}}
\put(7,-1.4){\color{white}\circle*{3}}
\put(7.5,-1.4){\color{white}\circle*{3}}
\put(7,-1.2){\color{white}\circle*{3}}
\put(7.5,-1.2){\color{white}\circle*{3}}
\put(6.8,-1.4){\large $\delta Z_c$}
\end{picture}
\end{minipage}
\begin{minipage}{0.35\linewidth}
\caption{In the upper panel the flow towards the ultraviolet in the $(G, \delta Z_c)$-plane for $\lambda=0.1$ shows that, due to the large difference in the size of the (real part) of the critical exponents, the flow stays in planes $\delta Z_c \approx \rm const$, approximately. The irrelevant direction of the fixed point at $\delta Z_c=1$ becomes visible only around the separatrix between the two fixed points. In the lower panel the flow towards the ultraviolet in the $(\lambda, \delta Z_c)$-plane for $G=0.1$ again exhibits the same behaviour.\label{flowdeltaZG}}
\end{minipage}
\end{figure}

Let us now study the full flow: For slices with $\delta Z_c=
\rm const$ through our three-dimensional theory space we
observe that the flow pattern resembles the flow in the
simple Einstein-Hilbert truncation to a very high degree
(see fig.~\ref{EHdeltaZflow}). This can also be observed by
setting $G= \rm const$ (see fig.~\ref{flowdeltaZG}) or
$\lambda = \rm const$ (see fig.~\ref{flowdeltaZG}), where
clearly in most regions the flow is approximately confined
to a lower-dimensional subspace. We could have inferred such
a flow pattern from observing the real part of the critical
exponents, which are more than twice as large in the
Einstein-Hilbert sector than in the new direction.
Accordingly the flow in the Einstein-Hilbert plane is
faster, and a flow in orthogonal directions occurs mostly in
the vicinity of the Einstein-Hilbert fixed point at fixed
$\delta Z_c$.\\

$\phantom{x}$\newline

\noindent Since we observe two NGFPs, one of which has one irrelevant direction, an RG trajectory exists that connects the two fixed points. Thus the fixed point at $\delta Z_c \approx 1.562$ can be interpreted as an \emph{infrared} fixed point. This allows to construct a complete RG trajectory, i.e. one without singularities in the flow which can be extended both to $k \rightarrow \infty$ and $k \rightarrow 0$. However the trajectory between these two fixed points does not support a long classical regime (cf. fig. \ref{flowdeltaZG} ) and therefore cannot constitute a complete RG trajectory that would describe our universe.

Interestingly, a similar structure has been observed in a bi-metric truncation \cite{Manrique:2010am}. It is a highly speculative possibility if a further extension of truncations including gauge-fixing-induced directions  will support the existence of such two fixed points. If this were the case, an extended truncation might also allow for an RG trajectory that features a long classical regime. This would allow to circumvent the singularity at $\lambda= \frac{1}{2}$ that shows up in the flow in the Einstein-Hilbert truncation in many regularisation schemes.

Having focussed on operators that are part of a simple Faddeev-Popov operator, we will now broaden our view a bit, and explain why we should generically expect the existence of further non-zero couplings in the ghost sector.

\subsection{Non-Gau\ss{}ian fixed point for ghost couplings}\label{ghostNGFPs}
Starting from a perturbative gauge-fixing technique, one
only needs to consider a Faddeev-Popov operator in the ghost
sector. We have already stressed that this presumably breaks
down in a non-perturbative setting, one indication of which
is the Gribov problem. Therefore one is lead to think about
further ghost operators. Within the setting used here, one
can easily see, that a non-trivial flow in many ghost
operators will presumably be induced even if one starts with
a simple Faddeev-Popov term only. The NGFP for the Newton
coupling generically implies that, beyond the possibility
that ghost couplings may also show NGFPs, even the GFP will
be shifted and become interacting for $G = G_{\ast} \neq 0$.

Let us consider a specific example here in order to exemplify this, namely a truncation with an Einstein-Hilbert and a gauge-fixing term and an extended ghost sector of the form
\begin{equation}
\Gamma_{k\,\rm ghost}= - \sqrt{2} Z_c \int d^4 x \sqrt{\bar{g}}\bar{c}^{\mu} \mathcal{M}_{\mu \nu}c^{\nu}+ g_{\rm ghost} \int d^4x \sqrt{\bar{g}} \left(\bar{c}^{\mu}\mathcal{M}_{\mu \nu}c^{\nu}\right)\cdot \left(\bar{c}^{\mu}\mathcal{M}_{\mu \nu}c^{\nu}\right),
\end{equation}
where $\mathcal{M}_{\mu \nu}$ is the Faddeev-Popov operator. Then, the flow will be given by the $\tilde{\partial}_t$ derivatives of the following diagrams:
\begin{figure}[!here]
\begin{minipage}{0.23\linewidth}
 \includegraphics[scale=0.07]{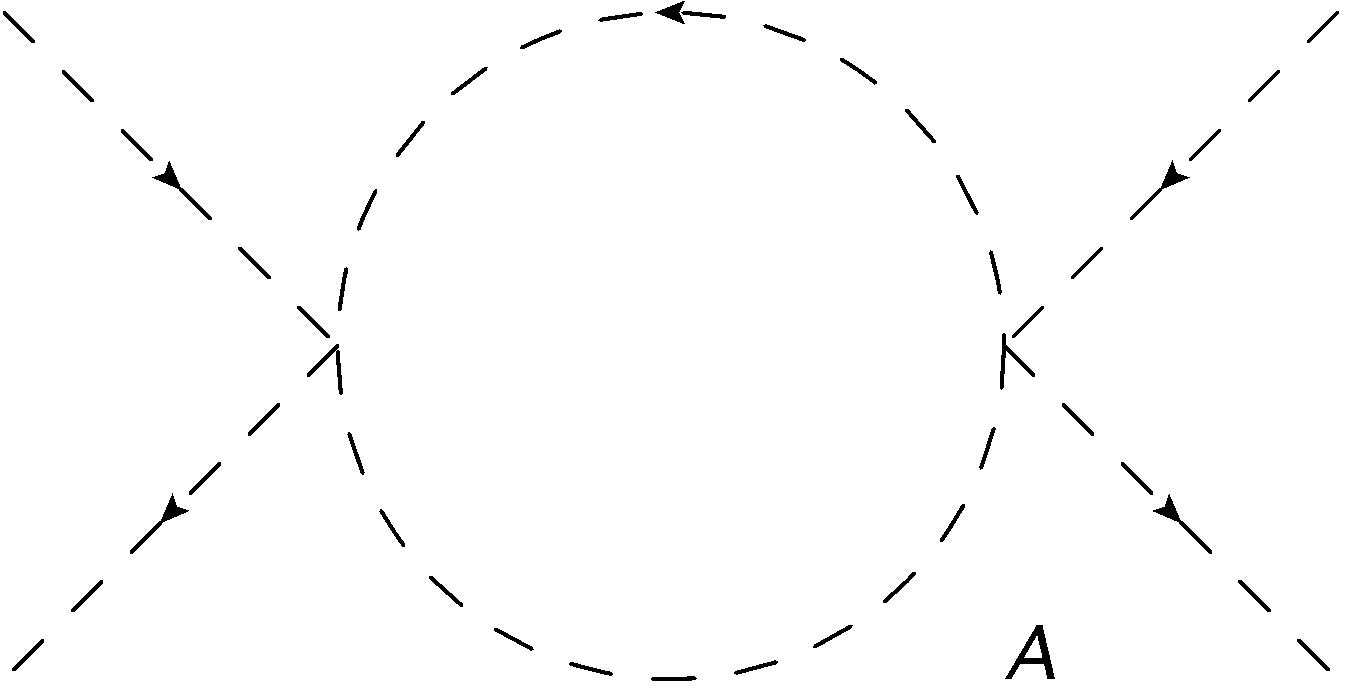}
\end{minipage}
\begin{minipage}{0.18\linewidth}
\begin{flushright}
 \includegraphics[scale=0.08]{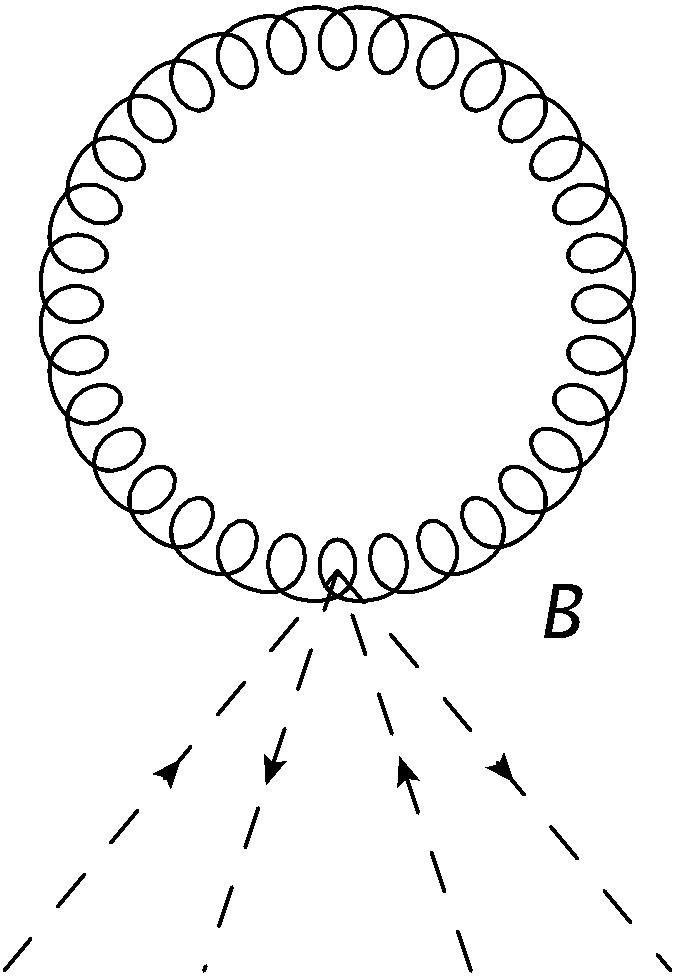}
\end{flushright}
\end{minipage}
\begin{minipage}{0.25\linewidth}
 \includegraphics[scale=0.08]{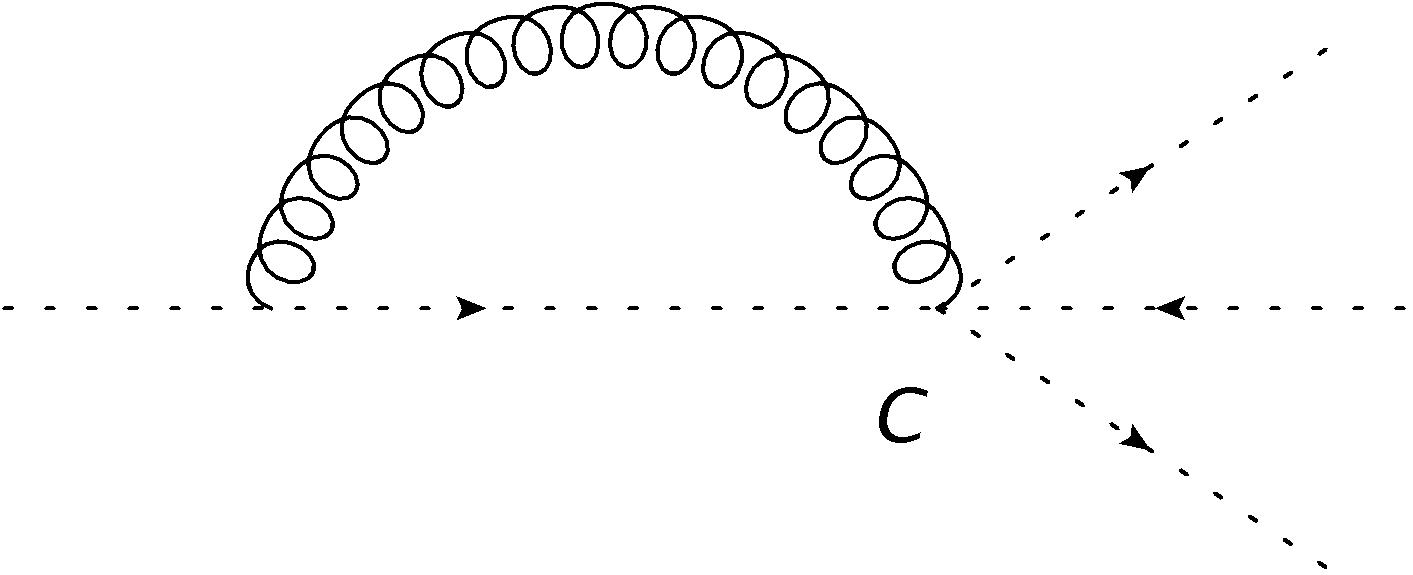}
\end{minipage}
\begin{minipage}{0.22\linewidth}
\begin{flushright}
 \includegraphics[scale=0.08]{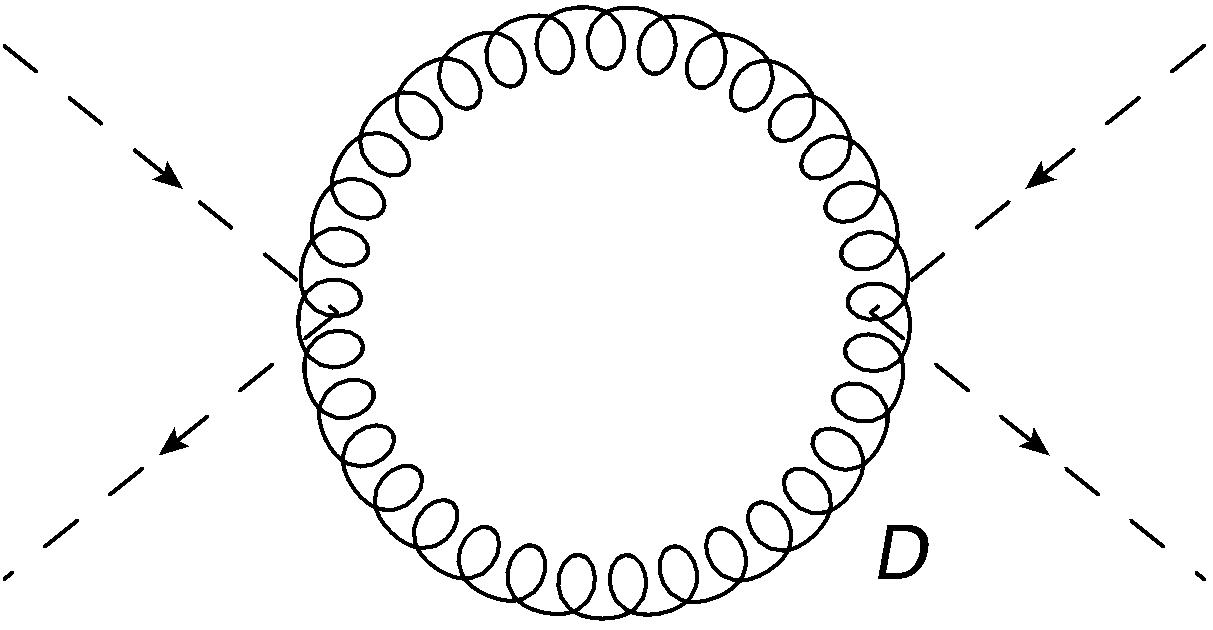}
\end{flushright}
\end{minipage}\newline
\begin{minipage}{0.2\linewidth}
\begin{flushleft}
 \includegraphics[scale=0.08]{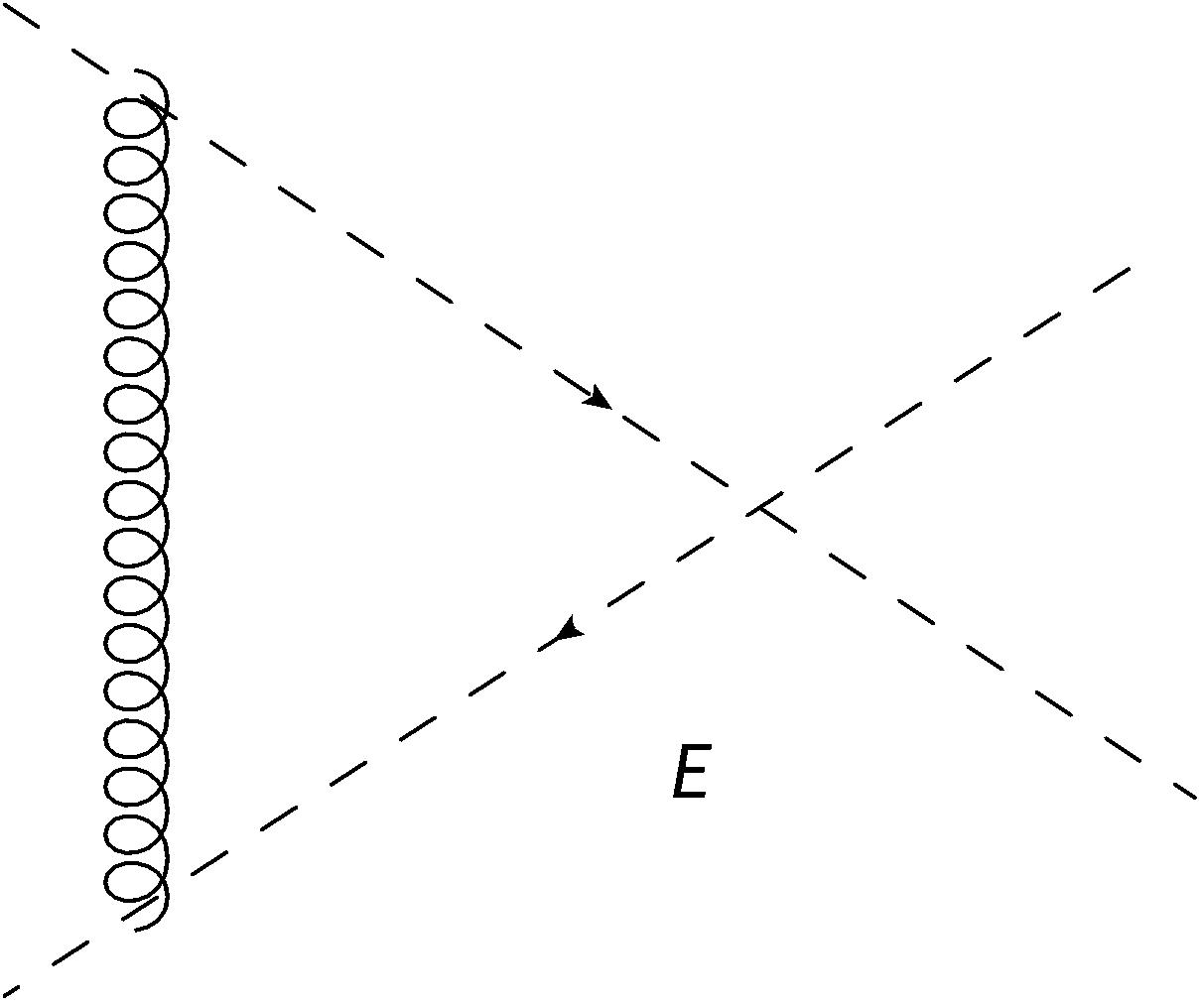}
\end{flushleft}
\end{minipage}
\begin{minipage}{0.2\linewidth}
\begin{flushright}
 \includegraphics[scale=0.33]{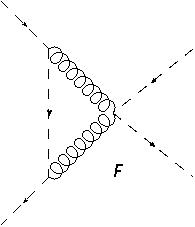}
\end{flushright}
\end{minipage}
\begin{minipage}{0.24\linewidth}
\begin{flushright}
 \includegraphics[scale=0.25]{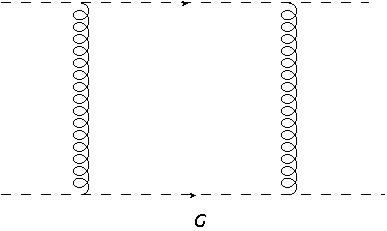}
\end{flushright}
\end{minipage}
\begin{minipage}{0.24\linewidth}
\begin{flushright}
 \includegraphics[scale=0.06]{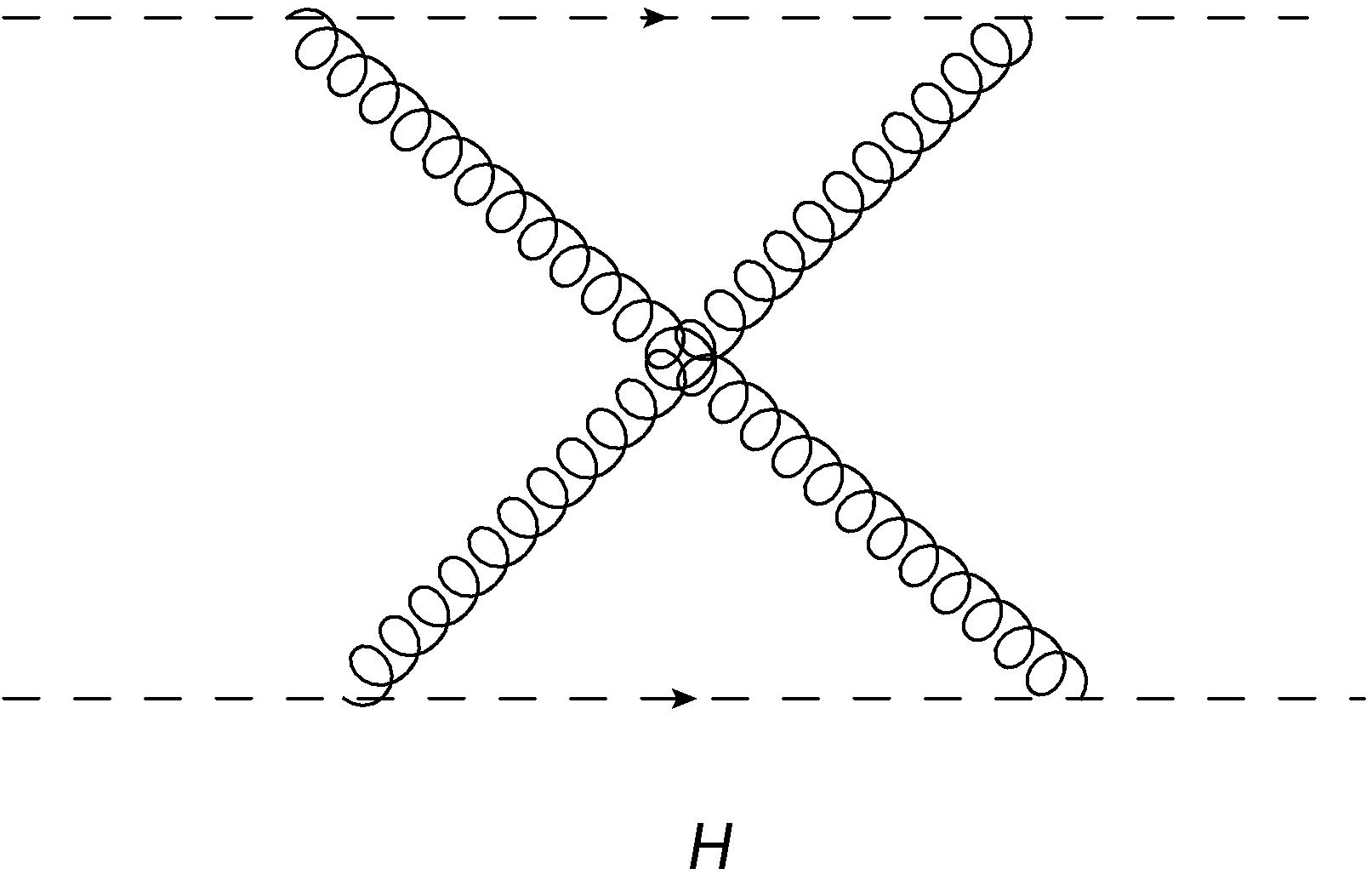}
\end{flushright}
\end{minipage}
\caption{Within a truncation involving a Faddeev-Popov-term and the above four-ghost term, these diagrams drive the flow of the new coupling. A closer inspection of the vertices reveals that due to the ghost-antighost-two-metric vertex being zero (see app.~\ref{tadpole}), diagrams D and F vanish.
In addition, diagrams A, B, C and E are proportional to at least one power of the four-ghost coupling and therefore vanish if this coupling is set to zero by hand. Thereby the two box diagrams generate a non-trivial flow of this coupling even if it is set to zero.\label{fourghostflow}}
\end{figure}

\noindent In particular we observe that the non-vanishing
diagrams A,B,C and E are $\sim g_{\rm ghost} G$, and
accordingly cannot generate a non-vanishing flow of $g_{\rm
ghost}$, if this coupling is set to zero initially. In
contrast, the ladder and the crossed-ladder diagrams are
$\sim Z_c^4 G^2$, and therefore exist even if $g_{\rm
ghost}=0$. 
These will presumably induce a non-zero flow $\beta_{g_{\rm ghost}}\sim G^2$, unless a cancellation between the two occurs. We observe a cancellation mechanism for such box diagrams in sect.~\ref{Wetteqfourfermion}, which relies on a scalar coupling of the metric to the vertex. Since this does not hold in this case, as in particular the transverse traceless metric modes couples to the momentum-dependent vertex, we do not expect a cancellation between the two diagrams. In this case $g_{\rm ghost}=0$ would not be a fixed point of $\beta_{g_{\rm ghost}}$, but the Gau\ss{}ian fixed point, that exists if $G=0$, would be shifted and become interacting. From this simple example we see that one should generically not expect the RG trajectories to be confined to a hypersurface defined by vanishing ghost couplings in the non-perturbative regime. 

We therefore conclude, that, even starting from a simple Faddeev-Popov term, the flow will in general induce further non-zero ghost couplings, which are non-vanishing at the NGFP for the Newton coupling. Here, further investigations are necessary to determine the structure of terms in this sector and discuss the Gribov problem.

\subsection{Ghost curvature couplings}\label{ghostcurv}
As explained above, a negative anomalous
dimension suggests the emergence of further relevant
directions in the ghost sector. Here a large number of couplings exist, which is only restricted by the Grassmannian nature of the ghost fields.

We start with a simple truncation of the type \cite{Eichhorn:2009ah}
\begin{equation}
\Gamma_k = \Gamma_{\rm EH}+ S_{gf}+ S_{gh} + \bar{\zeta}(k) \int
d^4x \sqrt{g}\, \bar{c}^{\mu}R c_{\mu}.
\end{equation}
For simplicity, we work in the TT approximation, where only the fluctuations of the transverse traceless tensor modes induce the RG flow. In the Einstein-Hilbert sector we have observed that the leading contribution to the critical exponents, which decides about the (ir)relevance of a coupling, is stable under this approximation. For simplicity we shall assume that this property also holds in the above truncation.

We further assume that the new coupling will be at the (presumably shifted) Gau\ss{}ian fixed point, and are interested in its critical exponent.

\begin{figure}[!here]
\begin{minipage}{0.55\linewidth}
 \includegraphics[scale=0.7]{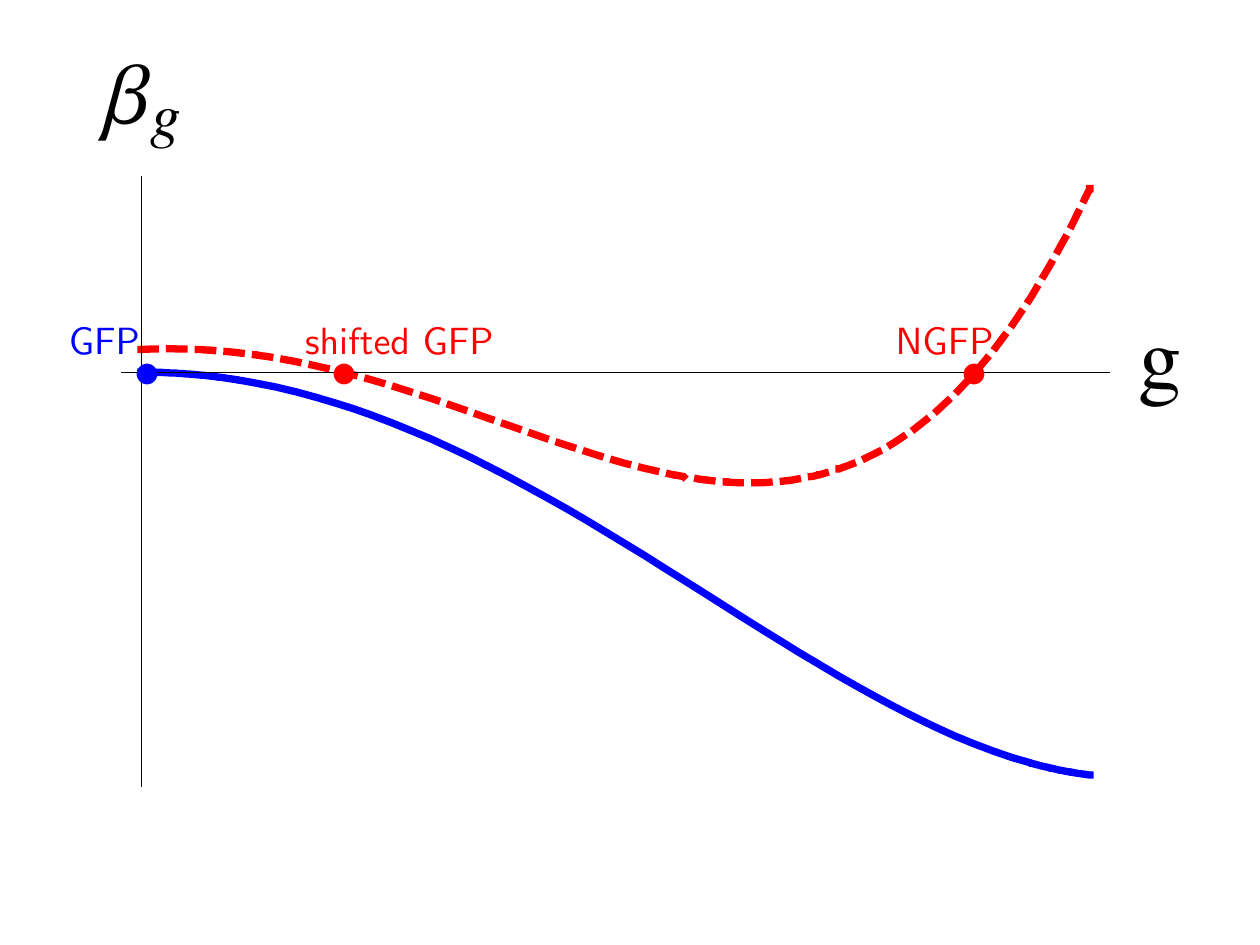}
\end{minipage}
\begin{minipage}{0.4\linewidth}
\caption{Sketch of a $\beta$ function without metric
fluctuations in blue, where the GFP is
assumed to be the only fixed point. Switching on metric
fluctuations will generically shift the GFP point to a NGFP,
without however changing its stability properties although
typically the value of the critical exponent will change,
but not necessarily its sign. Furthermore a second NGFP may
be induced.\label{shiftedGFP1}}
\end{minipage}
\end{figure}

The shift of the GFP should typically not change its stability properties, as
is exemplified in fig.~\ref{shiftedGFP1}.

\begin{figure}[!here]
\begin{minipage}{0.68\linewidth}
 \includegraphics[scale=0.38]{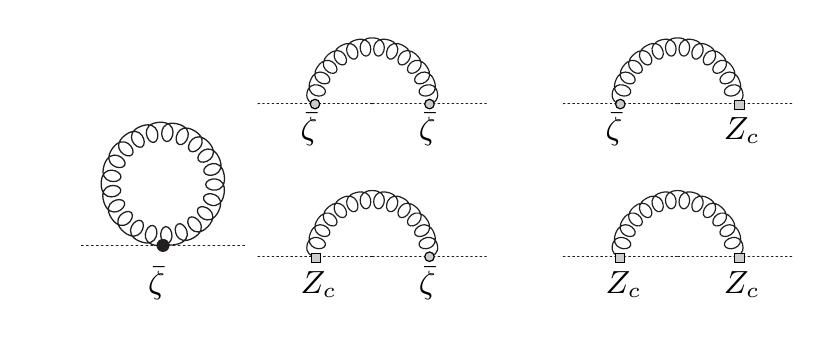}
\end{minipage}
\begin{minipage}{0.27\linewidth}
\caption{The flow of $\bar{\zeta}$ is driven by a tadpole diagram, and four different self-energy diagrams, since we can form two types of ghost-antighost-metric vertices $\sim Z_c$ and $\sim \bar{\zeta}$, respectively. \label{ghostbusters}}
\end{minipage}
\end{figure}

\noindent Diagrammatically, the complete flow in our truncation is given by the set of diagrams in fig.~\ref{ghostbusters}.

\noindent Now we have to carefully consider which diagram will be important to determine whether $\bar{\zeta}$ will be a relevant coupling at the (shifted) GFP. As explained, we do not expect that the shift of the GFP changes its stability properties. Therefore we neglect the self-energy diagram $\sim Z_c^2$ and work at the unshifted and therefore truly Gau\ss{}ian fixed point for $\bar{\zeta}$.
Clearly the tadpole and two of the self-energy diagrams are $\sim \bar{\zeta}$ and therefore contribute to the critical exponent at the GFP. The diagram $\sim \bar{\zeta}^2$ can induce a NGFP for $\bar{\zeta}$, but does not contribute to the critical exponent at the GFP.

We observe that within the TT approximation, the ghost-antighost-metric vertex which is required to build three of the sunset diagrams, vanishes on a spherical background. Since this effect is background-dependent, one might at this point worry how the background-independence of the UV part of the $\beta$ function is to be preserved if a specific class of contributions vanishes for a certain choice of backgrounds. In fact, the contribution to the $\beta$ function from these two sunset diagrams is of the form $\sim G\, \zeta$, as is also the contribution from the tadpole diagram. Thus these add up to one joint contribution to the $\beta$ function, which should not be background-dependent. We therefore conclude, that the weight of the two contributions is background-dependent, and the spherical background is a particularly efficient choice for the evaluation of the $\beta$ function. In other words, choosing a different background also alters the ghost-antighost-two-metric vertex, and hence the numerical factor that results from the tadpole diagram. Adding the non-vanishing contribution from the sunset diagrams would then result in the same contribution as from the tadpole alone on a spherical background.

Therefore we remain with the tadpole diagram to decide about the (ir)relevance of $\zeta$.

Accordingly, when investigating the stability properties of
the shifted GFP we focus on the tadpole
diagram only.

Since we have chosen a spherical background, the flow of $\zeta$ is mixed with the flow of a coupling of the type $\xi \bar{c}^{\mu}R_{\mu \nu}c^{\nu}$. We would like to disentangle these as far as possible. We therefore determine our vertex without at this stage specifying a background.
We first observe that if we set $\bar{D}^{\mu}c_{\kappa}=0$
then also,
\begin{equation}
 [\bar{D}_{\mu}, \bar{D}_{\nu}]c^{\kappa}=
\bar{R}^{\kappa}_{\, \alpha \mu \nu}c^{\alpha}=0.
\end{equation}
Therefore also the contraction $\bar{R}_{\mu \nu}c^{\nu}=0$
for this choice. 
Accordingly in our determination of the vertex necessary to build the tadpole, we drop all contributions that would vanish for $\bar{D}^{\mu}c_{\kappa}=0$, since these would feed into the flow of $\xi$, but not of $\zeta$ \footnote{Note that this argument clarifies why the approximations in \cite{Eichhorn:2009ah} are justifiable.}.

The necessary vertex requires the variation of $\sqrt{g}R$
at constant external (anti)ghost fields, which is the same
as in the pure Einstein-Hilbert term. Using the technique
outlined in sec.~\ref{EHtrunc} for the evaluation of the
right-hand side of the flow equation, we find for the
dimensionless renormalised coupling
$\zeta= \bar\zeta/Z_{\mathrm{c}}$, that
\begin{eqnarray}
  \partial_t \zeta &=& \eta_{c} \zeta
 +\frac{25G_{\ast} \zeta}{6\pi} f(\lambda_{\ast}),\nonumber\\
f(\lambda)&=&e^{4 \lambda}\left(2 \lambda+\frac{1}{2}\right)-e^{2 \lambda}\left(4\lambda+2\right)+8\lambda^2\bigl({\rm Ei}(2 \lambda)- {\rm Ei}(4 \lambda)\bigl).
\end{eqnarray}
Setting $\eta_c=0$ we observe that the critical exponent is positive for the physical regime $G_{\ast}>0$, since $f(\lambda)<0$ for all $\lambda$. 
As explained, there might be two possible values for the ghost anomalous dimension: The negative one renders the critical exponent for $\zeta$ even more positive. For the second value, $\eta_c \approx 0.241$, the value of $G_{\ast}f(\lambda_{\ast})$ is decisive: For the fixed-point values in the Landau-deWitt gauge, $\zeta$ still corresponds to a relevant direction. Interestingly, it can be turned into an irrelevant coupling by including a large number of minimally coupled fermions into the truncation. As observed in \cite{Percacci:2002ie}, these will induce a shift in $\lambda_{\ast}$ towards large negative values. This in turn results in $f(\lambda)\rightarrow 0$, thus the contribution from the tadpole diagram will be suppressed. Then the (ir)relevance of the coupling will be determined purely by the anomalous dimension, since the canonical dimension is zero. We thus observe that the number of relevant couplings in the ghost sector might also depend on the number of matter fields, even if we do not include terms coupling matter and ghosts directly into our truncation.

Hence we have found a first coupling that we expect to be relevant even beyond our simple approximation. The physical interpretation of the related free parameter requires the solution of the Ward identities.

The same investigation can be applied to a more generalised
truncation of the form
\begin{equation}
\Gamma_k = \Gamma_{\rm EH}+ S_{gf}+ S_{gh} + \sum_{i=1}^n \zeta_i(k) \int d^4x \sqrt{g}\, \bar{c}^{\mu}R^i c_{\mu}.
\end{equation}
As discussed in \cite{diplomathesis_Albrecht}, a similar result holds here: Depending on the precise value of $\lambda$, a growing number of curvature-ghost couplings may become relevant. Demanding the irrelevance of couplings beyond $\bar{c}^{\mu}R c_{\mu}$ would then result in restrictions on the possible fixed-point values of $G$ and $\lambda$. If this result holds beyond the approximations involved, we have found a region in theory space, where the fixed point comes endowed with a very large number of relevant directions. Avoiding such a large number of free parameters would imply that the coordinates of the fixed point have to satisfy bounds in the $(G_{\ast}, \lambda_{\ast})$ plane, beyond the physical bound $G_{\ast}>0$.

\section{Summary and Outlook: Towards a non-perturbative understanding of the ghost sector}
In this chapter we have introduced a specific scenario for the UV completion of gravity, namely the asymptotic-safety scenario, which may allow to construct a local quantum field theory of the metric invoking a non-perturbative notion of renormalisability. It is founded upon the existence of a NGFP with a finite number of relevant directions. 

We have reinvestigated the well-understood Einstein-Hilbert
truncation in a new combination of gauge choice and
regularisation scheme, and found further evidence for the
stability of the universal quantities describing the fixed
point under a variation of gauge and regularisation. On the
technical side we have explained how an evaluation of
diagrams contributing to the RG flow is possible without
relying on heat-kernel methods. This is of advantage when
considering flows of Faddeev-Popov ghost couplings, and also
matter couplings, see chap.~\ref{ferminAS}.

The use of the Wetterich equation in the investigation of this scenario necessitates the application of the background field method.
Therefore theory space is enlarged by directions
corresponding to Faddeev-Popov ghosts. In this larger theory
space all directions have to be treated on an equal footing,
meaning that a priori there is no possibility to decide,
what the coordinates of a NGFP and what its relevant
directions may be. We have discussed four scenarios for the
ghost sector, which differ in two respects: The NGFP may be
compatible with a simple form of Faddeev-Popov gauge fixing,
or necessitate further non-zero ghost couplings. This second
possibility could have interesting implications for a
possible solution of the Gribov problem, i.e. the
non-uniqueness of gauge fixing in some gauges.  Furthermore
the NGFP may have relevant directions that correspond to
ghost directions or mixtures of ghost and metric directions.
The challenging task is then to establish a connection
between these relevant couplings and  quantities that are
accessible to experiment. Here, the Ward-identities will
play a crucial role.

In this chapter we have started investigations of this
previously unexplored region of theory space and performed a
first step in analysing these properties of the ghost sector
of asymptotically safe quantum gravity in more detail. In
particular, we have considered a truncation with a running
ghost wave-function renormalisation. It further adds to the
evidence for a NGFP, and results in a negative value for the
anomalous dimension for all gauge parameters under
investigation. This implies in particular that further
ghost couplings could be shifted towards relevance.

In an extended truncation we have considered the separate
running of different tensor structures within the
Faddeev-Popov operator. Here, we have found two NGFPs, which
differ in that at the first one the ghost anomalous
dimension is negative, and a new relevant direction exists.
At the second one the anomalous dimension is positive, and a
further irrelevant direction exists. Thus this fixed point
may potentially allow for the construction of a UV
completion for gravity with a higher predictive power.
This presents the possibility to construct a complete RG
trajectory, which, however, within our truncation, fails to
show a long classical regime.

Furthermore we have presented an argument why further couplings beyond those in the Faddeev-Popov operator should be expected to be non-zero. Accordingly the non-perturbative regime of gravity is presumably characterised by a more complicated ghost sector. This is in accordance with the expectation, that a simple Faddeev-Popov gauge fixing does not hold beyond the perturbative regime. The implications for a possible solution of the Gribov problem remain to be investigated.

Finally we have considered a specific class of ghost-curvature couplings, which turn out to be relevant within our approximation. The interpretation of these directions and their connection to measurable quantities remains to be studied further.

To conclude, the ghost sector of asymptotically safe quantum
gravity may be more complex than one might think from a
perturbative point of view. Non-zero couplings beyond the
Faddeev-Popov operator and relevant directions may
characterise the NGFP. Hence further investigations have to
show explicitly, what the structure of the ghost sector is,
and how it can be reconciled with a unique gauge fixing and
also with a connection of all relevant couplings to
measurable parameters.

Let us again emphasise that also the couplings in the ghost sector all have to admit a fixed point in order for the asymptotic-safety scenario to be realised. Furthermore, since we have argued that the fixed-point value of ghost couplings will typically be shifted to non-zero values, a non-trivial backcoupling into the $\beta$ functions of metric couplings exists. Thereby these further ghost directions can also considerably influence the fixed-point properties in the metric sector. Beyond this question, several interesting questions remain to be analysed in the future:
\begin{itemize}
 \item What is the status of terms beyond the Faddeev-Popov operator at the NGFP? Which ghost couplings acquire a non-zero fixed-point value? What does this imply for the uniqueness of gauge-fixing?
\item Which ghost couplings contribute to relevant directions? How can these be related to measurable quantities?
\end{itemize}

The methods that we have introduced here, pave the way for a
more detailed investigation of these questions. This opens
up the possibility to gain a more detailed understanding of
this aspect of asymptotically safe quantum gravity and shed
light on the structure of an interacting UV fixed point for
this particular gauge theory.

\chapter{Light fermions in quantum gravity} \label{ferminAS}
\section{Matter in asymptotically safe quantum gravity}
Quantum gravity phenomenology has two main goals: One is to predict experimentally accessible phenomena which are genuine outcomes of quantum gravity and thus in some way result from the quantum nature of space-time. This goal is a highly challenging one, since the energy scale determining the realm of quantum gravity is many orders of magnitude above present earth-bound accelerator experiments. A possible access to quantum gravity phenomena may be given by astrophysical observations, since tiny effects can accumulate over large distances\footnote{As an example, consider the high-energy end of the spectrum of high-energetic cosmic showers, which might be explained by some property of quantum gravity, see, e.g. \cite{Dowker:2003hb}. Further, some approaches to quantum gravity may be connected to Lorentz-symmetry violations or deformations, see, e.g. \cite{Magueijo:2001cr}. This is currently being constrained experimentally, see, e.g. \cite{Reimer:2010zzb}.}. 
  
On the other hand the phenomenology of quantum gravity also consists of showing that everything that \emph{has} already been observed is consistent with a candidate theory for quantum gravity. 
In particular the theory has to support a semi-classical regime where an effective description in terms of the Einstein equations holds. Furthermore it has to allow for the inclusion of (quantised) matter without changing the observed low-energy properties of the matter content. Here one specific requirement is that it has to be possible to support masses much below the Planck scale, i.e. there has to be a separation of scales between the regime of strong gravity fluctuations (around the Planck scale), and the scale of chiral symmetry breaking, which is responsible for the generation of the fermion masses in the Standard Model.

\begin{figure}[!here]
 \begin{minipage}{0.15\linewidth}
\begin{flushright}
  $\partial_t \Gamma_k=$
\end{flushright}
 \end{minipage}
\begin{minipage}{0.8\linewidth}
 \includegraphics[scale=0.08]{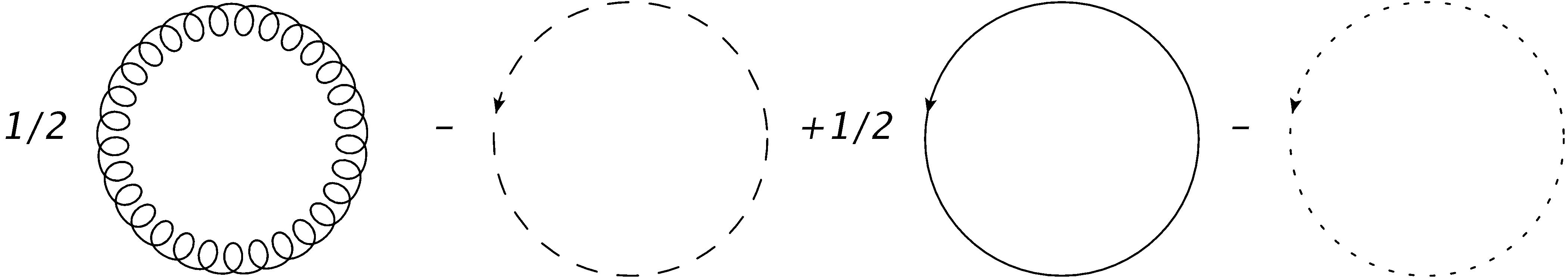}
\end{minipage}
\caption{Including matter, the flow equation contains contributions by metric (curly line), ghost
(dashed line), bosonic matter (full line) and fermionic
matter (dotted line) loops.\label{floweqwithmatter}}
\end{figure}

Within the asymptotic-safety scenario for gravity the inclusion of matter is straightforward, in contrast to some other approaches to quantum gravity\footnote{For example basing the theory on a discrete nature of space-time prohibits the use of most methods of quantum field theory in the microscopic regime, thus rendering the inclusion of matter a highly challenging task.}.
 As the asymptotic-safety scenario stays within the framework of local QFT, matter actions are simply built along the lines of quantum field theory on curved space-times, since these lead to generally covariant equations of motion. The framework naturally differs in the fact that metric fluctuations are also taken into account, not just matter fluctuations on a fixed given background. 
Accordingly in a functional RG approach both matter and
metric loops exist, see fig.~\ref{floweqwithmatter}.

\noindent Thus running couplings in the matter sector
receive contributions from metric fluctuations, and running
couplings in the gravitational sector are driven by matter
fluctuations. In such a way
a unified picture emerges, that treats gravity and matter
(meaning fermionic matter as well as gauge bosons) in the
same way, quantising an action containing both fields in the
path-integral framework by assuming the existence of a
NGFP\footnote{This does not preclude that some couplings may
be zero at the fixed point, accordingly this picture is
compatible with, e.g. asymptotic freedom of QCD.}. Let us emphasise that the RG flow will generically also generate non-minimal
matter-gravity couplings, which may have
interesting phenomenological consequences, e.g. in cosmology
or in regimes where curvature invariants become large.

In the following, we will also employ the framework of the FRG in the context of effective field theories. Here, the main idea is that on scales presumably below $M_{\rm Planck}$ it is not necessary to know the UV completion for gravity and its true microscopic degrees of freedom. Instead, the theory can be studied as an effective field theory with effective degrees of freedom, namely the metric. The underlying microscopic theory then determines the initial conditions for the RG flow within the effective field theory. Thus, the compatibility of other UV completions for gravity with the observed low-energy properties of matter can also be studied here.

Thereby our studies provide a well-defined framework to investigate a much larger class of UV completions for gravity than asymptotic safety, since the effective-field-theory framework is also expected to be applicable to UV completions which rely on other assumptions\footnote{Of course an effective description in terms of the metric would break down for a UV completion for gravity that included further massless degrees of freedom, since then the decoupling assumption underlying the effective field theory framework would not apply. Since we do not observe additional massless gravitational degrees of freedom at low energies, their existence and their couplings are severely constrained.}. Thus the RG framework may also be used to study implications of quantum gravity which hold generically, and can thus establish connections between different approaches to quantum gravity.

The investigation of the compatibility of asymptotically safe quantum gravity with quantised matter was pioneered in \cite{Percacci:2002ie,Percacci:2003jz}, where the back-reaction of minimally coupled fermionic and bosonic matter onto the Einstein-Hilbert sector was investigated. Constraints on the matter content of the universe can be derived in this way. Most importantly, the matter content of the Standard Model is compatible with asymptotically safe quantum gravity within the investigated truncation \cite{Percacci:2002ie}, where not all possible combinations of fermion number and boson number support a NGFP with a positive Newton coupling.
Non-minimally coupled scalar fields have been studied in \cite{Percacci:2003jz, Narain:2009gb}, which may be interesting for scenarios of inflation.

These studies focus on the question, how matter fluctuations
influence the existence and the properties of the fixed
point. On the other hand the effect of metric fluctuations
and the NGFP on the matter sector allows to investigate
which properties of, e.g. the Standard Model are altered by
gravity. 

Here investigations so far have focussed on Yang-Mills theories and the Higgs sector.
Within the functional RG setting, calculations in \cite{Daum:2009dn, Daum:2010bc, Folkerts:2011jz} show that asymptotically free gauge theories keep this property when fluctuations in the metric tensor are included.
Most interestingly, a U(1) gauge theory may also be rendered asymptotically free \cite{Harst:2011zx}, thus potentially allowing for a solution of the triviality problem. On the other hand the QED-QEG system may also allow for a NGFP where the QED coupling corresponds to an irrelevant coupling.
In principle, such a mechanism, where the NGFP renders perturbatively relevant couplings irrelevant is of course a very attractive, albeit highly speculative, possibility to reduce the large number of free parameters of the Standard Model.

The Higgs sector of the Standard Model is likely to inherit the triviality problem from scalar $\phi^4$ theory. A NGFP in the Higgs sector itself may solve this problem \cite{Gies:2009hq,Scherer:2009wu,Gies:2009sv}, or the coupling of the Higgs to gravity may render the theory fundamental \cite{Zanusso:2009bs,Vacca:2010mj}.

\subsection{Chiral symmetry breaking through metric fluctuations}
A sector where metric fluctuations may be expected to have interesting consequences is the fermionic sector of the Standard Model. It contains light fermions, since their mass is protected by chiral symmetry. Mass generation in the Standard Model is then linked to chiral symmetry breaking, which results from strong correlations between fermions, which are induced either in a Yukawa-type fashion in the Higgs sector and also through gluonic interactions in QCD. The associated mass of the fermions is then naturally related to the typical mass scale of this sector. In the case of gravity this scale would be around the Planck scale, making fermions so massive as to remove them from currently accessible energies. This is of course in most severe contradiction to observations.

Within the framework of the FRG, the onset of chiral symmetry breaking can be accessed within a purely fermionic truncation. In a fermionic language, chiral symmetry breaking is signalled by a diverging four-fermion coupling. In particular, we consider a Fierz-complete basis of SU$({\rm N}_f)_L \, \times$ SU$({\rm N}_f)_R$ symmetric four fermion couplings here, since any of the possible channels might become critical, with the others only following as a consequence. Thus the use of a Fierz-complete basis is expedient. In our case, it is given by
\begin{eqnarray}
 \Gamma_{k\, 4-\rm fermion}&=&\frac{1}{2} \int d^4x \, \sqrt{g}\, \left(\bar{\lambda}_- \left(V-A \right)+ \bar{\lambda}_+ \left(V+A \right) \right), \,\mbox{ where}\nonumber\\
V&=& \left( \bar{\psi}^i \gamma_{\mu}\psi^i \right)\left( \bar{\psi}^j \gamma^{\mu}\psi^j \right) \phantom{xx}
A=- \left( \bar{\psi}^i \gamma_{\mu}\gamma^5\psi^i \right)\left( \bar{\psi}^j \gamma^{\mu}\gamma^5\psi^j \right)\label{fourfermion}
\end{eqnarray}
Here the brackets indicate expressions with fully contracted
Dirac indices and the Latin indices $i,j=1,...N_f$ denote
flavour indices. The four-fermion couplings $\bar{\lambda}_{\pm}$
should not be confused with the cosmological constant
$\lambda$. For our notation regarding the vierbein and the
spin connection, see app.~\ref{fermvertsandconvs}.

Let us now explain how four-fermion couplings are linked to chiral symmetry breaking, before we motivate, how a coupling to gravity might modify the picture.

Chiral symmetry breaking is expected to become manifest in the scalar-pseudo-scalar channel, which is actually related to the $\lambda_+$ channel by a Fierz transformation
\begin{equation}
\bar{\lambda}_+ \big[(\bar\psi^i\gamma_\mu \psi^i)^2-(\bar\psi^i\gamma_\mu \gamma_5
  \psi^i)^2\big] 
= \bar{\lambda}_\sigma\big[(\bar\psi^i \psi^j)^2-(\bar\psi^i\gamma_5
  \psi^j)^2\big],
\label{eq:Fierz}
\end{equation}
where $(\bar\psi^i \psi^j)^2 \equiv \bar\psi^i \psi^j \bar\psi^j \psi^i$, and
similarly for the pseudo-scalar channel. Here $\bar{\lambda}_\sigma = -\frac{1}{2} \bar{\lambda}_+$
has to hold for an exact Fierz-identity.

Introducing a composite boson for a fermion bilinear, we can schematically rewrite the path integral
\begin{equation}
 \int \mathcal{D}\bar{\psi}\,\mathcal{D}\psi\,
e^{-\int \left(\bar{\psi}i \slashed{D}\psi+\lambda(\bar{\psi}\psi)^2\right)} = \int
\mathcal{D}\bar{\psi}\,\mathcal{D}\psi\, \mathcal{D}\phi\,
e^{-\int \left(\bar{\psi}i \slashed{D}\psi -\frac{2}{\lambda}\phi^2-h \phi \bar{\psi}\psi\right)}.\label{Hubbard}
\end{equation}
Since chiral symmetry breaking implies the formation of a fermion bilinear condensate, this is related to the generation of a condensate for the boson. For a simple bosonic potential with a global symmetry we can read off spontaneous symmetry breaking easily: It corresponds to a Mexican-hat-type potential. The transition between the symmetric and the symmetry-broken phase occurs when the boson mass vanishes. As is clear from \Eqref{Hubbard}, the inverse four-fermion coupling is related to the boson mass. Accordingly we find that the criterion for chiral symmetry breaking within a purely fermionic truncation is a diverging four-fermion coupling.
In order to determine when this criterion can be satisfied, let us introduce the
dimensionless renormalised couplings $\lambda_\pm$ and the fermionic anomalous
dimension $\eta_\psi$:
\begin{equation}
\lambda_\pm = \frac{k^2 \bar\lambda_\pm}{Z_\psi}, \quad 
\eta_\psi = -\partial_t \ln Z_\psi, 
\label{eq:dimrenlambda}
\end{equation}
where $Z_{\psi}$ denotes the fermionic wave-function renormalisation.
The one-loop form of the Wetterich equation determines the possible terms in the $\beta$ functions for
$\lambda_\pm$:
\begin{equation}
 \beta_{\lambda_{\pm}}= (2+\eta_{\psi})\lambda_{\pm}+a \,\lambda_{\pm}^2 +b\, 
\lambda_{\pm}\lambda_{\mp} + c \,\lambda_{\mp}^2+d \lambda_{\pm}+e.
\label{eq:betalambda}
\end{equation}
Herein the first term arises from dimensional (and anomalous) scaling. The only purely fermionic diagram that can be constructed from a truncation containing a kinetic term and four-fermion terms is obviously a two-vertex diagram, which results in the quadratic contribution.
The coupling to further fields can result in a tadpole contribution $\sim
d\lambda_{\pm}$, as well as a $\lambda_{\pm}$-independent part. The numerical values for $a,b$ and $c$ depend on the
regulator; the contributions $d$ and $e$ will also depend on further
couplings.

\begin{figure}[!here]
\begin{minipage}{0.5\linewidth}
 \includegraphics[scale=0.6]{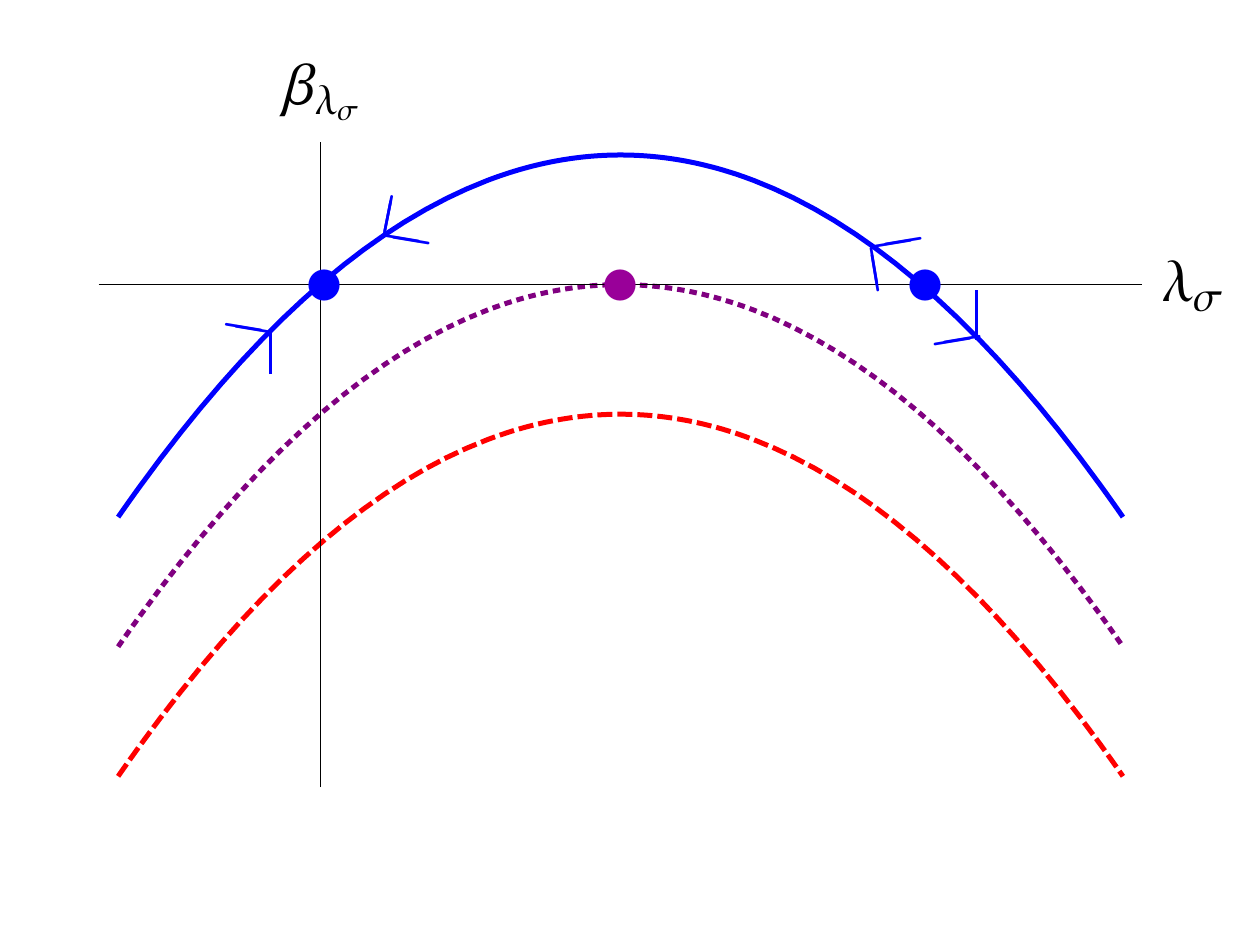}
\end{minipage}
\begin{minipage}{0.47\linewidth}
\caption{Sketch of the $\beta$ function for
$\lambda_{\sigma}$: The upper (blue, full) curve shows the
generic situation in a purely fermionic truncation, where
two real fixed points exist. Arrows indicate the flow
towards the infrared. The purple dotted curve indicates the
situation, where the Gau\ss{}ian and the non-Gau\ss{}ian
fixed point annihilate. For the red dashed curve the zeros
of the $\beta$ function are all
complex.\label{betalambdasketch}}
\end{minipage}
\end{figure}

Within a purely fermionic truncation, the $\beta$ function for $\lambda_{\sigma}$ is then a parabola, which generically admits two real fixed points, a Gau\ss{}ian and a non-Gau\ss{}ian one. As shown in fig.~\ref{betalambdasketch}, any initial condition to the left of the interacting fixed point results in a flow that is attracted towards the trivial fixed point towards the infrared. Accordingly fermions will be weakly correlated on macroscopic scales and chiral symmetry will remain intact. On the other hand an initial condition to the right of the NGFP results in a diverging flow at a finite scale, and hence chiral symmetry breaking.

\noindent This picture gets modified in a crucial way when
further interactions are allowed. In particular, in QCD,
non-Abelian gauge boson fluctuations lead to chiral symmetry
breaking and the formation of massive bound states when the
gauge coupling $g$ exceeds a critical value
\cite{Gies:2002hq, Gies:2005as, Braun:2006jd,Braun:2005uj}.
This effect is due to contributions $\sim g^4$ in the
$\beta$ functions (cf. \Eqref{eq:betalambda}). Since the
sign of this contribution is such as to shift the parabola
downwards, the two fixed points annihilate for $g= g_{\rm
crit}$. For larger values of the gauge coupling no fixed
points exist and chiral symmetry breaking occurs for any
initial condition.

Due to the similarity between non-Abelian gauge theories and
gravity a similar mechanism might be expected to cause
chiral symmetry breaking in the coupled fermion-metric
system, induced by metric fluctuations. As such a process
would be incompatible with our observation that light
fermions exist, this might allow to place  restrictions on
the asymptotic-safety scenario or the allowed number of
fermions. In particular, since the running Newton coupling
plays a similar role to the running gauge coupling, the
requirement of unbroken chiral symmetry might result in an
upper bound on the allowed fixed point value for the Newton
coupling.

The question that we are interested in here is actually not restricted
to the asymptotic-safety scenario: Our calculation applies
to any scenario which allows to parameterise gravity
fluctuations in terms of metric fluctuations within an
effective field-theory framework. 
Here we use that, regardless of the nature of the UV completion for gravity, one can expect a regime where quantum fluctuations of space-time 
can be parameterised in terms of metric fluctuations. This description presumably holds on scales $k_0 \lesssim M_{\rm Planck}$. Then, the microscopic theory in principle determines the initial conditions for the flow. Every quantum theory of
gravity has to allow for the possibility to incorporate
light fermions. Here we will show that this observation in principle may allow to restrict other UV completions for gravity, since the initial conditions determined from the microscopic theory for the RG flow within the effective theory may actually lie in a region of theory space which is not compatible with the existence of light fermions. 
Any candidate theory for quantum gravity which in such a way precludes the existence of light fermions, may be an internally consistent theory, but is
experimentally found to be incompatible with our universe.
On the other hand a quantum theory of gravity which
naturally allows for the existence of light fermions,
without requiring, e.g. a high degree of fine-tuning,
receives non-trivial experimental support here.

To study chiral symmetry breaking, we thus need to evaluate the $\beta$ functions of the four-fermion couplings $\lambda_{\pm}$. 
Since in particular NGFPs play a crucial role
to stabilise the fermionic as well as the gravitational
sector, it does not suffice to evaluate the $\beta$
functions perturbatively around a GFP.
We therefore need to access genuinely non-perturbative
information on the $\beta$ functions. Here the Wetterich equation is a well-suited tool.

Of course a purely fermionic truncation is not suited for
studies of the
symmetry-broken regime, but allows to study the onset of symmetry breaking. In QCD, this strategy allows to determine
the critical temperature for chiral symmetry breaking \cite{Braun:2005uj,Braun:2006jd}.

It is well-known, and can indeed be seen directly from the canonical dimension, that four-fermion interactions are perturbatively non-renormalisable. This does of course not preclude their study in this setting here. The perturbative non-renormalisability translates into the fact that these couplings are irrelevant at the GFP. Accordingly they have to be set to zero at the UV scale, when one uses them to construct a fundamental theory within perturbation theory. Even if set to zero initially, such couplings are generated in the context of QCD, or also when coupled to gravity. Their RG flow will then show if chiral symmetry is broken in this context. 

Furthermore NGFPs may exist which may allow to construct a non-perturbatively renormalisable theory with non-vanishing four-fermion interactions.

Note that our study differs from investigations of chiral symmetry breaking by a classical background curvature. In such a system the fermion propagator contains a term proportional to the curvature. Then, a classical background will screen
or enhance fermionic fluctuations that lead to chiral
criticality. For large positive curvatures, e.g. in a deSitter-type space-time, long-range fluctuations are effectively screened, which averts chiral symmetry breaking.  Screening mechanisms for chiral symmetry breaking of this type
have already been studied in various chiral models, such as two- and
three-dimensional (gauged) Thirring models
\cite{Sachs:1993ss,Sachs:1995dm,Geyer:1996yf} or the four-dimensional gauged
NJL model \cite{Geyer:1996kg}.
This effect should not be confused with the setting that we
will investigate in the following, where the classical
background curvature is not taken into account and instead
large metric fluctuations, i.e. the \emph{quantum} nature
of gravity in contrast to its classical properties may break
chiral symmetry.

\subsection{Wetterich equation for four-fermion couplings}\label{Wetteqfourfermion}

Since a minimally coupled kinetic term and four-fermion
terms of the type above (see \Eqref{fourfermion}) suffice to
study the onset of the chiral phase transition in QCD
\cite{Gies:2002hq,Gies:2005as,Braun:2005uj,Braun:2006jd,
Braun:2009gm} with quantitatively good results, we study an
analogous truncation here. As coupling fermions to gravity
allows for a larger number of operators, a different
mechanism for chiral symmetry breaking may apply here, see sec.~\ref{conclusions}.
We also disregard momentum-dependent four-fermion interactions, which could conveniently be included by performing a Hubbard-Stratonovich transformation and including a kinetic term for the resulting bosonic fields. Furthermore the flow will also generate fermion-ghost couplings which we set to zero here.

On a general (curved) space-time our truncation then reads:
\begin{eqnarray}
 \Gamma_k = \Gamma_{k\, \rm EH}+ S_{\rm gf}+S_{\rm ghost}+ \int d^4x \sqrt{g}\, i Z_{\psi} \bar{\psi}^i \gamma^{\mu}\nabla_{\mu}\psi^i + \Gamma_{k\, 4-\rm fermion},
\end{eqnarray}
with the Einstein-Hilbert term  \Eqref{GEH} and the gauge-fixing term \Eqref{Ggf}.
The covariant derivative $\nabla_{\mu}$ is given by
$\nabla_{\mu}\psi= \partial_{\mu}\psi +
\frac{1}{8}[\gamma^a,\gamma^b]\omega_{\mu\, ab}\psi$, where
Latin indices refer to the tangent space and $\gamma^{\mu}=
e^{\mu}_a \gamma^a$. $\omega_{\mu}^{ab}$ denotes the spin
connection, which can be determined in terms of the vierbein
and the Christoffel connection by requiring that
$\nabla_{\mu}e^{\nu}_a=0$, for details on the coupling of
fermions to gravity see, e.g. \cite{deWitt1963, ParkerToms}.
For the vierbein we work in the symmetric vierbein 
gauge \cite{Woodard:1984sj, vanNieuwenhuizen:1981uf} such
that $O(4)$ ghosts do not occur. This gauge also allows to
re-express vierbein fluctuation purely in terms of metric
fluctuations. Details on the second functional derivative of the effective action can be found in app.~\ref{fermvertsandconvs}.

Note that even in the Landau deWitt gauge $\rho \rightarrow
\alpha \rightarrow 0$ the wave-function renormalisation
$Z_{\psi}$ receives a non-trivial contribution, which is a first difference between gravity and Yang-Mills theory \cite{Gies:2003dp}. Within a
first study we do not evaluate $\eta_{\psi} = - \partial_t
\ln Z_{\psi}$ directly. Instead we simply keep $Z_{\psi}\neq
1$ in our calculation, which allows us to test the possible
effects of a non-zero anomalous dimension.

In our calculation we apply the following strategy: As the four-fermion interaction cannot couple directly into
the Einstein-Hilbert sector, nor
modify the wave-function renormalisation in the pointlike limit, the fixed-point
structure in the Einstein-Hilbert sector in the truncation $\eta_{\psi}=0$ is exactly given by
the calculation taking into account only a minimally coupled
kinetic fermion term \cite{Percacci:2002ie}.
Therefore our new task is to evaluate the $\beta$ functions
in the fermionic sector.
Accordingly a flat background
$\bar{g}_{\mu \nu}= \delta_{\mu \nu}$ is fully sufficient,
and technically highly favourable over backgrounds with
non-trivial curvature invariants.

\begin{figure}[!here]
\begin{minipage}{0.58\linewidth}
 \includegraphics[scale=0.34]{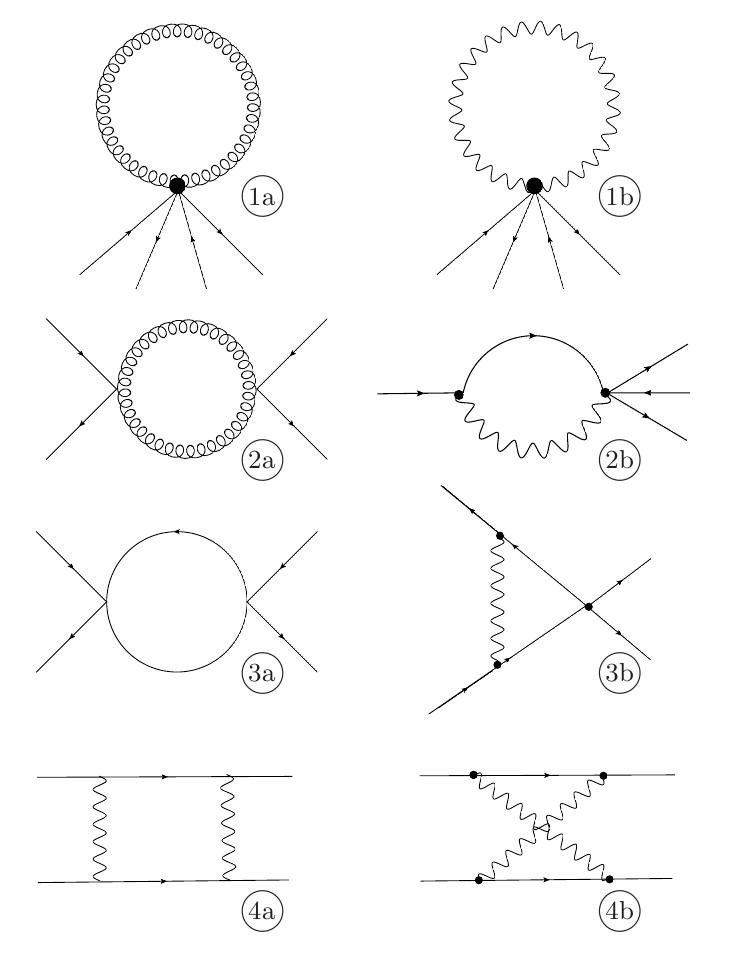}
\end{minipage}
\begin{minipage}{0.39\linewidth}
 \caption{Contributions to the running of the four-fermion
coupling, sorted according to the number of vertices they
contain. The diagrams containing wavy lines receive
contributions from the trace mode, only (see below), the
diagrams with curly lines exist only for the TT mode.
Diagrams 3a, 3b, 4a and 4b also exist if fermions are
coupled minimally to Yang-Mills theory. The additional
diagrams 1a, 1b, 2a and 2b can be traced back to the fact
that the $\sqrt{g}$ in the volume element generates
additional graviton-fermion-couplings and the covariant
derivative in the kinetic term generates not only one- but
also two-graviton fermion couplings.
\label{diagrams_4f}
}
\end{minipage}
\end{figure}

Diagrammatically the $\beta$ function can be represented by the $\tilde{\partial}_t$-derivatives of the set of diagrams in fig.~\ref{diagrams_4f}, where we have performed a York decomposition of the metric fluctuation (as in \Eqref{York}) and work in Landau-deWitt gauge (i.e. $\alpha =0$), so that only the transverse traceless and the trace contributions exist.

\begin{figure}[!here]
 \begin{minipage}{0.25\linewidth}  
\includegraphics[scale=0.11]{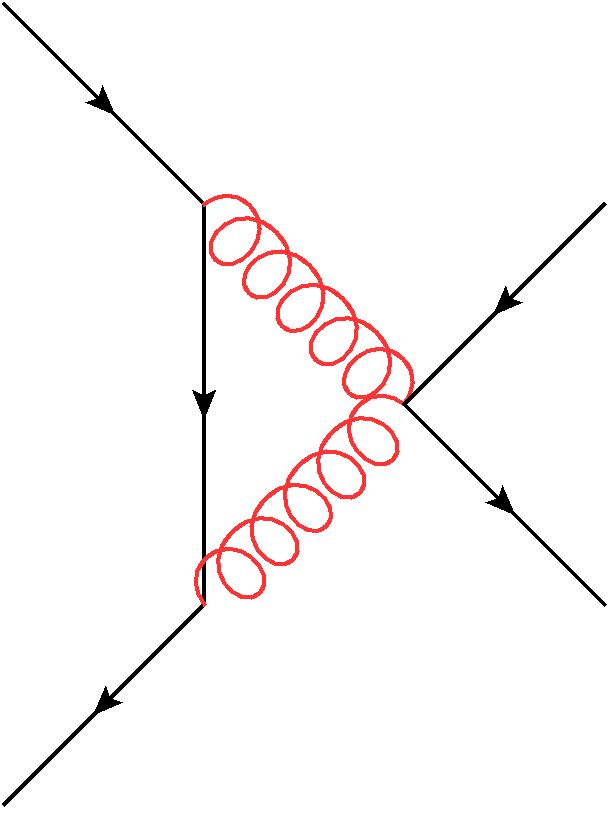}
 \end{minipage}
 \begin{minipage}{0.7\linewidth}  
\caption{The two vertices to the left exist only for the trace mode, whereas the vertex to the right only exists for the transverse traceless mode. The required propagator for the metric (the curly red line) would be required to be non-diagonal in the metric modes. There exists no choice of gauge parameters $\alpha$ and $\rho$, for which the corresponding off-diagonal matrix entry exists.\label{vanishingdiag}}
 \end{minipage}
\end{figure}

\noindent In order to construct these diagrams we have used some properties of the vertices, which are derived in app.~\ref{Gamma2}: The vertex that couples a fermion, antifermion and two gravitons\footnote{Note that again by gravitons we refer to metric fluctuations which are not necessarily small, i.e. we mean either $h_{\mu \nu}^T$ or $h$ resp. by this term.} only exists for the transverse traceless component, whereas the vertex coupling a fermion, antifermion and graviton only exists for the trace mode. 
In particular this leads to the fact that the second three-vertex diagram that one can draw using fermion-antifermion-graviton and fermion-antifermion-two-graviton vertices vanishes identically, cf. fig.~\ref{vanishingdiag}. 

\noindent As a first result we observe a cancellation between the two box diagrams, which is actually not restricted to gravity: The condition under which these diagrams cancel is that the coupling should be scalar and not chiral. In particular, a Yukawa coupling of the type $\phi \bar{\psi}\psi$ also leads to a cancellation, whereas a chiral Yukawa coupling does not \cite{Gies:2003dp}. 
For a detailed calculation of this cancellation see app.~\ref{cancellation}.

In Yang-Mills theory the box-diagrams are the only 
contribution that generates the four-fermion interaction even
if it is set to zero initially. As gravity allows for a
larger number of vertices from a
minimally-coupled kinetic fermion term, the two-vertex
diagram 2a in fig.~\ref{diagrams_4f} will create this
contribution here. Thus the $\beta$ functions will contain
similar terms as in Yang-Mills theory, altough their
diagrammatic origin is different.\\
To determine the structure of the $\beta$ functions note
that each metric propagator is $\sim G$, and each
four-fermion-vertex $\sim \lambda_{\pm}$. We observe that
the gravitational diagrams do not lead to mixing between
$\lambda_{+}$ and $\lambda_-$ in the respective $\beta$
functions, hence, e.g. no term $\sim G \lambda_-$
contributes to $\beta_{\lambda_+}$.
The $\beta$ functions for the dimensionless couplings
are then given by
\begin{equation}
 \partial_t \lambda_{\pm}= (2+\eta_{\psi})\lambda_{\pm}+ a
\,\lambda_{\pm}^2 +b\, 
\lambda_{\pm}\lambda_{\mp} + c \,\lambda_{\mp}^2 + d \,G
\,\lambda_{\pm} f_1(\lambda)+ e\, G^2
f_2(\lambda).\label{betalambda}
\end{equation}
Here the details of the functions $f_1(\lambda)$ and 
$f_2(\lambda)$ as well as the precise numerical values of
the coefficients $a,...,e$ depend on the choice of
regulator. 
Eq.~\ref{betalambda} shows that the $\beta$ functions are
structurally analogous to Yang-Mills theory, in that they
contain terms $\sim G^2$ (cf. $\sim g^4$) and $\sim G
\lambda_{\pm}$ (cf. $\sim g^2 \lambda_{\pm}$). We might
therefore expect a similar mechanism as in Yang-Mills theory
to lead to chiral symmetry breaking through metric
fluctuations.

\begin{figure}[!here]
\begin{minipage}{0.6\linewidth}
 \includegraphics[scale=0.7]{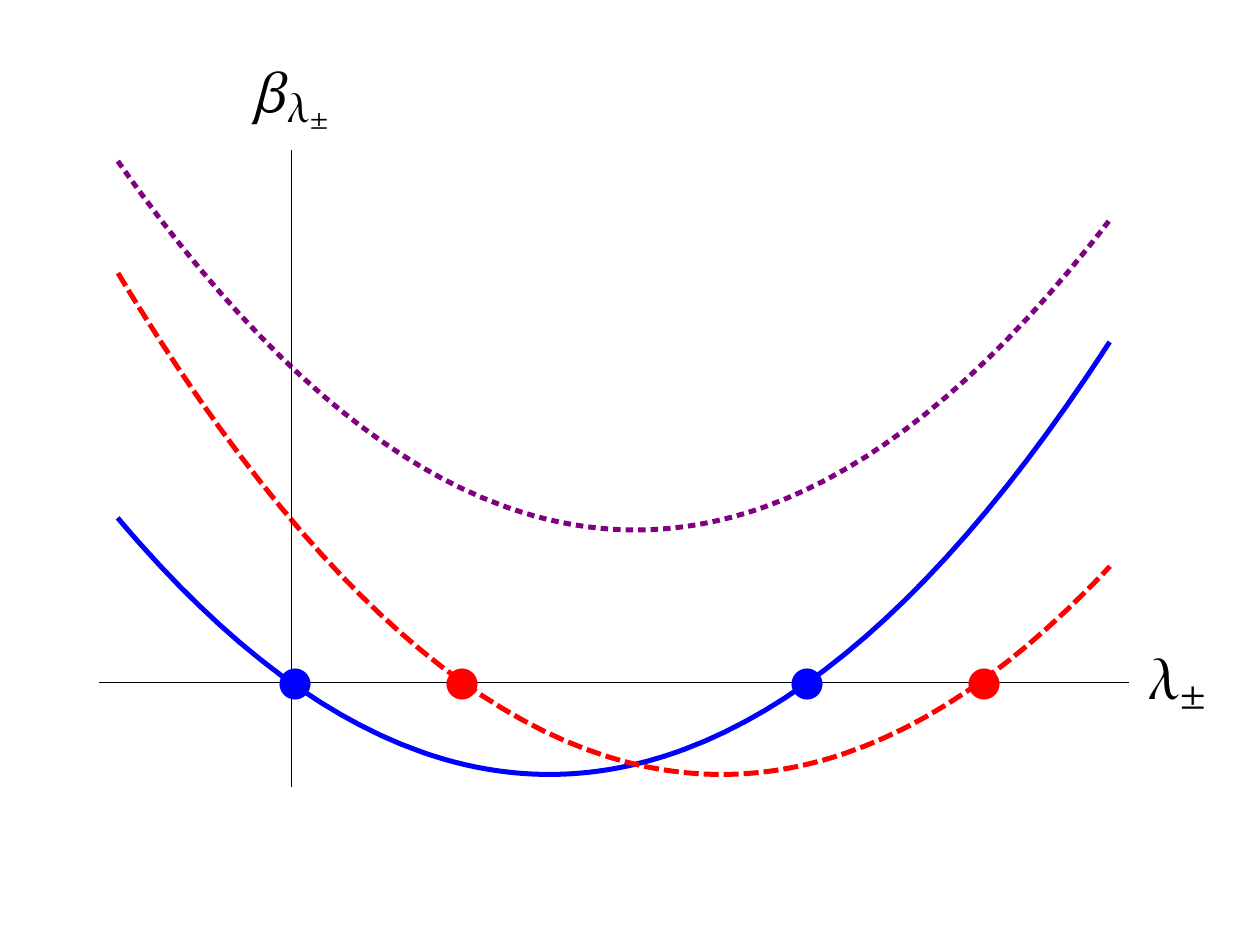}
\end{minipage}
\begin{minipage}{0.37\linewidth}
\caption{Sketch of the $\beta$ function of one four-fermion coupling: Without gravitational interactions the parabola for the four-fermion coupling admits a GFP and a NGFP (blue full line). Metric fluctuations may shift it (red dashed line), thereby inducing two interacting fixed points, or lead to a destabilisation (purple dotted line).\label{betalambdaGsketch}}
\end{minipage}
\end{figure}

Clearly the coupled system of $\beta$ functions
for $\tilde{\lambda}_{\pm}$ will admit $2^2$ fixed points
which need not all be real. If no real fixed points exist
for some value of $(G, \lambda, \rm N_f)$ then chiral
symmetry is
broken for this choice. If fixed points exist then it may
depend on the choice of initial conditions, if chiral
symmetry breaking is possible, see
fig.~\ref{betalambdaGsketch}. In particular it may happen
that the fixed point is found at negative values for
$\lambda_{\pm}$. This does not pose any stability problems
in a fermionic setting. 

\noindent From the vertices that exist within our truncation it is possible to construct a number of diagrams with six external fermion/antifermion legs, which accordingly seem to generate an explicitly symmetry-breaking 6-fermion interaction. A careful inspection of these diagrams reveals that the loop integral in each of these is over an uneven power of momenta and hence vanishes. This follows, as the fermion propagator is linear, the metric propagator quadratic, and the fermion-antifermion-$n$-graviton vertex linear in momentum.
Accordingly, although one may naively draw diagrams that lead to explicit chiral symmetry breaking, these diagrams vanish identically if the regulator respects the symmetry, and the symmetry can only be broken spontaneously. This constitutes an explicit example of how the flow equation respects symmetries of the theory as long as no explicit breaking is introduced by the regulator.

\subsection{Results: Existence of light fermions}\label{results}

Evaluating the trace over space-time and Dirac indices to
arrive at the $\beta$ functions, we employ the identities
for $\gamma$-matrices given in app.~\ref{gammamat}.
The $\beta$ functions for the dimensionful couplings for a
general regulator $R_k(p^2)= \Gamma_k^{(2)}r(y)$ with $y
=\frac{p^2}{k^2}$ are then given by: 
\begin{eqnarray}
 \beta_{\bar{\lambda}_\pm}&=& \partial_t \bar{\lambda}_{\pm}= 2 \Bigl[
\bar{\lambda}_{\pm} \frac{1}{32}
I[0,0,1]-\bar{\lambda}_{\pm}\frac{5}{8}I[0,1,0] \mp
\frac{15}{4096} I[0,2,0] -\bar{\lambda}_{\pm}
\frac{3}{16}I[1,0,1] \nonumber\\
&{}&+\bar{\lambda}_{\pm} 
\frac{27}{256} I[2,0,1] + {\rm fermionic\, contr.}\Bigr].
\label{generalbetas}
\end{eqnarray}
The fermionic contribution that we have not written out
can be found in \cite{Gies:2003dp}, and details on the integral $I[f, {\rm TT}, h]$ can be found in app.~\ref{threshold}.

Specialising to the case of a cutoff of the type
\begin{eqnarray}
R_{k\, \rm grav}(p^2)&=& \left(\Gamma_k^{(2)}(k^2)-\Gamma_k^{(2)}(p^2) \right) \theta (k^2 -p^2)\nonumber\\
R_{k\, \rm ferm}(p^2)&=& Z_{\psi} \slashed{p}
\left(\sqrt{\frac{k^2}{p^2}}-1 \right)\theta(k^2-p^2)
\end{eqnarray}
the pair of $\beta$ functions for the dimensionless
variables is given by
\begin{eqnarray}
 \partial_t \lambda_{-}&=& 2 \lambda_- +2 \eta_{\psi}
\lambda_-+2 \Bigl[ -\frac{5G (\eta_N-6)}{24 \pi (1-2
\lambda)^2}\lambda_- - \frac{G (-6+\eta_N)}{4 \pi (3- 4
\lambda)^2}\lambda_- -\frac{5 G^2 (\eta_N-8)}{128
(-1+2\lambda)^2} \nonumber\\
&{}&+ \frac{G \left(36 \eta_N -7
\left(54-24 \lambda + \eta_{\psi}(-3+4 \lambda) \right)
\right)}{25 \pi (3-4 \lambda)^2}\lambda_-\\
&{}& - \frac{9G
\left(21 \eta_N + 24 (-14+\eta_{\psi})-32
(-7+\eta_{\psi})\lambda \right)}{448 \pi (3-4
\lambda)^2}\lambda_-\Bigr]+ (-5+\eta_{\psi})
\frac{\lambda_-^2-{\rm N_f} \lambda_-^2-{\rm N_f}
\lambda_+^2}{40
\pi^2}\nonumber
\end{eqnarray}
\begin{eqnarray}
 \partial_t \lambda_{+}&=& 2 \lambda_++2 \eta_\psi
\lambda_+ +2 \Bigl[-\frac{5G (\eta_N-6)}{24 \pi (1-2
\lambda)^2}\lambda_+- \frac{G (-6+\eta_N)}{4 \pi (3- 4
\lambda)^2}\lambda_+ +\frac{5 G^2 (\eta_N-8)}{128
(-1+2\lambda)^2}\nonumber\\
&{}&+ \frac{G \left(36 \eta_N -7
\left(54-24 \lambda + \eta_{\psi}(-3+4 \lambda) \right)
\right)}{25 \pi (3-4 \lambda)^2}\lambda_+ \\
&{}&- \frac{9G \left(21
\eta_N + 24 (-14+\eta_{\psi})-32 (-7+\eta_{\psi})\lambda
\right)}{448 \pi (3-4 \lambda)^2}\lambda_+\Bigr]\nonumber\\
&{}&+
(\eta_{\psi}-5)\frac{-2 \lambda_- \lambda_+ -2 N_f
\lambda_- \lambda_+ -3 \lambda_+^2}{40 \pi^2}.\nonumber
\end{eqnarray}
Herein, the single terms correspond to the diagrams in
fig.~\ref{diagrams_4f} in the following sequence: The first
two
terms
 are due to the dimensionality of $\lambda_{\pm}$. The third and fourth
term correspond to the transverse traceless tadpole (1a) and the conformal tadpole (1b), respectively. The next term is represented
by the two-vertex diagram with internal gravitons only (2a). The
mixed two-vertex diagram (2b) results in the sixth term. Finally
the three-vertex diagram (3b) corresponds to the second last term
and the purely fermionic contributions (3a) are represented in
the two differing last terms.

We find four pairs of NGFPs for
$\lambda_{\pm}$ for $G= 0$, as in \cite{Gies:2003dp}. The first is the Gau\ss{}ian one, characterised by two critical exponents $\theta_{1,2}=-2$, corresponding to the canonical dimension of the couplings. Two further fixed points have one relevant direction with $\theta_1=2$, and one irrelevant direction. A last fixed point has two relevant directions.
This structure persists under the inclusion of metric
fluctuations. 
We cannot find values $(G_{\ast} \in (-10,10),
\lambda_{\ast}\in (-10,1/2), {\rm N_f}\in(2,10^3))$ for $\eta_{\psi}=0$, such that less than four real fixed points exist. 

This surprising difference to Yang-Mills theory can be understood from the fact that in the case of gravity terms $\sim G \lambda_{\pm}$ outweigh the effect of terms $\sim G^2$, in contrast to Yang-Mills theory. The naive expectation, that gravity, mediating an attractive interaction, will also lead to the formation of bound states in our setting, is contradicted (within our simple truncation). Instead metric fluctuations mainly modify the anomalous scaling of the interaction through terms of the form $\sim G \lambda_{\pm}$. 
In more physical terms, the $\sim G^2$ terms can be identified with the attractive interaction mediated by
gravity, whereas the terms $\sim G\lambda_\pm$ constitute the gravity contribution to the (anomalous) scaling of the fermion couplings,
$\partial_t \lambda_\pm = + 2(1+\eta_\psi+ \dots G)\lambda_\pm+ \dots$. 
The mechanism for the generation of fixed points in the fermionic flows is a balancing between the dimensional and anomalous scaling on the one hand, and the fermionic fluctuations on the other hand. Gauge-field fluctuations generate the four-fermion couplings even if they are set to zero, hence they support the fermionic fluctuation channels. In contrast, metric
fluctuations also take a strong influence on the anomalous dimensional scaling
which counteracts the general attractive effect of gravity.

This viewpoint is further supported by other technical observations: whereas
gravity is channel blind with respect to the scaling terms,
i.e. $\partial_t
\lambda_i \sim G\lambda_i$, gauge boson fluctuations with coupling $g$ also
give rise to terms $\partial_t \lambda_i \sim g^2 \lambda_j$ with $i\neq j$ that
act rather like the above mentioned fluctuation terms.

Let us comment on the question, if metric fluctuations can in this way counteract gluonic fluctuations and prevent chiral symmetry breaking in QCD. Here it is important to realise that the scale at which metric fluctuations are strong is a regime, where due to asymptotic freedom gluonic fluctuations can be neglected. On the other hand, quantum gravity effects are expected to be negligible on scales where the QCD coupling becomes large and gluonic fluctuations lead to chiral symmetry. This separation of scales implies that the results concerning chiral symmetry breaking through gluonic fluctuations can be expected to remain unaltered.

\subsubsection{Universality classes for the fermionic system}
For $G_{\ast} \neq 0$ the
GFP is shifted and becomes an
interacting fixed point (see fig.~\ref{betalambda}).
This effect can be traced back to the terms $\sim G^2$ in
the $\beta$ function. Since the terms $\sim G
\lambda_{\pm}$ dominate, the NGFP
also experiences a considerable shift, such that for $G>0$
the fixed points move further apart. 
If this structure persists beyond our simple truncation,
the system has four different universality classes
available in the UV.

\begin{figure}[!here]
\begin{minipage}{0.45\linewidth}
 \includegraphics[scale=0.75]{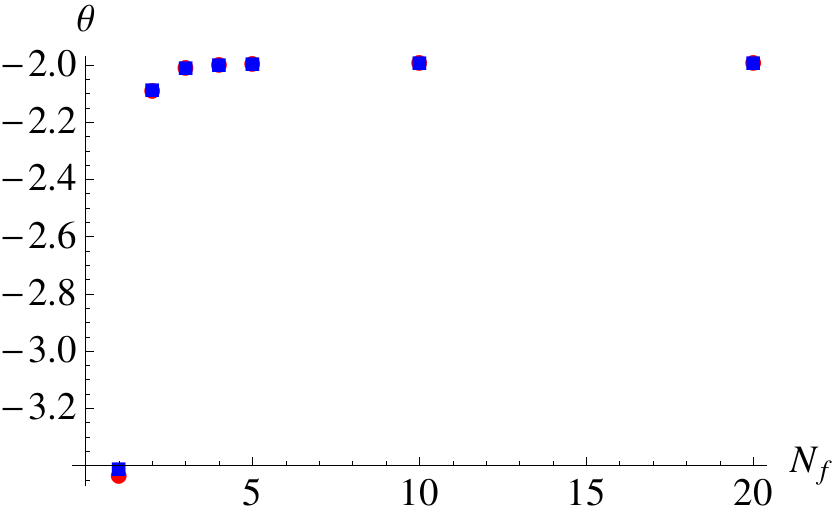}\newline\\
 \includegraphics[scale=0.75]{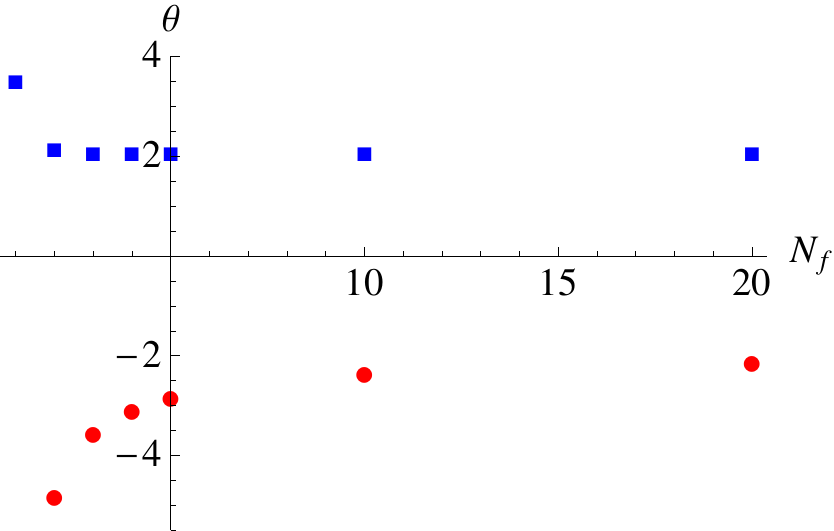}
\end{minipage}
\begin{minipage}{0.45\linewidth}
 \includegraphics[scale=0.75]{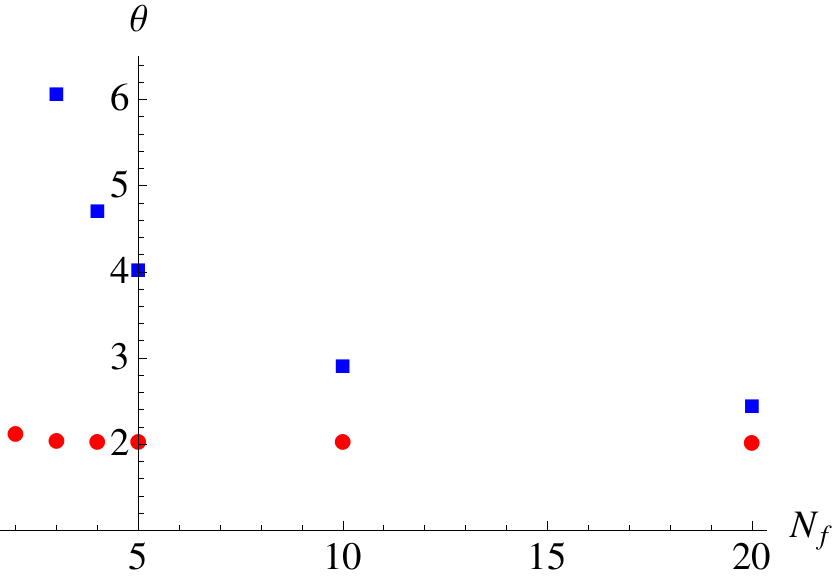}\newline\\
\includegraphics[scale=0.75]{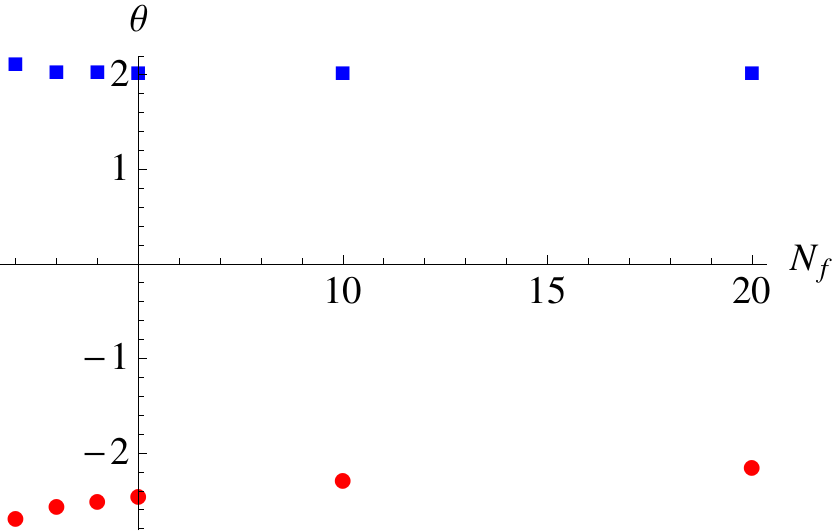}
\end{minipage}
\caption{Critical exponents in the fermionic subsector as a
function of $\rm N_f$, where we always plot one critical
exponent with red dots and the second one with blue
squares. The upper left panel corresponds to the shifted GFP and
therefore has irrelevant directions only.\label{thetaplots}}
\end{figure}

Since the four-fermion couplings do not couple back into
the flow of the Einstein-Hilbert sector, the stability
matrix has a $2\times 2$ block of zeros off the diagonal.
Therefore the eigenvalues are determined by the eigenvalues
in the Einstein-Hilbert, and the four-fermion subsector.
Accordingly the critical exponents are given by the
two relevant directions in the Einstein-Hilbert sector and the two real critical
exponents in the fermionic subsector. The dependence
of the critical exponents on $\rm N_f$ at each of the four
fixed points is shown in fig.~\ref{thetaplots}. 
Here we show values which are determined by inserting a
triplet $(N_f, G=G_{\ast},\lambda= \lambda_{\ast})$  as
determined in \cite{Percacci:2002ie} within
asymptotically safe quantum gravity. The difference of \cite{Percacci:2002ie} to our regularisation scheme adds to our truncation error, but will not change our findings qualitatively.

\subsubsection{Decoupling of metric fluctuations within asymptotic safety}
The critical exponents tend to their values from the purely
fermionic system for large $N_f$. This is due to the
following mechanism being at work here:
A negative value for $\lambda$ acts
similar to a "mass term" for the metric, since the metric propagator is schematically given by $\frac{1}{p^2-2\lambda}$. Therefore it
suppresses the metric contribution to
$\beta_{\lambda_{\pm}}$. As shown in
\cite{Percacci:2002ie}, the backcoupling of a minimally
coupled fermion sector into the Einstein-Hilbert sector
shifts $\lambda$ to increasingly negative values as a
function of $N_f$. Therefore the interaction of both
sectors, the fermionic and the metric one, leads to a regime
where the large "graviton mass" suppresses the contribution
of metric fluctuations to the running of fermionic
couplings. This decoupling mechanism ensures that the properties of the
matter sector will not be strongly altered by metric
fluctuations.\\
This feature distinguishes gravity from Yang-Mills theory,
where no analogous mechanism exists to suppress gluonic
fluctuations.

Let us stress that this decoupling mechanism is a feature of
asymptotically safe quantum gravity, and does not naturally
occur in the effective field-theory setting, where the
couplings can have values unrestricted by any fixed-point
requirement.
Within the asymptotic-safety scenario it is the interplay
between fermionic and metric fluctuations in combination
with the fixed point, that results in the
observed decoupling mechanism.

We observe that the decoupling mechanism is only at work in
theories with a larger number of fermionic degrees of
freedom, as it is the case for the Standard Model.
Since minimally coupled scalars shift the fixed-point
value for the cosmological constant towards $\lambda_{\ast}\rightarrow
\frac{1}{2}$, see \cite{Percacci:2002ie}, a larger number of scalars even results in an enhancement of metric fluctuations.
As a consequence, even at the shifted
GFP the fermionic system can develop strong correlations,
since the fixed point values for $\lambda_{\pm}$ can then become quite large.
Accordingly, in theories with a supersymmetric
matter content and low-scale supersymmetry breaking, such a decoupling
mechanism might not occur or only in a much weaker fashion. Supersymmetric theories with a quantum
gravity embedding may thus have to satisfy stronger constraints as far as the
initial conditions of their RG flow are concerned.

\begin{figure}[!here]
\begin{minipage}{0.45\linewidth}
 \includegraphics[scale=0.65]{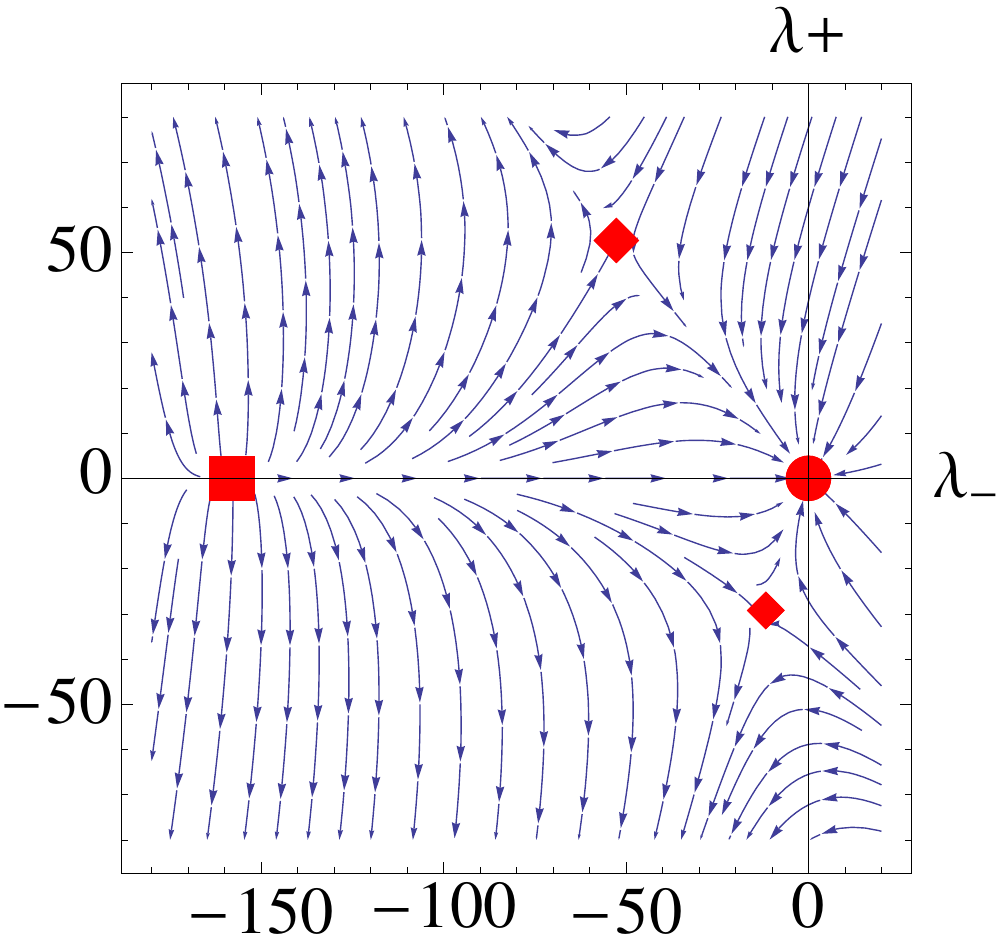}
 \end{minipage}
\begin{minipage}{0.45\linewidth}
\includegraphics[scale=0.65]{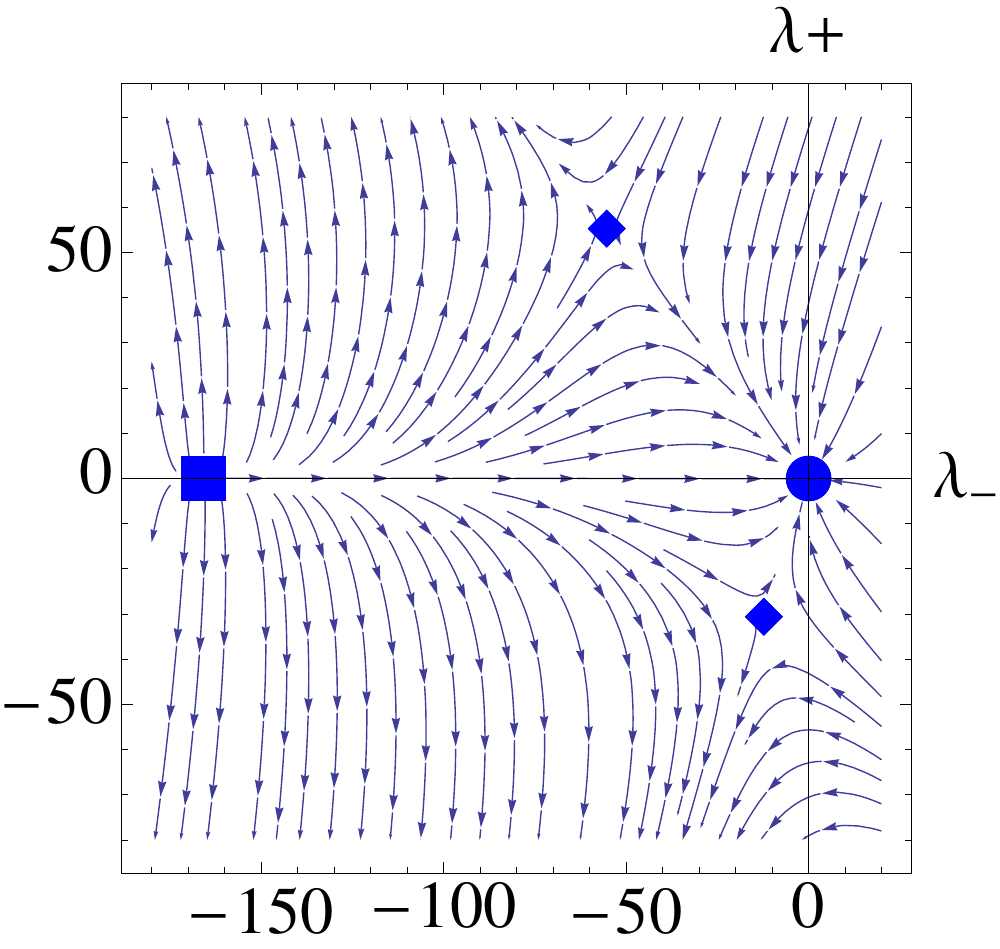}
\end{minipage}
\caption{RG trajectories towards the infrared in the $(\lambda_+,
\lambda_-)$ plane. The left panel shows the flow with
$G=0=\lambda$, the right one with $G\approx 2.3$ and
$\lambda \approx -1.38$, which are the fixed-point values for $N_f=2$ from \cite{Percacci:2002ie}.\label{flows}}
\end{figure}

We exemplify the above by showing the RG trajectories (directed
towards the infrared) in the $(\lambda_+, \lambda_-)$ plane
(see fig.~\ref{flows}). We set $N_f=2$ and $\eta_{\psi}=0$
and then show the flow without the metric contribution, and
with the metric contribution with $G, \lambda$ taking their
fixed-point values for $N_f=2$ according to \cite{Percacci:2002ie}.

\subsubsection{Effective field theory: Restrictions on UV completions of gravity}
Let us now analyse the system in the setting of effective
field theories, where no fixed-point requirement
 determines $G,\lambda$ and
$\eta_N$. Instead, any microscopic UV completion of gravity
should allow for a regime at $k_0 \lesssim M_{\rm Planck}$,
where our framework is applicable. The microscopic theory
then in principle determines the initial conditions for the
flow at $k_0$.  

\begin{figure}[!here]
\begin{minipage}{0.5\linewidth}
 \includegraphics[scale=0.6]{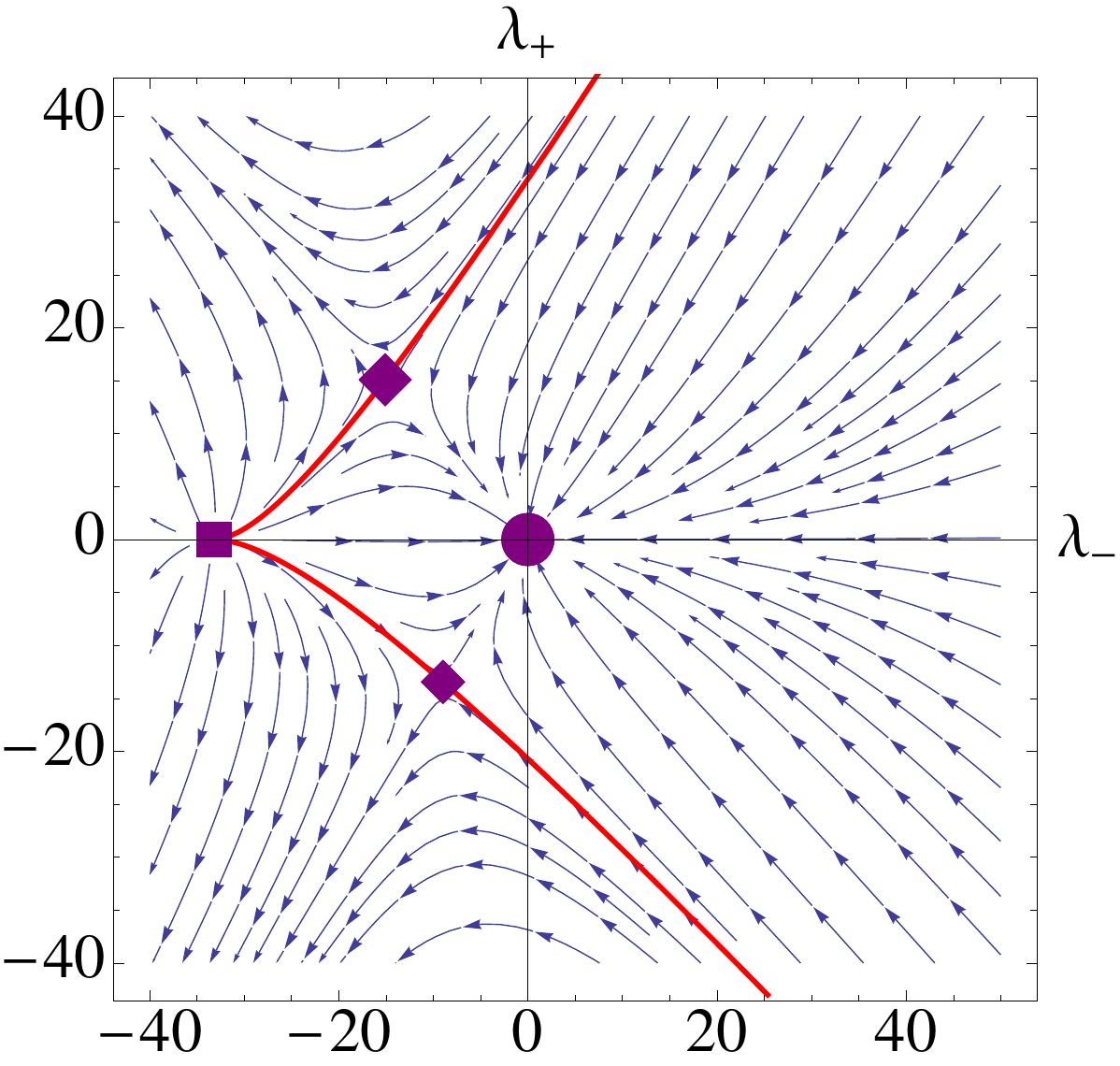}
\end{minipage}
\begin{minipage}{0.45\linewidth}
\caption{Flow towards the infrared in the $\lambda_{+},
  \lambda_{-}$-plane for $\eta_N =0, \eta_{\psi}=0, G=0.1,
\lambda=0.1$ and
  $N_f=6$. For initial values to the right of the red lines
the fermionic system
  is in the universality class of the (shifted) Gau\ss ian
fixed point. Any
  microscopic theory that would put the effective quantum
field theory to the
  left of the red lines would generically not support light
fermions.\label{efftheoryplot}}
\end{minipage}
\end{figure}

Within standard scenarios for matter, the fermionic system
should be in the vicinity of the GFP on
scales where metric fluctuations die out and matter and
gauge boson fluctuations dominate the picture. Thereby the allowed initial conditions for the flow at
$k_0$ are restricted to lie within the
basin of attraction of the (shifted) GFP (see fig.~\ref{efftheoryplot}). 

\noindent As is clear from fig.~\ref{efftheoryplot}, it is generically possible for a UV completion for gravity to determine initial conditions for the RG flow that are compatible with the existence of light fermions without a high degree of finetuning. Thereby our calculation suggests that the existence of light fermions might generically be compatible with quantum gravity. Naturally it has to be checked within any specific proposal for quantum gravity if this possibility is indeed realised, which requires to determine the values of the effective theory from the microscopic theory.   

\begin{figure}[!here]
\begin{minipage}{0.6\linewidth}
\includegraphics[width=1\linewidth]{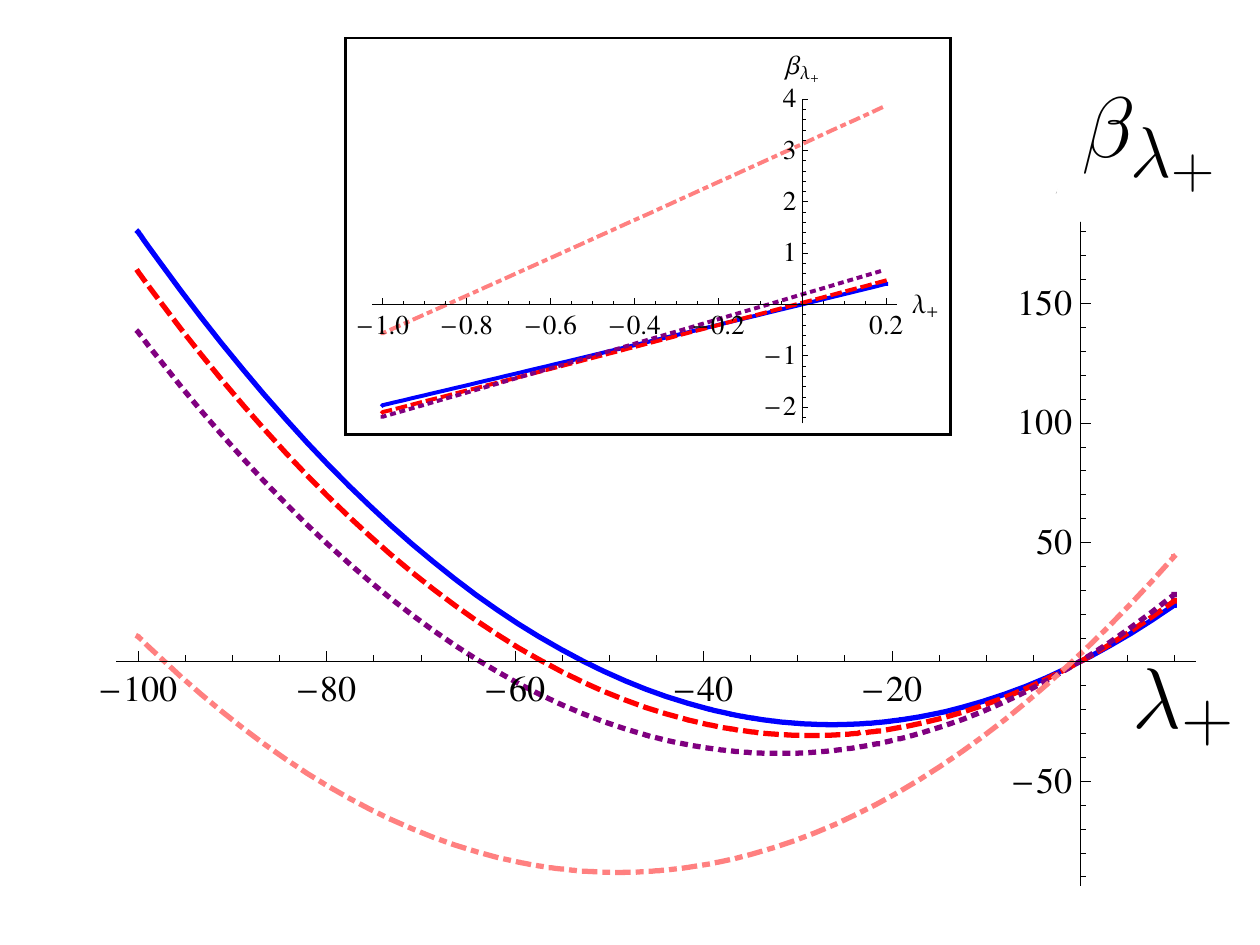}
\end{minipage}
\begin{minipage}{0.35\linewidth}
\caption{$\beta_{\lambda_{+}}$ as a function of
$\lambda_{+}$ for different values of $G$, where
$\lambda_-$ is set on the shifted GFP-value. The full blue
line corresponds to $G=0$, the red dashed one to $G=0.2$,
the purple dotted one to $G=0.5$ and the pink dotted-dashed
one to $G=2$. The inlay shows the region around the
Gau\ss{}ian fixed point, where its shift is clearly
visible.\label{betalambdaaa}}
\end{minipage}
\end{figure}

We also observe that for generic values of $G$ and
$\lambda$
the system can be
altered considerably.
As a first observation, the parabolas broaden as a
function of $G$ for a fixed value of $\lambda$, as shown
in fig.~\ref{betalambdaaa}.
As an example, we depict the $\beta$ function for
$\lambda_{+}$, for fixed
$N_f=2$, $\eta_N=-2$, $\eta_{\psi}=0$ and $\lambda=0$. We
set $\lambda_{-}$ on the shifted Gau\ss{}ian fixed point
value.\newline

\begin{figure}[!here]
\begin{minipage}{0.55\linewidth}
 \includegraphics[scale=0.5]{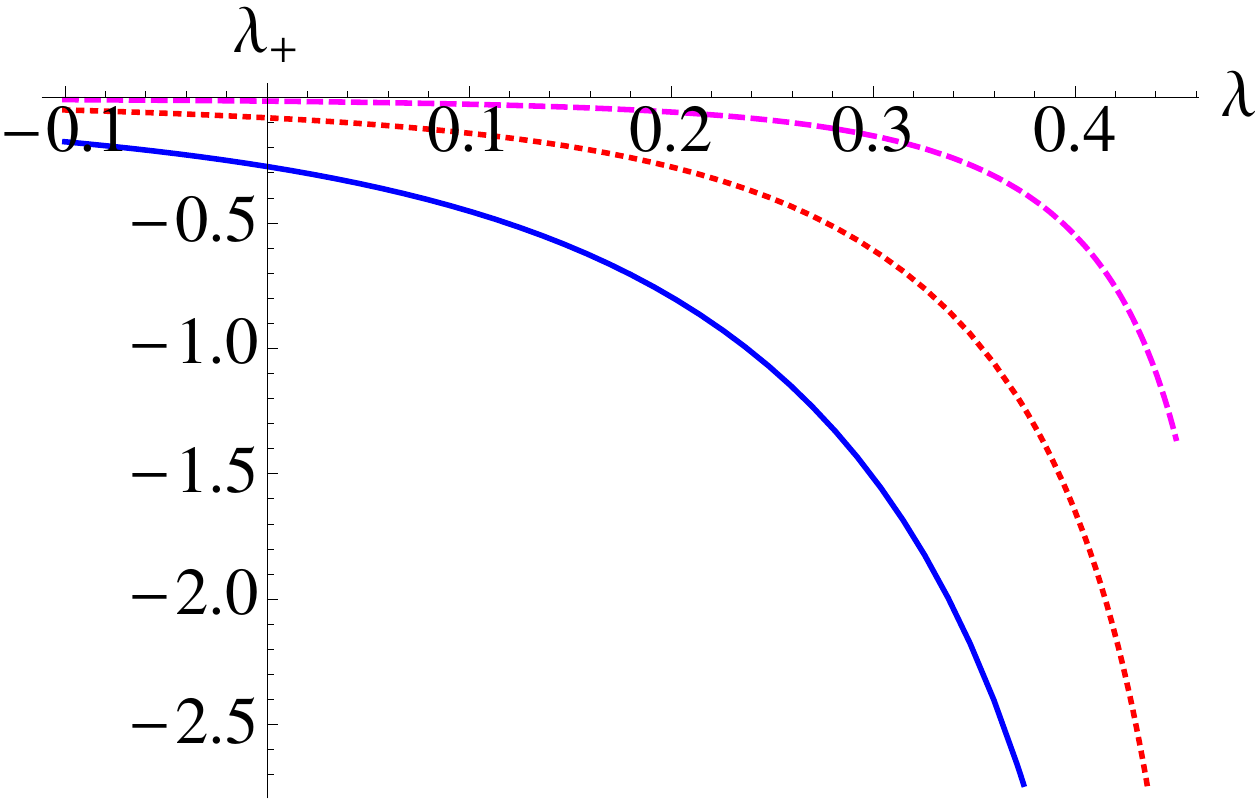}
\end{minipage}
\begin{minipage}{0.42\linewidth}
\caption{Fixed-point value of $\lambda_+$ at the shifted Gau\ss ian fixed
  point for $N_f=2$, $\eta_{\psi}=0$ and $\eta_N=0$ as a function of $\lambda$
  for $G=1$ (full blue line), $G=0.5$ (red dotted line) and $G=0.2$ (magenta
  dashed line).\label{lambdaplusplot}}
\end{minipage}
\end{figure}

\noindent A particularly strong effect can be observed for positive
values of $\lambda$.
Here, the contribution from the metric sector is further
enhanced for $\lambda >0$, and indeed the $\beta$ functions
show the well-known divergence for $\lambda = \frac{1}{2}$,
which presumably should be attributed to a breakdown of the
simple Einstein-Hilbert truncation in the infrared.\newline
As an example, we depict the value of $\lambda_+$ at the
shifted Gau\ss{}ian fixed point in fig.~\ref{lambdaplusplot}.

\noindent We observe that the GFP can be shifted
to considerably larger values of $(\lambda_-, \lambda_+)$.
This implies that the system will be strongly-interacting in
this sector even at the shifted GFP, which may alter its
physical behaviour.

\begin{figure}[!here]
\begin{minipage}{0.55\linewidth}
 \includegraphics[scale=1]{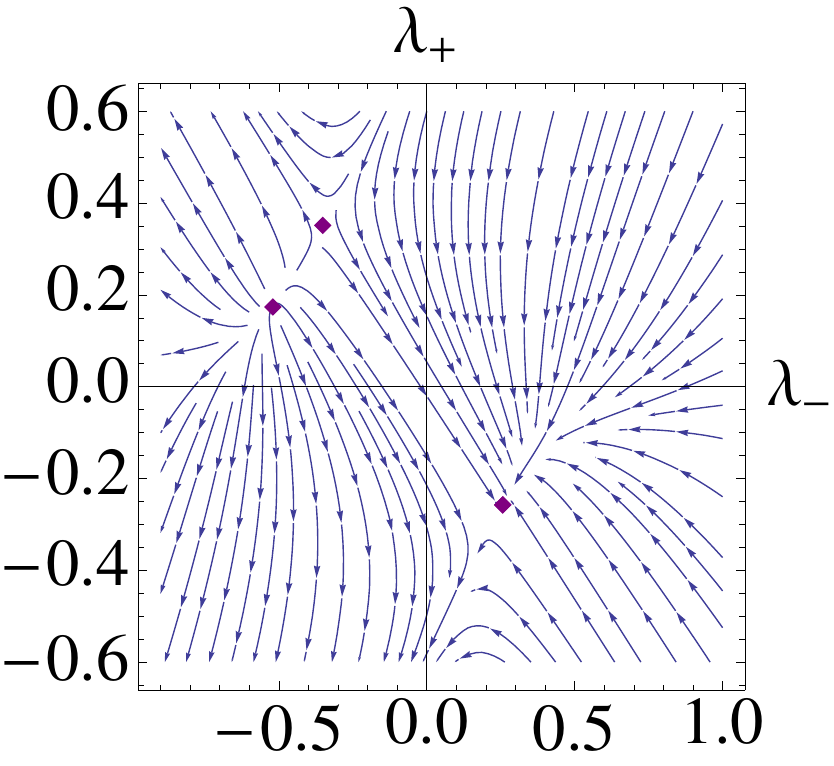}
\end{minipage}
\begin{minipage}{0.4\linewidth}
\caption{At $\eta_{\psi} \approx -1.0592$ (for the specific parameter
    values $N_f=6, G=0.1, \lambda=0.1, \eta_N =-2$) the Gau\ss{}ian fixed
  point and a fixed point with one relevant direction fall on top of each
  other (in the lower right quadrant). \label{degenerateflow}}
\end{minipage}
\end{figure}

We may now study to which extent an extension of the
truncation
with $\eta_{\psi}\neq 0$ can change our findings. Here we do not determine the value of $\eta_{\psi}$ from the Wetterich equation, but simply check the effect of non-zero values for $\eta_{\psi}$ on the fixed-point structure in the fermionic sector.
Interestingly a negative value of $\eta_{\psi}$ can lead to
a crucial change in the flow: For $\eta_{\psi}= \eta_{\psi
\, \rm crit}$ the shifted Gau\ss{}ian and one fixed point
with one relevant direction fall on top of each other (cf.
fig.~\ref{degenerateflow}), exhibiting one marginal direction with
a zero eigenvalue at this point. Since gravity has been
observed to induce a negative anomalous dimension for the
fermions \cite{Vacca:2010mj}, this critical value might be
assumed at high energies. What one naively would suppose to
be the basin of attraction for the GFP might then end up in
a different region in theory space under the flow, when
$\eta_{\psi}$ crosses the critical value. A more detailed
investigation is straightforwardly possible here with the methods outlined in
this section. 

\subsection{Outlook: Spontaneous symmetry breaking in gravity}\label{conclusions}
We have investigated the compatibility of light fermions
(i.e. unbroken chiral symmetry) with the asymptotic-safety
scenario for quantum gravity and also studied a setting
within the
effective-field-theory framework. In contrast to Yang-Mills
theory, where gluonic fluctuations break chiral symmetry if a
critical value of the coupling is exceeded, metric
fluctuations induce no such effect within our truncation.

As no combination of values $(G, \lambda, N_{\rm f}, \eta_N, \eta_{\psi})$ can be found for which chiral symmetry is directly broken, the asymptotic-safety scenario is compatible with light fermions within our truncation. In the case of a different UV completion for gravity the initial conditions for the flow equations that hold within an effective description, have to fall into a certain region in theory space, in order to avoid chiral symmetry breaking. 

In Yang-Mills theory, a truncation of FRG equations
containing only minimally coupled fermions and gluons
suffices to discover the effect of chiral symmetry
breaking. In gravity an extension of the truncation might
be necessary in order to observe chiral symmetry breaking.

At dimension six (for two-fermion terms) and eight (for
four-fermion-terms) we encounter a variety of new terms that
are not forbidden by explicit chiral symmetry breaking,
for instance
\begin{eqnarray}
\text{dim 6:} &\quad&\int d^4x \, \sqrt{g} \,R\, \bar{\psi}
\slashed{\nabla}\psi,\quad \quad\int d^4x \, \sqrt{g}\, R_{\mu \nu}\, \bar{\psi}
\gamma^{\mu}\nabla^{\nu}\psi, \label{extended_trunc}
\\
\text{dim 8:}&\quad&\int d^4x \, \sqrt{g} \, R \left(V^2 \pm
A^2 \right),\nonumber\\
&{}&\int d^4x \, \sqrt{g}\, R_{\mu \nu} \Bigl(\left(
\bar{\psi}^i
\gamma^{\mu}\psi^i \right)\left( \bar{\psi}^j
\gamma^{\nu}\psi^j \right) - \left( \bar{\psi}^i
\gamma^{\mu}\gamma^5\psi^i \right)\left( \bar{\psi}^j
\gamma^{\nu}\gamma^5\psi^j \right)\Bigr).
\label{extended_trunc2}
\end{eqnarray}
At higher dimensionalities the number of terms increases
considerably, as then
also, e.g. contractions involving the Riemann tensor will be
possible. Furthermore couplings involving $\slashed{\nabla}$
or higher
  powers of the curvature are possible.  Distinguishing
between the
background and the fluctuation metric leads to an even
larger "zoo" of
possible operators.

\begin{figure}[!here]
 \begin{minipage}{0.55\linewidth}
\begin{center}
  \includegraphics[scale=0.1]{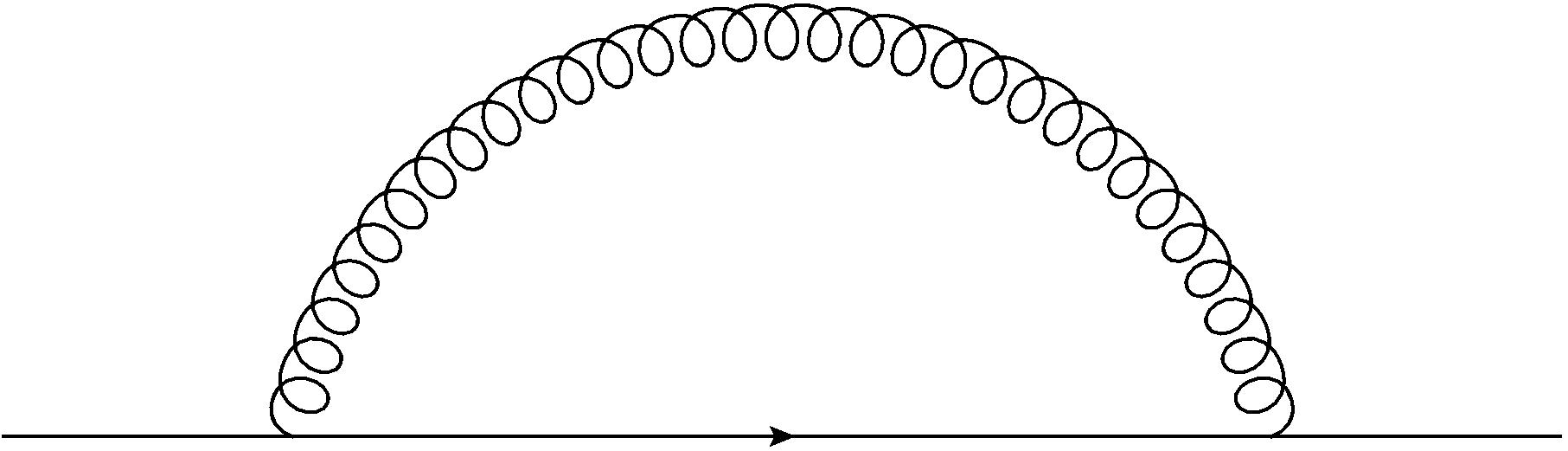}\newline\\
  \includegraphics[scale=0.1]{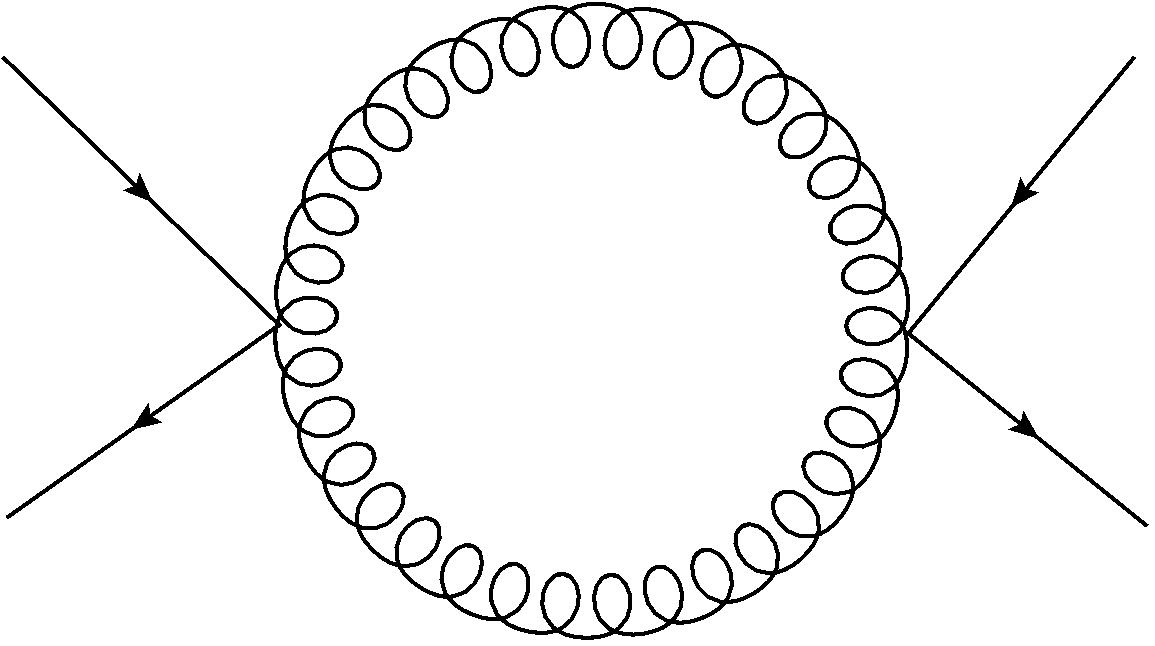}
\end{center}
 \end{minipage}
 \begin{minipage}{0.4\linewidth}
 \caption{The $\tilde{\partial}_t$ derivatives of these diagrams generate contributions to the couplings in \Eqref{extended_trunc2} (couplings to external metric structures result from corresponding derivatives of the diagrams). The upper diagram also contributes to the fermion anomalous dimension. Both terms are generated from vertices arising from the minimally coupled kinetic term. \label{furtherterms}}
 \end{minipage}
\end{figure}

Several comments are in order here: 
As a first requirement we demand the existence of a fixed
point for the extended truncation. Here, the above
couplings do not only alter the fermionic flow, but, e.g.
the non-minimal kinetic terms couple back into the flow of the
Einstein-Hilbert sector and further metric operators.
The effect of metric
fluctuations implies that none of these couplings will have a
GFP, as the
antifermion-fermion-two-graviton vertex generically
generates these couplings even if they are set to zero
(see fig.~\ref{furtherterms}).

\noindent Hence these couplings will typically assume
non-zero fixed point values, thus constituting
a non- vanishing contribution to the $\beta$ functions in
the Einstein-Hilbert as well as the fermionic sector.

In particular, non-minimal kinetic terms give a
contribution $\sim G^2$ to the $\beta$ functions for
$\lambda_{\pm}$. As the absence of chiral symmetry breaking
in our truncation follows from the dominance of anomalous
scaling of the fermionic interactions (i.e. terms $\sim G
\lambda_{\pm}$) over fluctuations that induce attractive
interactions and drive the fermionic system to criticality
(i.e. terms $\sim G^2$), we expect further terms of the
second type to play a crucial role. Clearly an extension of
our truncation in this direction is of particular interest.
Note that furthermore such non-minimal terms may be
interesting in the context of cosmological applications,
such as chiral symmetry breaking in the history of our
universe. Here, finite temperature effects as well as such
non-minimal terms may play a crucial role and demand for
further investigation.

Let us stress again that the property, that metric fluctuations within the asymptotic-safety scenario, but also within an effective field theory framework generically shift a Gau\ss{}ian to a non-Gau\ss{}ian fixed point is of interest beyond the question of chiral symmetry breaking. The same mechanism may apply for generic matter couplings, thus inducing (potentially strong) interactions of the matter theory at high energies. These may potentially play a role in early-universe cosmology and thus are of interest for further studies.

Let us now broaden our view a bit, concerning the so far
unaddressed question of gravity-induced symmetry-breaking
patterns. Here, we have
assumed an SU$({\rm
N_f})_{\mathrm{L}}$ $\times$
SU$({\rm N_f})_{\mathrm{R}}$ symmetry (with additional U(1)
factors of particle number
and axial symmetry), and implicitly assumed its breaking in
a QCD-like
fashion, i.e. to a remaining mesonic SU($\rm N_f$) symmetry.
It may now be possible that the pattern of symmetry
breaking is different in gravity. One may, e.g. consider a
scenario including an originally larger symmetry that may
break to the standard chiral symmetry upon large metric
fluctuations. 
In particular, such an extension of the symmetry group may
induce a rich structure in the corresponding phase diagram
of the theory: Different symmetry-breaking patterns may be
possible, corresponding to different phases with appropriate
condensates and excitations on top of these.
 If, e.g. a
gravitationally-stimulated
symmetry breaking transition with a remnant standard chiral
symmetry occurred
near the Planck scale, stable bound states (analogously to
hadrons in QCD) may
have remained and (if equipped with the right quantum
numbers) could
contribute to the dark matter in the universe.  

Furthermore, in analogy to recent ideas in QCD, where a
quarkyonic phase with
confinement but intact chiral symmetry supports a spectrum
of bound states,
bound states may form that correspond to bosonised
operators, e.g. of the
form \Eqref{extended_trunc2}. These might be generated at a scale
where quantum
gravity is strongly interacting, and may then become massive
at the much lower
scale of chiral symmetry breaking. Supporting a stable bound
state over such
a large range of scales requires, of course, a highly
non-trivial
interplay between gravity and matter.

Let us also point out that studies concerning the spontaneous breaking of global symmetries of matter theories through metric fluctuations are generically of interest. Here, we have focussed on studying the spontaneous breaking of chiral symmetry, but of course metric fluctuations may potentially induce the formation of non-trivial VEVs that also break other global symmetries of matter theories, such as, e.g. baryon number or similar. Depending on the phenomenological implications of such mechanisms, these may serve to further restrict generic UV completions for gravity, as well as the asymptotic-safety scenario.

\chapter{Conclusions: Relating micro- and macrophysics in high-energy physics}

In this thesis we have studied how the Functional
Renormalisation Group connects microscopic and macroscopic
degrees of freedom and descriptions of physics. We have
introduced the Renormalisation Group flow, whose natural
direction is from high to low momenta, i.e. from the ultraviolet to the infrared.
The flow successively takes into account the effect of
quantum fluctuations scale by scale and works not only in
the perturbative regime, but most importantly also beyond.

In particular we have been interested in fundamental theories. These are theories which are valid on all (momentum) scales, and do not show any unphysical divergences at any scale. In a more physical language a fundamental theory can only be one, where we have identified the correct UV degrees of freedom. In contrast effective theories are only valid on a limited range of scales and use effective degrees of freedom adapted to these scales. As an illustration, consider a theory where the UV degrees of freedom form bound states in the macroscopic regime. The effective low-energy theory can then be formulated with the help of these macroscopic bound states as effective degrees of freedom.

As an example, we have considered QCD, where the microscopic action is determined by the property of asymptotic freedom, i.e. the theory has a Gau\ss{}ian fixed point with one relevant direction. Towards the infrared, the theory becomes strongly interacting, and the relevant degrees of freedom change, since quarks and gluons become confined inside hadrons. We have studied the property of confinement at zero and at finite temperature in Yang-Mills theory. In both parts, i.e. for the evaluation of the gluon condensate at zero temperature as well as the study of the deconfinement phase transition, we have relied on the background field method, where we have used a non-trivial reconstruction of the corresponding fluctuation field propagators, thereby going beyond former approximations in this framework.\newline
In the first part we have focussed on the vacuum structure of Yang-Mills theory, and deduced the existence of a condensate $\langle F^2 \rangle$ from a full evaluation of the effective potential, which can be interpreted as a condensation of gluons in the vacuum. Our value for the condensate agrees rather well with estimates from other methods.
Interestingly, the functional form of the 
effective potential supports the leading-log model for the
ground state of Yang-Mills theory, which allows us to deduce
a value for the string tension between a static
quark-antiquark pair from our calculation. Here, a
straightforward extension of our work will allow to resolve
approximations that we have applied. In addition, an
implementation of our calculation at finite temperature and
with the effects of quark fluctuations taken into account is
also possible along the same lines, thus establishing a
connection to the QCD phase diagram. 
Furthermore we have related the critical exponents
$\kappa_{A,c}$, characterising the scaling behaviour of the
ghost and gluon fluctuation propagator in the deep infrared,
to the asymptotic form of the $\beta$ function of the
background running coupling. The requirement of a
strongly-interacting regime in the infrared, which is at the
heart of both confinement as well as chiral symmetry
breaking, imposes constraints on the form of the $\beta$
function in the vicinity of the Gau\ss{}ian fixed point.
These translate into an upper bound on the critical
exponents. This confinement criterion is fulfilled for critical exponents
evaluated from functional methods.

In a second part we have gone beyond the zero-temperature limit of the theory, and studied the deconfinement phase transition. For the physical case of the gauge group SU(3) we find good agreement with results from lattice gauge theory. Let us stress that since we do not use the standard order parameter, the Polyakov loop, but a related quantity instead, we can deduce a potential encoding confinement for our order parameter from the sole knowledge of the ghost and the gluon propagator. This is a crucial advantage in contrast to the Polyakov loop, the knowledge of which requires information on all $n$-point correlation functions of the gluon.
We have then extended our study to different gauge groups, in particular SU(N) with $2\leq {\rm N}\leq 12$, Sp(2) and E(7), which allows to shed light on the question what determines the order of the phase transition. Here our method allows to consider gauge groups which for technical reasons are currently unaccessible to lattice gauge theory.
We have found evidence suggesting that not the center, but the size of the gauge group is the decisive quantity. Physically, this is due to a large mismatch in the number of dynamical degrees of freedom on both sides of the phase transition. We have further discussed how the group structure enters the order of the phase transition through the eigenvalues of the Cartan generators in the adjoint representation. We have developed a picture of constructive and destructive interference of SU(2)-potentials for the order parameter, which can induce or prevent a second order phase transition, respectively. In the future, e.g. thermodynamic properties of the deconfined phase can be studied with our method. 

Using the RG flow in a less intuitive way, namely towards the ultraviolet, allows to search for UV completions of effective theories, thereby establishing a connection from a known macroscopic regime to a possible microscopic description. Here we use that UV completions can be found with the help of non-interacting, i.e. Gau\ss{}ian, or interacting, i.e. non-Gau\ss{}ian fixed points in theory space. In gravity a fundamental quantum field theory of the metric can only be defined at the interacting fixed point. The physical assumption underlying this scenario is that the description of gravity with the help of the metric field is indeed valid on all scales. In a first study we have examined the Einstein-Hilbert truncation for a new combination of regularisation scheme and gauge fixing, and used a method independent of heat-kernel techniques to evaluate the right-hand side of the Wetterich equation.
We have then studied a particular sector of the theory, namely the Faddeev-Popov ghost sector, which arises after gauge fixing. We have discussed several scenarios for this sector which we discriminate by the existence of relevant directions and an interacting fixed point in this sector. The first has potentially challenging implications for the relation between relevant couplings and measurable free parameters, whereas the second is closely related to the possible non-uniqueness of gauge-fixing in the non-perturbative regime.

Our first extension of formerly studied truncations in this sector contains a non-trivial ghost anomalous dimension. Our findings constitute further evidence for the existence of a physically admissible, interacting fixed point. 

In contrast to a perturbative setting, the properties of the ghost sector in the non-perturbative regime are largely unclear. Firstly, relevant couplings can be related to operators containing ghost fields, which makes the connection between relevant couplings and measurable quantities highly non-trivial. Our evaluation of the ghost anomalous dimension suggests the existence of such relevant couplings. As a specific example we have considered a ghost-curvature coupling, that turns out to be relevant at its Gau\ss{}ian fixed point. A second issue is closely related to the non-perturbative non-uniqueness of gauge-fixing, the so-called Gribov problem. Going beyond the perturbative regime, and in particular requiring the existence of an interacting fixed point for the Newton coupling induces non-zero couplings in the ghost sector. In a diagrammatic expansion of the flow equation it is easy to see that non-zero values of couplings are generated even if these are initially set to zero in a truncation.
Thereby terms beyond a simple perturbative Faddeev-Popov operator will be generated, and might even result in a unique gauge-fixing beyond the perturbative regime. As a first study in this direction we have focussed on a resolution of different running tensor structures in the Faddeev-Popov operator. Here we observe two physically admissible fixed points, where one has a new relevant direction. The other one is characterised by a further irrelevant direction and a positive ghost anomalous dimension, which shifts further ghost operators towards irrelevance. This second fixed point can be interpreted as an infrared fixed point, which allows to construct an RG trajectory with a well-defined IR as well as UV limit. 
We have further considered an explicit example for terms beyond a simple Faddeev-Popov operator and suggested the existence of non-vanishing four-ghost couplings.
Further investigations of the ghost sector along the lines outlined in this thesis are possible in the future, thereby further clarifying the structure of the interacting fixed point for gravity in this particular sector.

Finally we have applied the RG flow to a system containing fermionic fields coupled to the metric. Specifically we have been interested in the compatibility of massless fermions with a quantum gravity regime which can be parametrised by (strongly-coupled) metric fluctuations, presumably around the Planck scale. This question establishes a connection between the microscopic regime of quantum gravity and the macroscopic regime, where we observe the existence of light (compared to the Planck scale) fermions. In this setting, we have not only focussed on a fundamental theory of gravity, but also shed light on this question within the effective field theory framework. We deduce chiral symmetry breaking from the fixed-point structure of the $\beta$ functions of four-fermion couplings: If these do not show any real fixed points, the fermionic system becomes strongly correlated and chiral symmetry is broken, thereby endowing fermions with a mass.
Our findings suggest that asymptotically safe quantum gravity favours universes in which light fermions exist. Here, we have also discussed a physical decoupling mechanism: Fermionic fluctuations induce a negative fixed-point value for the cosmological constant, which in turn suppresses metric fluctuations, whereby their influence on the properties of the matter sector is reduced. 

In the case of effective theories we have shown that the requirement of unbroken chiral symmetry in principle restricts any UV completion of gravity. Here we use that, within the effective field theory framework, gravity fluctuations can be parametrised as metric fluctuations in a regime presumably below the Planck scale, even if the microscopic theory supposes a different nature of quantum gravity fluctuations, or even a discrete space-time. Then, the microscopic theory determines the values of couplings in the effective theory at some initial scale. Since these have to fulfill certain requirements in order for light fermions to naturally exist, this allows -- in principle -- to restrict the microscopic theory.
  
Our findings of unbroken chiral symmetry are in contrast to Yang-Mills theory, where in a similar truncation gluonic fluctuations induce chiral symmetry breaking. This can be traced back to the fact that metric fluctuations strongly alter the scaling behaviour of the fermionic couplings, thereby preventing the breaking of chiral symmetry. Within our study we have also identified terms beyond our truncation that may potentially induce chiral symmetry breaking in gravity. Here, the methods used in our investigation also allow for a further extension of the truncation and a detailed study of possible mechanisms of chiral symmetry breaking in gravity.

Let us stress, that applying the framework of effective field theories to quantum gravity, as in this thesis, could provide for a very useful window into the quantum gravity realm: Testing the compatibility of low-energy properties of matter with quantum gravity can be used to restrict theories of quantum gravity. In the absence of experimental insights into quantum gravity such tests provide for a non-trivial testing ground for candidate theories for quantum gravity.

To summarise, let us say that the formalism of the functional RG, used in this thesis, allows to access, in a qualitatively as well as quantitatively meaningful way, the non-perturbative regime of large variety of  theories. In particular, structural questions as, e.g. the properties of the sector of a theory arising after gauge fixing, as well as more physical questions such as the deconfinement phase transition or the existence of light fermions in our universe, can be studied further within this formalism.

Since the existence of a fundamental quantum field theory relies on the existence of a well-defined UV limit, candidates for fundamental theories show either a Gau\ss{}ian or a non-Gau\ss{}ian fixed point. In the first case, the microscopic regime of the theory is accessible with perturbative tools, but non-perturbative physics will emerge in the low-energy limit, such as in QCD. If we on the other hand construct a fundamental theory with the help of an interacting fixed point, the microscopic physics will generically be non-perturbative. We therefore find that in order to understand fundamental theories, be they non-Abelian gauge theories as used to construct the Standard Model, or be they candidates for a quantum gravity theory, we are bound to use non-perturbative tools. As discussed in this thesis, the functional Renormalisation Group can be applied to gain further understanding of these theories and might allow to show if the framework of quantum field theory is rich enough to incorporate not only low-energy phenomena such as confinement in QCD, but also a microscopic theory of quantum gravity.

\newpage
\pagestyle{empty}
\begin{center}
 {\large{\emph{Acknowledgements}}}\newline\\
\end{center}
I would like to express my gratitude towards Holger Gies for the possibility to work on the topics presented in this thesis, and for providing for a very stimulating and motivating atmosphere during this time. I especially would like to convey my thanks for many instructive discussions, from which I have benefited very much. I have been extremely motivated by his taking my ideas seriously and always taking a lot of time to discuss my projects with me.\newline

I thank Jan M. Pawlowski for very interesting collaborations on Yang-Mills theory and many helpful and insightful discussions on topics from both QCD as well as quantum gravity, which have helped my understanding of both topics a lot.\newline

It is a pleasure to thank Jens Braun for a very rewarding collaboration and many useful and interesting discussions and also many helpful suggestions during the course of my PhD-thesis. I very much appreciate his careful proof-reading of this thesis.\newline

I would like to thank Michael Scherer, and also Albrecht Werner for a collaboration on the "ghostbusters" project, which has been very challenging and interesting.\newline

I particularly thank the research training group "quantum and gravitational fields", as well as its speaker Andreas Wipf for providing  me with ample opportunities to attend workshops and conferences, as well as providing the financial support necessary to organise a seminar for the graduate students. The possibility to decide on possible topics and invite speakers has been very rewarding and has allowed me to learn more about many interesting aspects of physics.\newline
In particular I am highly grateful for the financial support during my research stay at the Perimeter Institute in Waterloo, Ontario.\newline

I would like to thank the Perimeter Institute for its hospitality during my research stay. I am particularly grateful to the members of the quantum gravity group for numerous
discussions, which have considerably added to my understanding of the topics presented in this thesis, and have also allowed me to get an insight into several topics or points of view, that were new to me. In particular I thank Joseph Ben Geloun, Razvan Gurau, Tim Koslowski, Lee Smolin and Rafael Sorkin for inspiring discussions.
Tim Koslowski also deserves many thanks for suggesting an exciting research project on matrix models in the context of quantum gravity.\newline

Furthermore I would also like to thank all past and present members of the Heidelberg-imported ``JCJ'' for numerous discussions, talks and presentations, and for a (nearly) inexhaustible amount of questions, that have
allowed me to understand many topics better and have always helped to remind me of the fun about physics. \newline

I particularly thank Roberto Percacci, Reinhard Alkofer and Frank Saueressig for their readiness to referee this thesis.\newline

I gratefully acknowledge helpful discussions with Fay Dowker, Axel Maas, Emil Mottola, Martin Reuter and Frank
Saueressig.\newline

\newpage
\begin{appendix}
\chapter{Appendix}

\section{Yang-Mills theory}
\subsection{Notation in Yang-Mills theory}\label{notation_YM}

The gauge field carries a Lorentz (denoted by greek letters) and an adjoint colour index (denoted by latin letters), and an explicitly indicated spacetime dependence: $A_{\mu}^a(x)$. At zero temperature the spacetime coordinates will be collectively denoted by $x$, whereas at non-zero temperature we will sometimes write $(x^0, \vec{x})$. The field strength tensor is given by
\begin{equation}
F_{\mu \nu}= \partial_{\mu}A_{\nu}- \partial_{\nu}A_{\mu}+ [A_{\mu}, A_{\nu}] = F_{\mu \nu}^a T^a= (\partial_{\mu}A_{\nu}^a- \partial_{\nu}A_{\mu}^a- i g f^{abc} A_{\mu}^b A_{\nu}^c)T^a 
\end{equation}
The dual field strength tensor is defined according to
\begin{equation}
\widetilde{F}_{\mu \nu}^a=\frac{1}{2}\epsilon^{\mu \nu \kappa \lambda}F_{\kappa \lambda},
\end{equation}
where $\epsilon^{0123}=1$.

Here we have introduced the Lie algebra generators which satisfy
\begin{equation}
 [T^a, T^b]= i f^{abc}T^c,
\end{equation}
with the structure constants $f^{abc}$ which define the adjoint representation
\begin{equation}
 \left(T^a\right)^{bc}= -i f^{abc}.
\end{equation}
In the fundamental representation they satisfy the normalisation
\begin{equation}
 {\rm Tr}(T^a_{\rm fund} T^b_{\rm fund})=
\frac{1}{2}\delta^{ab},
\end{equation}
whereas in the adjoint representation
\begin{equation}
  {\rm Tr}(T^a_{\rm adj} T^b_{\rm adj})=N \delta^{ab}
\end{equation}
for SU(N).

\subsection{Isospectrality relation on self-dual backgrounds}\label{selfdualspec}

To evaluate the heat-kernel traces on a self-dual background we use that defining $\Tr'$ as the trace without the zero mode, we make the following
useful observation for the trace over some function $\mathcal F$:
\begin{eqnarray}
&{}&\Tr_{x\text{cL}}' \mathcal F(\mathcal{D}_{\,\text{T}})
\nonumber\\
&=& 2
\sum_{l=1}^{{\rm N}_c^2-1}\, \left( \frac{f_l}{2\pi}\right)^2 \left\{
\sum_{n,m=0}^\infty \mathcal F \big( 2f_l(n+m+2)\big) + 
\sum_{n=0}^{\infty} \sum_{m=1}^{\infty} \mathcal F  \big( 2f_l(n+m)\big)
+ \sum_{n=1}^{\infty} \mathcal F  \big( 2f_l n\big)\right\} \nonumber\\
&=& 4
\sum_{l=1}^{{\rm N}_c^2-1}\, \left( \frac{f_l}{2\pi}\right)^2 
\sum_{n,m=0}^\infty \mathcal F \big( 2f_l(n+m+1)\big) \nonumber\\
&=& 4 \Tr_{x\text{c}} \mathcal F (-D^2), \label{eq:iso}
\end{eqnarray}
where the trace subscripts denote traces over coordinate space ``$x$'', color
space ``c'' and Lorentz indices ``L''.  In other words, there exists an
isospectrality relation between $-D^2$ and the non-zero eigenvalues of
$\mathcal {D}_{\,\text{T}}$.
\subsection{Symplectic group Sp(2)}\label{SP2}
The symplectic groups Sp(N) in the convention that we 
apply here can be defined in the following way: Group
elements $U \in$ SU(2N) that satisfy
\begin{equation}
U^{\ast}= J U J^{\dagger},\label{Sp2def}
\end{equation}
where the matrix $J$ is defined by $J = i \sigma_2 \otimes
\mathbf{1}$. 
The requirement \Eqref{Sp2def} clearly allows the elements
$U$ to form a group, as closure and the existence of an
inverse as well as a unit element can be shown
straightforwardly. Eq.~\ref{Sp2def} also implies that Sp(1)
is
actually SU(2). To find the center of Sp(N), realise that
\Eqref{Sp2def} implies that any $U \in$ Sp(2) can be written
as
\begin{equation}
U = (W,X),(-X^{\ast},W^{\ast})
\end{equation} 
with complex N $\times$ N matrices $X$ and $W$. 
As a center element must be a multiple of the unit matrix,
it follows that $W = W^{\ast}$ for center elements. This
clearly restricts the center to be $\mathbf{Z}_2$ for all
Sp(N).
Our represenation of the generators of $Sp(2)$ in the
fundamental representation is
as follows:

\begin{displaymath}
 C_1=\left(
\begin{array}{cccc}
 0 & i & 0 & 0 \\
 i & 0 & 0 & 0 \\
 0 & 0 & 0 & 0 \\
 0 & 0 & 0 & 0
\end{array}
\right),
\phantom{xxx}
C_2=\left(
\begin{array}{cccc}
 0 & 1 & 0 & 0 \\
 -1 & 0 & 0 & 0 \\
 0 & 0 & 0 & 0 \\
0 & 0 & 0 & 0
\end{array}
\right),
\phantom{xxx}
C_3=\left(
\begin{array}{cccc}
 i & 0 & 0 & 0 \\
 0 & -i & 0 & 0 \\
 0 & 0 & 0 & 0 \\
 0 & 0 & 0 & 0
\end{array}
\right),
\end{displaymath}

\begin{displaymath}
C_4=\left(
\begin{array}{cccc}
 0 & 0 & 0 & 0 \\
 0 & 0 & 0 & 0 \\
 0 & 0 & 0 & i \\
 0 & 0 & i & 0
\end{array}
\right),
\phantom{xxx}
C_5=\left(
\begin{array}{cccc}
 0 & 0 & 0 & 0 \\
 0 & 0 & 0 & 0 \\
 0 & 0 & 0 & 1 \\
 0 & 0 & -1 & 0
\end{array}
\right),
\phantom{xxx}
C_6=\left(
\begin{array}{cccc}
 0 & 0 & 0 & 0 \\
 0 & 0 & 0 & 0 \\
 0 & 0 & i & 0 \\
 0 & 0 & 0 & -i
\end{array}
\right),
\end{displaymath}
\begin{displaymath}
C_7=\left(
\begin{array}{cccc}
 0 & 0 & \frac{i}{\sqrt{2}} & 0 \\
 0 & 0 & 0 & -\frac{i}{\sqrt{2}} \\
 \frac{i}{\sqrt{2}} & 0 & 0 & 0 \\
 0 & -\frac{i}{\sqrt{2}} & 0 & 0
\end{array}
\right),
\phantom{xxx}
C_8=\left(
\begin{array}{cccc}
 0 & 0 & \frac{1}{\sqrt{2}} & 0 \\
 0 & 0 & 0 & \frac{1}{\sqrt{2}} \\
 -\frac{1}{\sqrt{2}} & 0 & 0 & 0 \\
 0 & -\frac{1}{\sqrt{2}} & 0 & 0
\end{array}
\right),
\end{displaymath}
\begin{displaymath}
C_9=\left(
\begin{array}{cccc}
 0 & 0 & 0 & \frac{i}{\sqrt{2}} \\
 0 & 0 & \frac{i}{\sqrt{2}} & 0 \\
 0 & \frac{i}{\sqrt{2}} & 0 & 0 \\
 \frac{i}{\sqrt{2}} & 0 & 0 & 0
\end{array}
\right),
\phantom{xxx}
C_{10}=\left(
\begin{array}{cccc}
 0 & 0 & 0 & \frac{1}{\sqrt{2}} \\
 0 & 0 & -\frac{1}{\sqrt{2}} & 0 \\
 0 & \frac{1}{\sqrt{2}} & 0 & 0 \\
 -\frac{1}{\sqrt{2}} & 0 & 0 & 0
\end{array}
\right).
\end{displaymath}
\section{Asymptotically safe quantum gravity}
\subsection{Conventions and variations in gravity}\label{gravvariations}
We define the Riemann tensor by
\begin{equation}
 [D_{\mu}, D_{\nu}]V^{\kappa}= R^{\kappa}_{\, \, \alpha \mu \nu}V^{\alpha}.
\end{equation}
The Ricci tensor and Ricci scalar are then given by
\begin{eqnarray}
 R_{\mu\nu}&=& R^{\kappa}_{\, \, \mu \kappa \nu}\\
R&=& g^{\mu \nu}R_{\mu \nu}.
\end{eqnarray}
For the evaluation of variations, we use the following symbolic notation:
\begin{equation}
 \Gamma_k= \Gamma_k[h=0]+ \delta \Gamma_k+ \frac{1}{2}\delta^2 \Gamma_k +...
\end{equation}
We have that
\begin{equation}
 \delta g_{\mu \nu}= h_{\mu \nu}
\end{equation}
and for the inverse metric
\begin{equation}
 \delta g^{\mu \nu}= - h^{\mu \nu}.
\end{equation}
Here, a second variation is also non-zero, since $\delta^2 (g_{\mu \kappa}g^{\kappa \lambda})= \delta^2\delta_{\mu}^{\lambda}=0$ and hence
\begin{equation}
 \delta^2g^{\mu \nu}= 2 h^{\mu}_{\lambda}h^{\lambda \nu}.
\end{equation}
Further we have that
\begin{equation}
 \delta \sqrt{g}= \sqrt{g}\frac{1}{2}g^{\mu \nu}h_{\mu \nu}
\end{equation}
and
\begin{equation}
 \delta^2 \sqrt{g}= \sqrt{g}(-\frac{1}{2}h^{\mu \nu}h_{\mu \nu}+ \frac{1}{4}h^2),
\end{equation}
where $h= g^{\mu \nu}h_{\mu \nu}$.
We also have that
\begin{eqnarray}
 \delta \Gamma^{\kappa}_{\mu \nu}= \frac{1}{2}g^{\kappa \lambda}\left(D_{\mu}h_{\lambda \nu}+ D_{\nu}h_{\lambda \mu}-D_{\lambda}h_{\mu \nu} \right),\label{firstchristoffel}
\end{eqnarray}
from which we deduce that
\begin{equation}
 \delta^2 \Gamma^{\kappa}_{\mu \nu}= -h^{\kappa \lambda}\left(D_{\mu}h_{\lambda \nu}+ D_{\nu}h_{\lambda \mu}-D_{\lambda}h_{\mu \nu} \right).\label{secondchristoffel}
\end{equation}
This structure arises, as the variation of the terms $D_{\mu}h_{\lambda \nu}$ etc. can be summarised to be of the same form as the variation of the metric $g^{\kappa \lambda}$ multiplied by the covariant derivatives of the fluctuation metric.
Finally we have the first variation of the Riemann tensor
\begin{equation}
 \delta R^{\rho}_{\,\,\mu\lambda \nu }= D_{\lambda}\delta \Gamma^{\rho}_{\mu \nu}- D_{\nu}\delta \Gamma^{\rho}_{\lambda \mu}.
\end{equation}
From here, all further variation of metric invariants can be constructed by using the above first and second variations.

In particular, we have that
\begin{equation}
 \delta^2 g^{\mu \nu}R_{\mu \nu}\rightarrow -h D_{\kappa}D^{\mu}h_{\mu}^{\kappa} + \frac{1}{2}h D^2 h + h_{\mu}^{\lambda}R^{\mu}_{\, \, \alpha \rho \lambda}h^{\alpha \rho}+ h_{\mu}^{\lambda}R_{\alpha \lambda}h^{\alpha \mu}+ h_{\mu \lambda}D^{\lambda}D_{\rho}h^{\mu \rho}- \frac{1}{2}h_{\mu \rho}D^2 h^{\mu \rho},
\end{equation}
where by the arrow we indicate that we neglect all terms that will turn into total derivatives if this expression is integrated over.

Furthermore
\begin{equation}
 \delta R \rightarrow -h^{\mu \nu}R_{\mu \nu}+ D^{\lambda}D_{\mu}h^{\mu \lambda}- D^2 h.
\end{equation}
The second variation of the Ricci scalar is given by
\begin{equation}
 \delta^2 R \rightarrow 2 h^{\mu \lambda}h_{\lambda}^{\nu}R_{\mu \nu}+ h^{\mu \nu}D^2 h_{\mu \nu}+ \frac{1}{2}h D^2 h - h_{\mu}^{\lambda}R^{\mu}_{\, \, \alpha \rho \lambda}h^{\alpha \rho}- h_{\mu}^{\lambda}R_{\alpha \lambda} h^{\alpha \mu}- h_{\mu \lambda}D^{\lambda}D_{\rho}h^{\mu \rho}- \frac{1}{2} h_{\mu \rho}D^2 h^{\mu \rho}.
\end{equation}
Finally we have the second variation of the Einstein-Hilbert term
\begin{eqnarray}
 \delta^2 \int d^4x \sqrt{g}\left(R- 2\lambda\right)&=& \int d^4x \sqrt{g}\Bigl(\left(-\frac{1}{2}h^{\mu\nu}h_{\mu \nu}+ \frac{1}{4}h^2 \right)(R- 2\lambda) - h R_{\mu \nu}h^{\mu \nu}+h D^{\lambda}D_{\mu}h^{\mu \lambda}\nonumber\\
&{}&- \frac{1}{2}h D^2 h + h^{\mu \lambda}R_{\mu \nu}h^{\nu}_{\lambda}- h_{\mu}^{\lambda}R^{\mu}_{\, \, \alpha \rho \lambda}h^{\alpha \rho}- h_{\mu \lambda}D^{\lambda}D_{\rho}h^{\mu \rho}+ \frac{1}{2}h^{\mu \nu}D^2 h_{\mu \nu}\Bigr).\nonumber\\
&{}&
\end{eqnarray}
Applying a York-decomposition and going over to a spherical background finally leads to the form from which the inverse propagators for the metric modes can be directly read off. Note that we set $g_{\mu \nu}= \bar{g}_{\mu \nu}$ here, but for simplicity do not write the bars over the background quantities.
\begin{eqnarray}
\delta^2 2 \bar{\kappa}^2\int d^4x \sqrt{g}(-R)&=&-2 \bar{\kappa}^2 \Bigl[ h_{\mu \nu}^T\left(\frac{1}{2}D^2+\frac{-d(d-3)-4}{2d(d-1)}R \right)h^{\mu \nu\, T}\nonumber\\
&{}&+v^{\mu}\left(R D^2 \frac{d-2}{d}+R^2\frac{d-2}{d^2} \right)v_{\mu} \nonumber\\
&{}&+\sigma \left(\frac{-(d-1)(d-2)}{2d^2}D^6 + \frac{2-d}{2d^2}R^2 D^2+\frac{2-d}{2d}R D^4  \right)\sigma\nonumber\\
&{}&+h \left(\frac{d-2}{d^2}D^2 R+\frac{(d-1)(d-2)}{d^2}D^4  \right)\sigma\nonumber\\
&{}&+h \left(R \frac{(d-2)(d-4)}{4d^2} + D^2 \frac{-(d-1)(d-2)}{2d^2} \right)h\Bigr]\label{delta2GammaEH}
\end{eqnarray}
Further the decomposition of the gauge-fixing term takes the following form
\begin{eqnarray}
 \Gamma_{k\, gf}&=& \frac{Z_N\kappa^2}{\alpha}\int d^4x \sqrt{g}\Bigl[v_{\mu}\left(\frac{R}{d}+D^2 \right)^2 v^{\mu} \nonumber\\
&{}&- \sigma \frac{1}{d^2}\left((d-1)D^2+R^2 \right)^2 \sigma\nonumber\\
&{}&+ h \left(\frac{2 \rho}{d^2}D^2 R + D^4 \frac{2(d-1)\rho}{d^2} \right)\sigma\nonumber\\
&{}&- \frac{\rho^2}{d^2}h D^2 h \Bigr]\label{delta2Gammagf}
\end{eqnarray}
Note that in these expression the rescaling \eqref{rescaling_jaco} remains to be completed.

\subsection{Hyperspherical harmonics}\label{hypersphericals}
A basis for symmetric transverse traceless tensor functions on a spherical background is given by the tensor hyperspherical harmonics $T_{\mu \nu}^{lm}(x)$, which are eigenfunctions of the covariant Laplacian:
\begin{equation}
-\bar{D}^2 T_{\mu \nu}^{lm}(x)=\Lambda_{l,2}(d) T_{\mu \nu}^{lm}(x).
\end{equation}
Similarly a basis of transverse vector functions is given by the vector hyperspherical harmonics
\begin{equation}
-\bar{D}^2 T_{\mu}^{lm}(x)=\Lambda_{l,1}(d) T_{\mu}^{lm}(x).
\end{equation}
Finally a basis for scalar functions exists, the scalar hyperspherical harmonics:
\begin{equation}
 -\bar{D}^2 T^{lm}(x)= \Lambda_{l,0}(d)T^{lm}(x).
\end{equation}
The eigenvalues depend on the curvature scalar $\bar{R}$, the dimensionality $d$, and the eigenvalue $l$:
\begin{eqnarray}
 \Lambda_{l,2}(d)&=& \frac{l(l+d-1)-2}{d(d-1)}\bar{R}\nonumber\\
\Lambda_{l,1}(d)&=& \frac{l(l+d-1)-1}{d(d-1)}\bar{R}\nonumber\\
\Lambda_{l,0}(d)&=&\frac{l(l+d-1)}{d(d-1)}\bar{R}.
\end{eqnarray}
As expected, the eigenvalue does not depend on $m$, therefore the degeneracy factors $D_l(d,s)$ (with $s=2,1,0$) are non-trivial and read:
\begin{eqnarray}
 D_{l}(d,2)&=&\frac{(d+1)(d-2)(l+d)(l-1)(2l+d-1)(l+d-3)!}{2(d-1)!(l+1)!}\nonumber\\
D_{l}(d,1)&=& \frac{l(l+d-1)(2l+d-1)(l+d-3)!}{(d-2)!(l+1)!}\nonumber\\
D_l(d,0)&=& \frac{(2l+d-1)(l+d-2)!}{l!(d-1)!}.
\end{eqnarray}

Since the hyperspherical harmonics are basis functions, they satisfy completeness and orthogonality relations as follows:
\begin{eqnarray}
 \frac{\delta^d(x-x')}{\sqrt{\bar g}}\frac{1}{2}\left(\bar{g}_{\mu \rho}\bar{g}_{\nu \sigma}+\bar{g}_{\mu \sigma}\bar{g}_{\nu \rho}\right)&=&
 \sum_{l=2}^{\infty}\sum_{m=1}^{D_l(d,2)}T_{\mu \nu}^{lm}(x)T_{\rho
   \sigma}^{lm}(x')\label{complete} \\
\delta^{lk}\delta^{mn}&=&\int d^d x \, \sqrt{\bar{g}}\, \frac{1}{2}\left(\bar{g}^{\mu
  \rho}\bar{g}^{\nu \sigma}+\bar{g}^{\mu \sigma}\bar{g}^{\nu \rho}\right)T^{lm}_{\mu \nu}(x)T^{kn}_{\rho \sigma}(x) \label{tortho}\\
 \frac{\delta^d(x-x')}{\sqrt{\bar g}}\bar{g}_{\mu \rho}&=&
 \sum_{l=1}^{\infty}\sum_{m=1}^{D_l(d,1)}T_{\mu}^{lm}(x)T_{\rho}^{lm}(x')\qquad\label{completev} \\
\delta^{lk}\delta^{mn}&=&\int d^d x \, \sqrt{\bar{g}}\, \bar{g}^{\mu
  \rho}T^{lm}_{\mu}(x)T^{kn}_{\rho}(x) \label{vortho}\\
 \frac{\delta^d(x-x')}{\sqrt{\bar g}}&=&
\sum_{l=0}^{\infty}\sum_{m=1}^{D_l(d,0)}T^{lm}(x)T^{lm}(x')\qquad\label{completes} \\
\delta^{lk}\delta^{mn}&=&\int d^d x \, \sqrt{\bar{g}}\, T^{lm}(x)T^{kn}(x), \label{sortho}\\
\end{eqnarray}

\section{Faddeev-Popov ghost sector of asymptotically safe quantum gravity}
\subsection{Vanishing of the tadpole diagram}\label{tadpole}

The vanishing of the graviton tadpole contribution to the running of $Z_c$
can be shown by making use of the second variation of the Christoffel symbol, see \Eqref{secondchristoffel}.
Varying the ghost kinetic term twice with respect to the metric produces the
following type of terms:
\begin{eqnarray}
 \delta^2 (g_{\kappa \nu}D_{\rho}c^{\nu})&=& 
\delta^2 (g_{\kappa \nu}\Gamma_{\rho \lambda}^{\nu})c^{\lambda}\nonumber\\
&=& (2 h_{\kappa \nu} \delta \Gamma^{\nu}_{\rho \lambda}
+ g_{\kappa \nu}\delta^2 \Gamma^{\nu}_{\rho \lambda})c^{\lambda}.
\end{eqnarray}
Inserting the first and second variation of the Christoffel symbol from \Eqref{firstchristoffel} and \Eqref{secondchristoffel} leads to a cancellation between the two
terms. Accordingly, the second variation of the ghost kinetic term with
respect to the metric vanishes (for all choices of $\rho$). Hence, there is
no gauge in which a graviton tadpole can contribute to the running of the
ghost wave function renormalisation. 

\subsection{Details on the
$\tilde{\partial}_t$-derivative}\label{tilde_partial}

After having evaluated the derivative with respect to the external momentum, the right-hand side of the Wetterich equation contains the shape function $r(y)$ as well as its derivative $r'(y)$. Then we have that
\begin{equation}
 \tilde{\partial}_t f(r_j,r_j')(p)=\sum_i \int \frac{d^dp'}{(2\pi)^d} \partial_t R_i(p')\frac{\delta}{\delta R_i(p')} f(r_j, r_j')(p), 
\end{equation}
where by the sum over Latin indices we denote a summation
over different types of fields.
Since
\begin{equation}
 R_i = Z_i(p^2-\lambda_i)r_i(p),
\end{equation}
where $\lambda_i$ may be, e.g. a cosmological constant or a
mass-term,
we can write
\begin{equation}
\frac{\delta}{\delta R_i(p')}= \frac{1}{Z_i(p'^2-\lambda_i)}\frac{\delta}{\delta r_i(p')}
\end{equation}
Thus
\begin{eqnarray}
&{}&  \tilde{\partial}_t f(r_j,r_j')(p)\nonumber\\
&=&\sum_i \int \frac{d^dp'}{(2\pi)^d}\frac{\partial_t R_i(p')}{Z_i(p'^2 -\lambda_i)} \Bigl[f^{(1,0)}(r_j,r_j') \delta(p,p')+f^{(0,1)}(r_j,r_j')\frac{k^2 p'_{\mu}}{2p^2}\partial_{p'_{\mu}}\delta_{ij}\delta(p,p') \Bigr]\nonumber\\
&=& \sum_i \Bigl[ \left( -2y r_i'- \eta_i r_i- \frac{\partial_t \lambda_i}{k^2}\frac{1}{y}(r_i + y r_i') \right)f^{(1,0)}(r_j,r_j')\nonumber\\
&{}& + \left( -2(r_i'+y r_i'')-\eta_i r_i'+ \frac{\partial_t \lambda_i}{k^2} \frac{1}{y^2}(r_i-yr_i'-y^2 r_i'')\right)f_i^{(0,1)}(r_j,r_j')\Bigr],
\end{eqnarray}
where $y= \frac{p^2}{k^2}$ and $\eta_i = - \partial_t \ln Z_i$.

\subsection{Vertices for the diagrams contributing to $\eta_c$}\label{etacvertices}

In this appendix, we derive the building blocks for the expansion of the flow equation in terms of the quantities $\mathcal
P$ and $\mathcal F$. In the following, we always aim at a Euclidean flat
  background. Here, our conventions for 2-point functions are given by
\begin{equation}
\Gamma_{k,ij}^{(2)}(p,q)= \frac{\overset{\rightarrow}{\delta}}{\delta \phi_i(-p)} 
\Gamma_{k} \frac{\overset{\leftarrow}\delta}{\delta \phi_j(q)},
\end{equation}
where $\phi(p)= \left(h_{\mu \nu}^{\text{T}}(p), v_{\mu}(p), \sigma(p), h(p), c_{\mu}(p),
\bar{c}_{\mu}(-p) \right)$ and $i,j$ label the field components. Here, we have
chosen the momentum-space conventions for the anti-ghost opposite to those of
the ghost, i.e. if $c^\mu(p)$ denotes a ghost with {\em
incoming} momentum
$p$ then $\bar{c}^\mu(q)$ denotes an anti-ghost with {\em outgoing} momentum
$q$. The ghost propagator is an off-diagonal matrix,
\begin{eqnarray}
\mathcal{P}_{\text{gh}}^{-1}&=&\left( \begin{array}{cc}
                 0& \Gamma^{(2)}_{k, c \bar{c}}(p,q)+R_k\\
 \Gamma^{(2)}_{k,\bar{c} c }(p,q))+R_k & 0
                    \end{array} \right)^{-1}\\
                    &=&\left( \begin{array}{cc}
                 0& \left( \Gamma^{(2)}_{k,\bar{c} c}(p,q)+R_k\right)^{-1}\\
\left( \Gamma^{(2)}_{k,c \bar{c}}(p,q)+R_k\right)^{-1}&0
                    \end{array} \right)\nonumber\\
&=&\left( \begin{array}{cc}
                 0& \mathcal{P}_{c\bar c}^{-1}\\
\mathcal{P}_{\bar c c}^{-1}&0
                    \end{array} \right),
\end{eqnarray}
where
\begin{eqnarray}
 \Gamma^{(2)}_{k,c \bar{c} , \mu \nu}(p,q)&=&
 \frac{\overset{\rightarrow}{\delta}}{\delta c_{\mu}(-p)}\Gamma_{k\, \rm
   gh}\frac{\overset{\leftarrow}{\delta}}{\delta \bar{c}_{\nu}(-q)}= -
 \Gamma^{(2)}_{k,\bar{c} c , \mu \nu}(p,q)).\nonumber\\
&{}& 
\end{eqnarray}
Within our truncation, the ghost propagator reads explicitly
\begin{eqnarray}
&{}&\left(\Gamma_{k,\bar{c} c}^{(2)}+R_k\right)_{\mu \nu}^{-1}= \frac{1}{\sqrt{2}Z_c p^2}\left(\delta_{\mu \nu}+ \frac{\rho-1}{3-\rho}\frac{p_{\mu}p_{\nu}}{p^2}\right)\frac{1}{(1+r(y))}\delta^{4}(p-q).\nonumber
\end{eqnarray}
The graviton propagators are obtained from the second variation of the
Einstein-Hilbert and the gauge-fixing action. Setting $g_{\mu
  \nu}=\bar{g}_{\mu \nu}=\delta_{\mu \nu}$ after the functional variation yields the following
expression in Fourier space:
\begin{eqnarray}
\delta^2\Gamma_{k\, \rm EH + gf}
&=& \kappa^2 \int \frac{d^4p}{(2\pi)^4} h^{\alpha \beta}(-p)\Bigl[\frac{1}{4}\left(\delta_{\alpha \mu}\delta_{\beta \nu} +\delta_{\alpha \nu}\delta_{\beta \mu}\right)p^2-\frac{1}{2}\delta_{\alpha \beta}p^2 \delta_{\mu \nu} + \delta_{\alpha \beta}p_{\mu}p_{\nu}\nonumber\\
&{}& -\frac{1}{2}\left(p_{\beta}p_{\mu}\delta_{\nu \alpha}+p_{\alpha}p_{\nu}\delta_{\mu \beta} \right) +\lambda_k \left(\frac{1}{2}\delta_{\alpha \beta}\delta_{\mu \nu} -\frac{1}{2}\left(\delta_{\alpha \mu}\delta_{\beta \nu}+\delta_{\alpha \nu} \delta_{\beta \mu} \right) \right)\\
&{}&+\frac{1}{\alpha}\Bigl(\frac{1}{2}\left(p_{\alpha}p_{\mu}\delta_{\beta \nu}+ p_{\beta}p_{\mu} \delta_{\beta \mu}\right)-\frac{1+\rho}{2}p_{\alpha}p_{\beta}\delta_{\mu \nu}+\frac{(1+\rho)^2}{16} \delta_{\alpha \beta}p^2 \delta_{\mu \nu}\Bigr)  \Bigr]h_{\mu \nu}(p).\label{Gamma2EHgf}\nonumber
\end{eqnarray}
Inserting the York decomposition \Eqref{York} into \Eqref{Gamma2EHgf} then
results in the following expression, from which the inverse propagators follow
directly by functional derivatives:

\begin{eqnarray}
\delta^2\Gamma_{k\, \rm EH + gf}&=& Z_\text{N} \kappa^2 \int
\frac{d^4p}{(2\pi)^4}
\Bigl[h^{\text{T}\,\alpha \beta} (-p) \frac{1}{2}\left(p^2-2\lambda_k \right)
h_{\alpha \beta}^T(-p)+ v^{\beta}(-p)\left(\frac{p^2}{\alpha}-2 \lambda_k
\right)v_{\beta}(p)\nonumber\\
&+& \sigma(-p)\frac{3}{16}\left(p^2 \frac{3-\alpha}{\alpha}-4 \lambda_k
\right)\sigma(p)
+ h(-p)3\frac{\rho-\alpha}{8 \alpha}p^2\sigma(-p)\nonumber\\
&{}&+\frac{1}{16}h(-p)\left( p^2 \frac{\rho^2 -3 \alpha}{\alpha}+4 \lambda
\right)h(p) \Bigr]. \label{Gamma2EHYork}
\end{eqnarray}

In this work, we confine ourselves to the gauge choice $\rho \rightarrow
\alpha$ where the propagator matrix becomes diagonal in the graviton modes. The
vector and transverse traceless tensor propagators go along with transverse and
transverse traceless projectors, respectively. In $d$-dimensional Fourier
space, these projectors read
\begin{eqnarray}
 {P}_{\text T\, \mu \nu}(p)&=& \delta_{\mu \nu} -
 \frac{p_{\mu}p_{\nu}}{p^2},\nonumber\\ 
{P}_{\text{TT}\, \mu \nu \kappa \lambda}(p)&=&
\frac{1}{2}\left({P}_{\text T\, \mu \kappa}{P}_{\text T\, \nu \lambda}+
  {P}_{\text T\, \mu \lambda}{P}_{\text T\, \nu \kappa}\right)-
\frac{1}{d-1}{P}_{\text T\, \mu \nu} {P}_{\text T\, \kappa \lambda}, 
\end{eqnarray}
where the last term in the transverse traceless projector ${P}_{\text{TT}}$
removes the trace part.

The resulting propagators together with the regulator $R_k$ constitute the
$\mathcal P$ term in the expansion of the flow equation
\Eqref{eq:flowexp}.

 The $\mathcal F$ term carries the dependence on the ghost
fields that couple via vertices to the fluctuation modes. 
To obtain these vertices, we vary the ghost action once with respect to
the metric and then proceed to a flat background,
yielding:

\begin{eqnarray}
\delta \Gamma_{k \, \rm gh}
&=& -\sqrt{2} Z_c \int d^4x \sqrt{\bg} 
\bar{c}^{\mu} \Bigl(\bar{D}^{\rho}h_{\mu \nu}\bar{D}_{\rho}
+ \bar{D}^{\rho}\left[\bar{D}_{\nu}h_{\mu \rho}\right] 
+\bar{D}^{\rho}h_{\rho \nu}\bar{D}_{\mu}
- \frac{1}{2}(1+\rho)\bar{D}_{\mu}h^{\rho}_{\nu}\bar{D}_{\rho}\nonumber\\
&{}&-\frac{1}{4}(1+\rho)\bar{D}_{\mu}[\bar{D}_{\nu}h_{\lambda}^{\, \lambda}]
\Bigr)c^{\nu}\nonumber\\ 
&\rightarrow& -\sqrt{2}Z_c \int \frac{d^4p}{(2 \pi)^4}\frac{d^4 q}{(2 \pi)^4}
 \bar{c}^{\mu}(p+q)h_{\rho \sigma}(p)c^{\kappa}(q)
\Bigl(-q\cdot(p+q)\frac{1}{2}\left(\delta_{\mu}^{\rho}\delta_{\kappa}^{\sigma}
+\delta^{\sigma}_{\mu}\delta^{\rho}_{\kappa}\right)\nonumber\\
&{}&-\frac{p_{\kappa}}{2}\left((p^{\rho}+q^{\rho})\delta^{\sigma}_{\mu}
+(p^{\sigma}+q^{\sigma})\delta_{\mu}^{\rho}\right)-\frac{q_{\mu}}{2}\Bigl((p^{\sigma}
+q^{\sigma})\delta_{\kappa}^{\rho}\nonumber\\
&{}&+(p^{\rho}+q^{\rho})\delta^{\sigma}_{\kappa}\Bigr)+\frac{1}{2}\frac{1+\rho}{2}(p_{\mu}+q_{\mu})
\left(\delta_{\kappa}^{\rho}q^{\sigma}+\delta^{\sigma}_{\kappa}q^{\rho}\right)
+\frac{1+\rho}{4}p_{\kappa}\delta^{\rho \sigma}(p_{\mu}+q_{\mu})\Bigr)\nonumber\\
&=&-\sqrt{2}Z_c \int \frac{d^4p}{(2 \pi)^4}\frac{d^4 q}{(2 \pi)^4}
\bar{c}^{\mu}(p+q)c^{\kappa}(q) \Bigl(V^{(\text{T})}_{\kappa \mu}{}^{\rho
  \sigma}(p,q)h_{\rho \sigma}^{\text T}(p)
+V^{(v)}_{\kappa \mu}{}^{\rho}(p,q)v_{\rho}(p)\nonumber\\
&{}&
+V^{(\sigma)}_{\kappa \mu}(p,q)\sigma(p)        
+V^{(h)}_{\kappa \mu}(p,q)h(p) \Bigr).
\end{eqnarray}

Here, we introduced the York decomposition \Eqref{York} for the graviton
fluctuation, such that we can read off the corresponding vertices connecting
ghost and anti-ghost with the graviton components:

\begin{eqnarray}
 V^{(\text{T})}_{\kappa \mu}{}^{\rho \sigma }(p,q)&=& -q\cdot(p+q)\frac{1}{2}
\left(\delta_{\mu}^{\rho}\delta_{\kappa}^{\sigma}
+\delta^{\sigma}_{\mu}\delta^{\rho}_{\kappa}\right)
-\frac{p_{\kappa}}{2}\left(q^{\rho}\delta^{\sigma}_{\mu}
+q^{\sigma}\delta_{\mu}^{\rho}\right)-\frac{q_{\mu}}{2}\left(q^{\sigma}
\delta_{\kappa}^{\rho}+q^{\rho}\delta^{\sigma}_{\kappa}\right)\nonumber\\
&{}&+\frac{1}{2}\frac{1+\rho}{2}(p_{\mu}+q_{\mu})
\left(\delta_{\kappa}^{\rho}q^{\sigma}+\delta^{\sigma}_{\kappa}q^{\rho}\right),\nonumber
\end{eqnarray}
\begin{eqnarray}
V^{(v)}_{\kappa \mu}{}^{\rho }(p,q)&=& \frac{2 i}{\sqrt{p^2}} p_{\sigma}
\Bigl(-q\cdot(p+q)\frac{1}{2}\left(\delta_{\mu}^{\rho}\delta_{\kappa}^{\sigma}
+\delta^{\sigma}_{\mu}\delta^{\rho}_{\kappa}\right)
-\frac{p_{\kappa}}{2}\left(q^{\rho}\delta^{\sigma}_{\mu}
+(p^{\sigma}+q^{\sigma})\delta_{\mu}^{\rho}\right)\nonumber\\
&{}&-\frac{q_{\mu}}{2}\left((p^{\sigma}+q^{\sigma})\delta_{\kappa}^{\rho}
+q^{\rho}\delta^{\sigma}_{\kappa}\right)+\frac{1}{2}\frac{1+\rho}{2}(p_{\mu}
+q_{\mu})\left(\delta_{\kappa}^{\rho}q^{\sigma}+\delta^{\sigma}_{\kappa}
q^{\rho}\right) \Bigr) \label{V_terms}\nonumber
\end{eqnarray}
\begin{eqnarray}
V^{(\sigma)}_{\kappa \mu}(p,q)&=&- \frac{1}{p^2}\Bigl[p_{\kappa}p_{\mu}
\left(\frac{3}{4}p^2+q^2+\frac{3-\rho}{2}q\cdot p \right)
+q_{\mu}q_{\kappa}\frac{1}{4}p^2\left(\frac{\rho-1}{2} \right)\nonumber\\
&{}&-\frac{1}{4}p^2\delta_{\mu \kappa}\left(q^2+q\cdot p \right)+p_{\kappa}q_{\mu}\left(\frac{1}{2}p^2+\frac{1-\rho}{2}q\cdot p \right)
+p_{\mu}q_{\kappa}\frac{1+\rho}{8}p^2\Bigr]\nonumber
\end{eqnarray}
\begin{eqnarray}
V^{(h)}_{\kappa \mu}(p,q)&=&\frac{1}{4}\Bigl(-q\cdot(p+q)
\delta_{\mu \kappa}+p_{\kappa}(p_{\mu}+q_{\mu})(\rho -1) \nonumber\\
&{}&
+ p_{\mu}p_{\kappa}-q_{\mu}q_{\kappa}\frac{1-\rho}{2}
+p_{\mu}q_{\kappa}\frac{1+\rho}{2}\Bigr).
\end{eqnarray}

From this, the four possible fluctuation matrix entries contributing to the
quantity $\mathcal F$ in the expansion \Eqref{eq:flowexp} can be evaluated:\\
\begin{eqnarray}
\Gamma^{(2)}_{h^{\text{T}} c}(q,p)&=& 
\frac{\overset{\rightarrow}{\delta}}{\delta h^{\text{T}}_{\mu \nu}(-q)}
\Gamma_{k\, \rm gh}\frac{\overset{\leftarrow}{\delta}}{\delta c_{\kappa}(p)}= -\sqrt{2}Z_c\bar{c}^{\tau}(p-q)V_{\kappa \tau \mu \nu}^{(\text{T})}(-q,p)\nonumber\\
\Gamma^{(2)}_{h^{\text{T}} \bar{c}}(q,p)&=& 
\frac{\overset{\rightarrow}{\delta}}{\delta h^{\text{T}}_{\mu \nu}(-q)}
\Gamma_{k\, \rm gh}\frac{\overset{\leftarrow}{\delta}}{\delta
  \bar{c}_{\kappa}(-p)}
=\sqrt{2} Z_c c^{\tau}(q-p)V^{(\text{T})}_{\tau \kappa \mu \nu}(-q,q-p)\nonumber\\
\Gamma^{(2)}_{c h^{\text{T}}}(q,p)&=& 
\frac{\overset{\rightarrow}{\delta}}{\delta c_{\kappa}(-q)}
\Gamma_{k\, \rm gh}\frac{\overset{\leftarrow}{\delta}}{\delta
  h^{\text{T}}_{\mu \nu}(p)}
= \sqrt{2}Z_c \bar{c}^{\lambda}(p-q)V^{(\text{T})}_{\kappa \lambda \mu
  \nu}(p,-q)
\nonumber\\
\Gamma^{(2)}_{\bar{c} h^{\text{T}}}(q,p)&=& 
\frac{\overset{\rightarrow}{\delta}}{\delta \bar{c}_{\kappa}(q)}
\Gamma_{k\, \rm gh}\frac{\overset{\leftarrow}{\delta}}{\delta h^{\text{T}}_{\mu \nu}(p)}= -\sqrt{2}Z_c c^{\lambda}(q-p)V^{(\text T)}_{\lambda \kappa \mu \nu}(p,q-p),
\end{eqnarray}
and similarly for the other graviton modes. 
\newpage 

\subsection{$\beta$ functions for $G$, $\lambda$ and
$\eta_c$}\label{betafunctionsghost}
Setting $\rho = \alpha$ and using a spectrally adjusted
exponential regulator function, we arrive at the following
$\beta$ functions, where all couplings are dimensionless:

\begin{eqnarray}
&{}&\partial_t \lambda+ 2\lambda \nonumber\\
&=&\frac{1}{36\pi} \Biggl(G \lambda \Biggl\{-150 (\partial_t\lambda+2 \lambda ) \text{Ei}(2 \lambda )-\frac{12 (\partial_t\lambda+2 \lambda ) \text{Ei}\left(-\frac{4 \lambda }{\alpha -3}\right)}{\alpha -3}-150
   \text{Li}_2\left(e^{2 \lambda }\right)\nonumber\\
&{}&+6 \text{Li}_2\left(e^{-\frac{4 \lambda }{\alpha -3}}\right)+36 \text{Li}_2\left(e^{2 \alpha  \lambda }\right)\nonumber\\
&{}&+\frac{6}{(\alpha -3)^2} \Biggl(-3 \alpha  (3 \alpha -2) (\partial_t\lambda+2 \lambda ) \text{Ei}(2 \alpha  \lambda ) (\alpha -3)^2-9 \alpha  \text{Li}_2\left(e^{2 \alpha  \lambda }\right) (\alpha -3)^2\nonumber\\
&{}&+(7 \alpha -9) \Bigl[\text{Li}_2\left(e^{-\frac{4 \alpha  \lambda }{\alpha
   -3}}\right) (\alpha -3)-2 \alpha  (\partial_t\lambda+2 \lambda ) \text{Ei}\left(-\frac{4 \alpha  \lambda }{\alpha -3}\right)\Bigr]\nonumber\\
&{}&-2 i \pi\lambda  \Bigl[\alpha  (39 \alpha -166)+219\Bigr]\nonumber\\
&{}&+\partial_t\lambda \Biggl[25 \ln \left(-1+e^{2 \lambda }\right) (\alpha -3)^2+2 \ln \left(-1+e^{-\frac{4 \lambda }{\alpha -3}}\right) (\alpha -3)\nonumber\\
&{}&+\alpha  \Big[3 (3 \alpha -2) \ln \left(-1+e^{2 \alpha  \lambda }\right)
   (\alpha -3)^2+2 (7 \alpha -9) \ln \left(-1+e^{-\frac{4 \alpha  \lambda }{\alpha -3}}\right)\Bigr]\Biggr]\Biggr)\nonumber\\
&{}&+\frac{1}{\alpha -3}e^{-\frac{4 (\alpha +1) \lambda }{\alpha -3}} \Biggl(-2 e^{\frac{4
   (\alpha +1) \lambda }{\alpha -3}} \Bigl(18 i \pi  (\alpha -3) \alpha  (3 \alpha -2) \lambda+(7 \alpha -27) \left(\pi ^2-3 \eta_c\right)\Bigr)\nonumber\\
&{}&-3 e^{\frac{4 \alpha  \lambda }{\alpha -3}} (\alpha -3) \eta_{\text{N}}+75 e^{\frac{2 (3 \alpha -1) \lambda }{\alpha -3}} (\alpha -3) \eta_{\text{N}}+9 e^{\frac{2 ((\alpha -1) \alpha +2) \lambda }{\alpha -3}} (\alpha -3) (3 \alpha -2) \eta_{\text{N}}\nonumber\\
&{}&-3 e^{\frac{4 \lambda }{\alpha -3}} (7
   \alpha -9) \eta_{\text{N}}\Biggr)\Biggr\}\quad \quad \quad \nonumber\\
&{}&+9 G \Biggl[\frac{8 \alpha  \lambda  \text{Li}_2\left(e^{-\frac{4 \alpha  \lambda }{\alpha -3}}\right)}{\alpha -3}+\frac{4\alpha  e^{\frac{-4 \alpha  \lambda }{\alpha -3}}}{(\alpha -3)^2}
   (\partial_t\lambda+2 \lambda ) \Biggl\{\alpha +e^{\frac{4 \alpha  \lambda }{\alpha -3}} \Biggl(4 \alpha  \lambda  \left(\text{Ei}\left(\frac{-4 \alpha 
   \lambda }{\alpha -3}\right)+i \pi \right)\nonumber\\
&{}&-(\alpha -3) \text{Li}_2\Bigl[e^{\frac{-4 \alpha  \lambda }{\alpha -3}}\Bigr]\Biggr)-3\Biggr\} + 6 \Biggl[2 \lambda  (\partial_t\lambda+2 \lambda )
   (\text{Ei}(2 \alpha  \lambda )+i \pi ) \alpha ^2+\partial_t\lambda \text{Li}_2\left(e^{2 \alpha  \lambda }\right) \alpha \nonumber\\
&{}&+2 \text{Li}_3\left(e^{2 \alpha  \lambda }\right)\Biggr]+4
   \text{Li}_3\left(e^{-\frac{4 \alpha  \lambda }{\alpha -3}}\right)-e^{-\frac{4 \alpha  \lambda }{\alpha -3}} \eta_{\text{N}}-3 e^{2 \alpha  \lambda } \left(2 \partial_t\lambda \alpha +4 \lambda  \alpha +\eta
   _N\right)\nonumber
\end{eqnarray}
\begin{eqnarray}
&{}&+5 \Biggl[4 \lambda  (\partial_t\lambda+2 \lambda ) (\text{Ei}(2 \lambda )+i \pi )+2\partial_t\lambda \text{Li}_2\left(e^{2 \lambda }\right)+4 \text{Li}_3\left(e^{2 \lambda }\right)-e^{2 \lambda
   } \left(2 \partial_t\lambda+4 \lambda +\eta_{\text{N}}\right)\Biggr]\nonumber\\
&{}&+\frac{4}{(\alpha -3)^2}\Biggl( \text{Li}_3\left(e^{-\frac{4 \lambda }{\alpha -3}}\right) (\alpha -3)^2-4 \partial_t\lambda \text{Li}_2\left(e^{-\frac{4 \lambda
   }{\alpha -3}}\right) (\alpha -3)\nonumber\\
&{}&+e^{-\frac{4 \lambda }{\alpha -3}} \left(4 \partial_t\lambda+8 \lambda -(\alpha -3) \eta_{\text{N}}\right) (\alpha -3)+16 i \pi  \lambda  (\partial_t\lambda+2 \lambda )\nonumber\\
&{}&+16 \lambda 
   (\partial_t\lambda+2 \lambda ) \text{Ei}\left(-\frac{4 \lambda }{\alpha -3}\right)\Biggr)+8 \left(\eta_c-4 \zeta (3)\right)\Biggr]\Biggr), \quad \mbox{ for $\lambda>0$.}\label{dtlambda}
\end{eqnarray}

We find the following expression for $\eta_N$ within the gauge choice $\rho =\alpha$ and a spectrally and RG adjusted regulator with exponential shape function:
\begin{eqnarray}
\eta_{\text{N}}&=&\frac{G}{36 \pi } \Biggl[54 e^{2 \alpha  \lambda } (\partial_t \lambda+2 \lambda ) \alpha ^2-\frac{144 \lambda  \ln \left(1-e^{-\frac{4 \alpha  \lambda }{\alpha -3}}\right) \alpha ^2}{(\alpha -3)^2}-36 e^{2 \alpha 
   \lambda } (\partial_t \lambda+2 \lambda ) \alpha\nonumber\\
&{}&-18 (3 \alpha -2) (\partial_t \lambda+2 \lambda ) \text{Ei}(2 \alpha  \lambda ) \alpha \nonumber\\
&{}&-18 (3 \alpha -2) \left(2 i \pi  \lambda +e^{2 \alpha  \lambda }
   (\partial_t \lambda+2 \lambda )-\partial_t \lambda \ln \left(-1+e^{2 \alpha  \lambda }\right)\right) \alpha\nonumber\\
&{}&-\frac{24 \lambda  \ln \left(1-e^{-\frac{4 \alpha  \lambda }{\alpha -3}}\right) \alpha }{\alpha
   -3}+\frac{144\lambda }{(\alpha -3)^2}  \ln \left(1-e^{-\frac{4 \alpha  \lambda }{\alpha -3}}\right) \alpha-\frac{12 (\partial_t \lambda+2 \lambda ) \text{Ei}\left(-\frac{4 \lambda }{\alpha -3}\right)}{\alpha -3}\nonumber\\
&{}&+\frac{12 (7 \alpha -9)}{(\alpha -3)^2} (\partial_t \lambda+2 \lambda ) \left(\ln \left(-1+e^{-\frac{4 \alpha 
   \lambda }{\alpha -3}}\right)-\text{Ei}\left(-\frac{4 \alpha  \lambda }{\alpha -3}\right)\right) \alpha +27 e^{2 \alpha  \lambda } \eta_{\text{N}} \alpha\nonumber\\
&{}&-\frac{3 e^{-\frac{4 \lambda }{\alpha -3}}
   \eta_{\text{N}} \alpha }{\alpha -3}-150 (\partial_t \lambda+2 \lambda ) \text{Ei}(2 \lambda )-300 \lambda 
   \ln \left(1-e^{2 \lambda }\right)+150 \partial_t \lambda \ln \left(-1+e^{2 \lambda }\right)\nonumber\\
&{}&+300 \lambda  \ln \left(-1+e^{2 \lambda }\right)-\frac{24 \lambda  \ln \left(1-e^{-\frac{4 \lambda }{\alpha
   -3}}\right)}{\alpha -3}+\frac{24 \lambda  \ln \left(-1+e^{-\frac{4 \lambda }{\alpha -3}}\right)}{\alpha -3}\nonumber\\
&{}&+\frac{12 \partial_t \lambda \ln \left(-1+e^{-\frac{4 \lambda }{\alpha -3}}\right)}{\alpha
   -3}-150 \text{Li}_2\left(e^{2 \lambda }\right)+6 \text{Li}_2\left(e^{-\frac{4 \lambda }{\alpha -3}}\right)+18 (2-3 \alpha ) \text{Li}_2\left(e^{2 \alpha  \lambda }\right)\nonumber\\
&{}&+\frac{6 (7 \alpha -9)
   \text{Li}_2\left(e^{-\frac{4 \alpha  \lambda }{\alpha -3}}\right)}{\alpha -3}-\frac{2 (\alpha -9) \left(\pi ^2-3 \eta_c\right)}{\alpha -3}-12 \left(\pi ^2-3 \eta_c\right)+75 e^{2 \lambda } \eta_{\text{N}}
   \nonumber\\
   &{}&-18 e^{2
   \alpha  \lambda } \eta_{\text{N}}+\frac{9 e^{-\frac{4 \lambda }{\alpha -3}} \eta_{\text{N}}}{\alpha -3}+\frac{3 e^{-\frac{4 \alpha  \lambda }{\alpha -3}} (9-7 \alpha ) \eta_{\text{N}}}{\alpha -3}\Biggr]\mbox{ for $\lambda>0$.} \label{etaN}
\end{eqnarray}
\newpage
In the Landau-deWitt gauge ($\alpha=0$), we find the following expression for
$\eta_c$, where the three lines are the transverse traceless contribution
and the last lines are due to the trace mode:

\begin{eqnarray}
\eta_{c\, \text{L}}
&=& -\frac{35}{162 \pi } e^{-4 \lambda } G \Biggl(e^{6 \lambda } \bigr[6 \partial_t\lambda+\eta_c-3 \eta_{\text{N}}+6\bigl]+e^{8 \lambda } \eta_{\text{N}}+3 e^{4 \lambda } \bigl[-\eta_c+8 \lambda +4\bigr]\nonumber\\
&{}&+12 \Bigl(e^{2
   \lambda } \bigl[\partial_t\lambda (6 \lambda -1)+\lambda  (-\eta_{\text{N}}+12 \lambda -2)\bigr]-\eta_c \lambda \Bigr) E_1(-6 \lambda )\nonumber\\
&{}&+12 e^{2 \lambda } \Bigl(\partial_t\lambda e^{2 \lambda } (4
   \lambda -1)+\lambda  \Bigl[-\eta_c+4 \lambda +e^{2 \lambda } (-\eta_{\text{N}}+8 \lambda -2)\Bigr]\Bigr) E_1(-2 \lambda )\nonumber\\
&{}&+12 \Bigl(\eta_c \lambda +e^{2 \lambda } \Bigl(-6 \lambda 
   \partial_t\lambda+\partial_t\lambda+(\eta_c+\eta_{\text{N}}-16 \lambda +2) \lambda \Bigr)\nonumber\\
&{}&+e^{4 \lambda } (-4 \lambda  \partial_t\lambda+\partial_t\lambda+\eta_{\text{N}}-8 \lambda +2) \lambda
   \Bigr) \Gamma (0,-4 \lambda )\Biggr)\nonumber\\
&{}&+\frac{2}{243 \pi } e^{-8 \lambda /3} G \Biggl(-8 e^{4 \lambda /3} \Bigl(\lambda  (8 \lambda -3 \eta_c)+e^{4 \lambda /3} \bigr[\partial_t\lambda (8 \lambda -3)+\lambda  (-3 \eta_{\text{N}}+16 \lambda -6)\bigl]\Bigr)
   \text{Ei}\left(\frac{4 \lambda }{3}\right)\nonumber\\
&{}&+8 \Bigl[-3 \eta_c \lambda +e^{8 \lambda /3} \bigl[\partial_t\lambda (8 \lambda -3)+\lambda  (-3 \eta_{\text{N}}+16 \lambda -6)\bigr]+e^{4 \lambda /3} \bigl[3
  \partial_t\lambda (4 \lambda -1)\nonumber\\
&{}&+\lambda  (32 \lambda -3 (\eta_c+\eta_{\text{N}}+2))\bigr])\Bigr] \text{Ei}\left(\frac{8 \lambda }{3}\right)\nonumber\\
&+&3 \Biggl[e^{8 \lambda /3} \Bigl(-3 \eta_c+e^{4
   \lambda /3} \Bigl(4 \partial_t\lambda+\eta_c+\left(-3+e^{4 \lambda /3}\right) \eta_{\text{N}}+6\Bigr)+16 \lambda +12\Bigr)\nonumber\\
&+&8 \left(\eta_c \lambda +e^{4 \lambda /3} (-4 \lambda 
   \partial_t\lambda+\partial_t\lambda+(\eta_{\text{N}}-8 \lambda +2) \lambda )\right) \text{Ei}(4 \lambda )\Biggr]\Biggr).\quad \mbox{ for $\lambda>0$.}
\label{etac_Landau}
\end{eqnarray}

The exponential factors result from the spectral adjustment of the
regulator. The expression is linear in $G$, as each contributing diagram contains
exactly one graviton propagator.\newline\\

In the deDonder or harmonic gauge ($\alpha=1$), we have to take contributions
from all modes into account. Accordingly, we arrive at the following
expression, which decomposes into transverse traceless \Eqref{etacTT}, vector
\Eqref{etacv}, scalar \Eqref{etacsc} and trace \Eqref{etacphi}
contributions:

\begin{eqnarray}
\eta_{c\, \text{dD}}&=&-\frac{5}{18 \pi } e^{-4 \lambda } G \Biggl(e^{6 \lambda } (6 \partial_t\lambda+\eta_c-3 \eta_{\text{N}}+6)+e^{8 \lambda } \eta_{\text{N}}+3 e^{4 \lambda } (-\eta_c+8 \lambda +4)\nonumber\\
&{}&+12 \Bigl(e^{2 \lambda
   } (\partial_t\lambda (6 \lambda -1)+\lambda  (-\eta_{\text{N}}+12 \lambda -2))-\eta_c \lambda \Bigr) E_1(-6 \lambda )\nonumber\\
&{}&+12 e^{2 \lambda } \Bigl(\partial_t\lambda e^{2 \lambda } (4 \lambda
   -1)+\lambda  \left(-\eta_c+4 \lambda +e^{2 \lambda } (-\eta_{\text{N}}+8 \lambda -2)\right)\Bigr) E_1(-2 \lambda )\nonumber
\end{eqnarray}
\begin{eqnarray}
&{}&+12 \Bigl(\eta_c \lambda +e^{2 \lambda } ((-6 \lambda+1)  \partial_t\lambda+(\eta_c+\eta_{\text{N}}-16 \lambda +2) \lambda )\nonumber\\
&{}&+e^{4 \lambda } ((-4 \lambda +1) \partial_t\lambda+(\eta_{\text{N}}-8 \lambda +2) \lambda )\Bigr) \Gamma (0,-4
   \lambda )\Biggr)\label{etacTT}\\
&{}&+\frac{1}{72 \pi }e^{-4 \lambda } G \Biggl(12 e^{8 \lambda } \left(24 \lambda ^2-2 (-6 \partial_t \lambda+\eta_{\text{N}}+2) \lambda -2 \partial_t \lambda-\eta_{\text{N}}\right)\nonumber\\
&{}&-e^{6 \lambda } \Bigl(576 \lambda ^2+12 (18
  \partial_t \lambda+\eta_c-3 (\eta_{\text{N}}+2)) \lambda +30 \partial_t \lambda+7 \eta_c-33 \eta_{\text{N}}+66\Bigr)\nonumber\\
&{}&+6 e^{4 \lambda } (\eta_c+16 \lambda  (3 \lambda -2)+16
 \partial_t \lambda (4 \lambda -1)+\lambda  (-\eta_{\text{N}}+8 \lambda -2)) (\text{Ei}(2 \lambda )-\text{Ei}(4 \lambda ))-4)\nonumber\\
&{}&-96 \eta_c \lambda  (3 \lambda -1) (\text{Ei}(4 \lambda )-\text{Ei}(6
   \lambda ))+48 e^{2 \lambda } \Bigl(\lambda  (2 (7-6 \lambda ) \lambda +\eta_c(3 \lambda -2)) \text{Ei}(2 \lambda )\nonumber\\
&{}&+\Bigl(2\partial_t \lambda (9 (\lambda -1) \lambda +1)+\lambda  \left(48 \lambda
   ^2-(3 \eta_c+3 \eta_{\text{N}}+50) \lambda +2 (\eta_c+\eta_{\text{N}}+2)\right)\Bigr) \text{Ei}(4 \lambda )\nonumber\\
&{}&-\Bigl(2 \partial_t \lambda (9 (\lambda -1) \lambda +1)+\lambda  \bigl(36
   \lambda ^2-3 (\eta_{\text{N}}+12) \lambda +2 (\eta_{\text{N}}+2)\bigr)\Bigr) \text{Ei}(6 \lambda )\Bigr)\Biggr)\label{etacv}\\
&{}&+\frac{1}{432 \pi }e^{-4 \lambda } G \Biggl(4 e^{8 \lambda } \left(72 \lambda ^2-6 (-6 \partial_t \lambda+\eta_{\text{N}}+2) \lambda -6 \partial_t \lambda-7 \eta_{\text{N}}\right)\nonumber\\
&{}&-e^{6 \lambda } \Bigl(576 \lambda ^2+12 (18
   \partial_t\lambda+\eta_c-3 (\eta_{\text{N}}+2)) \lambda +126 \partial_t \lambda+23 \eta_c-81 \eta_{\text{N}}+162\Bigr)\nonumber\\
&{}&+18 e^{4 \lambda } (3 \eta_c+4 (4 (\lambda -2) \lambda
   -3)+16 (\partial_t \lambda (4 \lambda -1)+\lambda  (-\eta_{\text{N}}+8 \lambda -2)) (\text{Ei}(2 \lambda )-\text{Ei}(4 \lambda )))\nonumber\\
&{}&-288 \eta_c (\lambda -1) \lambda  (\text{Ei}(4 \lambda
   )-\text{Ei}(6 \lambda ))-144 e^{2 \lambda } \Bigl(\lambda  \left(4 \lambda ^2-(\eta_c+10) \lambda +2 \eta_c\right) \text{Ei}(2 \lambda )\nonumber\\
&{}&+\Bigl(-16 \lambda ^3+(-6 \partial_t \lambda+\eta_c+\eta_{\text{N}}+38) \lambda ^2-2 (-7 \partial_t \lambda+\eta_c+\eta_{\text{N}}+2) \lambda -2 \partial_t \lambda \Bigr) \text{Ei}(4 \lambda )\nonumber\\
&{}&+\Bigl(12 \lambda ^3-(-6
  \partial_t \lambda+\eta_{\text{N}}+28) \lambda ^2+2 (-7 \partial_t \lambda+\eta_{\text{N}}+2) \lambda +2 \partial_t \lambda\Bigr) \text{Ei}(6 \lambda )\Bigr)\Biggr)\label{etacsc}\\
&{}&-\frac{1}{432 \pi }e^{-4 \lambda } G \Biggl(2 e^{8 \lambda } \left(144 \lambda ^2-12 (-6 \partial_t \lambda+\eta_{\text{N}}+2) \lambda -12 \partial_t \lambda+\eta_{\text{N}}\right)-e^{6 \lambda } (54 \partial_t \lambda (4
   \lambda -1)\nonumber\\
&{}&-9 (\eta_{\text{N}}-16 \lambda -2) (4 \lambda -1)+\eta_c (12 \lambda -7))-36 e^{4 \lambda } \Bigl(\eta_c-4 \left(2 \lambda ^2+\lambda +1\right)+2 (\partial_t \lambda (4 \lambda
   -1)\nonumber\\
&{}&+\lambda  (-\eta_{\text{N}}+8 \lambda -2)) (\text{Ei}(2 \lambda )-\text{Ei}(4 \lambda ))\Bigr)-72 \eta_c \lambda  (4 \lambda +1) (\text{Ei}(4 \lambda )-\text{Ei}(6 \lambda ))\nonumber\\
&{}&-72 e^{2 \lambda }
   \Bigl(-\lambda  \left(-8 \lambda ^2+2 \eta_c \lambda +\text{$\eta $c}\right) \text{Ei}(2 \lambda )\nonumber\\
&{}&+\left(-32 \lambda ^3+2 (-6 \partial_t \lambda+\eta_c+\eta_{\text{N}}-2) \lambda ^2+(-2
 \partial_t \lambda+\eta_c+\eta_{\text{N}}+2) \lambda +\partial_t \lambda\right) \text{Ei}(4 \lambda )\nonumber\\
&{}&+(\partial_t \lambda (2 \lambda  (6 \lambda +1)-1)+\lambda  (-\eta_{\text{N}} (2 \lambda +1)+4
   \lambda  (6 \lambda +1)-2)) \text{Ei}(6 \lambda )\Bigr)\Biggr).\quad \label{etacphi}\\
&{}&\mbox{ for $\lambda>0$.}
 \label{etac_deDonder}
\end{eqnarray}
\newpage
\subsection{Extended truncation in the ghost sector}\label{vertices_deltaZc}
The first variation with respect to the full metric $g_{\mu \nu}$ of our truncation \Eqref{trunc} yields the following expression, where we drop the bar on the background metric and the covariant derivative, as we now have identified $g_{\mu \nu}=\bar{g}_{\mu \nu}$:
\begin{eqnarray}
&{}& \delta \Gamma_{k\, \rm gh}\\
&=& -\sqrt{2}Z_c \int d^4 x \sqrt{\bar{g}}\bar{c}_{\mu}\Bigl[ D^{\rho}\left(h^{\mu}_{\, \, \nu}D_{\rho} \right)+D^{\rho}\left( D_{\rho}h^{\mu}_{\, \, \nu}\right)\frac{1-\delta Z_c}{2}+D^{\rho}\left(D_{\nu}h^{\mu}_{\, \, \rho} \right)\frac{1+\delta Z_c}{2}\nonumber\\
&{}&-D^{\rho}\left(D^{\mu}h_{\rho \nu} \right)\frac{1-\delta Z_c}{2}+D^{\rho} h_{\rho \nu}D^{\mu}\delta Z_c-\frac{1+\rho}{2}\delta Z_c
\left[D^{\mu }h^{\sigma}_{\, \, \nu}D_{\sigma} +\frac{1}{2}D^{\mu}\left(D_{\nu}h\right)  \right]\Bigr]c^{\nu}\nonumber,
\end{eqnarray}
where the covariant derivatives act on everything to the right of them, unless they are found inside a round bracket, when they do not act beyond the bracket.

We now proceed to flat space where we also use a York decomposition of the ghost
\begin{eqnarray}
 c^{\mu}(p)&=&c^{T\, \mu}(p)+ \frac{i p^{\mu}}{\sqrt{p^2}}\eta(p)\nonumber\\
\bar{c}^{\mu}(p)&=& \bar{c}^{T\, \mu}(p)- \frac{i p^{\mu}}{\sqrt{p^2}}\bar{\eta}(p), 
\end{eqnarray}
where $p^{\mu}c^{T}_{\mu}=0$.
Now the vertices are given by a derivative with respect to the ghost and antighost fields.
In the following we indicate the longitudinal/ transversal ghost and antighost by $(L,T)$, or $(\bar{L},\bar{T})$, respectively. The vertex involving a transverse traceless metric mode carries a $\rm TT$, the scalar vertex an $h$.\\
Accordingly
\begin{eqnarray}
 V_{\kappa \mu\rho\sigma}^{{\rm TT},\, \bar{T},T}(p,q)&=&\frac{1}{2}\left(\delta_{\mu \rho}\delta_{\sigma \kappa}+\delta_{\mu \sigma}\delta_{\rho \kappa} \right)\left(-p\cdot q -q^2 -\frac{1-\delta Z_c}{2}(p^2+p\cdot q) \right)\\
&{}&-\frac{1+\delta Z_c}{2}\frac{p_{\kappa}}{2}\left(q_{\rho}\delta_{\sigma \mu}+q_{\sigma}\delta_{\rho \mu} \right)+\frac{1}{4}\left(q_{\rho}\delta_{\sigma \kappa}+q_{\sigma}\delta_{\rho \kappa} \right)\left(- q_{\mu} \right)\left(1-\rho \right)\delta Z_c\nonumber\\
V_{\mu \rho \sigma}^{{\rm TT},\, \bar{T},L}(p,q)&=& \frac{i}{\sqrt{q^2}}\Biggl(\left(q_{\sigma}\delta_{\mu \rho}+q_{\rho}\delta_{\mu \sigma} \right) \left(-p\cdot q -\frac{1}{2}q^2-\frac{1}{4}p^2+\frac{\delta Z_c}{4}p^2 \right)\nonumber\\
&{}&-\frac{1+\delta Z_c}{2}q_{\rho}q_{\sigma}q_{\mu}\Biggr)\\
V_{\kappa \rho \sigma}^{{\rm TT}, \, \bar{L},T}(p,q)&=&\frac{-i}{\sqrt{(p^2+2 p \cdot q +q^2)}}\Biggl(-\frac{1+\delta Z_c}{4}p_{\kappa}\left(q_{\rho}p_{\sigma}+2 q_{\rho}q_{\sigma}+p_{\rho}q_{\sigma} \right)\\
&{}&
\frac{1}{4}\left(p_{\rho}\delta_{\sigma \kappa}+p_{\sigma}\delta_{\rho \kappa} \right) \left(p\cdot q(-3+\delta Z_c) -2 q^2 -\frac{1-\delta Z_c}{2}p^2 \right)\nonumber\\
&{}&+\frac{1}{4}\left(q_{\rho}\delta_{\sigma \kappa}+q_{\sigma}\delta_{\rho \kappa} \right)\Bigl(-2 p \cdot q -2 q^2 +\delta Z_c q^2(\rho -1)+2 p \cdot q \delta Z_c \rho\nonumber\\
&{}&+p^2 \delta Z_c (1+\rho) \Bigr)\Biggr)\nonumber
\end{eqnarray}
\begin{eqnarray}
V_{\rho \sigma}^{({\rm TT}, \, \bar{L},L)}(p,q)&=&-\frac{1}{\sqrt{p^2+2 p\cdot q + q^2}\sqrt{q^2}}\Biggl(\frac{1}{4}\left(p_{\sigma}q_{\rho}+p_{\rho}q_{\sigma}\right)\cdot\nonumber\\
&{}&\cdot \Biggl[\left((-1+\delta Z_c)p^2 -4 p\cdot q -2 q^2 \right)+ \frac{1}{2}\Bigl(-3 p \cdot q -2 q^2 +\delta Z_c \left( p^2 - p\cdot q -q^2\right) \nonumber\\
&{}&+ \delta Z_c \rho \left(p^2 +2 p\cdot q + q^2 \right) \Bigr)\Biggr]\Biggr),\label{transvertices}
\end{eqnarray}
where in the transverse antighost terms we used $p_{\mu}= -q_{\mu}$. (Note that due to the transverse ghost propagator the two versions of the vertex that can be obtained by interchanging $p_{\mu} \leftrightarrow -q_{\mu}$ yield the same contribution to the $\beta$-functions.)

The trace mode couples to the ghosts via:
\begin{eqnarray}
 V_{\kappa \mu}^{h,\, \bar{T},T}&=&\frac{p_{\kappa}q_{\mu}}{8}\left(-1-\delta Z_c \right) +\frac{\delta_{\mu \kappa}}{8}\left(-p^2(1-\delta Z_c)-3 p \cdot q + \delta Z_c p \cdot q-2q^2 \right)\\
V_{\kappa}^{(h,\, \bar{L},T)}&=&\frac{i}{\sqrt{(p^2+2 p\cdot q +q^2)}} \Biggl(\frac{p_{\kappa}}{8}\Bigl(q^2 \left( -3-\delta Z_c +2 \delta Z_c \,\rho\right)+p^2 \left(2 \rho\, \delta Z_c -1+\delta Z_c \right)\nonumber\\
&{}&+p\cdot q \left( -4+4 \delta Z_c\, \rho\right) \Bigr)\Biggr)\\
V_{\mu}^{h,\, \bar{T}, L}(p,q)&=& \frac{i\,q_{\mu}}{8\sqrt{q^2}}\left(-4 p \cdot q -3 q^2 -\delta Z_c q^2 -p^2 (1-\delta Z_c) \right)\\
V^{(h, \, \bar{L},L)}(p,q)&=& \frac{1}{\sqrt{q^2\, \left(p^2+2 p \cdot q + q^2 \right)}}\Biggl(\frac{p \cdot q}{8}q^2 \left( 4 \delta Z_c\, \rho-\delta Z_c -5\right)\nonumber\\
&{}&+\frac{(p\cdot q)^2}{8}\left(4 \rho \,\delta Z_c-4\right) +\frac{p \cdot q}{8}p^2 \left( 2 \rho \delta Z_c +\delta Z_c -1\right)\nonumber\\
&{}&+\frac{\left(q^2\right)^2}{8}\left( \rho\, \delta Z_c -\delta Z_c -2\right)+\frac{p^2 q^2}{8}(1+\rho)\delta Z_c\Biggr).\label{confvertices}
\end{eqnarray}
In all expressions involving a transverse antighost we have again used transversality to substitute $p_{\mu}=-q_{\mu}$.
On the 4-sphere the decomposition
\begin{equation}
 c_{\mu}=c_{\mu}^T+(\bar{D}^2)^{(-1)}\bar{D}_{\mu}\eta
\end{equation}
leads to the transverse and longitudinal inverse propagators:
\begin{eqnarray}
 \Gamma_k^{TT}&=& -\sqrt{2}Z_c \int d^4x \sqrt{\bar{g}}\bar{c}^{\mu \, T}\left(\bar{D}^2 \bar{g}_{\mu \nu}+\delta Z \frac{\bar{R}}{4}\bar{g}_{\mu \nu} \right)c^{\nu \, T}\nonumber\\
\Gamma_k^{LL}&=&-\sqrt{2} Z_c \int d^4x \bar{\eta}\left( \bar{D}^2 \left(1+\delta Z\frac{1-\alpha}{2} \right)+\frac{\bar{R}}{4}\left(1+\delta Z \right)\right)\eta
\end{eqnarray}
To project on $\partial_t Z_c$, note that
\begin{equation}
 \partial_t Z_c = \frac{1}{\sqrt{2}}\frac{1}{4}\delta^{\alpha \gamma}\frac{\partial}{\partial \tilde{p}^2}\int \frac{d^4 \tilde{q}}{(2\pi)^4}\frac{\overset{\rightarrow}{\delta}}{\delta \bar{c}^{\alpha\,T}(\tilde{p})}\partial_t \Gamma_k \frac{\overset{\leftarrow}{\delta}}{\delta c^{\gamma\, T}(\tilde{q})}\Big|_{c,\bar{c}, \eta, \bar{\eta}=0}\label{projetacext}
\end{equation}
The projection onto the longitudinal ghosts allows to write
\begin{equation}
 \partial_t \delta Z = \frac{\sqrt{2}}{Z_c (1-\alpha)}\left(\frac{\partial}{\partial \tilde{p}^2}\int \frac{d^4 \tilde{q}}{(2\pi)^4}\frac{\overset{\rightarrow}{\delta}}{\delta \bar{\eta}(\tilde{p})}\partial_t \Gamma_k \frac{\overset{\leftarrow}\delta}{\delta \eta(\tilde{q})} +\sqrt{2}Z_c \eta_c \left(1+\frac{\delta Z}{2}(1-\alpha) \right)\right)\Big|_{c,\bar{c}, \eta, \bar{\eta}=0}\label{projdeltaZ}
\end{equation}

\subsection{$\beta$ functions in extended truncation}\label{deltaZc}
We now find the following $\beta$ functions for $\lambda>0$ with a spectrally and RG-adjusted regulator with exponential shape function. Here, we have specialised to Landau-deWitt gauge $\rho = \alpha =0$.
 
\begin{eqnarray}
 \eta_N&=&\frac{1}{72 \pi  \left(\text{$\delta $Z}_c+2\right){}^2}\Biggl\{G \Biggl(12 \text{Li}_2\left(e^{4 \lambda /3}\right) \left(\text{$\delta $Z}_c+2\right){}^2-300 \text{Li}_2\left(e^{2 \lambda }\right) \left(\text{$\delta $Z}_c+2\right){}^2\nonumber\\
&{}&+8 \text{Ei}\left(\frac{4
   \lambda }{3}\right) \left(2 \lambda +\beta _{\lambda }\right) \left(\text{$\delta $Z}_c+2\right){}^2-300 \text{Ei}(2 \lambda ) \left(2 \lambda +\beta _{\lambda }\right) \left(\text{$\delta
   $Z}_c+2\right){}^2+4 \Bigl(4 \lambda  \ln \left(1-e^{4 \lambda /3}\right)\nonumber\\
&{}&-2 \ln \left(-1+e^{4 \lambda /3}\right) \left(2 \lambda +\beta _{\lambda }\right)+75 \left(\ln \left(-1+e^{2 \lambda }\right)
   \left(2 \lambda +\beta _{\lambda }\right)-2 \lambda  \log \left(1-e^{2 \lambda }\right)\right)\Bigr) \left(\text{$\delta $Z}_c+2\right){}^2\nonumber\\
&{}&-2 \Bigl(\pi ^2 \left(\text{$\delta $Z}_c \left(9 \text{$\delta
   $Z}_c+20\right)-2\right)-3 \Bigl[\left(\text{$\delta $Z}_c \left(9 \text{$\delta $Z}_c+29\right)+16\right) \eta _c\nonumber\\
&{}&+\left(-9-e^{4 \lambda /3}+25 e^{2 \lambda }\right) \left(\text{$\delta $Z}_c+2\right) \eta
   _N\Bigr]\Bigr) \left(\text{$\delta $Z}_c+2\right)-3 \pi ^2 \beta _{\text{$\delta $Z}_c} \left(3 \text{$\delta $Z}_c \left(\text{$\delta $Z}_c+4\right)+14\right)\nonumber\\
&{}&+6 \beta _{\text{$\delta $Z}_c}
   \left(\text{$\delta $Z}_c \left(9 \text{$\delta $Z}_c+28\right)+32\right)\Biggr)\Biggr\}
\end{eqnarray}

\begin{eqnarray}
 \beta_{\lambda}&=&\frac{1}{72} \Biggl(\frac{1}{\pi  \left(\text{$\delta $Z}_c+2\right){}^2}\Biggl\{G \Bigl(12 \text{Li}_2\left(e^{4 \lambda /3}\right) \left(\text{$\delta $Z}_c+2\right){}^2-300 \text{Li}_2\left(e^{2 \lambda }\right) \left(\text{$\delta $Z}_c+2\right){}^2\nonumber\\
&{}&+8
   \text{Ei}\left(\frac{4 \lambda }{3}\right) \left(2 \lambda +\beta _{\lambda }\right) \left(\text{$\delta $Z}_c+2\right){}^2-300 \text{Ei}(2 \lambda ) \left(2 \lambda +\beta _{\lambda }\right)
   \left(\text{$\delta $Z}_c+2\right){}^2\nonumber\\
&{}&+4 \Bigl(4 \lambda  \ln \left(1-e^{4 \lambda /3}\right)-2 \ln \left(-1+e^{4 \lambda /3}\right) \left(2 \lambda +\beta _{\lambda }\right)\nonumber\\
&{}&+75 \left(\ln \left(-1+e^{2
   \lambda }\right) \left(2 \lambda +\beta _{\lambda }\right)-2 \lambda  \ln \left(1-e^{2 \lambda }\right)\right)\Bigr) \left(\text{$\delta $Z}_c+2\right){}^2\nonumber\\
&{}&-2 \Bigl(\pi ^2 \left(\text{$\delta $Z}_c \left(9
   \text{$\delta $Z}_c+20\right)-2\right)-3 \Bigl(\left(\text{$\delta $Z}_c \left(9 \text{$\delta $Z}_c+29\right)+16\right) \eta _c\nonumber\\
&{}&+\left(-9-e^{4 \lambda /3}+25 e^{2 \lambda }\right) \left(\text{$\delta
   $Z}_c+2\right) \eta _N\Bigr)\Bigr) \left(\text{$\delta $Z}_c+2\right)-3 \pi ^2 \beta _{\text{$\delta $Z}_c} \left(3 \text{$\delta $Z}_c \left(\text{$\delta $Z}_c+4\right)+14\right)\nonumber\\
&{}&+6 \beta _{\text{$\delta
   $Z}_c} \left(\text{$\delta $Z}_c \left(9 \text{$\delta $Z}_c+28\right)+32\right)\Bigr) \lambda \Biggr\}\nonumber\\
&{}&-144 \lambda +\frac{1}{\pi }\Biggl\{2 G \Bigl(36 \text{Li}_3\left(e^{4 \lambda
   /3}\right)+180 \text{Li}_3\left(e^{2 \lambda }\right)+12 \text{Li}_2\left(e^{4 \lambda /3}\right) \beta _{\lambda }+90 \text{Li}_2\left(e^{2 \lambda }\right) \beta _{\lambda }\nonumber\\
&{}&+16 \lambda 
   \text{Ei}\left(\frac{4 \lambda }{3}\right) \left(2 \lambda +\beta _{\lambda }\right)+180 \lambda  \text{Ei}(2 \lambda ) \left(2 \lambda +\beta _{\lambda }\right)+4 \lambda  \Bigl(4 \ln \left(1-e^{4 \lambda
   /3}\right)\nonumber\\
&{}&-4 \ln \left(-1+e^{4 \lambda /3}\right)+45 \left(\ln \left(1-e^{2 \lambda }\right)-\ln \left(-1+e^{2 \lambda }\right)\right)\Bigr) \left(2 \lambda +\beta _{\lambda }\right)+72 \eta _c-36 \eta
   _N\nonumber\\
&{}&-45 e^{2 \lambda } \left(4 \lambda +2 \beta _{\lambda }+\eta _N\right)-3 e^{4 \lambda /3} \left(8 \lambda +4 \beta _{\lambda }+3 \eta _N\right)\nonumber\\
&{}&-\frac{18 \beta _{\text{$\delta $Z}_c}}{\text{$\delta
   $Z}_c+2}-144 \zeta (3)\Bigr)\Biggr\}\Biggr)\quad \mbox{ for $\lambda>0$.}
\end{eqnarray}
Finally we give the two $\beta$ functions in the ghost sector. Here, the large number of possible interaction vertices from the coupling of the diverse metric modes to the ghost modes leads to these rather lengthy expressions.
\newpage
\setlength{\hoffset}{-18mm}
\parbox[c][14\baselineskip][t]{\textwidth}{%
\scriptsize 
\begin{eqnarray}
{\eta_c}
&=&\frac{1}{15552 \pi }\Biggl\{e^{-4 \lambda } G \Biggl(12 \Bigl[90 e^{6 \lambda } \left(\text{$\delta $Z}_c \left(19 \text{$\delta $Z}_c+15\right)-58\right)+18 e^{2 \lambda } \Bigl(-80 (\text{Ei}(2 \lambda )-\text{Ei}(4 \lambda ))
   \Bigl(\text{$\delta $Z}_c \left((4 \lambda +5) \text{$\delta $Z}_c+5\right)-2 (2 \lambda +9)\Bigr) \lambda ^2\nonumber\\
&+&20 e^{2 \lambda } \left((2 \lambda  (4 \lambda +7)+9) \text{$\delta $Z}_c^2-2 (4 \lambda 
   (\lambda +5)+11)+5 \left(2 \lambda  \text{$\delta $Z}_c+\text{$\delta $Z}_c\right)\right)-5 e^{4 \lambda } \left((8 \lambda  (2 \lambda +3)+7) \text{$\delta $Z}_c^2-4 (\lambda  (4 \lambda +19)+5)+5 \left(4
   \lambda  \text{$\delta $Z}_c+\text{$\delta $Z}_c\right)\right)\Bigr)\nonumber\\
&+&6 e^{2 \lambda } \left(2 \lambda +\beta _{\lambda }\right) \Bigl(40 e^{6 \lambda } (6 \lambda -1) \left(\text{$\delta
   $Z}_c^2-1\right)-60 e^{2 \lambda } (4 \lambda -1) (\text{Ei}(2 \lambda )-\text{Ei}(4 \lambda )) \left(\text{$\delta $Z}_c \left(7 \text{$\delta $Z}_c+5\right)-20\right)\nonumber\\
&-&30 e^{4 \lambda } \Bigl(-12 \lambda
   +\text{$\delta $Z}_c \left(4 (3 \lambda -2) \text{$\delta $Z}_c-5\right)+21\Bigr)-12 (\text{Ei}(4 \lambda )-\text{Ei}(6 \lambda )) \Bigl(-40 \lambda  (3 \lambda -1) \left(\text{$\delta $Z}_c^2-1\right)-5
   (6 \lambda -1) \left(\text{$\delta $Z}_c \left(7 \text{$\delta $Z}_c+5\right)-20\right)\Bigr)\Bigr)\nonumber\\
&+&e^{2 \lambda } \eta_N \Bigl(90 e^{4 \lambda } \left(-4 \lambda +\text{$\delta $Z}_c \left((4
   \lambda -6) \text{$\delta $Z}_c-5\right)+19\right)-10 e^{6 \lambda } \left(-24 \lambda +\text{$\delta $Z}_c \left((24 \lambda -17) \text{$\delta $Z}_c-15\right)+56\right)\nonumber\\
&+&72 \lambda  \Bigl(5 e^{2 \lambda }
   (\text{Ei}(2 \lambda )-\text{Ei}(4 \lambda )) \left(\text{$\delta $Z}_c \left(7 \text{$\delta $Z}_c+5\right)-20\right)-5 (\text{Ei}(4 \lambda )-\text{Ei}(6 \lambda )) \left(\text{$\delta $Z}_c \left((4
   \lambda +7) \text{$\delta $Z}_c+5\right)-4 (\lambda +5)\right)\Bigr)\Bigr)\nonumber\\
&+&10 \Bigl(-9 e^{4 \lambda } \left(\text{$\delta $Z}_c \left(9 \text{$\delta $Z}_c+5\right)-22\right)+e^{6 \lambda } \left(12
   \lambda +\text{$\delta $Z}_c \left(2 (11-6 \lambda ) \text{$\delta $Z}_c+15\right)-61\right)+36 e^{2 \lambda } \lambda  (\text{Ei}(2 \lambda )-\text{Ei}(4 \lambda )) \left(\text{$\delta $Z}_c \left((4
   \lambda +7) \text{$\delta $Z}_c+5\right)-4 (\lambda +5)\right)\nonumber\\
&-&36 \lambda  (\text{Ei}(4 \lambda )-\text{Ei}(6 \lambda )) \left(\text{$\delta $Z}_c \left((8 \lambda +7) \text{$\delta $Z}_c+5\right)-4 (2
   \lambda +5)\right)\Bigr) \eta _c\nonumber\\
&+&\frac{1}{\text{$\delta $Z}_c+2}\Biggl\{2 \Biggl(\frac{1}{\text{$\delta $Z}_c+2}\Bigl\{\Bigl(e^{6 \lambda } (7-12 \lambda )-36 e^{4 \lambda }+72 e^{2 \lambda } \lambda  (2 \lambda +1) (\text{Ei}(2 \lambda )-\text{Ei}(4 \lambda ))-72
   \lambda  (4 \lambda +1) (\text{Ei}(4 \lambda )-\text{Ei}(6 \lambda ))\Bigr) \beta _{\text{$\delta $Z}_c}\Bigr\}\nonumber\\
&-&270 e^{6 \lambda } \text{$\delta $Z}_c \left(\text{$\delta $Z}_c+1\right)+90
   e^{2 \lambda } \Bigl(16 (\text{Ei}(2 \lambda )-\text{Ei}(4 \lambda )) \lambda ^2-4 e^{2 \lambda } (2 \lambda +1)+e^{4 \lambda } (4 \lambda +1)\Bigr) \text{$\delta $Z}_c \left(\text{$\delta
   $Z}_c+1\right)\nonumber\\
&-&30 e^{2 \lambda }  \eta_N \left(e^{4 \lambda } \left(-3+e^{2 \lambda }\right)+12 \lambda  \left(e^{2 \lambda } \text{Ei}(2 \lambda )-\left(1+e^{2 \lambda }\right) \text{Ei}(4 \lambda
   )+\text{Ei}(6 \lambda )\right)\right) \text{$\delta $Z}_c \left(\text{$\delta $Z}_c+1\right)\nonumber\\
&+&180 e^{2 \lambda } \left(2 e^{2 \lambda } (4 \lambda -1) (\text{Ei}(2 \lambda )-\text{Ei}(4 \lambda ))-e^{4
   \lambda }-2 (6 \lambda -1) (\text{Ei}(4 \lambda )-\text{Ei}(6 \lambda ))\right) \left(2 \lambda +\beta _{\lambda }\right) \text{$\delta $Z}_c \left(\text{$\delta $Z}_c+1\right)\nonumber\\
&-&30 \left(e^{4 \lambda }
   \left(-3+e^{2 \lambda }\right)+12 \lambda  \left(e^{2 \lambda } \text{Ei}(2 \lambda )-\left(1+e^{2 \lambda }\right) \text{Ei}(4 \lambda )+\text{Ei}(6 \lambda )\right)\right) \text{$\delta $Z}_c
   \left(\text{$\delta $Z}_c+1\right) \eta _c\Biggr)\Biggr\}\Bigr]\nonumber\\
&+&\frac{1}{\left(\text{$\delta $Z}_c+2\right){}^2}\Biggl\{2 e^{4 \lambda /3} \Biggl(8 e^{4 \lambda /3} \text{Ei}\left(\frac{4 \lambda }{3}\right) \Bigl(3 e^{4 \lambda /3} (8
   \lambda -3) \beta _{\lambda } \left(\text{$\delta $Z}_c+2\right) \left(\text{$\delta $Z}_c \left(\text{$\delta $Z}_c+7\right) \left(5 \text{$\delta $Z}_c-1\right)-96\right)\nonumber\\
&+&\lambda 
   \Bigl(\left(\text{$\delta $Z}_c+2\right) \Bigl(3 e^{4 \lambda /3} (-3 \eta_N+16 \lambda -6) \left(\text{$\delta $Z}_c \left(\text{$\delta $Z}_c+7\right) \left(5 \text{$\delta
   $Z}_c-1\right)-96\right)+8 \lambda  \Bigl(16 \lambda  \left(\text{$\delta $Z}_c-1\right) \left(\text{$\delta $Z}_c+1\right) \left(\text{$\delta $Z}_c+3\right)\nonumber\\
&+&3 \left(\text{$\delta $Z}_c \left(\text{$\delta
   $Z}_c \left(\text{$\delta $Z}_c+22\right)-3\right)-84\right)\Bigr)+3 \left(48 (\lambda +6)-3 \text{$\delta $Z}_c \left(\text{$\delta $Z}_c+7\right) \left(5 \text{$\delta $Z}_c-1\right)-16 \lambda 
   \text{$\delta $Z}_c \left(\text{$\delta $Z}_c \left(\text{$\delta $Z}_c+3\right)-1\right)\right) \eta _c\Bigr)-6 (4 \lambda +3) \beta _{\text{$\delta $Z}_c}\Bigr)\Bigr)\nonumber\\
&-&3 \Bigl(e^{8 \lambda /3}
   \Bigl(\left(e^{4 \lambda /3} (7-8 \lambda )-36\right) \beta _{\text{$\delta $Z}_c}+\left(\text{$\delta $Z}_c+2\right) \Bigl(\Bigl(4 \left(64 \lambda ^2+60 \lambda +81\right)+e^{8 \lambda /3} \Bigl(256
   \lambda ^2-32 \left(\eta_N-4 \beta _{\lambda }+2\right) \lambda +7 \eta_N-32 \beta _{\lambda }\Bigr)\nonumber\\
&-&81 \eta _c+e^{4 \lambda /3} \left(-512 \lambda ^2-16 \left(-3 (\ \eta_N+2)+12
   \beta _{\lambda }+\eta _c\right) \lambda -27  \eta_N+84 \beta _{\lambda }+17 \eta _c+54\right)\Bigr) \text{$\delta $Z}_c^3\nonumber\\
&+&6 \Bigl(4 (4 \lambda  (8 \lambda +17)+69)+e^{8 \lambda /3} \left(128
   \lambda ^2-16 \left(\eta_N-4 \beta _{\lambda }+2\right) \lambda +13 \eta_N-16 \beta _{\lambda }\right)-69 \eta _c\nonumber\\
&+&2 e^{4 \lambda /3} \left(-128 \lambda ^2-4 \left(-3 (\eta_N+2)+12
   \beta _{\lambda }+\eta _c\right) \lambda -21 \eta_N+40 \beta _{\lambda }+9 \eta _c+42\right)\Bigr) \text{$\delta $Z}_c^2\nonumber\\
&-&\Bigl(4 \left(64 \lambda ^2+84 \lambda +99\right)+e^{8 \lambda /3}
   \left(256 \lambda ^2-32 \left(\eta_N-4 \beta _{\lambda }+2\right) \lambda +13 \eta_N-32 \beta _{\lambda }\right)-99 \eta _c\nonumber\\
&+&e^{4 \lambda /3} \left(-512 \lambda ^2-16 \left(-3 (\eta_N+2)+12 \beta _{\lambda }+\eta _c\right) \lambda -45 \eta_N+108 \beta _{\lambda }+23 \eta _c+90\right)\Bigr) \text{$\delta $Z}_c-6 \Bigl(8 (16 \lambda  (\lambda +6)+81)\nonumber\\
&+&4 e^{8 \lambda /3}
   \left(32 \lambda ^2-4 \left(\eta_N-4 \beta _{\lambda }+2\right) \lambda +11 \eta_N-4 \beta _{\lambda }\right)-162 \eta _c\nonumber\\
&{}&+e^{4 \lambda /3} \left(-256 \lambda ^2-8 \left(-3 ( \eta_N+2)+12 \beta _{\lambda }+\eta _c\right) \lambda -135  \eta_N+204 \beta _{\lambda }+49 \eta _c+270\right)\Bigr)\Bigr)\Bigr)\nonumber\\
&-&8 \text{Ei}(4 \lambda ) \Bigl(e^{4 \lambda /3} \beta _{\lambda }
   \left(\text{$\delta $Z}_c+2\right) \left(64 \left(\text{$\delta $Z}_c-1\right) \left(\text{$\delta $Z}_c+1\right) \left(\text{$\delta $Z}_c+3\right) \lambda ^2+4 \left(\text{$\delta $Z}_c+11\right) \left(7
   \text{$\delta $Z}_c^2+\text{$\delta $Z}_c-24\right) \lambda -3 \left(\text{$\delta $Z}_c \left(\text{$\delta $Z}_c+7\right) \left(5 \text{$\delta $Z}_c-1\right)-96\right)\right)\nonumber\\
&+&\lambda  \Bigl(-3
   \left(\text{$\delta $Z}_c+2\right) \left(\text{$\delta $Z}_c \left(\text{$\delta $Z}_c+7\right) \left(5 \text{$\delta $Z}_c-1\right)-96\right) \left(e^{4 \lambda /3} (\eta_N+2)+\eta _c\right)\nonumber\\
&-&2
   \Bigl((8 \lambda +3) \beta _{\text{$\delta $Z}_c}-4 \lambda  \left(\text{$\delta $Z}_c+2\right) \Bigl(e^{4 \lambda /3} \left(6 (\eta_N-8 \lambda -44)+\text{$\delta $Z}_c \left(2  \eta_N-16
   \lambda +\text{$\delta $Z}_c \left(-2 \text{$\delta $Z}_c  \eta_N-6  \eta_N+7 \text{$\delta $Z}_c+16 \lambda  \left(\text{$\delta $Z}_c+3\right)+78\right)-13\right)\right)\nonumber\\
&-&4
   \left(\text{$\delta $Z}_c-1\right) \left(\text{$\delta $Z}_c+1\right) \left(\text{$\delta $Z}_c+3\right) \eta _c\Bigr)\Bigr)\Bigr)\Bigr)\Bigr)-8 \text{Ei}\left(\frac{8 \lambda }{3}\right) \Biggl(3 e^{4
   \lambda /3} \beta _{\lambda } \left(\text{$\delta $Z}_c+2\right) \Bigl(64 \left(\text{$\delta $Z}_c-1\right) \left(\text{$\delta $Z}_c+1\right) \left(\text{$\delta $Z}_c+3\right) \lambda ^2\nonumber\\
&+&4
   \left(\text{$\delta $Z}_c+11\right) \left(7 \text{$\delta $Z}_c^2+\text{$\delta $Z}_c-24\right) \lambda +e^{4 \lambda /3} (8 \lambda -3) \left(\text{$\delta $Z}_c \left(\text{$\delta $Z}_c+7\right) \left(5
   \text{$\delta $Z}_c-1\right)-96\right)-3 \left(\text{$\delta $Z}_c \left(\text{$\delta $Z}_c+7\right) \left(5 \text{$\delta $Z}_c-1\right)-96\right)\Bigr)\nonumber\\
&+&\lambda  \Biggl[\left(\text{$\delta $Z}_c+2\right)
   \Bigl(3 e^{8 \lambda /3} (-3  \eta_N+16 \lambda -6) \left(\text{$\delta $Z}_c \left(\text{$\delta $Z}_c+7\right) \left(5 \text{$\delta $Z}_c-1\right)-96\right)-96 \lambda  \left(\text{$\delta
   $Z}_c-1\right) \left(\text{$\delta $Z}_c+1\right) \left(\text{$\delta $Z}_c+3\right) \eta _c\nonumber\\
&-&9 \text{$\delta $Z}_c \left(\text{$\delta $Z}_c+7\right) \left(5 \text{$\delta $Z}_c-1\right) \eta _c+864 \eta
   _c+e^{4 \lambda /3} \Bigl(512 \left(\text{$\delta $Z}_c-1\right) \left(\text{$\delta $Z}_c+1\right) \left(\text{$\delta $Z}_c+3\right) \lambda ^2-48 \Bigl(\text{$\delta $Z}_c \Bigl[- \eta_N-\eta
   _c\nonumber\\
&+&\text{$\delta $Z}_c \left(3  \eta_N+3 \eta _c+\text{$\delta $Z}_c \left( \eta_N+\eta _c-4\right)-50\right)+8\Bigr]-3 \left( \eta_N+\eta _c-58\right)\Bigr) \lambda -9
   \left(\text{$\delta $Z}_c \left(\text{$\delta $Z}_c+7\right) \left(5 \text{$\delta $Z}_c-1\right)-96\right) \left( \eta_N+\eta _c+2\right)\Bigr)\Bigr)\nonumber\\
&-&6 \left(8 \lambda +e^{4 \lambda /3} (4
   \lambda +3)+3\right) \beta _{\text{$\delta $Z}_c}\Biggr]\Biggr)\Biggr)\Bigg\}\Biggr)\Biggr\}\quad \mbox{ for $\lambda>0$.}
\end{eqnarray}
}
\newpage
\setlength{\hoffset}{2mm}
\parbox[c][2\baselineskip][t]{\textwidth}{%
\scriptsize
\begin{eqnarray}
 \beta_{\delta Z_c}&=&\frac{1}{7776 \pi  \left(\text{$\delta $Z}_c+2\right){}^2}\Biggl\{e^{-4 \lambda } G \Biggl(8 e^{8 \lambda /3} \text{Ei}\left(\frac{4 \lambda }{3}\right) \Biggl(3 e^{4 \lambda /3} (8 \lambda -3) \beta _{\lambda } \left(\text{$\delta $Z}_c-1\right) \left(\text{$\delta
   $Z}_c+2\right) \left(\text{$\delta $Z}_c \left(\text{$\delta $Z}_c \left(5 \text{$\delta $Z}_c+7\right)-136\right)-48\right)\nonumber\\
&+&\lambda  \Biggl(\left(\text{$\delta $Z}_c^2+\text{$\delta $Z}_c-2\right) \Biggl(8
   \lambda  \Bigl(16 \lambda  \left(\text{$\delta $Z}_c-1\right) \text{$\delta $Z}_c \left(\text{$\delta $Z}_c+4\right)+3 \text{$\delta $Z}_c \left(\left(\text{$\delta $Z}_c-5\right) \text{$\delta
   $Z}_c-120\right)-144\Bigr)-3 \Bigl(16 \lambda  \left(\text{$\delta $Z}_c-1\right) \text{$\delta $Z}_c \left(\text{$\delta $Z}_c+4\right)\nonumber\\
&+&3 \text{$\delta $Z}_c \left(\text{$\delta $Z}_c \left(5
   \text{$\delta $Z}_c+7\right)-136\right)-144\Bigr) \eta _c+3 e^{4 \lambda /3} \left(\text{$\delta $Z}_c \left(\text{$\delta $Z}_c \left(5 \text{$\delta $Z}_c+7\right)-136\right)-48\right) \left(16 \lambda
   -3 \eta _N-6\right)\Biggr)\nonumber\\
&-&6 (4 \lambda +3) \beta _{\text{$\delta $Z}_c} \left(\text{$\delta $Z}_c-4\right)\Biggr)\Biggr)-432 e^{2 \lambda } \text{Ei}(2 \lambda ) \left(\text{$\delta $Z}_c-1\right) \Bigl(5
   e^{2 \lambda } (4 \lambda -1) \beta _{\lambda } \left(\text{$\delta $Z}_c+1\right) \left(\text{$\delta $Z}_c+2\right) \left(7 \text{$\delta $Z}_c^2+\text{$\delta $Z}_c-22\right)\nonumber\\
&+&\lambda  \Bigl(5
   \left(\text{$\delta $Z}_c+1\right) \left(\text{$\delta $Z}_c+2\right) \Bigl(4 \lambda  \left(5 \text{$\delta $Z}_c^2-\text{$\delta $Z}_c+4 \lambda  \left(\text{$\delta $Z}_c^2+\text{$\delta
   $Z}_c-2\right)-18\right)-\left(7 \text{$\delta $Z}_c^2+\text{$\delta $Z}_c+4 \lambda  \left(\text{$\delta $Z}_c^2+\text{$\delta $Z}_c-2\right)-22\right) \eta _c\nonumber\\
&+&e^{2 \lambda } \left(7 \text{$\delta
   $Z}_c^2+\text{$\delta $Z}_c-22\right) \left(8 \lambda -\eta _N-2\right)\Bigr)-2 \left(2 \lambda  \beta _{\text{$\delta $Z}_c}+\beta _{\text{$\delta $Z}_c}\right)\Bigr)\Bigr)+3 \Biggl(e^{4 \lambda }
   \Biggl(\Bigl(5504 \lambda ^2+9840 \lambda -1539 \left(\eta _c-4\right)\nonumber\\
&+&e^{4 \lambda /3} \left(512 \lambda ^2+16 \left(12 \beta _{\lambda }+\eta _c-3 \eta _N-6\right) \lambda -84 \beta _{\lambda }-17 \eta
   _c+27 \eta _N-54\right)-40 e^{2 \lambda } \Bigl(288 \lambda ^2+6 \left(18 \beta _{\lambda }+\eta _c-3 \eta _N-6\right) \lambda -72 \beta _{\lambda }\nonumber\\
&-&11 \eta _c+27 \eta _N-54\Bigr)+20 e^{4 \lambda }
   \left(288 \lambda ^2-24 \left(-6 \beta _{\lambda }+\eta _N+2\right) \lambda -24 \beta _{\lambda }+17 \eta _N\right)\nonumber\\
&+&e^{8 \lambda /3} \left(-256 \lambda ^2+32 \left(-4 \beta _{\lambda }+\eta _N+2\right)
   \lambda +32 \beta _{\lambda }-7 \eta _N\right)\Bigr) \text{$\delta $Z}_c^5+4 \Bigl(4064 \lambda ^2+5256 \lambda -882 \left(\eta _c-4\right)\nonumber\\
&+&e^{4 \lambda /3} \Bigl(512 \lambda ^2+16 \left(12 \beta _{\lambda
   }+\eta _c-3 \eta _N-6\right) \lambda -60 \beta _{\lambda }-11 \eta _c+9 \eta _N-18\Bigr)-60 e^{2 \lambda } \Bigl(144 \lambda ^2+3 \left(18 \beta _{\lambda }+\eta _c-3 \eta _N-6\right) \lambda \nonumber\\
&-&27 \beta
   _{\lambda }-4 \eta _c+9 \eta _N-18\Bigr)+15 e^{4 \lambda } \left(288 \lambda ^2-24 \left(-6 \beta _{\lambda }+\eta _N+2\right) \lambda -24 \beta _{\lambda }+11 \eta _N\right)\nonumber\\
&-&e^{8 \lambda /3} \left(256
   \lambda ^2-32 \left(-4 \beta _{\lambda }+\eta _N+2\right) \lambda -32 \beta _{\lambda }+\eta _N\right)\Bigr) \text{$\delta $Z}_c^4-\Bigl(4992 \lambda ^2+32208 \lambda -3861 \left(\eta _c-4\right)-40 e^{2
   \lambda } \Bigl(288 \lambda ^2\nonumber\\
&+&6 \left(18 \beta _{\lambda }+\eta _c-3 \eta _N-6\right) \lambda -252 \beta _{\lambda }-41 \eta _c+117 \eta _N-234\Bigr)+3 e^{4 \lambda /3} \Bigl(512 \lambda ^2+16 \left(12
   \beta _{\lambda }+\eta _c-3 \eta _N-6\right) \lambda -580 \beta _{\lambda }\nonumber\\
&-&141 \eta _c+399 \eta _N-798\Bigr)+20 e^{4 \lambda } \left(288 \lambda ^2-24 \left(-6 \beta _{\lambda }+\eta _N+2\right) \lambda
   -24 \beta _{\lambda }+77 \eta _N\right)-3 e^{8 \lambda /3} \Bigl(256 \lambda ^2-32 \left(-4 \beta _{\lambda }+\eta _N+2\right) \lambda \nonumber\\
&-&32 \beta _{\lambda }+131 \eta _N\Bigr)\Bigr) \text{$\delta $Z}_c^3-2
   \Biggl(18880 \lambda ^2+37728 \lambda -5499 \left(\eta _c-4\right)-40 e^{2 \lambda } \Bigl(1008 \lambda ^2+21 \left(18 \beta _{\lambda }+\eta _c-3 \eta _N-6\right) \lambda -297 \beta _{\lambda }-46 \eta
   _c\nonumber\\
&+&117 \eta _N-234\Bigr)+e^{4 \lambda /3} \left(2560 \lambda ^2+80 \left(12 \beta _{\lambda }+\eta _c-3 \eta _N-6\right) \lambda -1308 \beta _{\lambda }-307 \eta _c+801 \eta _N-1602\right)\nonumber\\
&+&10 e^{4 \lambda
   } \left(2016 \lambda ^2-168 \left(-6 \beta _{\lambda }+\eta _N+2\right) \lambda -168 \beta _{\lambda }+149 \eta _N\right)+e^{8 \lambda /3} \left(-1280 \lambda ^2+160 \left(-4 \beta _{\lambda }+\eta
   _N+2\right) \lambda +160 \beta _{\lambda }-257 \eta _N\right)\Biggr) \text{$\delta $Z}_c^2\nonumber\\
&+&\Bigl(\left(e^{2 \lambda } (28-48 \lambda )+e^{4 \lambda /3} (8 \lambda -7)-108\right) \beta _{\text{$\delta
   $Z}_c}+16 \Bigl(4 \left(-32 \lambda ^2+282 \lambda +81\right)-81 \eta _c+75 e^{2 \lambda } \left(6 \beta _{\lambda }+\eta _c-3 \eta _N+6\right)+75 e^{4 \lambda } \eta _N\nonumber\\
&+&e^{4 \lambda /3} \left(256 \lambda
   ^2+8 \left(12 \beta _{\lambda }+\eta _c-3 \eta _N-6\right) \lambda -180 \beta _{\lambda }-43 \eta _c+117 \eta _N-234\right)\nonumber\\
&-&2 e^{8 \lambda /3} \left(64 \lambda ^2-8 \left(-4 \beta _{\lambda }+\eta
   _N+2\right) \lambda -8 \beta _{\lambda }+19 \eta _N\right)\Bigr)\Bigr) \text{$\delta $Z}_c\nonumber\\
&+&4 \Bigl(36 \left(4 (2 \lambda  (20 \lambda +51)+59)-59 \eta _c\right)+e^{2 \lambda } \left((12 \lambda -7) \beta
   _{\text{$\delta $Z}_c}+40 \left(36 \lambda +17 \eta _c-6 \left(18 (\lambda -1) \beta _{\lambda }+\lambda  \left(48 \lambda +\eta _c-3 \eta _N\right)\right)-45 \eta _N+90\right)\right)\nonumber\\
&-&e^{4 \lambda /3}
   \left((8 \lambda -7) \beta _{\text{$\delta $Z}_c}+72 \left(4 \beta _{\lambda }+\eta _c-3 \eta _N+6\right)\right)-72 e^{8 \lambda /3} \eta _N+20 e^{4 \lambda } \left(288 \lambda ^2-24 \left(-6 \beta
   _{\lambda }+\eta _N+2\right) \lambda -24 \beta _{\lambda }+29 \eta _N\right)\Bigr)\Biggr)\nonumber\\
&-&144 \text{Ei}(6 \lambda ) \left(\text{$\delta $Z}_c-1\right) \Bigl(5 e^{2 \lambda } \beta _{\lambda }
   \left(\text{$\delta $Z}_c+1\right) \left(\text{$\delta $Z}_c+2\right) \left(24 \left(\text{$\delta $Z}_c^2+\text{$\delta $Z}_c-2\right) \lambda ^2+2 \left(\text{$\delta $Z}_c \left(17 \text{$\delta
   $Z}_c-1\right)-58\right) \lambda -7 \text{$\delta $Z}_c^2-\text{$\delta $Z}_c+22\right)\nonumber\\
&+&\lambda  \Bigl(5 \left(\text{$\delta $Z}_c+1\right) \left(\text{$\delta $Z}_c+2\right) \Bigl(\left(16 \lambda
   -\text{$\delta $Z}_c \left(7 \text{$\delta $Z}_c+8 \lambda  \left(\text{$\delta $Z}_c+1\right)+1\right)+22\right) \eta _c+e^{2 \lambda } \Bigl(48 \left(\text{$\delta $Z}_c^2+\text{$\delta $Z}_c-2\right)
   \lambda ^2\nonumber\\
&-&4 \left(-2 \eta _N+\text{$\delta $Z}_c \left(\text{$\delta $Z}_c \left(\eta _N-17\right)+\eta _N+1\right)+58\right) \lambda +\left(-7 \text{$\delta $Z}_c^2-\text{$\delta $Z}_c+22\right)
   \left(\eta _N+2\right)\Bigr)\Bigr)-2 \left(4 \lambda  \beta _{\text{$\delta $Z}_c}+\beta _{\text{$\delta $Z}_c}\right)\Bigr)\Bigr)\nonumber\\
&+&8 \text{Ei}(4 \lambda ) \Biggl(e^{2 \lambda } \beta _{\lambda } \Bigl(90
   e^{2 \lambda } (4 \lambda -1) \left(\text{$\delta $Z}_c-1\right) \left(\text{$\delta $Z}_c+1\right) \left(\text{$\delta $Z}_c+2\right) \left(7 \text{$\delta $Z}_c^2+\text{$\delta $Z}_c-22\right)\nonumber\\
&+&e^{2
   \lambda /3} \left(\text{$\delta $Z}_c^2+\text{$\delta $Z}_c-2\right) \left(-576 \lambda +\text{$\delta $Z}_c \left(64 \left(\text{$\delta $Z}_c-1\right) \left(\text{$\delta $Z}_c+4\right) \lambda ^2+4
   \text{$\delta $Z}_c \left(7 \text{$\delta $Z}_c-3\right) \lambda -1504 \lambda -3 \text{$\delta $Z}_c \left(5 \text{$\delta $Z}_c+7\right)+408\right)+144\right)\nonumber
\end{eqnarray}}
\newpage
\setlength{\hoffset}{-18mm}
\parbox[c][14\baselineskip][t]{\textwidth}{%
\scriptsize 
\begin{eqnarray}
&+&90 \left(\text{$\delta $Z}_c-1\right)
   \left(\text{$\delta $Z}_c+1\right) \left(\text{$\delta $Z}_c+2\right) \left(24 \left(\text{$\delta $Z}_c^2+\text{$\delta $Z}_c-2\right) \lambda ^2+2 \left(\text{$\delta $Z}_c \left(17 \text{$\delta
   $Z}_c-1\right)-58\right) \lambda -7 \text{$\delta $Z}_c^2-\text{$\delta $Z}_c+22\right)\Bigr)\nonumber\\
&+&\lambda  \Biggl(-2 e^{4 \lambda /3} (8 \lambda +3) \beta _{\text{$\delta $Z}_c} \left(\text{$\delta
   $Z}_c-4\right)-36 e^{2 \lambda } (2 \lambda +1) \beta _{\text{$\delta $Z}_c} \left(\text{$\delta $Z}_c-1\right)-36 (4 \lambda +1) \beta _{\text{$\delta $Z}_c} \left(\text{$\delta
   $Z}_c-1\right)\nonumber\\
&+&\left(\text{$\delta $Z}_c^2+\text{$\delta $Z}_c-2\right) \Biggl(-90 \left(\text{$\delta $Z}_c+1\right) \left(7 \text{$\delta $Z}_c^2+\text{$\delta $Z}_c+8 \lambda  \left(\text{$\delta
   $Z}_c^2+\text{$\delta $Z}_c-2\right)-22\right) \eta _c-e^{4 \lambda /3} \Bigl(32 \lambda  \left(\text{$\delta $Z}_c-1\right) \text{$\delta $Z}_c \left(\text{$\delta $Z}_c+4\right)\nonumber\\
&+&3 \text{$\delta $Z}_c
   \left(\text{$\delta $Z}_c \left(5 \text{$\delta $Z}_c+7\right)-136\right)-144\Bigr) \eta _c+90 e^{4 \lambda } \left(\text{$\delta $Z}_c+1\right) \left(7 \text{$\delta $Z}_c^2+\text{$\delta $Z}_c-22\right)
   \left(8 \lambda -\eta _N-2\right)\nonumber\\
&+&e^{8 \lambda /3} \Bigl(128 \left(\text{$\delta $Z}_c-1\right) \text{$\delta $Z}_c \left(\text{$\delta $Z}_c+4\right) \lambda ^2-8 \left(\text{$\delta $Z}_c \left(-7
   \text{$\delta $Z}_c^2+3 \text{$\delta $Z}_c+2 \left(\text{$\delta $Z}_c-1\right) \left(\text{$\delta $Z}_c+4\right) \eta _N+376\right)+144\right) \lambda \nonumber\\
&-&3 \left(\text{$\delta $Z}_c \left(\text{$\delta
   $Z}_c \left(5 \text{$\delta $Z}_c+7\right)-136\right)-48\right) \left(\eta _N+2\right)\Bigr)+90 e^{2 \lambda } \left(\text{$\delta $Z}_c+1\right) \Bigl(64 \left(\text{$\delta $Z}_c^2+\text{$\delta
   $Z}_c-2\right) \lambda ^2\nonumber\\
&-&4 \left(\text{$\delta $Z}_c \left(\eta _c+\eta _N+\text{$\delta $Z}_c \left(\eta _c+\eta _N-22\right)+2\right)-2 \left(\eta _c+\eta _N-38\right)\right) \lambda +\left(-7
   \text{$\delta $Z}_c^2-\text{$\delta $Z}_c+22\right) \left(\eta _c+\eta _N+2\right)\Bigr)\Biggr)\Biggr)\Biggr)\Biggr)\Biggr)\Biggr\}\nonumber\\
&+&\frac{1}{7776 \pi  \left(\text{$\delta $Z}_c+2\right){}^2}\Biggl\{e^{-4 \lambda } G \Biggl(-8 e^{4 \lambda /3} \text{Ei}\left(\frac{8 \lambda }{3}\right) \Biggl(3 e^{4 \lambda /3}
   \beta _{\lambda } \left(\text{$\delta $Z}_c^2+\text{$\delta $Z}_c-2\right) \Bigl(64 \lambda ^2 \text{$\delta $Z}_c^3+28 \lambda  \text{$\delta $Z}_c^3-15 \text{$\delta $Z}_c^3+192 \lambda ^2 \text{$\delta
   $Z}_c^2\nonumber\\
&-&12 \lambda  \text{$\delta $Z}_c^2-21 \text{$\delta $Z}_c^2-256 \lambda ^2 \text{$\delta $Z}_c-1504 \lambda  \text{$\delta $Z}_c+408 \text{$\delta $Z}_c-576 \lambda +e^{4 \lambda /3} (8 \lambda -3)
   \left(\text{$\delta $Z}_c \left(\text{$\delta $Z}_c \left(5 \text{$\delta $Z}_c+7\right)-136\right)-48\right)+144\Bigr)\nonumber\\
&+&\lambda  \Bigl(\left(\text{$\delta $Z}_c^2+\text{$\delta $Z}_c-2\right) \Bigl(-3
   \left(32 \lambda  \left(\text{$\delta $Z}_c-1\right) \text{$\delta $Z}_c \left(\text{$\delta $Z}_c+4\right)+3 \text{$\delta $Z}_c \left(\text{$\delta $Z}_c \left(5 \text{$\delta
   $Z}_c+7\right)-136\right)-144\right) \eta _c\nonumber\\
&+&3 e^{8 \lambda /3} \left(\text{$\delta $Z}_c \left(\text{$\delta $Z}_c \left(5 \text{$\delta $Z}_c+7\right)-136\right)-48\right) \left(16 \lambda -3 \eta
   _N-6\right)+e^{4 \lambda /3} \Bigl(512 \left(\text{$\delta $Z}_c-1\right) \text{$\delta $Z}_c \left(\text{$\delta $Z}_c+4\right) \lambda ^2\nonumber\\
&-&48 \Bigl(\text{$\delta $Z}_c \Bigl(\text{$\delta $Z}_c \left(3
   \eta _c+3 \eta _N+\text{$\delta $Z}_c \left(\eta _c+\eta _N-4\right)+4\right)\nonumber\\
&-&4 \left(\eta _c+\eta _N-62\right)\Bigr)+96\Bigr) \lambda -9 \left(\text{$\delta $Z}_c \left(\text{$\delta $Z}_c \left(5
   \text{$\delta $Z}_c+7\right)-136\right)-48\right) \left(\eta _c+\eta _N+2\right)\Bigr)\Bigr)-6 \left(8 \lambda +e^{4 \lambda /3} (4 \lambda +3)+3\right) \beta _{\text{$\delta $Z}_c} \left(\text{$\delta
   $Z}_c-4\right)\Bigr)\Biggr)\Biggr)\Biggr\}\nonumber\\
&{}& \mbox{ for $\lambda>0$.} \label{deltaZflowloooong}
\end{eqnarray}
}
\newpage
\setlength{\hoffset}{0mm}
\section{Fermions in quantum gravity}\label{Gamma2}
\subsection{Vertices for fermion-graviton couplings}\label{fermvertsandconvs}
The relation between the vierbeins and the metric is given by
\begin{equation}
 g_{\mu \nu} =\eta_{ab}e^{\mu}_{\p a} e^{\nu}_{\p b}\label{metric1}
\end{equation}
 The inverse vielbeins are obtained via
\begin{eqnarray}
 e^{\mu}_{\p a}e_{\nu}^{\p a}&=& \delta^{\mu}_{\nu}\\
e_{\mu}^{\p a}e^{\mu}_{\p b}&=&\delta^a_b
\end{eqnarray}
In the following we expand the vierbein around a (flat) background:
\begin{equation}
 e_{\mu a}= \bare_{\mu a}+ \delta e_{\mu a},\label{vierbeinfluc}
\end{equation}
where higher orders are not needed in our calculation.
In the following we choose the Lorentz symmetric gauge with gauge-fixing functional \cite{Woodard:1984sj}\cite{vanNieuwenhuizen:1981uf}, as then all vierbein fluctuations can be rewritten in terms of metric fluctuations without ghosts due to the $O(4)$ gauge fixing:
\begin{equation}
 F_{ab}= e_{\mu a}\bar{g}^{\mu \nu}\bare_{\nu b}- e^{\mu b}\bar{g}^{\mu \nu}\bare_{\nu a} \label{vierbeingauge}.
\end{equation}
This allows to write
\begin{eqnarray}
\delta e_{\mu a}&=& \frac{1}{2}h_{\mu}^{\,\,\kappa}\bar{e}_{\kappa a}\label{deltavar}\\
\delta e^{\kappa b}&=& -\frac{1}{2} h_{\mu}^{\,\,
\kappa}\bar{e}^{\mu b}\label{deltainvvar}\\
\end{eqnarray}
The spin connection is determined from the requirement that the covariant differentiation commutes with the transition to the local orthonormal frame. As the vierbein is used in this transition, this requirement translates into
\begin{equation}
 \nabla_{\mu}e^{a}_{\nu}= \partial_{\mu}e^{a}_{\nu}- \Gamma^{\lambda}_{\nu \mu}e^{a}_{\lambda}+ \omega_{\mu\, \, b}^a e^{b}_{\nu}.
\end{equation}
This establishes the following relation between the spin connection and the Christoffel connection
\begin{equation}
 \omega^{\p \p ab}_{ \mu}=-(\partial_{\mu}e_{\nu}^{\p a})e^{\nu b}+ \Gamma_{\nu \mu \sigma}e^{\nu a}e^{\sigma b}, \label{spinchristoffel}
\end{equation}
which also implies that 
\begin{equation}
 \omega_{\mu\, ab}= - \omega_{\mu \, ba}.
\end{equation}
For the variation, we then have that
\begin{eqnarray}
\phantom{,}[\gamma^a,\gamma^b]\, 
\delta \omega_{\mu ab}&=& [\gamma^{\lambda}, \gamma^{\nu}]D_{\nu}h_{\lambda \mu}\label{omegavar}.
\end{eqnarray}

From \Eqref{omegavar} we can deduce for constant external fermions, where
total derivatives can be discarded, that
\begin{eqnarray}
\lbrack\gamma^a, \gamma^b\rbrack\delta^2 \omega_{\mu ab}&=& [\gamma^{\lambda}, \gamma^{\nu}]\Bigl(-h^{\sigma}_{\, \lambda}D_{\nu}h_{\mu \sigma}-h^{\sigma}_{\, \nu}D_{\sigma}h_{\mu \lambda}-\frac{1}{2}h_{\kappa \lambda}D_{\mu}h^{\kappa}_{\, \nu} \Bigr),\label{omega2var}
\end{eqnarray}
where we have set $g_{\mu \nu}= \bar{g}_{\mu \nu}$ and $e_{\mu a}= \bare_{\mu
  a}$, and then dropped the bar on the covariant derivative.

We then go over to Fourier space
\begin{eqnarray}
 \psi(x)&=& \int \frac{d^4p}{(2 \pi)^4}\psi(p)e^{-ipx}\nonumber\\
 h_{\mu \nu}(x)&=& \int \frac{d^4p}{(2 \pi)^4}h_{\mu \nu}(p)e^{-ipx}\nonumber\\
 \bar{\psi}(x)&=& \int \frac{d^4p}{(2 \pi)^4}\bar{\psi}(p)e^{ipx},\nonumber\\
\end{eqnarray}
where $\psi(x)$ and $\psi(p)$ denote Fourier transforms of each other.

Now we may evaluate the mixed fermion-graviton
vertices, where our conventions are
\begin{eqnarray}
\Gamma^{(2)}= \frac{\overset{\rightarrow}{\delta}}{\delta
\Phi^T(-p)}\Gamma \frac{\overset{\leftarrow}\delta}{\delta
\Phi(q)},
\end{eqnarray}
where the collective fields
\begin{eqnarray}
 \Phi^T(-q)&=& \left(h_{\kappa \lambda}^{\mathrm{TT}}(-q), h(-q),
 \psi^T_i(-q), \bar{\psi}_i(q) \right)\\ 
\Phi(q)&=&\left(h_{\mu \nu}^{\mathrm{TT}}(q), h(q),\psi_j(q),
\bar{\psi}^T_j(-q) \right). 
\end{eqnarray}
Here the second line should be read as a column vector. The symbol $T$ refers
to transposition in Dirac space and in field space. 
As we work in the Landau deWitt gauge, only the transverse
traceless and the trace mode can contribute.

The first variation of the kinetic fermion term with respect to the metric is given by
\begin{eqnarray}
 \delta \Gamma_{\rm kin}&=& i Z_{\psi}\int d^4 x \bar{\psi}^i\Bigl(\delta(\sqrt{g}) \gamma^{\mu}\nabla_{\mu}+ \sqrt{g}\delta \gamma^{\mu}\nabla_{\mu}+ \sqrt{g}\gamma^{\mu}\delta \nabla_{\mu} \Bigr)\psi^i. \label{varkin}
\end{eqnarray}
To read off the trace-mode-fermion-vertices we
Fourier-transform the first variation of the kinetic term
with respect to the metric to get (in agreement with
\cite{Zanusso:2009bs})

\begin{eqnarray}
 \delta \Gamma_{\rm kin}&=& Z_{\psi} \int
\frac{d^4p}{(2\pi)^4}
 \Bigl(\Bigl(\frac{3}{16}h(p)\bar{\psi}^i(p)\slashed{p}
\psi^i
-\frac{3}{16}\bar{\psi}^i\slashed{p}
\psi^i(-p)h(p)\Bigr)\Bigr).\label{firstvar}
\end{eqnarray}
In this notation, $\bar{\psi}$ and $\psi$ are the constant background fields, whereas the momentum-dependent fluctuation fields are distinguished by carrying an appropriate argument.
This allows to evaluate the following vertices:
\begin{eqnarray}
 V_{\rm kin}^{h\, \bar{\psi}^{i\, T}}&=&
\frac{\delta}{\delta h(-p)}
\,\Gamma_{\rm kin}\frac{\overset{\leftarrow}{\delta}}{\delta
\bar{\psi}^{i\, T}(-q)}= \frac{3}{16}Z_{\psi}\psi^{i\,
T}\slashed p^T\\
V_{\rm kin}^{h \psi^i}&=& \frac{3}{16}Z_{\psi}\bar{\psi}^i\slashed{p}\\
V_{\rm kin}^{\psi^{i\,T} h}&=& \frac{3}{16}Z_{\psi}\slashed{p}^T \bar{\psi}^{i\, T}\\
V_{\rm kin}^{\bar{\psi^i}h}&=& \frac{3}{16}Z_{\psi}\slashed{p}\psi^i,
\end{eqnarray}
where the momentum is always the momentum of the incoming graviton. 

The corresponding vertices with the TT mode vanish, as the
first term in \Eqref{varkin} contains only the trace
mode, the second term vanishes by transversality for
constant external fermion fields and the last term vanishes
as the contraction $\gamma^{\mu}[\gamma^{\nu},
\gamma^{\kappa}]D_{\kappa}h^{\mathrm{TT}}_{\mu \nu}=0$.

The second variation of the kinetic fermion term with
respect to the metric contains only a TT contribution, as
the trace contribution is always of the form
$\bar{\psi}^i h \slashed{D}h \psi^i$, which can be rewritten
as a total derivative for constant external fermions. 

From the fact that we have constant external fermions at
 least one of the variations has to hit the covariant
derivative $\nabla_{\mu}$ and hence produce a $[\gamma^a,
\gamma^b]\delta \omega_{\mu\, a b}$. Accordingly the second
variation will necessarily contain $\gamma^{\mu}
[\gamma^{\kappa}, \gamma^{\lambda}]$. As there is one
derivative in the kinetic term, the vertex has to be
proportional to the momentum of one of the gravitons. The
only possibly structure that cannot be rewritten into a
total derivative is then $\gamma^{\mu}[\gamma^{\kappa},
\gamma^{\lambda}] h_{\kappa
\sigma}D_{\mu}h_{\lambda}^{\sigma}$. Our explicit
calculation now only has to fix the sign and the numerical
factor of the vertex. 
From \Eqref{omega2var} and \Eqref{omegavar} we deduce that
\begin{equation}
 \delta^2 \Gamma_{\rm kin}= i Z_{\psi} \int d^4 x 
\sqrt{g}\bar{\psi}^i\left(\frac{-1}{16}\right)\left(h^{\mu}_
{\lambda}\gamma^{\nu}[\gamma^{\lambda},
\gamma^{\kappa}]D_{\nu}h_{\kappa \mu}\right)\psi^i.
\end{equation}

The vertex that results from this expression is given by
\begin{eqnarray}
V^{h^{\mathrm{TT}}h^{\mathrm{TT}}}_{\text{kin}\, \mu \nu \kappa \lambda}&=& \frac{-1}{128}
p_{\tau}\bar{\psi}
 [\gamma^{\rho},\gamma^{\alpha}]\gamma^{\tau}\psi
\Bigl(\delta_{\mu\rho}\delta_{\kappa \alpha} \delta_{\nu
\lambda}+\delta_{\mu 
\rho}\delta_{\kappa\nu} \delta_{\alpha \lambda}+\delta_{\mu \lambda}\delta_{\nu \rho}\delta_{\kappa
\alpha}\nonumber\\
&{}&
+\delta_{\mu \kappa}\delta_{\nu \rho}\delta_{\alpha
\lambda}-\delta_{\rho \kappa}\delta_{\lambda
\nu}\delta_{\alpha \mu}-\delta_{\rho \kappa}\delta_{\lambda \mu}\delta_{\alpha\nu}-\delta_{\rho
  \lambda}\delta_{\kappa \nu}\delta_{\alpha \mu}-\delta_{\rho
  \lambda}\delta_{\kappa \mu}\delta_{\alpha \nu}\Bigr). 
\end{eqnarray}
The variations of the four-fermion term with respect to
 the metric are very simple: Due to
\begin{equation}
 \delta(\gamma^{\mu}\gamma_{\mu})= \delta(4)=0,
\end{equation}
only the determinant factor can contribute, and not the various $\gamma$
matrices. They always appear with completely contracted spacetime indices,
such that the above identity applies.  Hence the vertices containing three
external (anti)-fermions, one internal (anti)-fermion and one internal
graviton only exist for the trace mode, as $\delta \sqrt{g}=
\frac{1}{2}\sqrt{g}h$.  The vertices are given by
f
\begin{eqnarray}
 V_{4f}^{h \bar{\psi}^{j\,T}}&=&- \frac{\bar{\lambda}_- + \bar{\lambda}_+}{2}\left(\bar{\psi}^i \gamma^{\mu}\psi^i \right)\psi^{j\, T}\gamma_{\mu}^T- \frac{\bar{\lambda}_- - \bar{\lambda}_+}{2}\left(\bar{\psi}^i \gamma^{\mu}\gamma^5\psi^i \right)\psi^{j\, T}\gamma^{5T}\gamma_{\mu}^T\nonumber\\
V_{4f}^{h \psi^j}&=&\frac{\bar{\lambda}_-+\bar{\lambda}_+}{2} \left(\bar{\psi}^i \gamma^{\mu}\psi^i \right) \bar{\psi}^j \gamma_{\mu} +\frac{\bar{\lambda}_- -\bar{\lambda}_+}{2} \left(\bar{\psi}^i \gamma^{\mu}\gamma^5\psi^i \right) \bar{\psi}^j \gamma_{\mu}\gamma^5\nonumber\\
V_{4f}^{\psi^{j\,T} h}&=& -\frac{\bar{\lambda}_-+\bar{\lambda}_+}{2}\left(\bar{\psi}^i \gamma^{\mu}\psi^i \right) \gamma_{\mu}^T\bar{\psi}^{j\, T}-\frac{\bar{\lambda}_--\bar{\lambda}_+}{2}\left(\bar{\psi}^i \gamma^{\mu}\gamma^5\psi^i \right) \gamma^{5T}\gamma_{\mu}^T\bar{\psi}^{j\, T}\nonumber\\
V_{4f}^{\bar{\psi^j}h}&=&\frac{\bar{\lambda}_-+\bar{\lambda}_+}{2}\left(\bar{\psi}^i \gamma^{\mu}\psi^i \right) \gamma_{\mu} \psi^j+\frac{\bar{\lambda}_--\bar{\lambda}_+}{2}\left(\bar{\psi}^i \gamma^{\mu}\gamma^5\psi^i \right) \gamma_{\mu} \gamma^5\psi^j.
\end{eqnarray}
The tadpole receives contributions from both the TT and 
the trace mode. The corresponding vertices are given by
\begin{eqnarray}
 V_{4f}^{hh}&=&\frac{1}{16}\left(\bar{\lambda}_- (V-A)+\bar{\lambda}_+(V+A)
 \right), \\
V_{4f\, \mu \nu \kappa \lambda}^{h^{\mathrm{TT}} h^{\mathrm{TT}}}&=&
-\frac{1}{8}\left(\delta_{\mu \kappa}\delta_{\nu \lambda}+\delta_{\mu
  \lambda}\delta_{\nu \kappa} \right)\left(\bar{\lambda}_-
(V-A)+\bar{\lambda}_+ (V+A) \right). 
\end{eqnarray}
The variations of the four-fermion terms with respect to the fermions read as follows:
\begin{eqnarray}
 V_{4f}^{\psi^{i\,T} \bar{\psi}^{j\,T}}&=& \frac{\bar{\lambda}_-+\bar{\lambda}_+}{2}\left[2 \left(\gamma^{\mu\, T}\bar{\psi}^{i\, T} \right)\left(\psi^{j\, T}\gamma^{\mu\, T}\right)-2 \gamma^{\mu\, T}\delta^{ij}\left(\bar{\psi}^k \gamma^{\mu}\psi^k \right)\right]\nonumber\\
&{}&+\frac{\bar{\lambda}_--\bar{\lambda}_+}{2}\left[2 \left(\gamma^5\gamma^{\mu\, T}\bar{\psi}^{i\, T} \right)\left(\psi^{j\, T}\gamma^5\gamma^{\mu\, T}\right)-2\gamma^5 \gamma^{\mu\, T}\delta^{ij}\left(\bar{\psi}^k \gamma^5\gamma^{\mu}\psi^k \right)\right]\nonumber\\
V_{4f}^{\psi^{i\,T} \psi^j}&=& -\frac{\bar{\lambda}_-+\bar{\lambda}_+}{2}2 \left(\gamma^{\mu\, T}\bar{\psi}^{i\, T} \right)\left(\bar{\psi}^j \gamma^{\mu}\right)-\frac{\bar{\lambda}_--\bar{\lambda}_+}{2}2 \left(\gamma^5\gamma^{\mu\, T}\bar{\psi}^{i\, T} \right)\left(\bar{\psi}^j \gamma^{\mu}\gamma^5\right)\nonumber\\
V_{4f}^{\bar{\psi}^{i}\bar{\psi}^{j\,T}}&=& -\frac{\bar{\lambda}_-+\bar{\lambda}_+}{2}2 \left(\gamma^{\mu}\psi^i \right)\left(\psi^{j\, T}\gamma^{\mu \, T} \right)-\frac{\bar{\lambda}_--\bar{\lambda}_+}{2}2 \left(\gamma^{\mu}\gamma^5\psi^i \right)\left(\psi^{j\, T}\gamma^5\gamma^{\mu \, T} \right)\nonumber\\
V_{4f}^{\bar{\psi}^{i}\psi^j}&=& \frac{\bar{\lambda}_-+\bar{\lambda}_+}{2}\left[2 \delta^{ij}\gamma^{\mu}\left( \bar{\psi}^k \gamma^{\mu}\psi^k\right)+2 \left(\gamma^{\mu}\psi^i \right)\left(\bar{\psi}^j\gamma^{\mu} \right)\right]\nonumber\\
&{}&+\frac{\bar{\lambda}_--\bar{\lambda}_+}{2}\left[2 \delta^{ij}\gamma^{\mu}\gamma^5\left( \bar{\psi}^k \gamma^{\mu}\gamma^5\psi^k\right)+2 \left(\gamma^{\mu}\gamma^5\psi^i \right)\left(\bar{\psi}^j\gamma^{\mu}\gamma^5 \right)\right].
\end{eqnarray}
Here we suppress the Dirac index structure; by round brackets 
we indicate the way in which the Dirac indices of the terms 
have to be contracted.

\subsection{Cancellation of box diagrams}\label{cancellation}
The two box diagrams are given by terms $\mathcal{O}(\mathcal{P}^{-1}\mathcal{F})^4$, where the vertices arise from variations of the kinetic term only. Using the notation
\begin{equation}
 I_{h \, \psi}= \frac{\delta}{\delta h(-q)}\Gamma_{k\, \rm kin} \frac{\delta}{\delta \psi(p)},
\end{equation}
and analogously for derivatives involving the antifermion, the fluctuation matrix (restricted to the entries relevant for us) reads
\begin{eqnarray}
 \mathcal{F}= \begin{pmatrix}0 & I_{h\, \psi}& I_{h \,\bar{\psi}^T}\\ I_{\psi^T\, h}&0&0\\ I_{\bar{\psi}\, h}&0&0\end{pmatrix}.
\end{eqnarray}
Multiplication with the propagator matrix
\begin{eqnarray}
 \mathcal{P}^{-1}= \begin{pmatrix}\mathcal{P}^{-1}_{h\,h}&0&0\\0&0&\mathcal{P}^{-1}{\psi^T \,\bar{\psi}^T}\\0 & \mathcal{P}^{-1}_{\bar{\psi}\, \psi}&0 \end{pmatrix},
\end{eqnarray}
taking the fourth power and the supertrace (not forgetting the negative signs in the fermionic entries) yields
\begin{eqnarray}
 {\rm STr} (\mathcal{P}^{-1}\mathcal{F})^4 &=& 2\, {\rm tr} \mathcal{P}^{-1}_{h\,h}I_{h \,\psi}\mathcal{P}^{-1}_{\bar{\psi}\, h}I_{\bar{\psi}\, h}\mathcal{P}^{-1}_{h\, h}I_{h\, \psi} \mathcal{P}^{-1}_{\psi^T \,\bar{\psi}^T}I_{\bar{\psi}\, h}\nonumber\\
 &{}&+ 2\,{\rm tr} \mathcal{P}^{-1}_{h\,h}I_{h \,\bar{\psi}^T}\mathcal{P}^{-1}_{\bar{\psi}\,\psi}I_{\psi^T\, h}\mathcal{P}^{-1}_{h\, h}I_{h\, \bar{\psi}^T}\mathcal{P}^{-1}_{\bar{\psi}\, \psi}I_{\psi^T\, h}\nonumber\\
&{}&+ 2\,{\rm tr} \mathcal{P}^{-1}_{h\, h} I_{h\, \psi} \mathcal{P}^{-1}_{\psi^T \,\bar{\psi}^T}I_{\bar{\psi}\, h}\mathcal{P}^{-1}_{h\, h} I_{h \,\bar{\psi}^T}\mathcal{P}^{-1}_{\bar{\psi}\,\psi}I_{\psi^T \, h}\nonumber\\
&{}&+ 2 {\rm tr}\,\mathcal{P}^{-1}_{h\, h}I_{h\, \bar{\psi}^T}\mathcal{P}^{-1}_{\bar{\psi}\, \psi}I_{\psi^T\, h} \mathcal{P}^{-1}_{h\, h}I_{h\, \psi}\mathcal{P}^{-1}_{\psi^T\, \bar{\psi}^T}I_{\bar{\psi}\, h}.\nonumber\\
&{}&
\end{eqnarray}
Herein the remaining trace is over Dirac, flavour and Lorentz indices, and of course contains the momentum-integration.

The scalar coupling of the trace mode to the fermions allows to schematically rewrite all four terms as
\begin{equation}
 (\mathcal{P}^{-1}_{h\, h})^2 \cdot \left(\rm fermionic part \right).
\end{equation}
We can now use that
\begin{equation}
{\rm tr}_{L\, f\, D} I_{h\, \bar{\psi}^T}\mathcal{P}^{-1}_{\bar{\psi}\psi}I_{\psi^T\, h}(p)= - {\rm tr}_{L\, f\, D} I_{h\, \psi}\mathcal{P}^{-1}_{\psi^T \, \bar{\psi}^T}I_{\bar{\psi}\, h}(p),
\end{equation}
as there are now external momenta, and all vertices have the same sign. The trace is over Lorentz, flavour and Dirac indices. Accordingly the negative sign is induced by the transposition, following from the Grassmannian nature of the fermions. This yields a complete cancellation of all terms.

\subsection{Identities for traces containing gamma-matrices}\label{gammamat}
We may use the following identities in order to simplify the traces occuring on the right-hand-side of the flow equation:
\begin{eqnarray}
\gamma_{\mu}\gamma_{\mu}&=& 4\\
\gamma_{\mu}\gamma_{\nu}\gamma_{\mu}&=& -2 \gamma_{\nu}\\
\left(\bar{\psi}\gamma_{\mu \gamma_{\nu}\gamma_{\kappa}}\psi \right)\left(\bar{\psi}\gamma_{\mu \gamma_{\nu}\gamma_{\kappa}}\psi \right)&=& 10 \left( \bar{\psi}\gamma_{\mu}\psi\right) \left( \bar{\psi}\gamma_{\mu}\psi\right)+6 \left( \bar{\psi}\gamma_{\mu}\gamma_5\psi\right) \left( \bar{\psi}\gamma_{\mu}\gamma_5\psi\right)\\
\left(\bar{\psi}\gamma_{\mu \gamma_{\nu}\gamma_{\kappa}}\psi \right)\left(\bar{\psi}\gamma_{\kappa \gamma_{\nu}\gamma_{\mu}}\psi \right)&=& 10 \left( \bar{\psi}\gamma_{\mu}\psi\right) \left( \bar{\psi}\gamma_{\mu}\psi\right)-6 \left( \bar{\psi}\gamma_{\mu}\gamma_5\psi\right) \left( \bar{\psi}\gamma_{\mu}\gamma_5\psi\right)\\
\end{eqnarray}

Furthermore
\begin{eqnarray}
\int d^4p \, p_{\mu}p_{\nu}f(p^2)&=& \frac{1}{4}\delta_{\mu \nu}\int d^4p p^2 f(p^2)\\
\int d^4p \, p_{\mu}p_{\nu}p_{\kappa}p_{\lambda}f(p^2)&=& \frac{1}{24}\left(\delta_{\mu \nu}\delta_{\kappa \lambda}+ \delta_{\mu \kappa}\delta_{\nu \lambda}+ \delta_{\mu \lambda}\delta_{\nu \kappa}\right)\int d^4p \left(p^2\right)^2 f(p^2).
\end{eqnarray}
These follow from the fact that $p_{\mu}p_{\nu}$ and $p_{\mu}p_{\nu}p_{\kappa}p_{\lambda}$ are completely symmetric tensors. The numerical factors on the right-hand side are then fixed by taking the trace on both sides.

\subsection{Fermionic $\beta$ functions}\label{threshold}
We define the shape-function dependent
integral
\begin{eqnarray}
 I[f, {\rm TT}, h]&=& \tilde{\partial}_t \int
\frac{d^4p}{(2 \pi)^4}\, \frac{1}{\left(Z_{\psi}p^2(1+r_f(y))\right)^f}
\cdot\frac{1}{\left(\Gamma_{k\,
TT}^{(2)}(1+_{TT}(u) \right)^{TT}} \frac{1}{\left(\Gamma_{k\,
conf}^{(2)}(1+r_{\rm h}(y) \right)^{\rm h}} (p^2)^n,\nonumber
\end{eqnarray}
where $n = TT+f+\rm h-1$. Herein $TT$ counts the number of transverse traceless metric propagators, $h$ does the same for the trace part, and $f$ for the fermions.
\end{appendix}


\begin{thebibliography}{99}
\bibitem{Gies:1999jt}
  H.~Gies,
  ``Probing the quantum vacuum. Perturbative effective
action approach in QED and QCD and its applications,''
\end{thebibliography}
\end{document}